\documentclass[a4paper,11pt,twoside]{ThesisStyle}

\usepackage[avantgarde]{quotchap}
\usepackage{url}

\usepackage{amsmath,amssymb}             
\usepackage[latin1]{inputenc}
\usepackage[T1]{fontenc}
\usepackage[left=1.1in,right=1.1in,top=1.1in,bottom=1.0in,includefoot,includehead,headheight=13.6pt]{geometry}

\usepackage[nottoc, notlof, notlot]{tocbibind}
\usepackage{minitoc}
\usepackage{aecompl}

\usepackage{nomencl}

\usepackage{ifpdf}

\ifpdf
  \usepackage[pdftex]{graphicx}
  \DeclareGraphicsExtensions{.jpg}
  \usepackage[a4paper,hyperindex=true]{hyperref}
\else
  \usepackage{graphicx}
  \DeclareGraphicsExtensions{.ps,.eps}
  \usepackage[a4paper,dvipdfm,hyperindex=true]{hyperref}
\fi

\graphicspath{{.}{images/}}

\usepackage{color}
\definecolor{linkcol}{rgb}{0,0,0.2} 
\definecolor{citecol}{rgb}{0,0,0.2} 

\hypersetup
{
bookmarksopen=true,
pdftitle="Supernovae in Dense Circumstellar Medium",
pdfauthor="Takashi Moriya", 
pdfsubject="Supernovae illuminated by shock interaction of SN ejecta and CSM", 
pdfmenubar=true, 
pdfhighlight=/O, 
colorlinks=true, 
pdfpagemode=None, 
pdfpagelayout=SinglePage, 
pdffitwindow=true, 
linkcolor=linkcol, 
citecolor=citecol, 
urlcolor=linkcol 
}

\usepackage{rotating}                    

\usepackage{fancyhdr}                    

\let\headruleORIG\headrule
\renewcommand{\headrule}{\color{black} \headruleORIG}

\usepackage{colortbl}
\arrayrulecolor{black}

\fancypagestyle{plain}{
  \fancyhead{}
  \fancyfoot{}
  
}

\usepackage{algorithm}
\usepackage[noend]{algorithmic}

\makeatletter

\def\cleardoublepage{\clearpage\if@twoside \ifodd\c@page\else%
  \hbox{}%
  \thispagestyle{empty}
  \newpage%
  \if@twocolumn\hbox{}\newpage\fi\fi\fi}

\makeatother
 
{%

\hrulefill
\vspace*{0.5cm}%
\end{minipage}
}

\let\minitocORIG\minitoc
\renewcommand{\minitoc}{\minitocORIG \vspace{1.5em}}

\usepackage{multirow}
\usepackage{slashbox}

{ \begin{list}%
	{$\bullet$}%
	{\setlength{\labelwidth}{25pt}%
	 \setlength{\leftmargin}{30pt}%
	 \setlength{\itemsep}{\parsep}}}%
{ \end{list} }

\renewcommand{\epsilon}{\varepsilon}

\usepackage{color} 
\usepackage{ifpdf}
\usepackage{slashed}

\usepackage{graphicx, graphics, hyperref, amsmath, amssymb, slashed, color, bbm,amsthm, array}

\usepackage{subfigure, url}
\usepackage{comment}
\usepackage{amsmath, amssymb}

\newcommand{\hata}{\hat{\alpha}}

\newcommand{\sigmabar}{\bar{\sigma}}
\newcommand{\thetabar}{\bar{\theta}}
\newcommand{\delfive}{\partial_{5}}
\newcommand{\no}{\nonumber}
\newcommand{\M}{{\cal M}}
\newcommand{\mSbar}{m_{\bar{S}}}
\newcommand{\Sbar}{\bar{S}}
\newcommand{\order}{{\cal O}}
\newcommand{\MeV}{{\ \rm MeV}}
\newcommand{\GeV}{{\ \rm GeV}}
\newcommand{\TeV}{{\ \rm TeV}}
\newcommand{\hc}{{\rm h.c.}}

\def\kesu#1{}

\newcommand{\GEV}{\mbox{ GeV}}
\newcommand{\TEV}{\mbox{ TeV}}

\newcommand{\mttwo}{M_{T2}}
\newcommand{\mttm}{M_{T2}^{\max}}

\newcommand{\ptvec}[1]{\mathbf{p}_T^{#1} }
\newcommand{\Ptvec}[1]{\mathbf{P}_T^{#1} }

\newcommand{\ptmiss}{\slashed{\mathbf p}_T}
\newcommand{\Etmiss}{E_T^{miss}}
\newcommand{\ttbar}{t\bar{t}}
\newcommand{\bbbar}{b\bar{b}}

\newcommand{\MG}{{\tt  \tt MadGraph/MadEvent 4.4} }
\newcommand{\PY}{{\tt PYTHIA 6.4}}

\newcommand{\fb}{fb$^{-1}$ }

\begin{document}

\thispagestyle{empty}
\begin{center}
\noindent {\large \textbf{}} \\
\vspace*{2.0cm}

\noindent {\LARGE \textbf{Aspects of Supersymmetry after LHC Run I} } \\
\vspace*{3.5cm}
{\it
\noindent \Large A dissertation submitted in partial satisfaction of \\
\noindent \Large the requirement for the degree of \\
}
\vspace*{0.3cm}
\noindent \Large Doctor of Philosophy \\
\vspace*{0.3cm}
\noindent \Large {\it in} \\
\vspace*{0.3cm}
\noindent \Large Physics \\
\vspace*{1.0cm}
\noindent \Large {\it by} \\
\vspace*{1.5cm}
\noindent {\LARGE \bf Kohsaku Tobioka} \\
\vspace*{0.3cm}
\vspace*{0.5cm}
\noindent \Large {Kavli Institute for the Physics and Mathematics of the Universe} \\
\noindent \Large {Department of Physics, Graduate School of Science} \\
\noindent \Large {University of Tokyo} \\
\vspace*{0.4cm}
\noindent \large March 2014 \\
\vspace*{0.5cm}
\end{center}
\sloppy

\cleardoublepage

\thispagestyle{empty}
\begin{center}
{\large\textbf{Abstract\\ \ \\}}
\end{center}

Supersymmetry (SUSY) is a prime candidate for physics beyond the Standard Model, and we  investigate aspects of supersymmetry in light of Large Hadron Collider (LHC) results. The main accomplishment of LHC Run I is discovery of Higgs boson, and it is a momentous step towards understanding electroweak symmetry breaking. The Standard Model Higgs sector, however, is not theoretically satisfactory. Since the Higgs mass scale is not protected by any symmetry, the electroweak scale is unstable to be at observed scale and hence  unacceptable fine-tuning is required.  
Unlike the Standard Model, low-energy supersymmetry stabilizes the Higgs mass avoiding fine-tuning and leads to natural electroweak symmetry breaking.  The natural scale of supersymmetric particles (sparticles) is below $\approx\TeV$. 
The minimal supersymmetric Standard Model (MSSM) has been studied, and it was found the lightest supersymmetric particle (LSP) is a good dark matter candidate. 

However, there are tensions between low-energy supersymmetry and the LHC results. Firstly, searches at the LHC for sparticles have not found any signal and give strong limits on mass of gluino and squark up to $\approx 1.8 \TeV$ for a conventional model, constrained minimal supersymmetric Standard Model (CMSSM). Next, the observed Higgs mass of 125 GeV is not easily accommodated in the MSSM, where one has to rely on the radiative corrections to boost the Higgs mass beyond the tree-level upper bound of $m_Z\simeq91\GeV$. This requirement push sparticles scale well beyond the TeV within CMSSM, leading to fine-tuning.  
In this thesis we particularly study two scenarios of supersymmetry originating the tensions. 
\vspace{7pt}

A  nearly degenerate (compressed)  spectrum ameliorates the bounds from the current searches at the LHC whereas the CMSSM typically generates a widely spread spectrum. For the lack of SUSY signal, the scenario with a compressed spectrum recently has more attentions, but motivations and explicit models for the scenario have not been discussed. Therefore we study this direction in detail.  

Supersymmetry broken geometrically in extra dimensions, by the Scherk-Schwartz mechanism, naturally leads to a compressed spectrum.
 We present a minimal such model with a single extra dimension, which we call ``Compact Supersymmetry,'' and show that it leads to viable phenomenology despite the fact that it essentially has two less free parameters than the CMSSM. The theory does not suffer from the supersymmetric flavor or CP problem because of universality of geometric breaking, and automatically yields near-maximal mixing in the scalar top sector with $|A_t|\approx 2m_{\tilde{t}}$ to boost the Higgs boson mass. Despite the rather constrained structure, the theory is less fine-tuned than many supersymmetric models. 
The LSP is Higgsino-like and can be a component of  dark matter. We find direct detection experiments will cover a large portion of parameter space. 
The collider constraint on the Compact Supersymmetry is certainly weaker than that on the CMSSM as gluino and squark mass bound is relaxed down to $\approx 1\TeV$.  In order to increase sensitivity to models with a compressed spectrum, we suggest a kinematic variable, $M_{T2}$, can be useful since the Standard Model background is systematically removed by requiring $M_{T2}>m_{\rm top}$.
%
\vspace{7pt}

Naturalness implies new dynamics beyond the minimal theory. 
There have been many attempts to extend the MSSM to accommodate the Higgs mass. In such extensions, new states interact with the Higgs, raising its mass by increasing the strength of the quartic interaction of the scalar potential. 
One possibility is a non-decoupling $F$-term, as in the NMSSM (MSSM plus a singlet $S$).
If the new states are integrated out supersymmetrically, their effects decouple and the Higgs mass is not increased. On the other hand, 
SUSY breaking can lead to non-decoupling effects that increase the Higgs mass.  However, in general, these extensions require new states at the few hundred GeV scale, so that the new sources of SUSY breaking do not spoil naturalness.

In this thesis, we have identified a new model where the Higgs couples to two singlet fields with a Dirac mass, which we call Dirac NMSSM,
	\begin{eqnarray}
	W\supset \lambda S H_u H_d +M S \bar{S}. \nonumber
	\end{eqnarray}
The non-decoupling $F$-term increases the Higgs mass while maintaining  naturalness even in the presence of large SUSY breaking in the singlet sector as $m_{\bar S}\gtrsim 10\TeV$.
The key feature  in the Dirac NMSSM is that $\bar S$ couples to the MSSM only through the dimensionful Dirac mass, $M$. We note that interactions between $\bar S$  and other new states are not constrained by naturalness, even if these states experience SUSY breaking. Therefore, the Dirac NMSSM represents a new type of portal, whereby our sector can interact with new sectors, with large SUSY breaking, without spoiling naturalness in our sector.

Collider signatures of the Dirac NMSSM are  discussed. The low-energy  phenomenology is that of a two Higgs doublet model.  We obtain constraints from direct searches for heavier Higgs boson and coupling measurements for the lightest Higgs boson at the LHC. We also study the future reach based on prospects of high-luminosity LHC and future international linear collider, and show large parameter space can be probed. 

\vspace{-10pt}
\thispagestyle{empty}


\cleardoublepage
\pagenumbering{roman}
\tableofcontents
\cleardoublepage

\thispagestyle{empty}

\mainmatter

\pagebreak

\chapter{Invitation}
\section{Last Piece of the Standard Model}
One of the greatest accompaniments of physics in the past few decades is discovery of a Higgs-like boson. It was reported on July 4, 2012 by the ATLAS and CMS collaborations of the Large Hadron Collider (LHC), CERN \cite{:2012gk}. It has been about 50 years since its theoretical prediction. 
Eventually, the mass is confirmed to be about $125\GeV$, and the properties of spin and coupling are compatible with ones predicted by the Standard Model of particle physics (SM). 
To begin with, let us briefly explain the importance of Higgs boson and its discovery.

The Higgs boson is the only scalar among the SM particles, and does the most important role that triggers the electroweak symmetry breaking (EWSB). In the SM,  electroweak symmetry of $SU(2)_L\times U(1)_Y$ is broken down to the electromagnetic symmetry of $U(1)_{EM}$. The Higgs boson with $SU(2)_L\times U(1)_Y$ charge forms wine-bottle (or mexican-hat) type potential, 
	\begin{eqnarray}
	V=\mu_H^2|H|^2 +\frac{\lambda_H}{4}|H|^4 	\label{eq:higgs}
	\end{eqnarray}
where curvature at origin is negative, $\mu_H^2<0$. Since the origin of potential is unstable, the vacuum shifts a stable vacuum with a finite vacuum expectation value (VEV), 
	\begin{eqnarray}
	\langle H \rangle = \sqrt{\frac{-2\mu_H^2}{\lambda_H}}\equiv v \ .
	\end{eqnarray}
\begin{figure}[htb]
\begin{center}
  \includegraphics[clip,width=.3\textwidth]{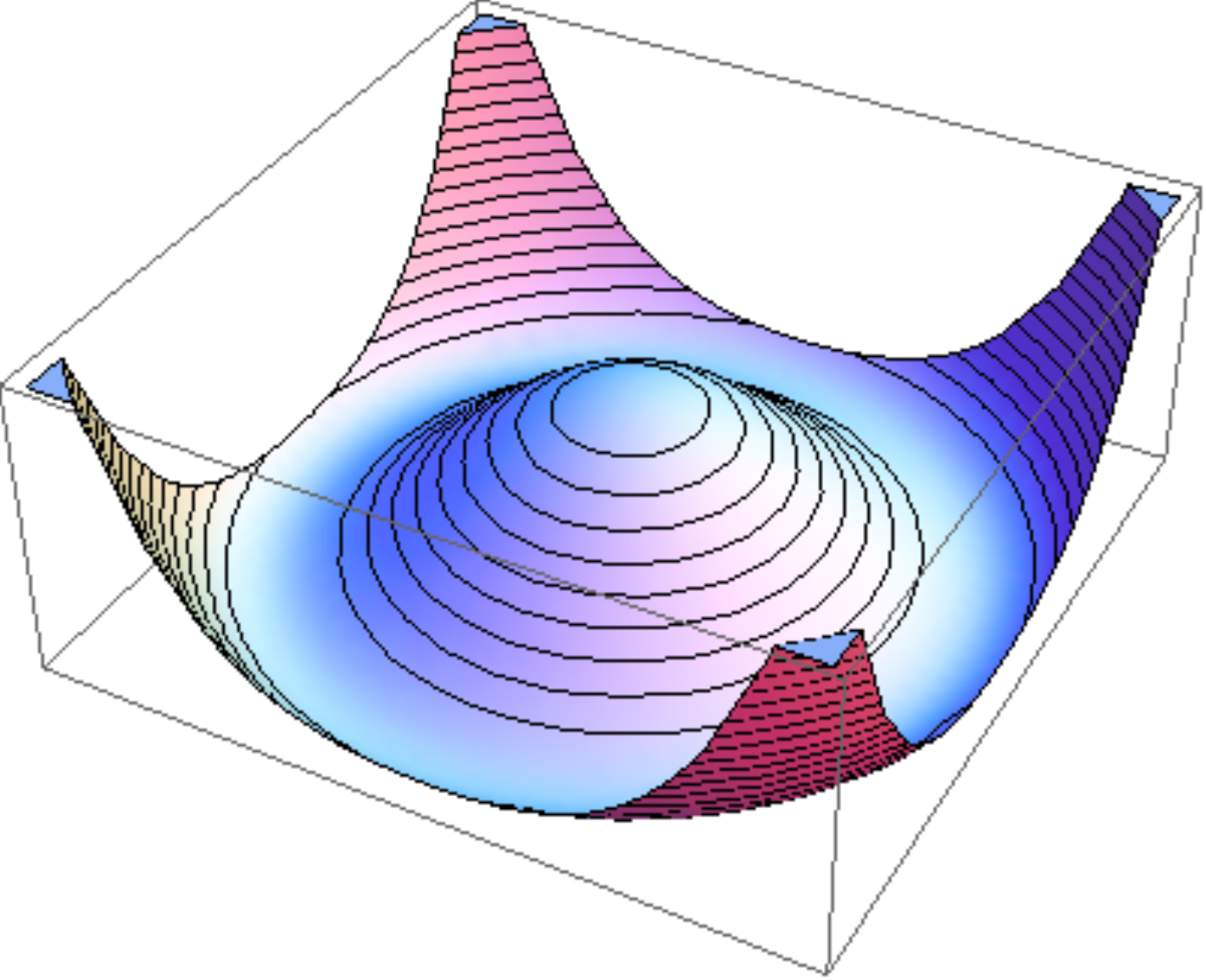}
\end{center}
  \vspace{-20pt}
\caption{Schematic view of Higgs potential}
\end{figure}
Thus the symmetry is {\it spontaneously broken} because the new vacuum respects only $U(1)_{EM}$ symmetry but the original $SU(2)_L\times U(1)_Y$ symmetry.
The Higgs field is re-parametrized as 
	\begin{eqnarray}
	H=v+\frac{h+i\pi}{\sqrt{2}} \ .
	\end{eqnarray}
In the case of global symmetries, Nambu-Goldstone (NG) boson corresponding to a broken symmetry is generated, while in the case of gauge symmetry the NG boson  $\pi$ is absorbed to a gauge boson making it massive. The radial mode of the potential, $h$, still remains as physical particle that is what is currently observed at the LHC experiment. The Higgs potential has only two parameters. The VEV, $v$, is well measured by electroweak precision studies, and the measurement of Higgs mass, 
	\begin{eqnarray}
	m_H^2=-2\mu_H^2 =\lambda_H v^2\approx 125\GeV, 
	\end{eqnarray}
determines the last remained parameter. 
This spontaneous breaking of gauge symmetry is called by Higgs mechanism, which is developed by R.~Brout, F.~Englert, P.~Higgs, G.~Guralnik, C.R.~Hagen, and T.~Kibble \cite{HiggsMech}.  \footnote{P.~Higgs and F.~Englert received the Nobel Prize in physics in 2013 for the theoretical discovery of Higgs mechanism.}

The Higgs boson in the SM also gives masses to quarks and leptons through Yukawa interactions. Yukawa interactions are bilinear of fermions plus Higgs field, and once Higgs field obtains a finite VEV, the interactions turn out to be the masses of quarks and leptons. The size of mass is proportional to the strength of each Yukawa interaction. The Higgs field couples to most of elementary particles, $Z, W,$ quarks, and leptons, giving their masses, so the Higgs is ubiquitous in our Universe. 

\begin{figure}[h!]
\begin{center}
  \includegraphics[clip,width=.35\textwidth]{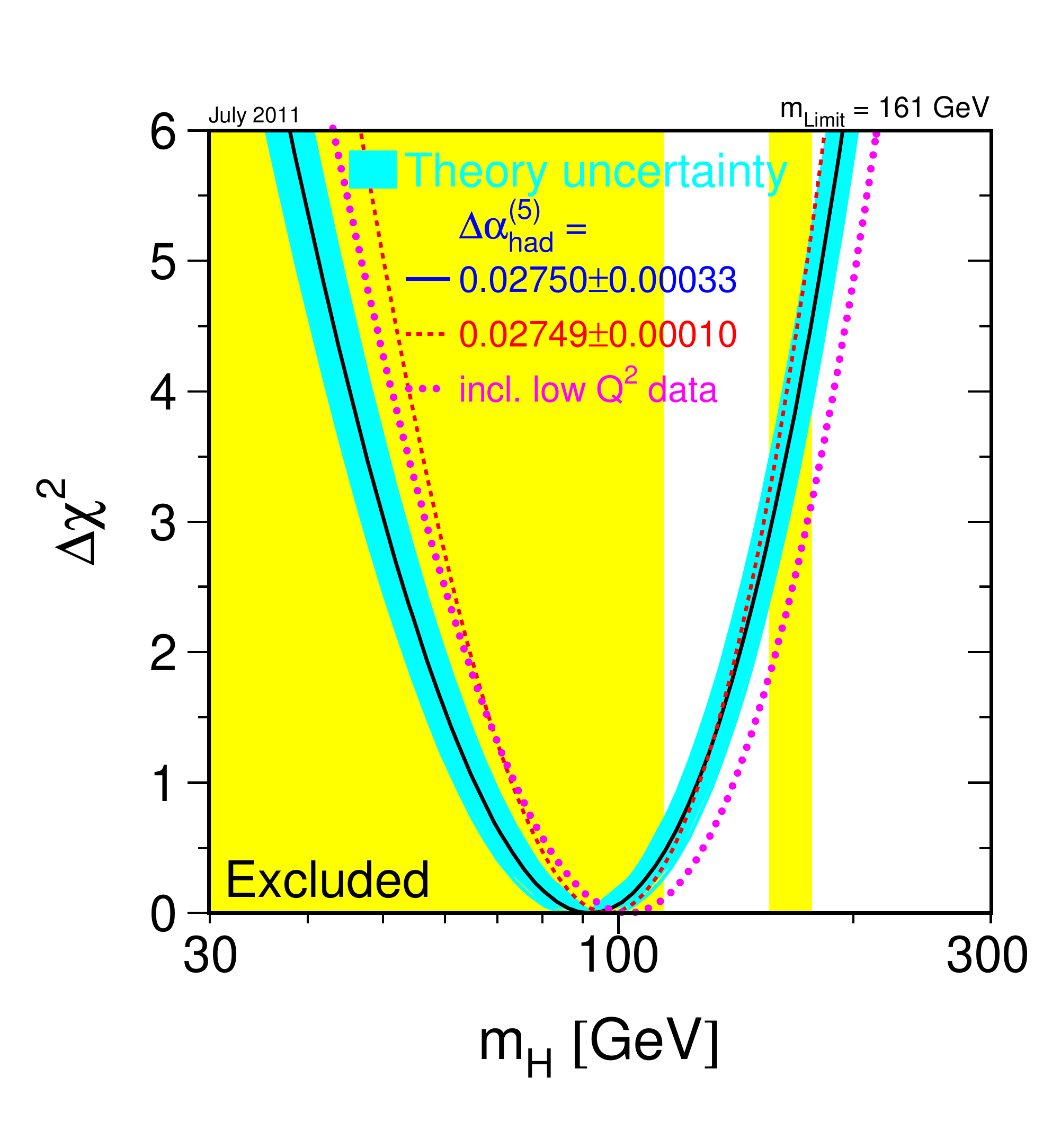}
  \vspace{-20pt}
\end{center}
\caption{$\Delta \chi^2$ for Higgs boson mass based on the electroweak precision measurements before its discovery from Ref.~\cite{lepweewg}. With 1$\sigma$ level, $m_H\lesssim 150\GeV$ was favored.}
\label{chi2p}
\end{figure}

The Higgs boson discovery is triumphs of theoretical prediction based on Quantum Field Theory and  of experiments with frontier technologies and international collaborations. The mass of Higgs boson seems arbitrary and might be very heavy, but it is suggested physics without Higgs boson does not guarantee the perturbative unitarity. Massive gauge boson's scattering amplitude grows with (energy)$^2$ due to their longitudinal modes and breaks perturbative  unitarity at energy about TeV unless the Higgs boson appears. This is a motivation to construct the collider which focuses on the TeV range. In addition, precise measurements of the electroweak sector favor Higgs mass $\lesssim 150 \GeV$ as in Fig.~\ref{chi2p}. 

The experimental efforts are tremendous. It is basically necessary to design the highest energy machine for search for undiscovered particles. The center of mass energy of LHC was $\sqrt{s}=7\TeV$ and 8 TeV for physics run I and the energy will be upgraded to 14 TeV in run II. Between 2010 and 2012, integrated luminosity is $\sim 30~\rm fb^{-1}$,\footnote{b (barn) is an unit of interacting cross section. ${\rm b}=10^{-28}{\rm m}^2=100~{\rm fm}^2$. } and in the high-luminosity upgrade the integrated luminosity will be $300~\rm fb^{-1}$ or more. 
In contrast, the previous strongest collider, Tevatron at Fermilab,  had the maximum energy of $\sqrt{s}=1.96\TeV$ and total integrated luminosity of $\sim10~\rm fb^{-1}$.  Developments occurred in the analysis level as well. For instance, since the backgrounds is well understood even in the proton-proton collisions where huge and various background usually exists, $h\to \gamma\gamma$ channel significantly contributed to the Higgs boson discovery even if the Higgs signal is very weak compared to the background  in this channel. 

Now we understand the electroweak symmetry breaking better; the Higgs potential actually exists. However, the potential in particular its mass term is unsatisfactory because no symmetry protects the mass scale of Higgs, and hence the electroweak scale is unstable against the radiative correction. Supersymmetry (SUSY)\cite{Martin:1997ns} is the leading theory beyond the SM (BSM) to deal with this issue.
In this article we study aspects of supersymmetry in light of LHC run I results. 

\section{Beyond the Standard Model}

\begin{figure}[t]
\begin{center}
  \includegraphics[clip,width=.35\textwidth]{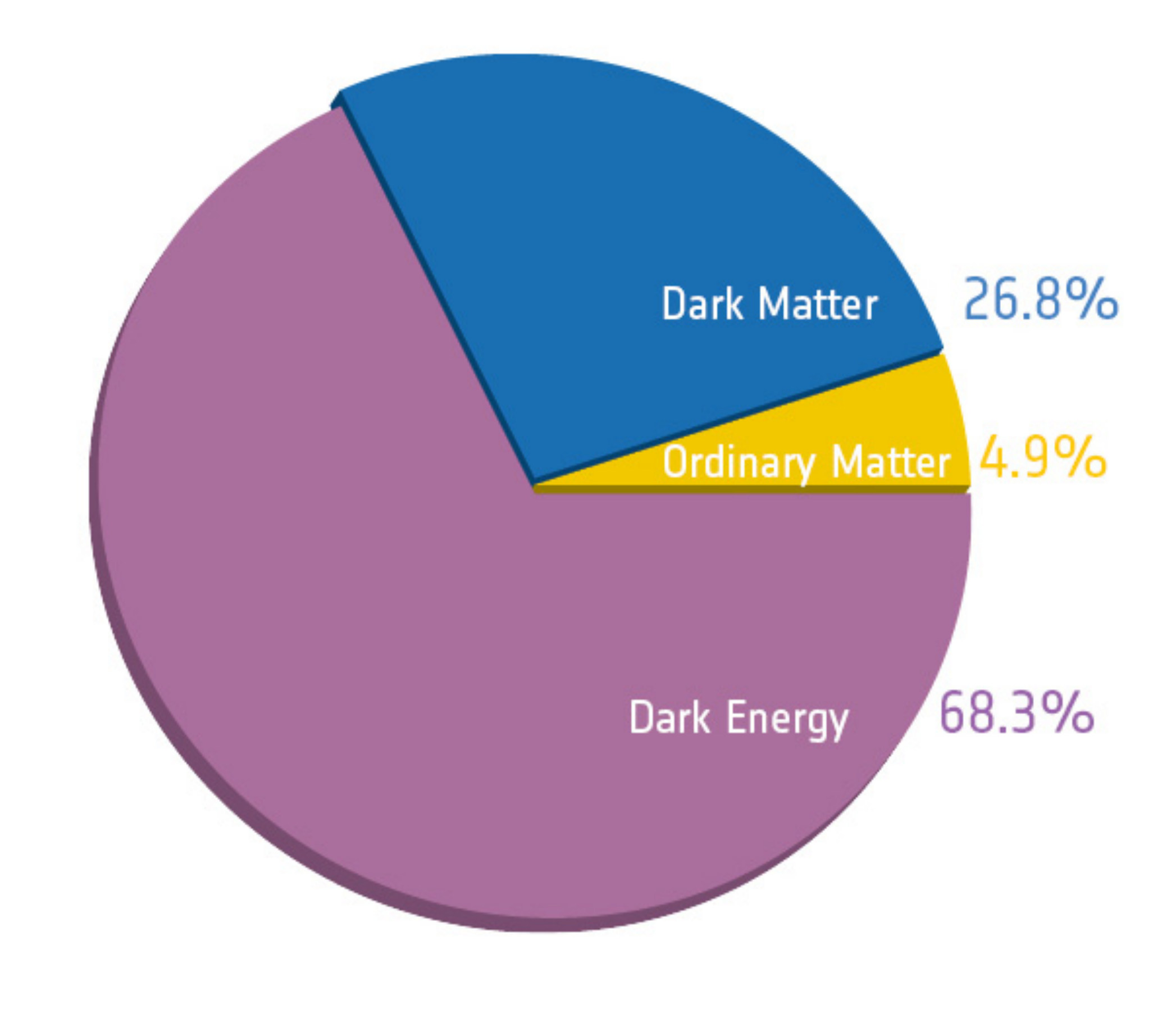}
\end{center}
  \vspace{-20pt}
\caption{Cosmic pie chart of the Universe's content by Planck satellite. [Credit: ESA/Planck Collaboration \cite{planck}]}
  \vspace{-0pt}
\label{cosmicpie}
\end{figure}

Observations at the ATLAS and CMS for a new boson is well consistent with the Higgs boson predicted by the SM. 
Then a question arises: Is particle physics at the end? The answer is clearly no since there are experimental results that cannot be explained within the SM, 
\begin{itemize}
\item 	Dark matter in the Universe
\item 	Dark energy in the Universe
\item 	Matter asymmetry in the Universe
\item 	Neutrino masses
\item 	Acausal correlation in the Universe
\end{itemize}
and  also strong CP problem could be categorized together with subjects above. The best example is seen in the pie chart of the Universe's energy content (Fig.~\ref{cosmicpie}) which is measured by the WMAP satellite and recently updated by the Planck satellite.

The dark energy, dominant composition, is unknown energy source which accelerate expansion of the Universe, and the dark matter is unknown relic abundance created in the early Universe. It is expected that the dark matter can be explained by a new elementary particle. A typical scenario is that the dark matter is a weakly interacting massive particle (WIMP), and many BSM models provide WIMP candidates. 
In WIMP scenario, experiments for direct and indirect detection of dark matter particle have discovery potential. Since the typical mass scale of WIMPs is order of TeV, there is a good chance to discover the dark matter at the LHC. On the other hand, the dark energy is still very mysterious. 

We do not understand an origin of ordinary matter (particle) of 4.9\% in the pie chart. This is because particles and antiparticles have almost the same nature, and in the early Universe they equally existed and are supposed to annihilate each other. However, the current Universe has only  matters. This occurs for the CP-violation, asymmetry between particle and antiparticle, but the size of CP-violation in the SM is not sufficient to explain the current abundance of matter. 

Neutrino masses are confirmed by observations of neutrino oscillation. The mass difference and mixing are measured by various experiments, and one neutrino at least has mass of $\gtrsim 50{\ \rm meV}$.  On the other hand, the SM predicts that all the neutrinos are massless. This clearly shows new physics must exist.

In our observable sky, there are correlation between points crossing the causal horizon. It is understood by inflation theory, an essential idea in the modern cosmology. Inflation leads to  exponential expansion in a very early stage of the Universe stretching small fluctuations to a large scale. 
A new scalar field that makes inflation dynamics is needed unless  Higgs field, a unique scalar in the SM, plays such a role of inflaton. This does not directly suggests that new physics in the electroweak sector but imply some new physics. 

Each of experimental results above requires some new physics beyond the Standard model. 
Another motivation to extend the SM is actually from the Higgs boson. 
Although the Higgs boson or mechanism plays a crucial role leading to the electroweak scale, the mass scale is not protected by any symmetry. Note that this is not theoretical contradiction, but thinking of the high-scale physics such as Grand Unification, quantum gravity or String Theory, the scale mass unlike the electron (fermion) mass is too sensitive to stabilize their scale. This is so-called the ``naturalness problem'' and quantified by ``fine-tuning.'' 
The leading candidate to solve the issue is supersymmetry. Supersymmety embeds a scalar in a multiplet with a fermion, and, roughly speaking, the multiplet is protected by sort of ``chiral symmetry'' thanks to the fermion. We will discuss the point in the next section. 

Another approach to the problem is to consider the Higgs boson as a pseudo-NG boson of some global symmetry like a pion.  The global symmetry which is broken and generates pseudo-NG bosons guarantees technical naturalness of Higgs boson mass.

\section{Supersymmetry}

\subsection{Motivation}
There is a question why electroweak scale, or vacuum expectation value, is at a scale we observe. Low-energy supersymmetry was introduced to answer such a question. 
We know at least there must be a new scale, planck scale ($M_{pl}$) at which quantum gravity appears,\footnote{There are scenarios that the planck scale itself is actually close to the weak scale. This happens when multiple extra dimensions or warped extra dimension exist where only gravity can propagate that is why the gravity force seems much weaker.} and the electroweak scale can receive  radiative corrections of the order of $M_{pl}$. 
In other words, this is because no symmetry forbids emergence of quadratic divergence to Higgs boson mass. Supersymmetry saves this issue leading the divergence to logarithmic one. 
Before going to the detail, let us give an example where this way of thinking such as naturalness worked. This analogy is given by H.~Murayama \cite{Murayama:1994kt}.

Back to 19th century, the electrodynamics had a similar problem that self-energy of the electron is linearly divergent.  In the nonreletivistic electrodynamics, the Coulomb part of the self-energy is just the expectation value of $\frac{e^2}{2}\frac{1}{4\pi|\bf x -\bf x'|}$ 
as $\bf x'$ approaches $\bf x$,
	\begin{eqnarray}
	E_{self}&=&\frac{e^2}{2}\int d^3x \ \psi_{NR}({\bf x})\left(
	\int \frac{d^3x'}{4\pi |\bf x - x'|}\delta^3({\bf x - x'})
	\right)\psi^*_{NR}({\bf x'})\no\\
	&=&\frac{e^2}{2}\int d^3x \ \psi_{NR}({\bf x})\left(
	\int d^3p \int \frac{d^3x'}{(2\pi)^3} \frac{e^{i{\bf p(x-x')}}}{4\pi |\bf x - x'|})
	\right)\psi^*_{NR}({\bf x})\no\\
	&=&\frac{e^2}{2(2\pi)^3}\int \frac{d^3 p}{{\bf p}^2}
	=\frac{\alpha}{\pi}\int^\Lambda d|\bf p|, 
	\end{eqnarray}
where $\Lambda$ is a cutoff scale of the theory. Then the correction above leads to the relation between bare mass and observed mass of electron, 
	\begin{eqnarray}
	m_{e,\rm obs} =m_{e,\rm 0}+E_{self}, 
	\end{eqnarray}
where $m_{e,0}$ is a bare electron mass.  If we substitute $\Lambda=10^3\times200 \MeV$ corresponding to observed electron size $10^{-3}$~fm, the bare mass  must be highly-tuned with three orders of magnitude to give the observed electron mass of $0.51 \MeV$. Furthermore, the perturbation theory does not work anymore below $r\approx \frac{\alpha}{\pi }/m_e \simeq 1$~fm.

The unnaturalness is cured by the discovery of positron. The relativistic theory of electrodynamics requires the existence of positron, and in the short distance, quantum fluctuation allows electron-positron pair creation and annihilation. 
It leads to a new correction of electron self-energy, and as a result cutoff dependence of the self-energy in the short-distance becomes logarithmic rather than linear,  
	\begin{eqnarray}
	E_{self}=\frac{3}{2\pi}\frac{e^2 }{4\pi }m_e\log \frac{\Lambda}{m_e } .
	\end{eqnarray}
The result is first calculated by V.F.~Weisskopf in 1939 \cite{Weisskopf:1939zz}.
\footnote{For more discussion and calculation, see for example Sec.~2-8, 4-7 of ``Advanced Quantum Mechanics'' by J.J.~Sakurai \cite{Sakurai}.} 
In the modern understanding by Quantum Field Theory, the cancelation is a consequence of chiral symmetry. The chiral symmetry is actually broken but only {\it softly} by the electron mass. In the exact limit of chiral symmetry, the mass of electron mass is not generated perturbatively, and therefore the radiative corrections to the electron mass must be proportional to the breaking parameter, the electron mass itself.
The chiral symmetry controls the radiative correction better. 

The motivation for supersymmetry is very similar to the self-energy of electron; simply replacing the electron self-energy in the electrodynamics and chiral symmetry with the Higgs mass in the Standard Model and supersymmetry, respectively. 
In the SM, the radiative correction from some high scale is expected to be 
	\begin{eqnarray}
	\mu_H^2 =\mu^2_{H,0} +\frac{g^2}{16\pi^2}\Lambda^2 \ . 
	\end{eqnarray}
where $\mu^2_{H,0}$ is a bare Higgs mass, and $g$ is ${\cal O}(1)$ coupling. The scalar mass term $|H|^2$ is singlet under any symmetry, and the coefficient cannot be a soft breaking parameter of some symmetry. Supersymmetry cures this problem in a way that chiral symmetry is promoted for bosons because fermion and boson are transformed each other by supersymmetry. 
However, since we have not observed supersymmetric particles in the low-energy, it must be broken {\it softly} by a breaking scale of $M_{SUSY}$. The self-energy of the Higgs boson is roughly 
	\begin{eqnarray}
	\mu_H^2 =\mu^2_{H,0} +\frac{g^2}{16\pi^2}M_{SUSY}^2\log\frac{\Lambda^2}{m_{weak}^2} \ . 
	\end{eqnarray}
The structure is same as the self-energy of electron: the correction must be proportional to the breaking parameter of $M_{SUSY}$ and the cutoff sensitivity is only logarithmic. Since the Higgs mass parameter is at weak scale, we expect the SUSY breaking scale is also in the same scale.  In the most natural case, 
	\begin{eqnarray}
	m_{SUSY}\sim m_{weak} \ . \label{eq:weak}
	\end{eqnarray}
However the na\"ive expectation of Eq.~(\ref{eq:weak}) is challenged by the observed Higgs mass and the direct search for supersymmetric particles at the LHC. We discuss these tensions in Sec.~\ref{direction}.  
In this article, we study possible scenarios of low-energy supersymmetry in light of these results. 

\subsection{Algebra}
 Supersymmetry is a continuous symmetry which transforms boson to fermion and vice versa. The supersymmet charge $Q$ is grassmanian, and its conjugate is denoted by $Q^\dag$,  
	\begin{eqnarray}
	\{Q,Q\}=\{Q^\dag,Q^\dag\}=0, 
	\end{eqnarray}
and $Q$ acts on a state such that
	\begin{eqnarray}
	Q|\rm Boson\rangle =|Fermion\rangle ,
	\end{eqnarray}
and vice verse. 
Supersymmetry is  non-trivial extension of Poincar\'e symmetry, and then it relates to the special transformation operator, $P_\mu$,
	\begin{eqnarray}
	&&{\{Q,Q^\dag\}\sim P_\mu, }
	\\
	&&\left[Q, P_\mu\right]=\small[Q^\dag, P_\mu\small] =0  . 
	\end{eqnarray}
The first line tells supersymmetric vacuum should be zero energy vacuum since $H=P_0\sim QQ^\dag$. In other words, the supersymmetry breaking vacuum $Q|0\rangle\neq0$ has non-zero energy, 
	\begin{eqnarray}
	\langle 0| H|0\rangle \sim \langle0|QQ^\dag|0\rangle\neq 0\ .
	\end{eqnarray}
The second line holds energies (masses) of boson and fermion same,
	\begin{eqnarray}
	E_F  |{\rm Fermion}\rangle =
	H |{\rm Fermion}\rangle =H Q|{\rm Boson}\rangle
	= E_B Q|{\rm Boson}\rangle \ .
	\end{eqnarray}

Let us briefly comment on construction of massless multiplet. For the minimal supersymmetric case in 4D, referred to as ${\cal N}=1, D=4$, have two kinds of multiplet for renormalizable theories. If we start with a state with zero helicity using $(Q^\dag)^2=0$, 
	\begin{eqnarray}
	&|h=0\rangle, \ \ Q^\dag|h=0\rangle \sim |h=\frac{1}{2}\rangle, &
	\end{eqnarray}
and their the CPT conjugate states form a supersymmetric multiplet, called chiral supermultiplet,\footnote{Strictly speaking, the condition for chiral multiplet is given by $\bar{\cal D}\Phi=0$ where $\Phi$ is a superfield and $\bar{\cal D}$ is super-covariant derivative. }
 consisting of a Weyl fermion (quark and lepton) and a complex scalar. Starting with a state with helicity $\frac{1}{2}$, 
	\begin{eqnarray}
	&|h=\frac{1}{2}\rangle, \  \ Q^\dag|h=\frac{1}{2}\rangle \sim |h=1\rangle ,& 
	\end{eqnarray}
and their CPT conjugate states form another supermultiplet consisting of a Weyl fermion and a vector. We call this multiplet vector supermultiplet.

In the extended supersymmetry, the multiplets become bigger because there are multiple supersymmetry charges. For example in $\cal N$=2 case, two kinds of charge $Q_1$ and $Q_2$ have properties, 
	\begin{eqnarray}
	Q_1^{\dag}Q_2^{\dag}\neq 0, \ (Q_1^{\dag})^2=(Q_2^{\dag})^2=0 .
	\end{eqnarray}
From a state with helicity $-\frac{1}{2}$, 
	\begin{eqnarray}
	&|h=-\frac{1}{2}\rangle,&\no\\
	&Q_1^\dag|h=-\frac{1}{2}\rangle , \ Q_2^\dag|h=-\frac{1}{2}\rangle,&\no\\
	&Q_1^\dag Q_2^\dag|h=-\frac{1}{2}\rangle,&
	\end{eqnarray}
and their CPT conjugate states form a multiplet, called hypermultiplet,  of two complex scalars and two Weyl fermions. A hypermultiplet corresponds to two chiral supermultiplets. 
On the other hand, a new vector supermultiplet is obtained by, 
	\begin{eqnarray}
	&|h=0\rangle,&\no\\
	&Q_1^\dag|h=0\rangle , \ Q_2^\dag|h=0\rangle, &\no\\
	&Q_1^\dag Q_2^\dag|h=0\rangle. &
	\end{eqnarray}
Hence there are a complex scaler, a vector and two Weyl fermions. This multiplet  corresponds to a chiral supermultiplet and a vector supermultiplet of $\cal N$=1. Note that there is a rotation symmetry which exchanges $Q_1$ and $Q_2$, called $SU(2)_R$ symmetry. Scalars of hypermultiplet and fermions of vector supermultiplet (gauginos) are doublets of $SU(2)_R$, and the others are singlet. When one considers 5D theory, minimal supersymmetry corresponds to ${\cal N}=2, D=4$ since Weyl fermions cannot exist because $\gamma_5$ is not 5D Lorentz invariant.

\subsection{The Minimal Supersymmetric Standard Model}
In order to build models, we use superfield and superspace for explicit supersymmetric invariance. Here we use the consequences,  for more detail see Refs.~\cite{Martin:1997ns, Wess:1992cp}. We describe the Minimal Supersymmetric Standard Model (MSSM)  in the following. 

Higgs, quark and lepton fields are contained in a chiral superfield (a superfield of chiral supermultiplet). Multiplication of chiral superfields is also a chiral superfield. Terms written in superpotential, $W(\Phi)$, are explicitly supersymmetric where $\Phi$ is a chiral superfied and superpotential is a holomorphic function. Yukawa-type interactions are in the superpotential,
	\begin{eqnarray}
	W = y_{U}^{ij} H_u Q_i U_j  + y_{D}^{ij} H_dQ_i D_j+y_{E}^{ij} H_dL_i E_j  + \mu H_u H_d .
	\end{eqnarray}
Note that there are two chiral superfields for Higgs, $H_u$ and $H_d$ for two reasons. 
Firstly, a fermionic partner of Higgs boson, Higgsino, gives a new anomaly of gauge symmetry, and then we need an additional Higgs field with opposite charge. Secondly, since the superpotential is holomorphic function, two Higgs bosons are necessary to give masses of up-type and down-type quarks. Scalar components of quark and lepton chiral superfield are called squark and slepton. 

Here, we have introduced an assumption, $R$-parity,
	\begin{eqnarray}
	R=(-1)^{3(B-L)+F} , 
	\end{eqnarray}
where $B$ and $L$ are baryon and lepton number, respectively, and $F=1(0)$ for fermions (bosons).  As a superfield, $Q, U, D,L,$ and $E$ are $-1$ while $H_u$ and $H_d$ are $+1$. This symmetry forbids $B$ and $L$ violating terms, $UDD$ and $DQL$,  which induces proton decay. As long as coupling constants of these terms are sufficiently small to be consistent with lower bound of proton lifetime, $R$-parity can be broken, and this small breaking is relevant to collider phenomenology. 

The gauge interactions and kinetic terms of chiral superfields appear in K\"ahlar potential $K(\Phi, e^{gT^aV^a}\Phi^\dag)$ where $K$ is a real function and $V^a$ is a vector superfield, a superfield of vector supermultiplet. Pure supersymmetric Yang-Mills theory is also given in K\"ahlar potential. 
A fermionic part of $V$ is called gaugino. There are three kinds of gaugino in the MSSM, gluino, Wino, and Bino, corresponding to $SU(3)_C$, $SU(2)_L$, and $U(1)_Y$ gauge symmetry, respectively. The gauginos are $R$-parity $-1$ while gauge bosons are $+1$.  

The new particles beyond the SM (and additional Higgs boson), Higgsinos, squarks, sleptons, and gauginos are all $R$-parity $-1$, and denoted as $\widetilde{X}$. More detail is given in Table~\ref{tab:MSSM}. 

 \begin{table}[htdp]
 \begin{center}\begin{tabular}{|c| c | c c c c   |}
 	\hline
	 &\small (Fermion, Boson)& $SU(3)_C$ & $SU(2)_L$ & $U(1)_Y$ &  $R$-parity   \\
     	\hline\hline
	$H_u$& $(\widetilde{H}_u, H_u)$ & \bf 1 & \bf 2 & 1/2  & $(-,+)$\\
	$H_d$& $(\widetilde{H}_d, H_d)$ & \bf 1 & \bf 2 & -1/2  &$(-,+)$\\
     	\hline
	$Q$&$(Q_i, \widetilde{Q}_i) $& \bf 3 & \bf 2 & 1/6 &  $(+,-)$\\
 	$U $&$(u^c_{Ri}, \widetilde{u}^c_{Ri}) $& $\bf \bar{3}$ & \bf 2 & -2/3&  $(+,-)$\\
	$D $&$(d^c_{Ri}, \widetilde{d}^c_{Ri}) $ & $\bf \bar{3}$ & \bf 1 & 1/3  & $(+,-)$\\
	$L $&$(L_i, \tilde{L}_i)$ & \bf 1 & \bf 2 & -1/2  & $(+,-)$\\
	$E$&$(e^c_{Ri}, \tilde{e}^c_{Ri}) $& \bf 1 & \bf 1 & 1  &  $(+,-)$\\
	\hline
	$V_C$&$(\widetilde{g}, G_\mu) $& \bf 8 & \bf 1 & 1  &  $(-,+)$\\
	$V_L$&$(\widetilde{W}, W_\mu) $& \bf 1 & \bf 2 & 1  &  $(-,+)$\\
	$V_Y$&$(\widetilde{B}, B_\mu) $& \bf 1 & \bf 1 & 1  &  $(-,+)$\\
	\hline
 \end{tabular} \caption{MSSM particle contents.}
\label{tab:MSSM}
\end{center}
\end{table}
Components of $SU(2)_L$ doublet are labeled by the electromagnetic charge $Q$  given by $Q=T^3_L+Y$ where $T^3_L$($Y$) is a generator of $SU(2)_L$ ($U(1)_Y$), and then,  
\begin{table}[h!]
 \begin{center}\begin{tabular}{ c c c }
 	$H_u=(H_u^+, H_u^0), $ &$H_d=(H_d^0, H_d^-),$&\\
	$Q_i=(u_L, d_L)_i, $&$L_i=(\nu, e_L)_i. $&
 \end{tabular} 
\vspace{-20pt}
\end{center}
\end{table}
\\The index $i$ represents generation of particles, 
\begin{table}[h!]
 \begin{center}\begin{tabular}{ c c c c   |}
 	$u_i=(u,c,t),$&$d_i=(d,s,b),$&\\
	$e_i=(e,\mu,\tau),$&$\nu_i=(\nu_e,\nu_\mu,\nu_\tau).$&\\
 \end{tabular} 
\vspace{-20pt}
\end{center}
\end{table}

The lightest sparticle (LSP) is stable for the $R$-parity conservation. If the LSP is nautralino, that is a mixture of $\widetilde{B}, \widetilde{W}^0, \widetilde{H_u^0}$, and $\widetilde{H_d^0}$, two important phenomenological consequences are led.
First, the LSP is a good dark matter candidate as WIMP, and the weak to TeV scale LSP gives a thermal relic abundance which is the same order of observed relic abundance of the dark matter. Next, the LSP cannot be detected at collider experiments, which leads to a distinct signal with $\slashed{E}_T$. 

There are soft supersymmetry breaking terms, 
	\begin{eqnarray}
	{\cal L}_{\rm soft}&=& -m_{H_u}^2|H_u|^2 -m_{H_d}^2|H_d|^2-(b H_uH_d+\hc)
	\no\\&&
	-m_{Qij}^2\widetilde{Q}_i^\dag \widetilde{Q}_j
	-m_{uij}^2 \widetilde{u}^{c\dag}_{Ri} \widetilde{u}^c_{Rj}
	-m_{dij}^2\widetilde{d}^{c\dag}_{Ri} \widetilde{d}^c_{Rj}
	-m_{Lij}^2\widetilde{L}_i^\dag \widetilde{L}_j
	-m_{eij}^2\widetilde{e}^{c\dag}_{Ri} \widetilde{e}^c_{Rj}
	\no\\&&
	-a_{U}^{ij} H_u \widetilde{Q}_i \widetilde{u}^c_{Rj}  
	- a_{D}^{ij} H_d \widetilde{Q}_i \widetilde{d}^c_{Rj} 
	-a_{E}^{ij} H_d\widetilde{L}_i \widetilde{e}^c_{Rj} +\hc
	\no\\&&
	+\frac{M_1}{2}\widetilde{B}\widetilde{B}+\frac{M_2}{2}\widetilde{W}\widetilde{W}
	+\frac{M_3}{2}\tilde{g}\tilde{g}+\hc
	\end{eqnarray}
where $b$ is sometimes written as $B\mu$ in the literature. The trilinear terms are often parametrized with an explicit factror of Yukawa coupling as $(a_U)_{ij}=(y_U A_U)_{ij}$.
The MSSM has many parameters in these soft breaking terms. And potentially there are a lot of dangerous terms leading to flavor-violation and/or CP-violation which are strongly constrained by the low-energy experiments. A simple solution to this is that  the soft-terms have a structure of Minimal Flavor Violation (MFV) where the flavor violation and CP-violation only come from the SM Yukawa couplings. This occurs, for instance, if the supersymmetry breaking is transferred to the MSSM sector through gauge couplings or geometry because either of them does not discriminate flavors. 
In constructing supersymmetry breaking models, we must address this issue. 

\begin{figure}[t!]
\begin{center}
  \includegraphics[clip,width=.43\textwidth]{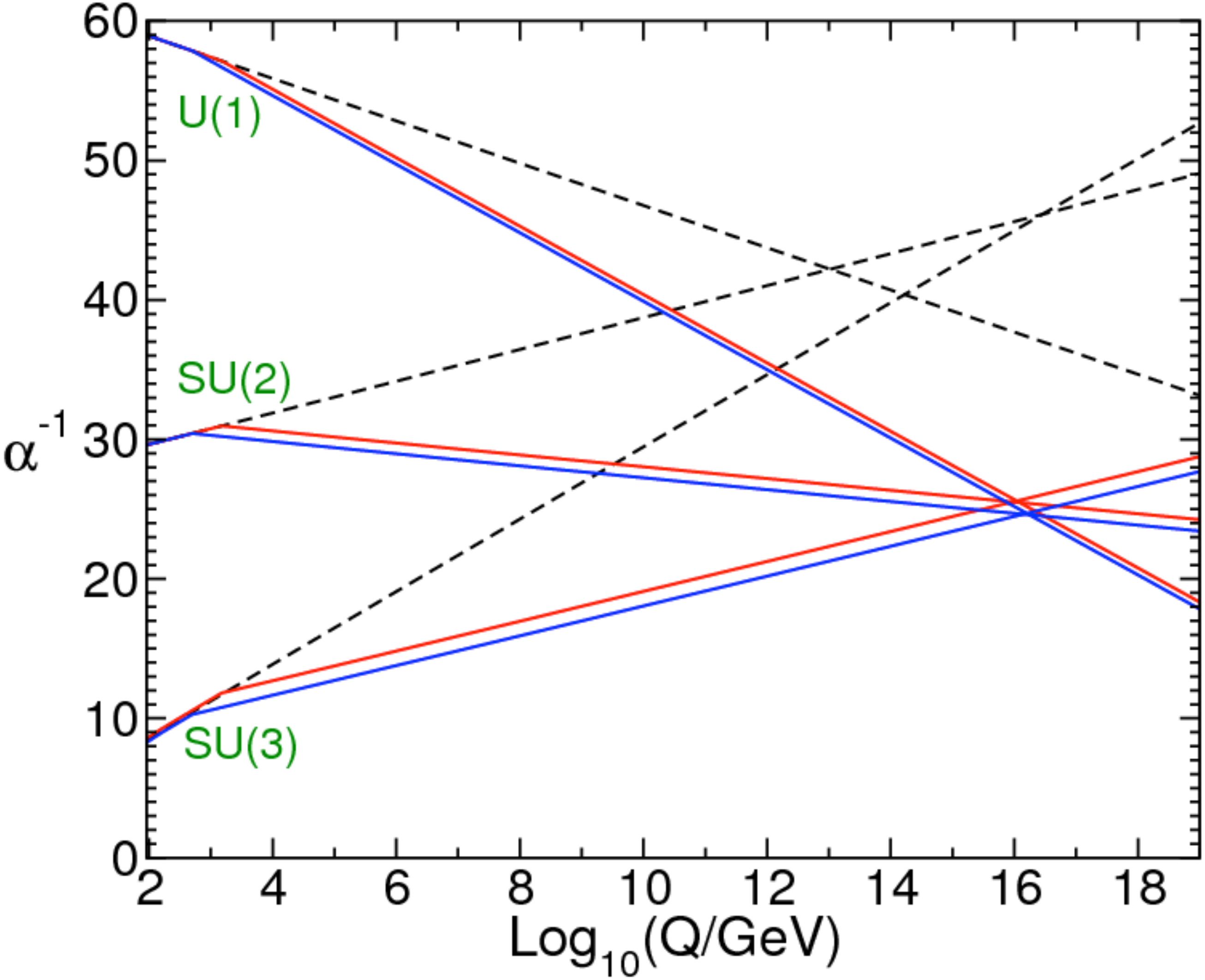}
\end{center}
\caption{Two-loop renormalization group evolution of the inverse gauge couplings $\alpha^{-1}(Q)$ in the Standard Model (dashed lines) and the MSSM (solid lines) from Ref.~\cite{Martin:1997ns}}
\label{gut}
\end{figure}

A nice feature of the MSSM is that quartic terms of Higgs potential are not free parameters but determined by the gauge couplings, 
	\begin{eqnarray}
	V(H_u^0, H_d^0)&=&(\mu^2+m_{H_u}^2)|H_u^0|^2+(\mu^2+m_{H_d}^2)|H_d^0|^2
	-(b H_u^0H_d^0+\hc)
	\no\\&&
	+\frac{g_1^2+g_2^2}{8}\left(|H_u^0|^2 -|H_d^0|^2 \right)^2 .
	\end{eqnarray}
where $g_1$and $g_2$ are $U(1)_Y$ and $SU(2)_L$ gauge coupling, respectively. $H_u^0$ and $H_d^0$ get VEVs,
	\begin{eqnarray}
	\langle H_u^0\rangle =v_u =v\sin\beta , \ 
	\langle H_d^0\rangle =v_d =v\cos\beta , \ 
	v\approx 174 \GeV . 	
	\end{eqnarray}
As in the SM, the lightest Higgs mass at tree-level is determined by the size of quartic coupling,
	\begin{eqnarray}
	m_h^2 \lesssim m_Z^2 \cos^2(2\beta)\leq (91.2\GeV)^2 \ ,
	\end{eqnarray}
 where $m_Z^2 =(g_1^2+g_2^2)v^2/8$. This upper bound is applied for the MSSM, and the Higgs mass can be larger if there are other matters which couple to the Higgs sector, for example, in an extension with a gauge singlet $S$,  a new interaction $W\supset \lambda S H_u H_d$ gives an additional quartic interaction at tree-level. 
  In the MSSM, radiative corrections boost the lightest Higgs mass, and an approximate formula at one-loop level is given by
  \cite{Ellis:1990nz}
 	\begin{eqnarray}
	m_{h,\rm MSSM}^2 \simeq m_Z^2 \cos 2\beta+ \frac{3m_t^4}{4\pi^2 v^2}\left(\log\frac{m_{\tilde{t}}^2}{m_t^2} +X_t^2\left(1-\frac{X_t^2}{12m_{\tilde{t}}^2} \right) \right), 
	\label{eq:1loopHiggs}
	\end{eqnarray}
where we use the common mass for stops, $m_{\tilde{t}}\equiv m_{\tilde{Q}_3}=m_{\tilde{u}_3}$, and $X_t=A_t -\mu \cot\beta$ is left-right mixing. 
The correction from $X_t$ is maximized at $X_t=\sqrt{6}m_{\tilde{t}}$.
The corrections are interpreted as effective Higgs quartic couplings generated at one-loop level. 
The large stop mass and/or large mixing are needed to accommodate $m_h=125 \GeV$, but they lead to a tension that the naturalness may be suffered from the large supersymmetry breaking in the MSSM.  
These are discussed in Sec.~\ref{direction}.

Supersymmetry, in particular conventional MSSM, is also motivated by the Grand Unification of gauge couplings. In presence of new sparticles, the beta functions of gauge couplings are changed and their couplings seem to unify for introducing supersymmetry in Fig.~\ref{gut}. The unification scale is about $10^{16}\GeV$.

\subsection{Naturalness in a nutshell}
Here we review a basic argument of naturalness and fine-tuning. The original discussion is given in Ref.~\cite{Barbieri:1987fn} and others are found in Refs.~\cite{SusyFineTuning, Kitano:2006gv, Hall:2011aa}. 

In some BSM models, there are various contributions to the quadratic terms of the Higgs potential, and their scale should be of order of weak scale $\cal O$(200 GeV) or less for the natural EWSB.  Since the relevant terms are particularly those in the direction of Higgs VEV, the discussion can be reduced to the one-dimensional potential problem as in the SM. 
Once again, the Higgs potential and the physical Higgs mass after the EWSB are
	\begin{eqnarray}
	&&V=\mu_H^2|H|^2 +\frac{\lambda_H}{4}|H|^4 	\label{eq:higgs}, 
	\\
	&&m_H^2=-2\mu_H^2 . 
	\end{eqnarray}
The quadratic term $\mu_H^2$ is in general a linear combination of various terms of scalar fields. Each contribution $\delta m_H^2$ to the Higgs mass should not be much greater than $m_H^2$, otherwise different parameters need to be tuned so that $m_H^2$ is at weak scale. Therefore, a ratio $\delta m_H^2/m_H^2$ should not be large. By using $\delta m_H^2=-2\delta \mu_H^2$, 
	\begin{eqnarray}
	\Delta\equiv \left|\frac{2\delta \mu_H^2}{m_H^2}\right|
	\end{eqnarray}
can be used as a fine-tuning measure. Before knowing the Higgs mass, $m_H$ is replaced with $m_Z$. Following the same philosophy, the commonly used measure to consider effects from dimensionless parameters \cite{Barbieri:1987fn} is
	\begin{eqnarray}
	\Delta_{} =\max_x \left| \frac{\partial \log m_{Z(h)}}{\partial \log x}\right|, 
	\end{eqnarray}
where $x$ is a parameter of the given theory. 

Let us give some exercises. Suppose the SM Higgs sector couples to a particle of $M_{pl}$ mass with $\cal O$(1) coupling, and  we estimate a radiative correction,
	\begin{eqnarray}
	\delta \mu_H^2 \sim \frac{1}{16\pi^2}M_{pl}^2\sim 10^{34} \GeV^2
	\end{eqnarray}
which leads to an extreme fine-tuning, 
	\begin{eqnarray}
	\Delta \sim \frac{2\times10^{34}\GeV^2}{125^2\GeV^2}\sim 10^{30}. 
	\end{eqnarray}
This tuning-level is  extremely high, and thus our EWSB is unlikely to happen. 
Next, let us consider a MSSM scenario where SUSY breaking effects are mediated at a scale of $\Lambda_{\rm mess}$. The important contribution is from top-stop sector for $y_t$ coupling. Renormalization group equation gives a  large correction to the soft-mass of $H_u$, 
	\begin{eqnarray}
	\delta m_{H_u}^2 =-\frac{3y_t^2}{8\pi^2}(m_{Q3}^2+m_{u3}^2+|A_{t}|^2) \log \frac{\Lambda_{\rm mess}}{m_{\rm weak}} \ , 
	\label{eq:1loopRGE}
	\end{eqnarray}
up to one-loop leading-log. For the low-scale mediation,  we have 
	\begin{eqnarray}
	\Delta \sim 10\left(\frac{m_0}{600\GeV} \right)^2\left(\frac{\log\frac{\Lambda_{\rm mess}}{m_{\rm weak}}}{6}\right),
	\end{eqnarray}
where we choose $m_0^2=m_{Q3}^2=m_{u3}^2=|A_{t}|^2$ for simplicity. We believe $\Delta\sim10$ tuning is acceptable. 

\section{LHC Run I: Higgs Boson}
\subsection*{Production and decay}

\begin{figure}[ht!]
\begin{center}
  \includegraphics[clip,width=.49\textwidth]{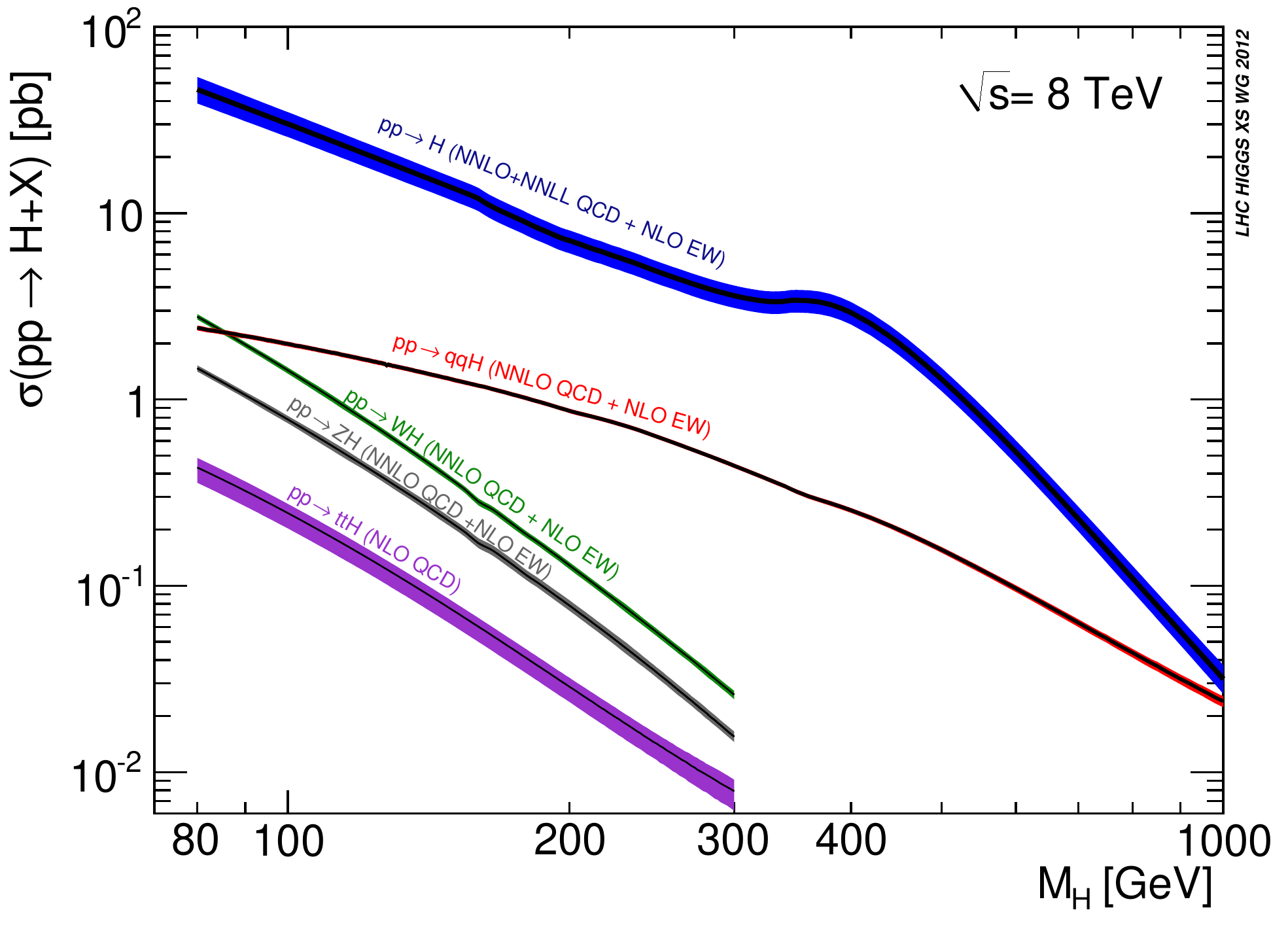}\hspace{20pt}
  \includegraphics[clip,width=.36\textwidth]{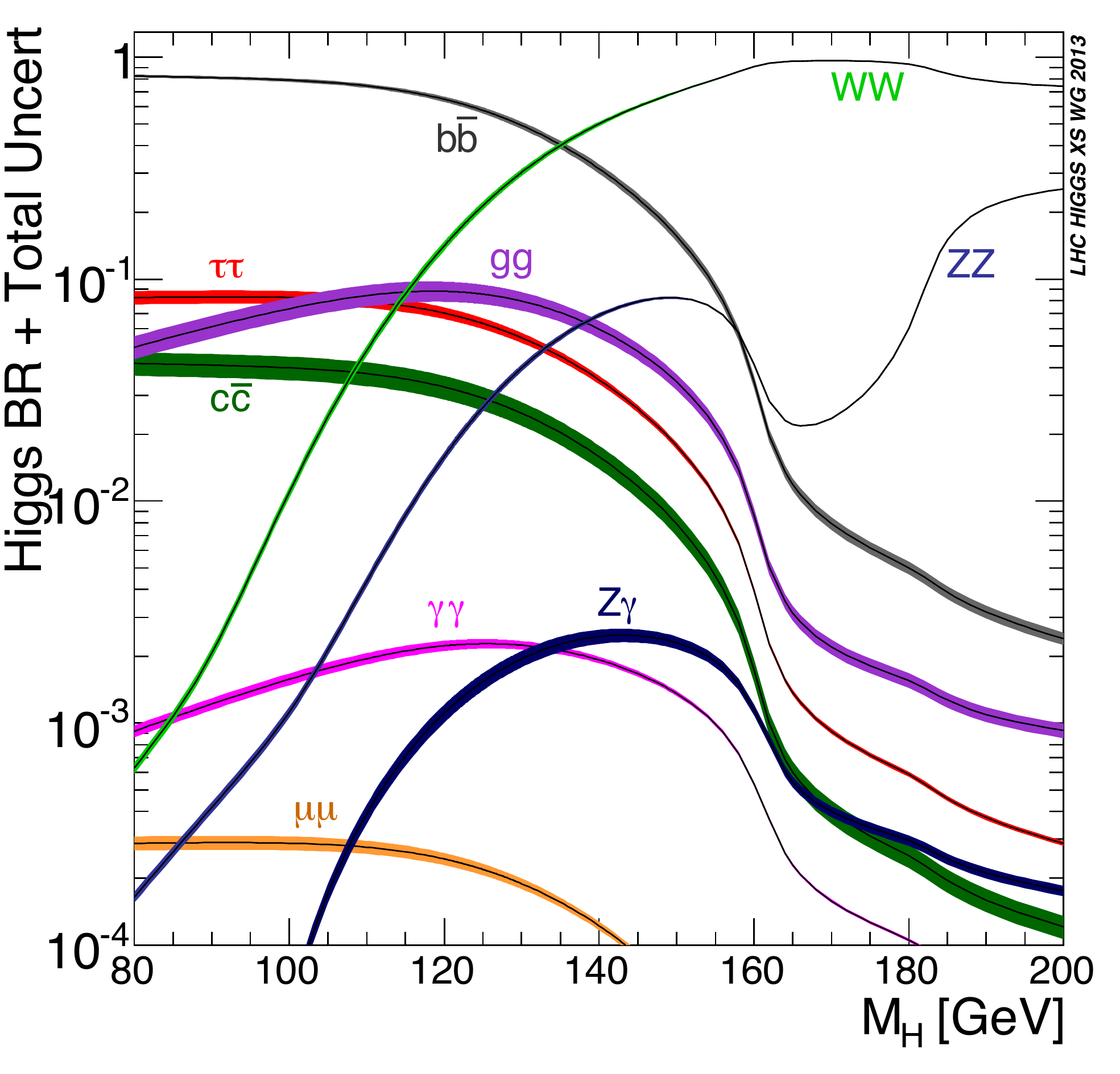}
\end{center}
\vspace{-20pt}
\caption{{\it Left}: cross section of Higgs boson as a function of its mass \cite{Dittmaier:2011ti}.   {\it Right}: branching Ratio of Higgs boson as a function of its mass \cite{Dittmaier:2012vm}. The band shows uncertainties. }
\label{fig:Dittmaier} 
\end{figure}

At the LHC, the Higgs boson is produced in several processes, and their cross sections including corrections of  higher order QCD and electroweak interactions are given in Ref.~\cite{Dittmaier:2011ti}. The result for $\sqrt{s}=8\TeV$ is seen in Fig.~\ref{fig:Dittmaier} left. 

The dominant production is a loop-induced channel, gluon-fusion (ggF), that is, two gluons go to Higgs boson via top quark.  
The cross section is about 19~pb with $+7.2\% \atop -7.8\%$ scale uncertainty\footnote{Scale uncertainty is due to choices of factorization scale and renormalization scale. } and $+7.5\% \atop -6.9\%$ PDF uncertainty at $m_H=125\GeV$ at $\sqrt{s}=8\TeV$. Another production process with top-Yukawa coupling is that a gluon splits into two top quarks and two top quarks from different gluons are fused into a Higgs boson. The final state has a Higgs boson and two top quarks, and this production is denoted by ttH. The cross section is small, but this is important because it is a direct information of the Higgs coupling to top quark. Scale and PDF uncertainties of ttH are similar to those of ggF.

Next, there are production processes by electroweak gauge interactions. Two weak gauge bosons ($W/Z$) which are radiated by initial quarks are  fused into a Higgs boson, called Vector Boson Fusion (VBF). To tag this process in the experiments, the Higgs boson should be produced with two forward jets. Another important process is associate production. A virtual $W/Z$ boson emits a Higgs boson, and the final state simply has a $W/Z$ boson and a Higgs boson, and they are denoted by WH and ZH, respectively (VH is referred  to both WH and ZH). Scale and PDF uncertainties of VBF and WH are about 0.2~\% and 2~\%, respectively, while those of ZH are about 2~\% and 3~\% at $\sqrt{s}=8\TeV$. 

The Higgs boson at mass of 125 GeV has many decay modes as shown in Fig.~\ref{fig:Dittmaier} right. The ATLAS and CMS detectors are designed to have good resolutions for $H\to \gamma\gamma$, and this diphoton decay occurs via top quark and $W$ boson loops. Two contributions are deconstructive and $W$ contribution  is larger. 
A typical process of Higgs production to decay at the LHC is $gg\to H\to \gamma\gamma$. Decays into $ZZ^*$ and  $WW^*$ are also important, and in particular there are less backgrounds in final states with more electrons and muons by $Z\to ee,\mu\mu,$ and $W\to e\nu,\mu\nu$. The golden channel is $H\to ZZ^*\to 4l \ (l=e,\mu)$.  
Although large portion of decay is in final states of $b\bar{b}$ and $gg$, they are not easily discriminated from QCD background. However, a process from VH production to $H\to b\bar{b}$ decay is be used  in the current analysis requiring $b$-jets and charged leptons.

\subsection*{Higgs mass}
Mass of Higgs boson has been measured in $H\to \gamma\gamma$ and $H\to ZZ^*\to 4l$ channels at both ATLAS and CMS. The combined results of each experiment are
	\begin{eqnarray}
	m_H &=& 125.5 \pm 0.2\ ({\rm stat}) {+0.5\atop -0.6}\ ({\rm sys}) \GeV  \ \ ({\rm ATLAS}~\cite{ATLAS:mass2013}), 
	\\
	m_H &=& 125.7 \pm 0.3\ ({\rm stat}) \pm 0.3\ ({\rm sys}) \GeV  \ \ ({\rm CMS}~\cite{CMS:mass2013strength}) . 
	\end{eqnarray}
The uncertainty is already within 1 GeV level. Note that there is a difference of measured mass in the two different channels at the ATLAS with $\approx2.5\sigma$ level. A mass of $m_H = 126.8 \pm 0.2\ ({\rm stat}) \pm 0.7\ ({\rm sys}) \GeV$ is found in the $H\to\gamma\gamma$ channel and a mass of $m_H = 124.3 {+0.6\atop -0.6}\ ({\rm stat}) {+0.5\atop -0.63}\ ({\rm sys}) \GeV$ is found in the $H\to ZZ^*\to 4l$ channel. However, for simplicity, we neglect this deviation so far and take the observed mass to be 125 GeV. 
\begin{figure}[ht!]
\begin{center}
  \includegraphics[clip,width=.38\textwidth]{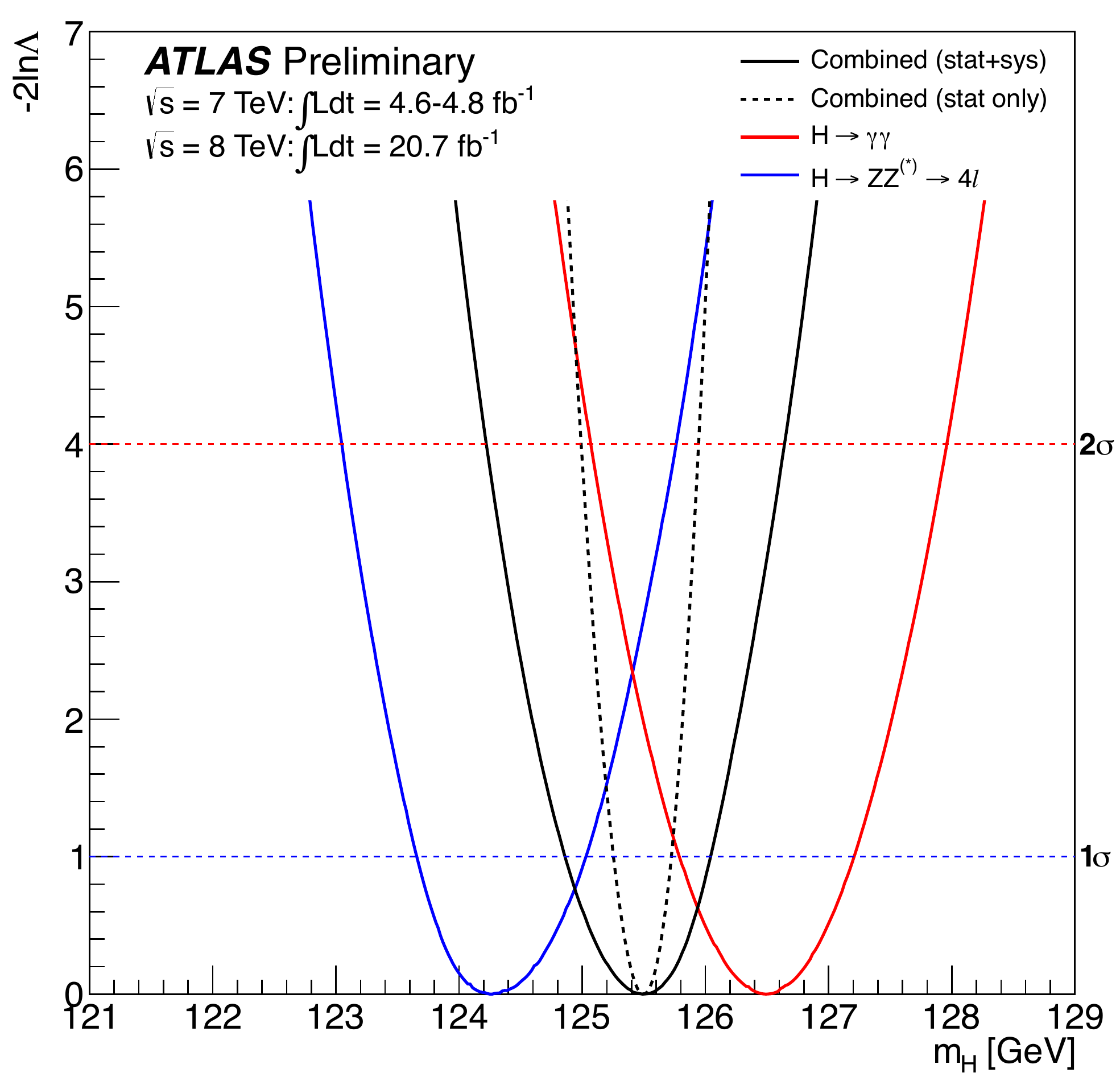}\hspace{20pt}
    \includegraphics[clip,width=.38\textwidth]{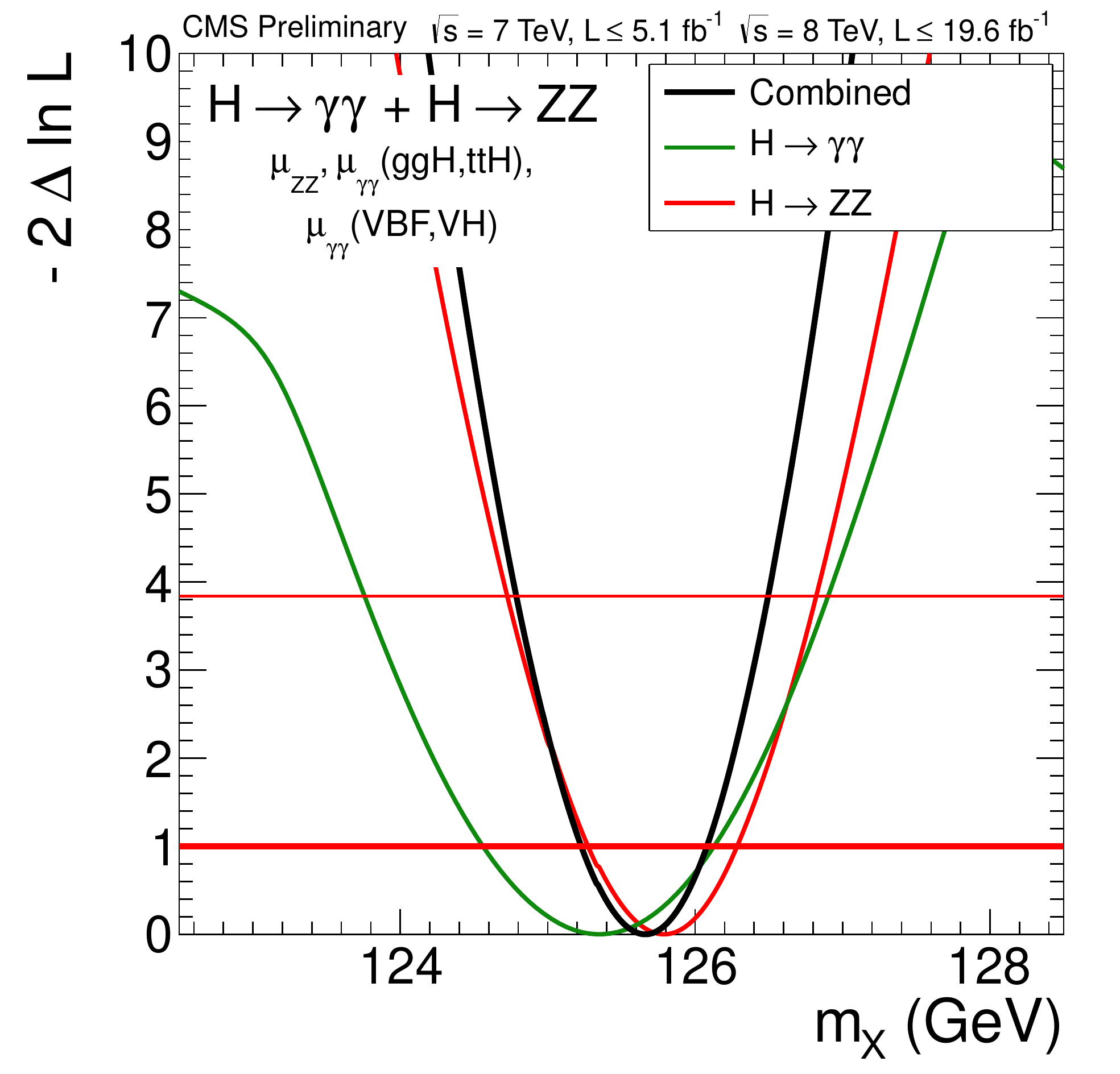}
\end{center}
\vspace{-10pt}
\caption{1D likelihood for measured Higgs mass. {\it Left}: ATLAS result \cite{ATLAS:mass2013}, and {\it right}: CMS result \cite{CMS:mass2013strength}}
\end{figure}

\subsection*{Spin and parity}
We have treated the observed particle as the Higgs {\it boson} which has spin 0 and parity even. Here, we briefly discuss other possibilities of spin and parity. 
The spin measurements are performed in  $H\to \gamma\gamma$,   $H\to ZZ^*\to 4l$, and $H\to WW^*\to l\nu l \nu$ channels \cite{Aad:2013xqa}. First of all, in $H\to \gamma\gamma$ channel, Landau-Yang theorem excludes spin 1 hypothesis and it is possible to test $J^P=0^+, 2^+$ where $J$ is spin and $P$ is parity.  The $H\to ZZ^*\to 4l$ channel has good sensitivies to $J^P=0^+, 0^-, 1^+, 1^-, 2^+$. For the spin and parity test, one can examine one hypothesis against another hypothesis using ratio of likelihood. The results of ATLAS and CMS disfavor $J^P=0^-, 1^+, 1^-, 2^+$ in comparison with $J^P=0^+$ with $\gtrsim 98$~\%C.L. 
 Therefore we conclude the observed particle is certainly Higgs boson like-particle of $J^P=0^+$. 


\subsection*{Signal Strength}
\begin{figure}[th!]
\begin{center}
  \includegraphics[clip,width=.37\textwidth]{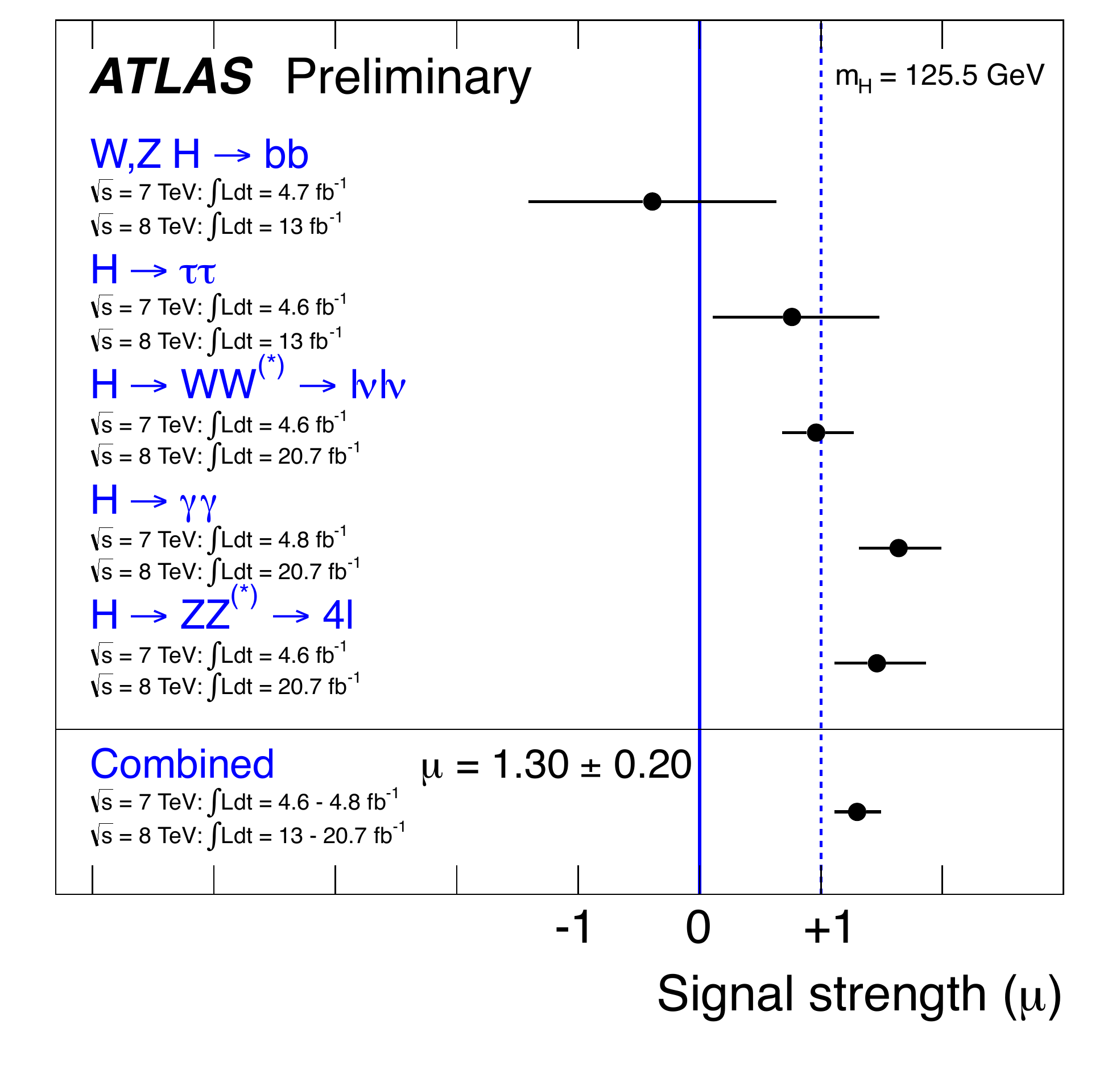}
    \includegraphics[clip,width=.45\textwidth]{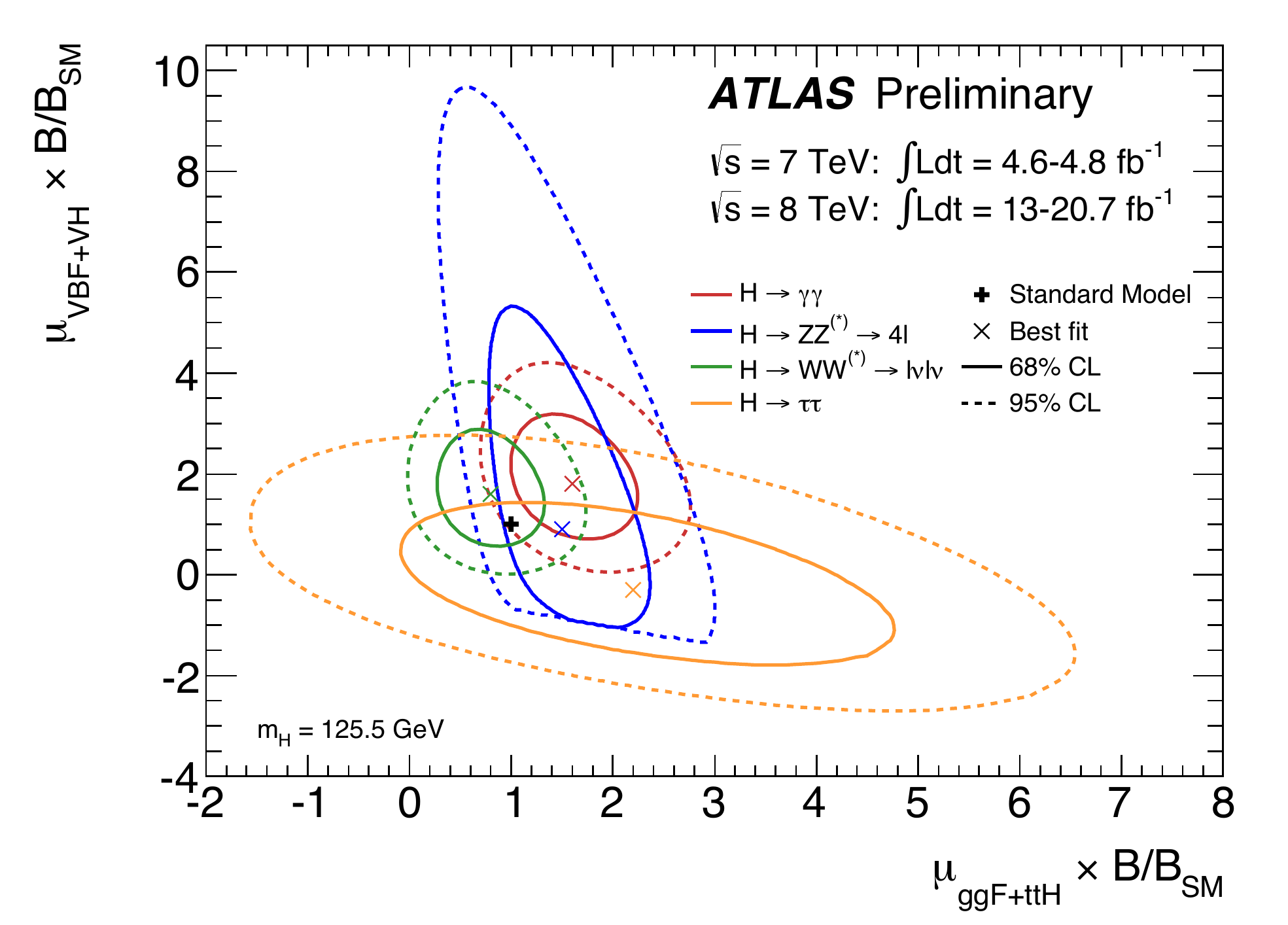}\vspace{-10pt}
\caption{{\it Left:} 1D signal strengths in each decay mode measured by the ATLAS.  {\it Right:} 2D signal strengths depending on production channels. Both plots from Ref.~\cite{ATLAS:strength2013}. }
\vspace{20pt}
  \includegraphics[clip,width=.4\textwidth]{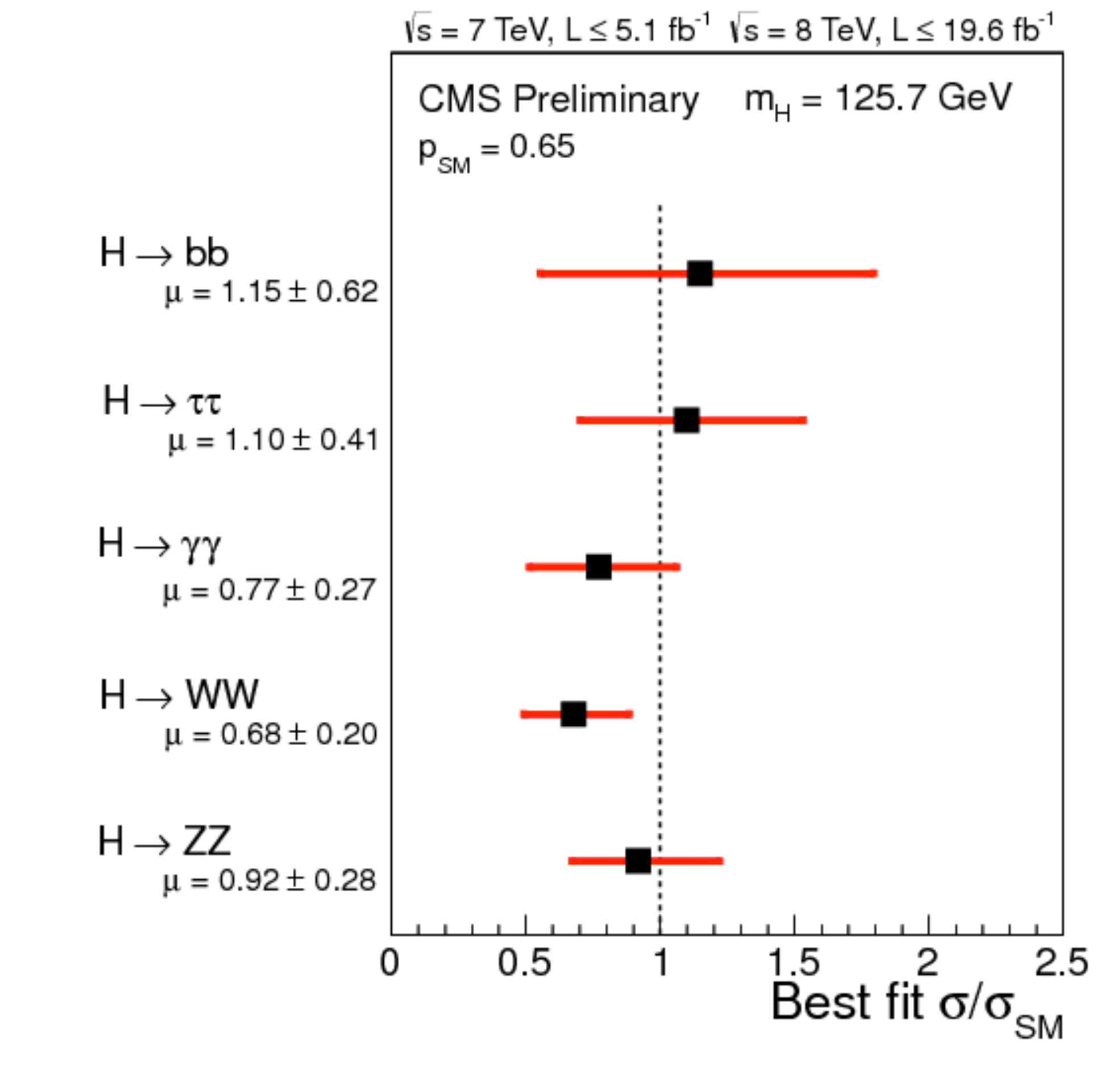}\hspace{10pt}
    \includegraphics[clip,width=.4\textwidth]{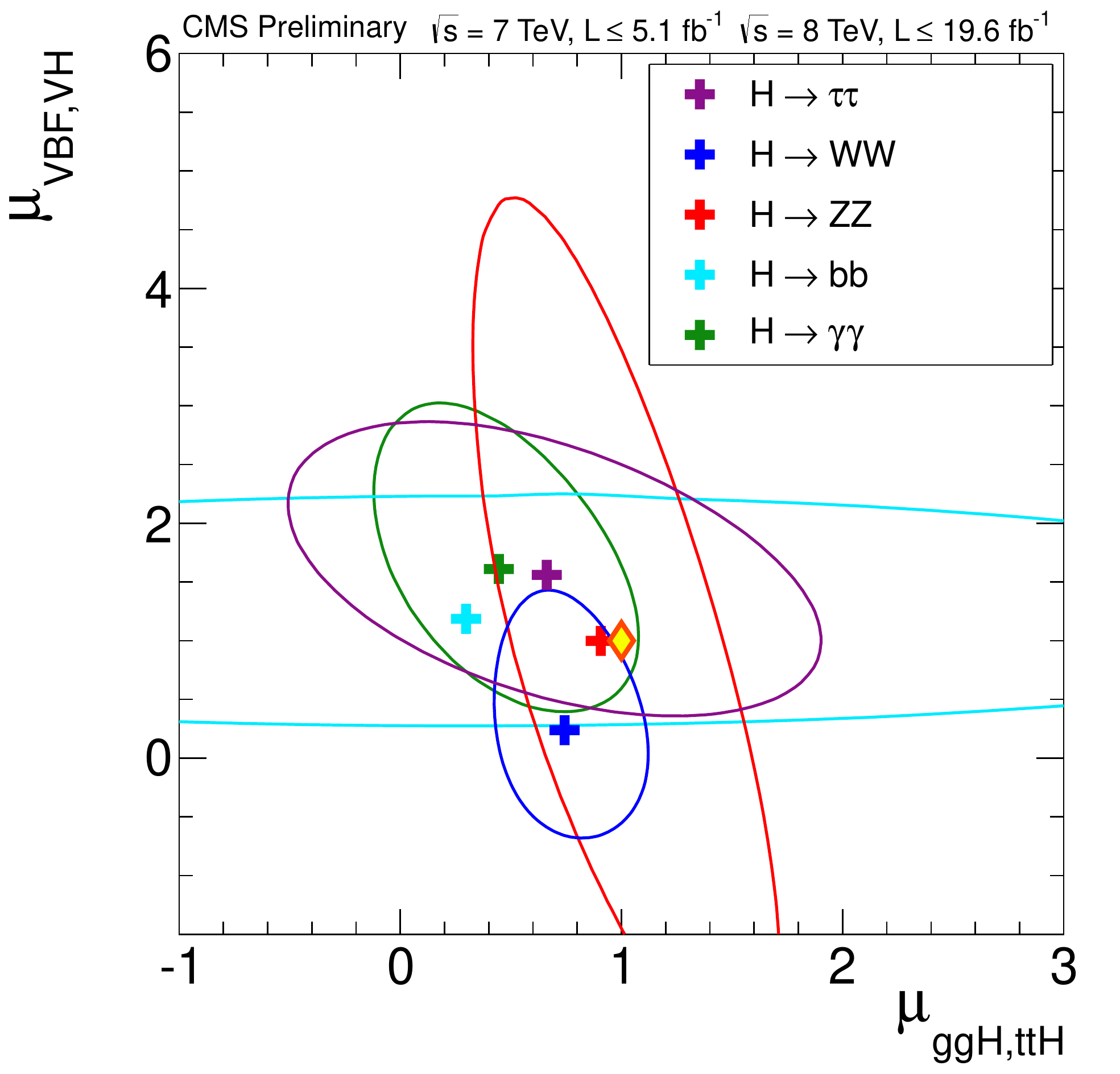}
\vspace{-10pt}
\caption{{\it Left:} 1D signal strengths in each decay mode measured by the CMS.  {\it Right:} 2D signal strengths depending on production channels. Both plots are from Ref.~\cite{CMS:mass2013strength}}
\vspace{-10pt}
\end{center}
\end{figure}
Signal strength is studied in various channels. The signal strength is roughly given by
	\begin{eqnarray}
	n^{\rm obs}_{{\rm signal},i\to f}= \mu_f B_{SM, f}\times \mu_i\sigma_{SM, i}\times ({\rm Integrated\ Luminosity}),
	\end{eqnarray}
where $\sigma_{SM, i}$ is SM production cross section of process $i$, $B_{SM, f}$ is SM Higgs boson production cross section of process $i$, the branching ratios $B_{SM,f}$ of the SM Higgs boson decays of $f$ final state, and $\mu_i$ and $\mu_f$ are relative strengths of production and decay, respectively. 
The number of signal events expected from each combination of production and decay modes is scaled by the corresponding product $\mu_i\mu_f$.

Results from the ATLAS and CMS are basically consistent with SM prediction. Although the errors of each channel is still large, it is possible to constraint BSM models which predicts large divination of signal strength. The $H\to \gamma\gamma$ channel at the ATLAS is slightly larger, but this is not conclusive.

\section{LHC Run I: Supersymmtric Particles}
Searches for supersymmetric particles (sparticles), heavy $Z/W$-bosons, fourth generation quarks, and many other phenomena predicted by BSM models have been carried out at the LHC. Here we present  results of sparticle searches assuming $R$-parity conservation. 

A typical (expected) scenario is that the colored sparticles are produced in pair and they decay into the neutral LSP emitting many jets and/or leptons\footnote{Leptons in the sense of collider physics mean electrons and muons, not tau leptons. Since tau leptons decay before reach to the detector, they behave like collimated QCD jets.} (cascade decay). The mass gap between those colored particles and the LSP are expected to be large such as $\gtrsim 500 \GeV$, which is certainly true for ``Minimal Supergravity''  or Constrained Minimal Supersymmetric Standard Model (CMSSM) scenarios \cite{Chamseddine:1982jx}, and then the LSP carries large momentum reflecting the mass gap. While the neutral LSPs themselves are not detected because they are stable for the $R$-parity, we can know their total transverse momentum thanks to momentum conservation.   The size of missing momentum ($\slashed{P}_T$) is usually referred to $\slashed{E}_T$. In the SM, neutrinos do the same role of LSP, and therefore high-energy SM processes emitting neutrinos such as $t\bar{t}$+jets and $W$+jets are background in the sparticle searches with $\slashed{E}_T$.

Particles emitted during cascade decays are highly model-dependent, but, in $R$-parity conserving case,  we can expect at least a few jets as well as $\slashed{E}_T$ because the produced parents are colored while the LSP are neutral.  If bottom quarks which form $b$-jets or energetic charged leptons are emitted, the signal becomes more distinct from SM background.

\begin{figure}[h]
\begin{center}
  \includegraphics[clip,width=.48\textwidth]{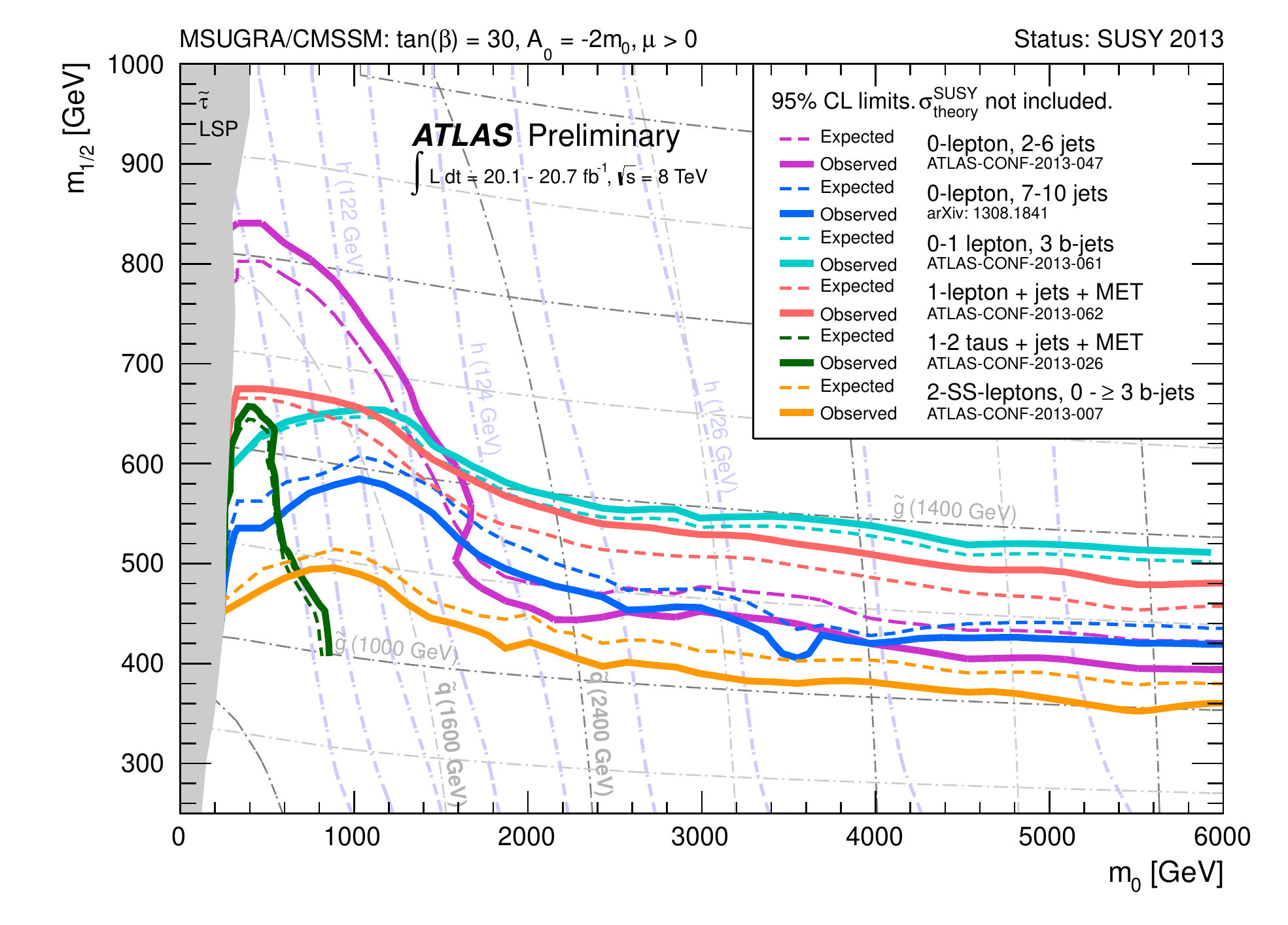}
  \includegraphics[clip,width=.50\textwidth]{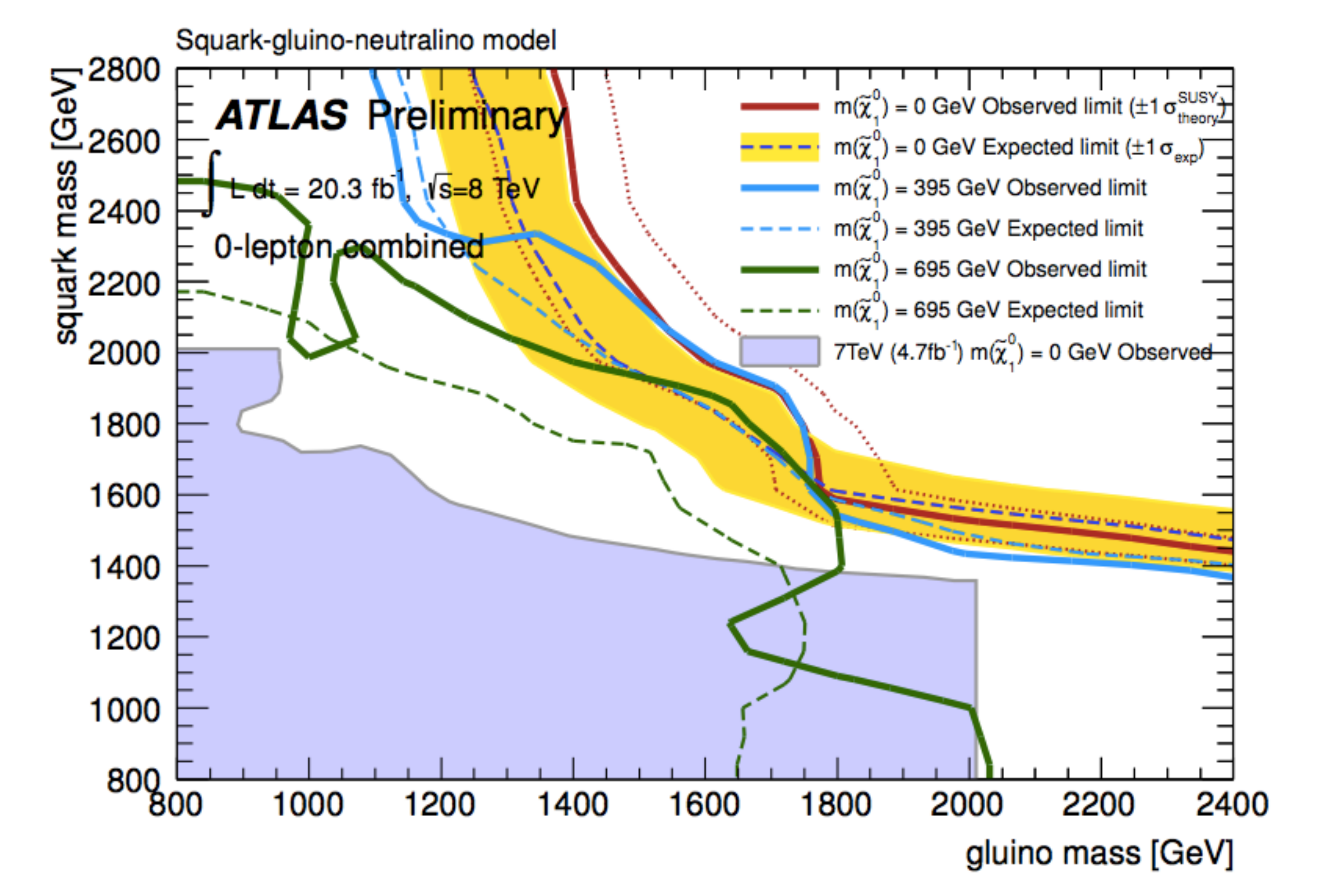}
\vspace{-10pt}
\end{center}
\caption{{\it Left:} constraints on CMSSM with large $A_t$ and large $\tan\beta$ by multiple analyses  \cite{ATLAS:multijet2013, ATLAS:summary2013}. {\it Right:} constraint on a simplified model of gluino, squarks, and neutralino in the right \cite{ATLAS:multijet2013}. }
\label{limit:MSSM}
\end{figure}

The dominant production of sparticles are $\tilde{g}\tilde{g}, \tilde{g}\tilde{q}$ and $\tilde{q}\tilde{q}$ pair productions where $\tilde{q}$ represents squarks of the first and second generations. Multijet+$\slashed{E}_T$ searches constrain many simple models of supersymmetry. For example, the CMSSM is excluded up to $m_{\tilde{q}}\approx 1.6\TeV$ (gluino decoupling), $m_{\tilde{g}}\approx 1.2\TeV$ (squark decoupling) as shown in Fig.~\ref{limit:MSSM}. 
Right panel of Fig.~\ref{limit:MSSM} shows multijet+$\slashed{E}_T$ searches constraint a simplified model of gluino, squarks, and neutralino where gluino and squark masses are varied independently with desecrate choices of neutralino mass.  The limit on this model is $m_{\tilde{g}}, m_{\tilde{q}}\gtrsim 1.4\TeV$ with massless LSP. When $m_{\tilde{g}}\simeq m_{\tilde{q}}$, the limit for simplified Model and CMSSM is extended up to about 1.7 TeV. 

\begin{figure}[h]
\begin{center}
  \includegraphics[clip,width=.49\textwidth]{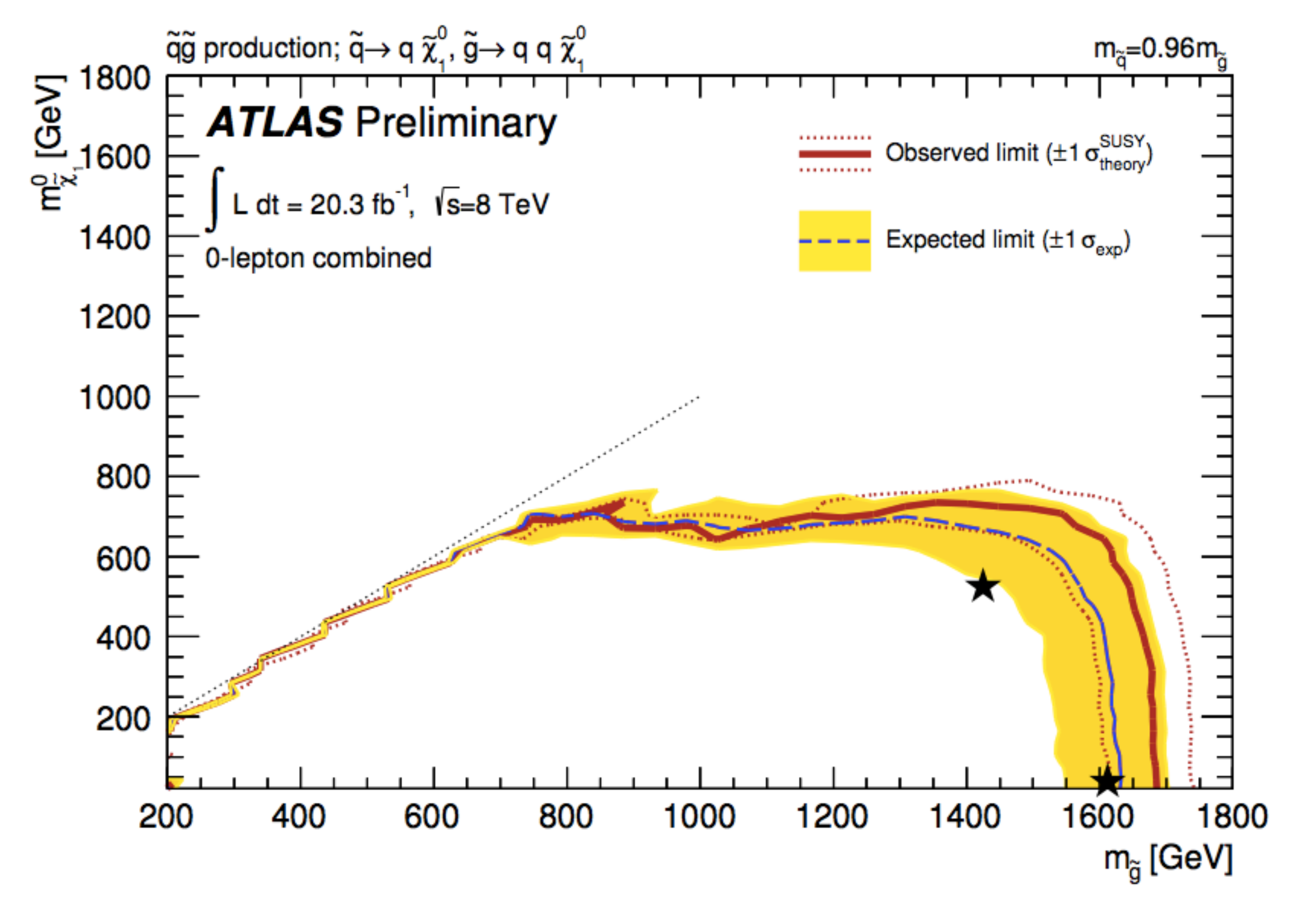}
\end{center}
\vspace{-20pt}
\caption{Squark-gluino-neutralino simplified model fixing squark mass as $m_{\tilde{q}}=0.96 m_{\tilde{g}}$ \cite{ATLAS:multijet2013}. }
\label{squark-gluino}
\end{figure}

Let us remind that $m_{SUSY}\gg m_{weak}$ gives rise to fine-tuning for the EWSB. The above results seem to disfavor low-scale supersymmetry.  However, very different consequences are led if the spectrum is nearly degenerate (compressed), {\it i.e.} the mass gap between the LSP and produced parent particles are relatively small, say $\lesssim 500\GeV$ (not necessarily as small as $\lesssim 50\GeV$).  This result shows much weaker constraints because the smaller mass gap lowers $\slashed{E}_T$.  For instance, for the previous simplified model with $m_{\tilde{g}}\approx m_{\tilde{q}}$, the gluino and squark masses are excluded up to 1.7 TeV with massless LSP, while the bound is relaxed down to 900 GeV with 200 GeV mass gap as seen Fig.~\ref{squark-gluino}. 
Therefore low-energy is still possible if the spectrum is compressed.  
In searches for this scenario, initial state radiation (ISR) will be important. An energy scale of ISR reflects the mass scale of parent colored particles, and thus we can expect energetic ISRs for the heavy particle productions. $\slashed{E}_T$ also gets larger because  LSPs are also recoiled by ISRs. 

\begin{figure}[ht]
\begin{center}
  \includegraphics[clip,width=.49\textwidth]{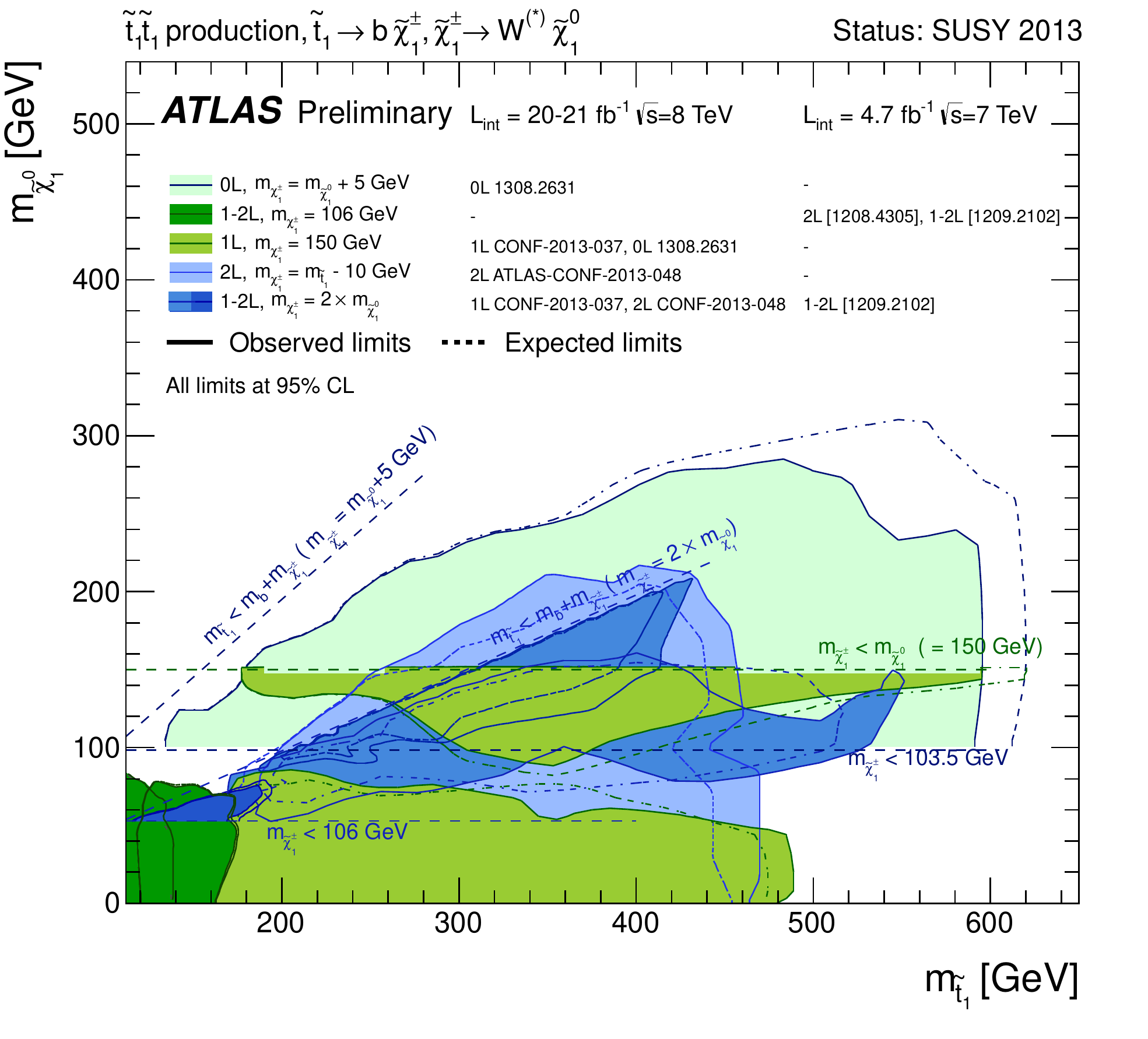}
  \includegraphics[clip,width=.49\textwidth]{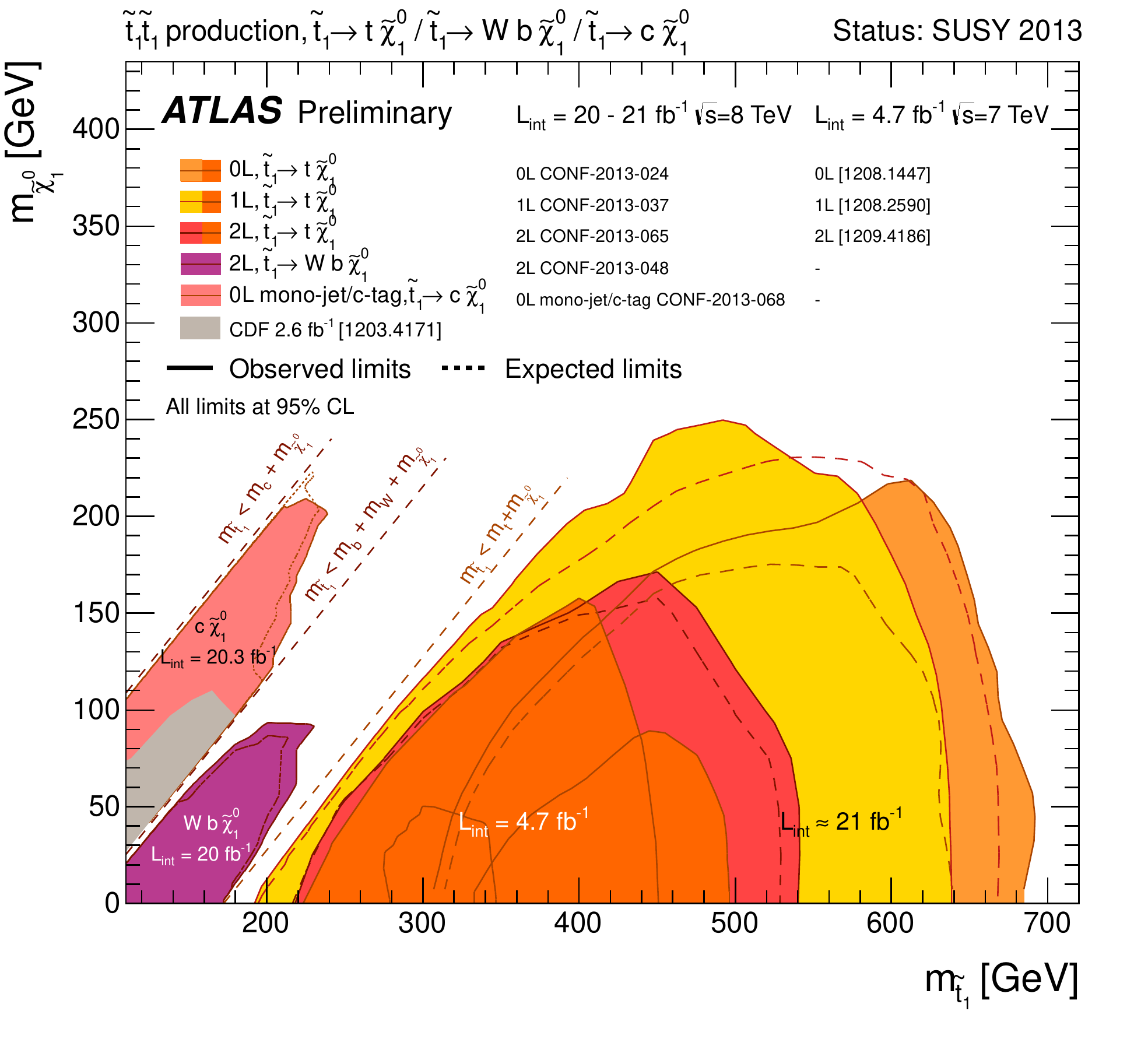}
\end{center}
\vspace{-20pt}
\caption{Direct stop searches for $\tilde{t}_1\to b\tilde{\chi}^\pm_1\to bW^{(*)}\tilde{\chi}^0_1$ in the left panel and for $\tilde{t}_1\to t\tilde{\chi}^0_1$, $\tilde{t}_1\to Wb \tilde{\chi}^0_1$, and $\tilde{t}_1 \to c\tilde{\chi}^0_1$ in the right panel \cite{ATLAS:summary2013}}.
\label{stoplimit}
\end{figure}

Constraints on the first and second generation squarks are indirect information for the naturalness. The third generation squarks, in particular stop, are more relevant to the naturalness. Of course, if one defines an explicit model, the first and second generation squarks bounds are relate to the stop mass bound. Stops are produced  from gluons initial state, and  there are basically  four decay patterns of the light stop: one via chargino, $\tilde{t}_1\to b\tilde{\chi}^\pm_1\to bW^{(*)}\tilde{\chi}^0_1$, and the others directly to the LSP, $\tilde{t}_1\to t\tilde{\chi}^0_1$, $\tilde{t}_1\to Wb \tilde{\chi}^0_1$, and $\tilde{t}_1 \to c\tilde{\chi}^0_1$. 
The former is studied with various assumptions of chargino mass, and the results are shown in left of Fig.~\ref{stoplimit}. The strongest limit is $m_{\tilde{t}_1}\gtrsim 600\GeV$. For the latter decays shown in right of Fig.~\ref{stoplimit}, the stop mass is excluded up to 700 GeV with massless LSP. 
However, these limits are also relaxed down to $m_{\tilde{t}_1}\sim 200\GeV$ when the mass gap is small. Note that production of stops is not only from direct stop production but also from gluino decays, and in this case limits becomes more stringent. Hence, the constraints are model-dependent. 

Finally let us briefly comment on other searches. 
Direct searches for EW-inos give bounds, sometimes up to 600 GeV, but the bound again depends on spectrum and decay patterns \cite{ATLAS:summary2013}. Scenarios with $R$-parity violation (RPV) are studied. Although baryonic RPV leads to no $\slashed{E}_T$ events, gluino mass is excluded up to about 900 GeV \cite{ATLAS:RPV2013}. 


\section{Directions for low-energy Supersymmetry}\label{direction}
\subsection{Tensions by LHC Run I  }
Here is a short summary of the LHC run I results relevant to supersymmetry, 
	\begin{itemize}
	\item New discovered particle is compatible with SM Higgs boson
	\item Observed mass of the new particle is 125 GeV, whereas the MSSM predicts the lightest Higgs mass is $m_h\leq91.2\GeV$ at tree-level
	\item The CMSSM and simplified model with massless LSP are excluded up to $m_{\tilde{q}}\simeq m_{\tilde{g}}\simeq 1.7\TeV$, while sub-TeV scenario is still possible when the spectrum is compressed
	\item Constraints on stop mass are model-dependent, and $m_{\tilde{t}_1}\sim 200\GeV$ is still allowed.
	\end{itemize}

Searches for sparticles based on $\slashed {E}_T$ constraint a lot of natural region of many supersymmetric models. As discussed above, if the spectrum is compressed, limits from LHC are much more relaxed. However, a motivation and an explicit model for low-energy supersymmetry with a compressed spectrum have not been discussed while collider studies are carried out by using simplified models. Also, after the Higgs discovery, it is necessary to discuss such a model with a compressed spectrum in the context of Higgs mass and fine-tuning. We give explicit such a model and study its phenomenology.  

	\begin{figure}[th!]
  	\centering
 	 \includegraphics[width=.9\textwidth]{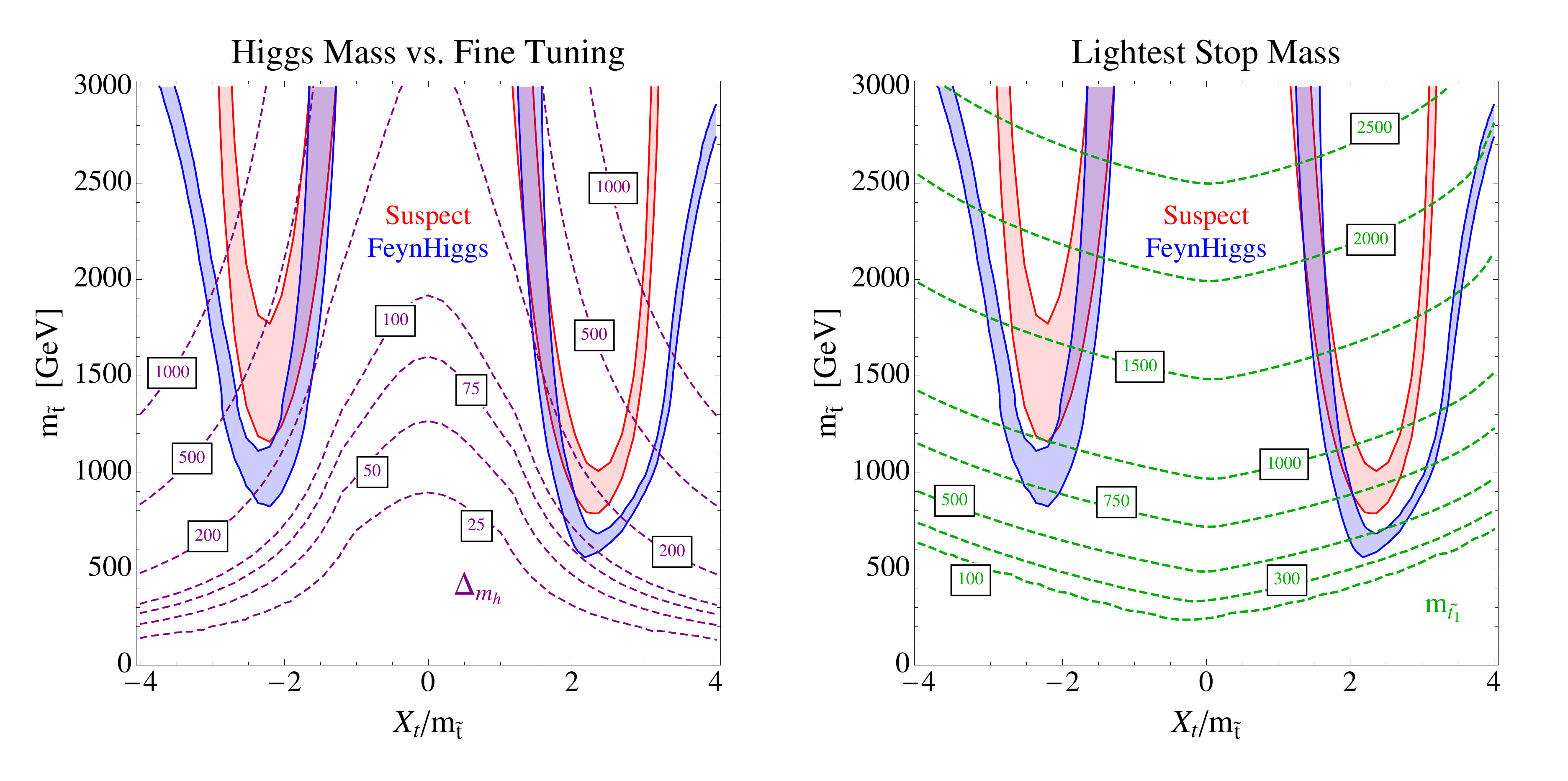}\vspace{-10pt}
	\caption{Plots for fine-tuning ({\it left}) and the lightest stop mass ({\it right}) as function of stop soft mass $m_{\tilde t}$ and mixing $X_t$ in Ref.~\cite{Hall:2011aa}. The SM-like Higgs mass is fixed to be 125 GeV calculated by two different softwares at two-loop level, {\tt FeynHiggs} and {\tt Suspect}. The tuning measure is $\Delta =|\partial \log m_h^2/\partial\log  x|$ with low mediation scale $\Lambda=10\TeV$.}
	\label{MSSM2D}
	\end{figure}

The collider limits are model-dependent and can be relaxed by a compressed spectrum or $R$-parity violation, or when sparticles unnecessary for naturalness are taken to be heavy. Even if sparticles evade detection for we must address the heavy Higgs mass.

In the MSSM, large radiative corrections by large stop mass and/or large mixing as in Eq.~(\ref{eq:1loopHiggs}) to the SM-like Higgs mass are required to accommodate $m_h=125\GeV$. 
However, the large stop mass and $X_t$ feed in $m^2_{H_u}$, and therefore the fine-tuning is caused. 
This tension is seen in Fig.~\ref{MSSM2D}. Assuming low mediation scale  $\Lambda=10\TeV$ it tells $\Delta \gtrsim 100$ ($\Delta^{-1} \lesssim 1\%$ ) is needed and the large mixing nearly $X_t\sim \sqrt{6}m_{\tilde t}$ is favored. 
The lightest stop mass should be $m_{{\tilde t}_1}\gtrsim 500\GeV$. 
For the CMSSM scenarios, the tension is worse \cite{Baer:2012mv}, $\it i.e.$ the fine-tuning is at best $\Delta \gtrsim 10^3$. 

Since we know the MSSM requires about 1\% tuning at least, possibilities beyond the minimal theory should be taken into account. 
For example, a singlet extension called Next-to-Minimal Supersymmetric Standard Model (NMSSM) has a new source of the lightest Higgs mass at tree-level, and therefore the tuning tend to be relaxed. Therefore, we also study an extended Higgs sector focusing on the tension of the Higgs mass and natural EWSB. 

\subsection{Organization of this thesis}

In this thesis we discuss low-energy supersymmetry in light of the LHC run I results, with particular concerns as follows.
	\begin{itemize}
	\item Compressed spectrum 
	\item Natural Higgs sector beyond the MSSM
	\end{itemize}

At first, we present an explicit model with a compressed spectrum in Ch.~\ref{compact:model}. We call it Compact Supersymmety.  The universality of gaugino, squark, and slepton masses is led by SUSY breaking from compactified extra dimensions, called Scherk-Schwarz mechanism. The model is explicitly testable since it has only three parameters, a size of extra dimension, SUSY breaking scale, and supersymmetric Higgs mass.  In Ch.~\ref{compact:pheno} we study Higgs mass, fine-tuning, dark matter nature, and collider constraint. The Higgs mass is generally large because large $A$-term is predicted. Chs.~\ref{compact:model}, \ref{compact:pheno} are based on work of Ref.~\cite{Murayama:2012jh}.
For future prospect, we discuss that a kinematic variable $M_{T2}$ is useful to search for BSM particles with a compressed spectrum in Appendix~\ref{ch:mt2}  based on work of Ref.~\cite{Murayama:2011hj}.

Next, we focus on the Higgs sector in particular the tension between Higgs mass and fine-tuning. We show that an extended Higgs sector with two gauge singlets  has a new window of natural theory of EWSB in Ch.\ref{dnmssm:model}. Unlike the MSSM and NMSSM favor lower mass scale of sparticles, in the new framework a singlet scaler is favored to be extremely heavy. This gives a non-decoupling effect without upsetting naturalness. 
In Ch.~\ref{dnmssm:pheno} we study constraints and future prospect of the model at the LHC and future $e^+e^-$ linear collider.  They are based on work of Ref.~\cite{Lu:2013cta}.
We finish with an overall summary in Ch.~\ref{ch:last}.

\part{Supersymmety, Extra Dimension and Compressed Spectrum}

\chapter{Compact Supersymmetry}\label{compact:model}

\section{Introduction}
Supersymmetry has widely been regarded as the 
prime candidate for physics beyond the standard model~\cite{Martin:1997ns}. 
It can explain the dynamical origin of electroweak symmetry breaking 
through renormalization group effects and provide a natural candidate 
for the cosmological dark matter in its simplest incarnations.  In 
particular, it stabilizes the large hierarchy between the electroweak 
scale $\approx~{\rm TeV}$ and the quantum gravity scale $\approx 
10^{15}~{\rm TeV}$ against radiative corrections to the Higgs mass 
parameter.  Barring a fine-tuning among parameters of the theory, 
this consideration strongly suggests the existence of sparticles 
below $\approx ~{\rm TeV}$.  The mass spectrum of sparticles 
has been mostly discussed within the CMSSM 
framework~\cite{Chamseddine:1982jx}, which typically generates 
a widely spread spectrum leading to experimentally identifiable 
large visible and missing energies.

However, no experimental hints have been seen so far at the Large Hadron 
Collider (LHC), which has led to substantial anxiety in the community. 
Moreover, the suggested mass of $125~{\rm GeV}$ for the Higgs boson by 
the LHC data~\cite{:2012gk, ATLAS:2012ae} is not easily accommodated in the MSSM, where one has to rely on radiative 
corrections to push the Higgs boson mass beyond the tree-level upper 
bound of $m_Z \simeq 91~{\rm GeV}$.  These requirements push the scalar 
quark masses well beyond the TeV within the CMSSM.

There are three main suggestions to allow for supersymmetry without 
the signal so far, within the context of $R$-parity conserving 
supersymmetry.  One is to simply accept a fine-tuning to maintain 
the hierarchy against radiative corrections, at a level significantly 
worse than a percent.  Quite often, the anthropic principle is brought 
in to justify this level of fine-tuning~\cite{ArkaniHamed:2004fb}. 
The second is to keep sparticles relevant to the Higgs mass 
parameter below TeV while to assume all other sparticles well 
beyond TeV~\cite{Dimopoulos:1995mi}.  The third is to assume that 
all sparticles are compressed (nearly degenerate) making them somewhat hidden 
from experimental searches due to low $Q$-values in visible and 
missing energies.  The last option, however, has been discussed 
only phenomenologically~\cite{LeCompte:2011fh}, lacking theoretical 
justifications based on simple and explicit models of supersymmetry 
breaking.

In this chapter, we point out that the third possibility of a compressed sparticle spectrum is quite automatic when supersymmetry 
is broken by boundary conditions in compact extra dimensions, the so-called 
Scherk--Schwarz mechanism~\cite{Scherk:1978ta,Pomarol:1998sd}.  With 
the simplest extra dimension---the $S^1/{\mathbb Z}_2$ orbifold---the 
mechanism has a rather simple structure~\cite{Barbieri:2001yz}.  In 
particular, locating matter and Higgs fields in the bulk and on a brane, 
respectively, and forbidding local-parity violating bulk mass parameters 
for the matter fields, the theory has only four parameters relevant 
for the spectrum of sparticles:\ the compactification scale $1/R$, 
the 5D cutoff scale $\Lambda$ ($> 1/R$), the supersymmetry breaking 
twist parameter $\alpha$ ($\in [0, \frac{1}{2}]$), and the supersymmetric 
Higgs mass $\mu$.

Using the common notation in the MSSM, the spectrum of sparticles 
is given at the compactification scale $\approx 1/R$ as
\begin{equation}
\begin{array}{c}
  M_{1/2} = \frac{\alpha}{R},
\quad
  m_{\tilde{Q},\tilde{U},\tilde{D},\tilde{L},\tilde{E}}^2 
  = \left( \frac{\alpha}{R} \right)^2,
\quad
  m_{H_u,H_d}^2 = 0,
\\[10pt]
  A_0 = -\frac{2\alpha}{R},
\quad
  \mu \neq 0,
\quad
  B = 0,
\end{array}
\label{eq:tree}
\end{equation}
at tree-level.  While these masses receive radiative corrections from 
physics at and above $1/R$, they are under control because of the 
symmetries in higher-dimensional spacetime, and thus can naturally 
be small.  Therefore, in this limit, the theory essentially has only 
three free parameters:
\begin{equation}
  \frac{1}{R},
\qquad
  \frac{\alpha}{R},
\qquad
  \mu.
\label{eq:free-param}
\end{equation}
This rather compact set of parameters gives all the sparticle as 
well as the Higgs boson masses.

Even though Eq.~(\ref{eq:free-param}) gives two less parameters than in 
the traditional CMSSM framework, we show that it still leads to viable 
phenomenology.  In addition, it solves the flavor problem that often 
plagues models of supersymmetry breaking, because the geometry is 
universal to all scalar particles and hence respect a large flavor 
symmetry.  The problem of accommodating a large enough Higgs boson 
mass is ameliorated by the near-degeneracy between $\tilde{t}_L$ and 
$\tilde{t}_R$, and $|A_t| \approx 2 m_{\tilde{t}}$.  And the compressed 
spectrum at tree-level automatically achieves a compact spectrum that 
allows sparticles to be hidden from the current searches even 
when they are below TeV.

This chapter is organized as follows.  We first review 5D supersymmetric gauge theory.
Then we investigate 
supersymmetry breaking by boundary conditions in the $S^1/{\mathbb Z}_2$ 
orbifold, and present the simplest model we study.  In Ch.~\ref{compact:pheno} we present 
the low-energy structure of the model and discuss its phenomenology. 

\section{5D Supersymmetric Lagrangian }
To begin with, we need to prepare 5D Lagrangian with a minimal supersymmetry. 
Constructing this Lagrangian is not trivial and actually quite complicated, but  there is a way to write down supersymmetric theories in higher dimensions within the familiar ${\cal N=} 1$, $D= 4$ superspace  \cite{ArkaniHamed:2001tb}. 
However, higher dimensional part of Lorentz symmetry is not clear within superspace. Lorentz symmetry is manifest after integrating out (or field redefinition of) auxiliary fields. 

The minimal supersymmetry of 5D corresponds to ${\cal N}= 2$, $D= 4$, and therefore $SU(2)_R$ global symmetry exists.  We start the theory in the superspace and derive component Lagrangian with explicit full Lorentz symmetry and $SU(2)_R$ global symmetry. Note that the superspace formalism is useful to interpret the Scherk-Schwarz supersymmetry breaking as Radion Mediation as we will see later in Sec.~\ref{sec:RM}. 

We basically follow notations of ``Supersymmetry and Supergravity'' by J.~Wess and J.~Bagger \cite{Wess:1992cp} for spinor formulae and so on.  
 The only difference is that we adopt a metric of  $g^{MN}=\rm diag(+,-,-,-,-)$ instead of  one used in the book, $g^{MN}=\rm diag(-,+,+,+,+)$,
where $M=\mu,5$ and $\mu=0,1,2,3$. To go to the metric of 
$g^{MN}=\rm diag(-,+,+,+,+)$, simply multiply  each contraction of Lorentz indices by $(-1)$, for example, 
$\partial^2 \to -\partial^2$ and $F^{MN}F_{MN}\to (-1)^2F^{MN}F_{MN}$. The spinor contractions and formulae are found in the Appendices A, B of the book.

The gamma matrices are given by
\begin{eqnarray}
  \Gamma^\mu = \gamma^\mu = 
   \left(
    \begin{array}{c c }
     0 & \sigma^\mu   \\
     \sigmabar^\mu &0 \\
    \end{array}
   \right) \ , \ 
     \Gamma^5 = i\gamma^5 = 
   \left(
    \begin{array}{c c }
     -i & 0   \\
     0 & i \\
    \end{array}
\right),
  \nonumber
\end{eqnarray}
 and we see $\{\Gamma^M,\Gamma^N \}=2g^{MN}$. Since $\gamma_5$ is incorporated in generators of 5D Lorentz symmetry, the 5D theory cannot have chiral fermions which are essential for the SM. 
 
\subsection{Non-Abelian gauge field}
Supersymmetric Yang-Mills theory in the 5D is given by
\begin{eqnarray}
{\cal L}_5^{\rm YM}= \frac{1}{16kg^2}{\rm Tr}\left[ \int d^2\theta {\cal W^\alpha W_\alpha} \ +{\rm h.c.} \right]
\hspace{4cm}  
  \nonumber\\\nonumber 
  +\frac{1}{4kg^2}   \int d^4\theta \ {\rm Tr}
  \Big[ (\sqrt{2}\partial_5+2g{\chi}^{\dag})e^{-2gV} (-\sqrt{2}\partial_5+2g{\chi})e^{2gV}
   \\ 
   +\partial_5 e^{-2gV}\partial_5 e^{2gV}+2g^2(\chi \chi+{\chi}^{\dag}{\chi}^{\dag})
  \Big],
  \label{213628_19Apr12}
\end{eqnarray}
where
\begin{eqnarray}
 {\cal W_\alpha} =\frac{1}{4}\bar{D}\bar{D}e^{2gV}D_\alpha e^{-2gV} 
  \ \mbox{and} \ \ {\rm Tr} [T^aT^b] =k\delta^{ab}.
\end{eqnarray}
$V=V(x,y)$ and $\chi=\chi(x,y)$ are adjoint vector and chiral superfields. The superfields are expanded as follows 
(with $V$ in the Wess-Zumino gauge, and $\chi$ in the $y$-basis), 
\begin{eqnarray}
 V= V^aT^a &=& -\theta\sigma^\mu {\thetabar} A  _\mu
+i\theta^2\thetabar\bar{\hat{\lambda}}_1- i\thetabar^2\theta\hat{\lambda}_1
+\frac{1}{2}\theta^2\thetabar^2D\label{165100_1May12}
\\
&\equiv&-\theta\sigma^\mu{\thetabar} A  _\mu
-\theta^2\thetabar\bar{\lambda}_1  -\thetabar^2\theta\lambda_1
+\frac{1}{2}\theta^2\thetabar^2D,	\label{165148_1May12}
\\
\chi=\chi ^aT^a&=&\frac{\Sigma+iA_5}{\sqrt{2}}+\sqrt{2}\theta\lambda_2 +\theta^2F_\chi,
\end{eqnarray}
where $\lambda\equiv i\hat{\lambda}$. In 5D, $\Sigma^a$ is a real scalar in adjoint representation of gauge group, and,  in  6D, it behaves as the sixth component of gauge field, $\Sigma \rightarrow A_6$.
The last two lines of Eq.~(\ref{213628_19Apr12}) are unconventional and describe terms of fifth component of gauge field, such as $F_{\mu5}$, and all the terms involving $\lambda_2$ and $\Sigma$ including the term 
combining two Weyl spinors through fifth dimensional derivative, $\lambda_1{\cal D}_5\lambda_2$. It certainly tells that gaugino in this formalism forms a Dirac fermion.  
Note that the last term $\chi \chi+{\chi}^{\dag}{\chi}^{\dag}$ in eq.(\ref{213628_19Apr12}) vanishes under $\int d^4\theta$ integral, but this remains in the presence of a supersymmetry breaking term and is  necessary for the gauge invariance (Appendix~\ref{gaugeinv5D}). 

The conventional part of the Lagrangian is
\begin{eqnarray}
 \frac{1}{16kg^2}{\rm Tr}\left[ \int d^2\theta {\cal W^\alpha W_\alpha}  +{\rm h.c.} \right]
=
  -\frac{1}{4}F^{a\mu\nu}F^a_{\mu\nu} 
  {+}i \bar{\lambda}^a_1 \sigmabar^\mu {\cal D}_\mu \lambda_1^a
  +\frac{1}{2}D^aD^a,
  \label{020040_20Apr12}
\end{eqnarray}
where
\begin{eqnarray}
 {\cal D}_M \lambda_i^a &=&\partial_M \lambda_i^a +gf^{abc}A_M^b \lambda_i^c,
  \\
 F_{MN}^a &=&\partial_M A^a_N -\partial_N A^a_M +gf^{abc} A^b_M A^c_N.
\end{eqnarray}
And the rest part contains extra components of gauge field strength, all the kinetic terms involving $\Sigma$ and $\lambda_2$, and it shows particularly mixing between $\lambda_1$ and $\lambda_2$ through 5D derivative. 
The second and third lines of Eq.~(\ref{213628_19Apr12}) are decomposed by using  $V^3=0$ in the Wess-Zumino gauge, 
\begin{eqnarray}
&&\hspace{-30pt}\int d^4\theta \ \frac{1}{4kg^2}{\rm Tr} \Big[(\sqrt{2}\partial_5+2g{\chi}^{\dag})e^{-2gV} (-\sqrt{2}\partial_5+2g{\chi})e^{2gV}
 \nonumber\\ &&
\hspace{50pt} +\partial_5 e^{-2gV}\partial_5 e^{2gV}+2g^2(\chi \chi+{\chi}^{\dag}{\chi}^{\dag})
    \Big]
    \\\nonumber
 &&=\int d^4\theta \ \frac{1}{k}{\rm Tr}\Big[ 
  \left\{ (\delfive V)^2 +\chi{\chi}^{\dag} -\sqrt{2}(\chi+ {\chi}^{\dag})\delfive V 
	 +\frac{\chi \chi+{\chi}^{\dag}{\chi}^{\dag}}{2} \right \}
	 \nonumber\\
 &&\hspace{20pt}+g  \Big\{ \sqrt{2}(\chi- {\chi}^{\dag}) [{(\delfive V)V -V(\delfive V)}] 
  +2({\chi}^{\dag} \chi -\chi {\chi}^{\dag})V \Big\} 
  \nonumber\\
 &&\hspace{20pt}+g^2  \Big\{-4 {\chi}^{\dag}  V \chi V  +2({\chi}^{\dag} \chi +\chi {\chi}^{\dag}){V^2  } \Big\} \Big].
 \ \\\nonumber
  &&=
-\frac{1}{4}(2F^{a\mu5}F^a_{\mu5})
-i {\lambda}^a_2 \sigma^\mu {\cal D}_\mu \bar{\lambda}_2^a 
  - {\lambda}^a_2 {\cal D}_5 {\lambda}_1^a 
  +\bar{\lambda}^a_1 {\cal D}_5 \bar{\lambda}_2^a 
\nonumber\\
&&\hspace{20pt}
 {+}\frac{1}{2k} \mbox{Tr}  \Big[ {\cal D}_\mu \Sigma {\cal D}^\mu \Sigma \Big]
 +igf^{abc}(\lambda_2^a \Sigma^b \lambda^c_1 
 +\bar{\lambda}_1^a \Sigma^b \bar{\lambda}^c_2)
\nonumber\\
&&\hspace{20pt}
+\frac{1}{2k} \mbox{Tr}  \Big[2({\cal D}_5 \Sigma )D  \Big] +| F^a_\chi|^2 ,\label{020149_20Apr12} 
\end{eqnarray}  
where $ {\cal D}_M (\Sigma^aT^a) ={\cal D}_M \Sigma =\partial_M \Sigma -ig [A_M, \Sigma]$.  We combine Eq.(\ref{020040_20Apr12}) and Eq.(\ref{020149_20Apr12}), 
\begin{eqnarray}
 {\cal L}_5^{\rm YM} \!\!\!\!\!
&=& \!\!\!\!
-\frac{1}{4}F^{a MN}F^a_{MN}
  {+}i \big[  \ 
  \bar{\lambda}^a_1 \sigmabar^\mu {\cal D}_\mu \lambda_1^a 
  + {\lambda}^a_2 \sigma^\mu {\cal D}_\mu \bar{\lambda}_2^a 
  {+}i{\lambda}^a_2 {\cal D}_5 {\lambda}_1^a 
  {-}i\bar{\lambda}^a_1 {\cal D}_5 \bar{\lambda}_2^a \ \big]
\nonumber\\
&& {+}\frac{1}{2} {\cal D}_\mu \Sigma^a {\cal D}^\mu \Sigma^a 
 +\frac{g}{2}f^{abc} \Sigma^a(i\lambda_1^b  \lambda^c_2 -i\lambda_2^b  \lambda^c_1
 -i\bar{\lambda}_1^b \bar{\lambda}^c_2 +i\bar{\lambda}_2^b \bar{\lambda}^c_1)
\nonumber\\
&& +\frac{1}{2}D^aD^a
+({\cal D}_5 \Sigma^a )D^a   +| F^a_\chi|^2 
\\\nonumber\\
&=& -\frac{1}{4}F^{a MN}F^a_{MN}
 +\frac{1}{2} \Big[  
  i\bar{\lambda}^a_i \sigmabar^\mu {\cal D}_\mu \lambda_i^a 
  + {\lambda}^a_i \epsilon_{ij}{\cal D}_5 {\lambda}_j^a 
  \ +\mbox{h.c.} \
  \Big]
\nonumber\\
&& {+}\frac{1}{2} {\cal D}_M \Sigma^a {\cal D}^M \Sigma^a 
 +\frac{g}{2}f^{abc} \Sigma^a(i\lambda_i^b \epsilon_{ij} \lambda^c_j  \ +\mbox{h.c.})
+\frac{1}{2}(D'^a)^2 
  +| F^a_\chi|^2,
  \label{141250_23Apr12}  
\end{eqnarray}
where $D'^{a}\equiv D^a+{\cal D}_5 \Sigma^a $ 
and $\epsilon_{ij} = (i\sigma_2)_{ij}$. Terms of gauginos are written in  $SU(2)_R$ invariant way.  
If we use Dirac fermions, $\lambda \equiv  (\lambda_{1\alpha},  \bar{\lambda}_2^{\dot{\alpha}} )^T$ and
${\bar{\lambda}} \equiv  (\lambda_2^{\alpha},  \bar{\lambda}_{1\dot{\alpha}} )$, 
the Lagrangian becomes explicitly Lorentz invariant,
\begin{eqnarray}
{\cal L}_5^{\rm YM}
&=& -\frac{1}{4}F^{a MN}F^a_{MN}
 {+}i \bar{ \lambda}^a\Gamma^M {\cal D}_M {\lambda}^a
 {+}\frac{1}{2} {\cal D}_M \Sigma^a {\cal D}^M \Sigma^a 
 +igf^{abc} \Sigma^a {\bar{\lambda}}^b  \lambda^c
\nonumber\\&& 
 +\frac{1}{2}(D'^a)^{2} 
  +| F^a_\chi|^2 \ .
\end{eqnarray}
 For simultaineous $SU(2)_R$ invariance and Lorentz invariance, we must use a symplectic Majorana fermion that has four spinors, 
	\begin{eqnarray}
	{\boldsymbol \lambda}_{sm}\equiv 
		\left(
	 \begin{array}{c}
	  \lambda_{sm,1}  \\
	  \lambda_{sm,2} \\
	 \end{array}
	\right),  \ \  
		  \lambda_{sm,1} =	
	  \left( \begin{array}{c}
	  \lambda_{1} \\
	  \bar\lambda_{2} \\
	 \end{array}
	\right), \ 
\lambda_{sm,2}=
	  \left( \begin{array}{c}
	 - \lambda_{2} \\
	  \bar\lambda_{1} \\
	 \end{array}
	\right). 
\label{symajorana}
	\end{eqnarray}
 Two Dirac fermions $(\lambda_{sm,1}, \lambda_{sm,2})$ transform as $SU(2)_R$ doublet.
 The kinetic term of gauginos become both Lorentz and $SU(2)_R$ invariant, 
	\begin{eqnarray}
	 i \bar{ \lambda}^a\Gamma^M {\cal D}_M {\lambda}^a
	 =\frac{i}{2} \bar{\boldsymbol \lambda}^a_{sm}\Gamma^M {\cal D}_M {\boldsymbol \lambda}^a_{sm}.	 
	\end{eqnarray}

\subsection{Hypermultiplet}
Now we consider the matter sector. Matters are in Hypermultiplet which consists of two chiral supermultiplets in ${\cal N}=1, D=4$ language. 
The Lagrangian in superspace is given by
\begin{eqnarray}
 {\cal L}_5^{\rm Hyper} 
&=&
\int d^4\theta \left\{
{H_2} e^{2gV}{H_2^\dag} +{H_1^\dag} e^{-2gV}{H_1}
\right\}
\nonumber\\
&&+\int d^2\theta \ H_2 (\partial_5  -\sqrt{2} g \chi) H_1 \ +\hc \ ,
\label{matter-insuperspace}
\end{eqnarray}
where $H_{1,2}=H_{1,2}(x,y)$ and $V=V^a T_R^a, \ \chi=\chi^a T_R^a$. $H_1 (H_2)$ belongs to a representation $R \ (\bar{R})$ of the gauge group.
The chiral superfields are defined as (in the $y$-basis)
	\begin{eqnarray}
	 H_1=\phi_1+\sqrt{2}\theta\psi_1 +\theta^2F_1 \ , \label{defH1}
	\\
	 H_2=\phi_2+\sqrt{2}\theta\psi_2 +\theta^2F_2 \ . \label{defH2}
	\end{eqnarray}
The second line of Eq.~(\ref{matter-insuperspace}) leads to terms of fifth dimensional covariant derivative, such as $F_2{\cal D}_5\phi_1$ and $\psi_2{\cal D}_5\psi_1$, and $\Sigma$ interactions with $\psi_{1,2}$ and $\phi_{1,2}$.
When the Lagrangian is written down in component, many terms appear,
\begin{eqnarray}
 {\cal L}^{\rm Hyper}_5
  &=&\ 
  ({\cal D}^\mu \phi_1)^\dag {\cal D}_\mu \phi_1 
 {+} ({\cal D}^\mu \phi_2^*)^\dag {\cal D}_\mu \phi^*_2
   {+}i \bar{\psi}_1 \sigmabar^\mu {\cal D}_\mu \psi_1
   {+}i {\psi}_2 \sigma^\mu {\cal D}_\mu \bar\psi_2
  \nonumber\label{140825_23Apr12}\\ 
  &&  -i\sqrt{2}g (  {\phi}_1^\dag T^a {\psi}_1 {\hat{\lambda}}^a_1 
     -  \bar{\hat{\lambda}}^a_1 \bar{\psi}_1 T^a {\phi}_1  )
     +i\sqrt{2}g (  {\hat{\lambda}}^a_1 {\psi}_2 T^a    {\phi}^*_2 
     -  {\phi_2} T^a \bar{\psi}_2 \bar{\hat{\lambda}}^a_1 )
     \nonumber\\ 
    && -gD^a(\phi_1^* T^a \phi_1)+gD^a(\phi_2 T^a \phi_2^*)
   \nonumber\\  &&
  + |F_1|^2+|F_2|^2 
   \nonumber\\ 
   && + F_2 \big[ \delfive -g(\Sigma^a+iA^a_5)T^a \big]\phi_1
    +\phi_2 \big[ \delfive -g(\Sigma^a+iA^a_5)T^a \big]F_1  \ +\mbox{h.c.}  
    \nonumber\\
   &&+\phi_2 \big[ -\sqrt{2} g F_\chi^a T^a \big]\phi_1     \ +\mbox{h.c.}  
    \nonumber\\
    && -\psi_2[\delfive -g(\Sigma^a +iA_5^a)T^a  ] \psi_1    \ +\mbox{h.c.}  
    \nonumber\\
    && +\sqrt{2}g(  \psi_2 \lambda_2^a  T^a \phi_1   + \phi_2 T^a \lambda_2^a \psi_1)
       \ +\mbox{h.c.}  
\end{eqnarray}
where the covariant derivatives are 
   \begin{eqnarray}
       {\cal D}_M X = (\delfive -ig A^a_M T^a ) X   \ , 
	\ \ X=\phi_1 , \ \phi^*_2, \ \psi_1 \mbox{ and } \bar\psi_2  \ .
   \end{eqnarray}

It is hard to see especially because $\partial_\mu$ and $\partial_5$ are placed in different terms. We organize Lagrangian in an explicit 5D Lorentz invariant form by redefining   
 auxiliary fields of $F$ and $D^a$, 
\begin{eqnarray}
   F'_1 &\equiv& F_1 +  \big[ -{\delfive} -g(\Sigma^a-iA^a_5)T^a \big] \phi_2^*  \ , 
   \label{redefineF1}
    \\
   F'^*_2 &\equiv& F_2^* +  \big[ \delfive -g(\Sigma^a+iA^a_5)T^a \big]\phi_1 \ , \\
   D'^a &\equiv& D^a + {\cal D}_5 \Sigma^a \label{redefineD},
\end{eqnarray}
 and use   $\lambda_1 = i\hat\lambda_1$.	
The Lagrangian becomes,
\begin{eqnarray}
	  {\cal L}^{\rm Hyper}_5 &=&
	   |F'_1|^2+ |F'_2|^2
	    {+} ({\cal D}^M \phi_1)^\dag {\cal D}_M \phi_1 
	    {+} ({\cal D}^M \phi_2^*)^\dag {\cal D}_M \phi^*_2
	\nonumber\\&& 
	   - g^2(\Sigma^aT^a \phi_1)^\dag 
	     (\Sigma^aT^a \phi_1)
	       - g^2(\Sigma^aT^a \phi_2^*)^\dag 
	       (\Sigma^aT^a \phi_2^*)
	\nonumber\\ &&
	{+}i\left(\psi_2^\alpha, \bar\psi_{1\dot\alpha} \right)
	\Gamma^M{\cal D}_M 
	\left(
	 \begin{array}{c}
	  \psi_{1\alpha} \\
	  \bar\psi_2^{\dot\alpha} \\
	 \end{array}
	\right) \ 
	\nonumber\\ &&
	   +g \Sigma^a (\psi_2   T^a   \psi_1 +\bar\psi_1    T^a   \bar\psi_2    ) \ 
	\nonumber\\ &&
    +\sqrt{2}g
    (\phi_1^* , \phi_2)T^a 
    \left\{
        \psi_1 
	\left(
	 \begin{array}{c}
	-  \lambda^a_{1} \\
	  \lambda^a_2 \\
	 \end{array}
	\right)
	+
	\bar\psi_2 
	\left(
	 \begin{array}{c}
	  \bar\lambda^a_{2} \\
	  \bar\lambda^a_1 \\
	 \end{array}
	\right)
	       \right\}
    	   \ +\mbox{h.c.}  	   
     \nonumber\\ && 
     -gD'^a(\phi_1^* T^a \phi_1)+gD'^a(\phi_2 T^a \phi_2^*)
     \nonumber\\ &&
     +\phi_2 \big[ -\sqrt{2} g F_\chi^a T^a \big]\phi_1     \ +\mbox{h.c.}  
     \label{LagrangianHyper}
	\end{eqnarray}
Furthermore, we lead this to 
	\begin{eqnarray}	
	{\cal L}^{\rm Hyper}_5&=&
   	|F'_i|^2
    	{+} ({\cal D}^M \Phi_i)^\dag {\cal D}_M \Phi_i
	- g^2(\Sigma^aT^a {\boldsymbol \Phi})^\dag 
     (\Sigma^aT^a {\boldsymbol \Phi})
	\nonumber\\ &&
	    {+}i\overline{\Psi}
	\Gamma^M {\cal D}_M         \Psi
	   +g \Sigma^a (\overline{\Psi}   T^a   \Psi)
	\nonumber\\ &&
    -\sqrt{2}g {\Phi}_i^\dag \epsilon_{ij}T^a \Psi{{\bar\lambda}}_{sm,j}^a  	   \ +\mbox{h.c.}  	   
     \nonumber\\ && 
     -gD'^a(\Phi_1^* T^a \Phi_1)+gD'^a(\Phi_2^* T^a \Phi_2)
     \nonumber\\ &&
     +\Phi^*_2 \big[ -\sqrt{2} g F_\chi^a T^a \big]\Phi_1     \ +\mbox{h.c.}  \ ,
     \label{L5hyper}
\end{eqnarray}
where 
	\begin{eqnarray}
	&&{\boldsymbol \Phi}=
		\left( \begin{array}{c}
      	\Phi_1  \\ \Phi_2 
    	\end{array} \right)
	\equiv 
	\left( \begin{array}{c}
      	\phi_1 \\ \phi_2^* 
    	\end{array} \right),
	\quad
	\Psi \equiv 
	\left( \begin{array}{c}
      	\psi_1  \\ \bar\psi_2 
    	\end{array} \right).
	\end{eqnarray}
Here, $i,j$ are indices of $SU(2)_R$ doublet, and $\lambda_{sum,i}$ is given in Eq.(\ref{symajorana}). 
$SU(2)_R$ symmetry becomes clear 
except for terms involving gauge auxiliary fields $D'$ and $F_\chi$ shown in the last two lines of Eq.~(\ref{L5hyper}). 
However, once these are combined with ${\cal L}_5^{\rm YM}$, $SU(2)_R$ is actually realized. For detail see Appendix~\ref{explicitSU2}. 
\section{Supersymmetry Breaking by Boundary Conditions}
\subsection{Scherk-Schwarz mechanism}
We consider 
a single compact extra dimension with the coordinate $y$ identified 
under ${\cal T}: y \rightarrow y + 2\pi R$ and ${\cal P}: y \rightarrow -y$. Let $\phi$ be a column vector representing all fields of the theory. We require the 5D action is invariant under both operations, which is called orbifold compactification. Invariance of ${\cal P}$ is needed so that the low-energy theory has chiral fermions as in the SM and the MSSM. 
The action of these transformations on the fields can be written as
	\begin{eqnarray}
	{\cal T}:\phi(y)&\to&T^{-1}\phi(y+2\pi R)=\phi(y)\ ,\\
	{\cal P}:\phi(y)&\to&P\phi(-y)=\phi(y) \ .
	\end{eqnarray}
For the simultaneous imposition of $\cal T$ and $\cal P$, there is a consistency condition that the space-time motion induced by $\cal PT$ is identical to that induced by $\cal T^{\rm -1}P$.
Then the  two operations satisfy the algebra 
	\begin{eqnarray}
	P T P &=& T^{-1} \ ,
	\\ P^2&=&1 \ ,
	\end{eqnarray}
and 
the resulting extra dimension is an interval $y \in [0, \pi R]$:\ the 
$S^1/{\mathbb Z}_2$ orbifold.\footnote{For practical calculations, we use a larger interval $y \in [0, 2\pi R]$ for convenience. }
For a simple example, let us consider 5D quark where $T$ and $P$ commute. We obtain $T=\pm1$ and $P=\pm\gamma_5$, and ignore $T=-1$ solutions which do not have massless modes.  The Kaluza-Klein (KK) expansion for quark with $T=1$ and $P=\gamma_5$ is given by 
	\begin{eqnarray}
	\Psi(x,y)=\frac{P_R}{\sqrt{2\pi R}}\Psi_0(x) 
	+\sum_{n=1}^{\infty}\frac{P_R}{\sqrt{\pi R}}\Psi_n(x) \cos\frac{ny}{R}+\sum_{n=1}^{\infty}\frac{P_L}{\sqrt{\pi R}}\Psi_n(x) \sin\frac{ny}{R}
	\end{eqnarray}
where $P_R=(1+\gamma_5)/2$ and $P_L=(1-\gamma_5)/2$.  A massless chiral fermion is certainly obtained as zero mode while non-zero modes have both left- and right-handed components which form a Dirac mass of $n/R$. 

In presence of supersymmetry, $T$ and $P$ are not necessarily commute but we can think of other boundary conditions using $SU(2)_R$ space.  The boundary 
conditions for  two dimensional representation such as $SU(2)_R$ doublets 
can be
\begin{eqnarray}
  &&P = \left(
    \begin{array}{cc}
      1 & 0 \\ 0 & -1
    \end{array} \right) \ ( \otimes \gamma_5 \ \mbox{ for fermion}),
\no\\
  &&T = \left(
    \begin{array}{cc}
      \cos (2\pi\alpha) & \sin(2\pi\alpha) \\ 
      -\sin(2\pi\alpha) & \cos(2\pi\alpha)
    \end{array} \right)=e^{i\sigma_2 (2\pi\alpha)  },
\label{eq:b-c}
\end{eqnarray}
under ${\cal P}$ and $\cal T$ (see Ref.~\cite{Barbieri:2001yz} for details). The twist parameter of $\alpha$ is real. 
For one dimensional representation such as  $SU(2)_R$ singlets, the condition is $T^2=1$ because $P$ and $T$ commute. 
 In this thesis we consider the case $\alpha \ll 1$.
 As explained later, the twist parameter $\alpha$ in the boundary conditions 
 is equivalent to an $F$-term vacuum expectation value of the radion 
 superfield~\cite{Marti:2001iw}, which can be generated dynamically 
 through a radion stabilization mechanism and hence can be naturally 
 small. The non-zero $\alpha$ leads to mass-splitting of KK modes between $SU(2)_R$ singlets and doublets, and therefore supersymmetry is broken. This supersymmetry breaking through the non-trivial boundary condition is generally called Scherk-Schwarz mechanism \cite{Scherk:1978ta}. 
In other words, for the orbifold boundary condition $SU(2)_R$ is broken down to $U(1)_R$ and thus ${\cal N}=1$ supersymmetry remains. Furthermore the remaining supersymmetry is broken by Scherk-Schwarz mechanism with non-zero $\alpha$. 

The boundary conditions of Eq.~(\ref{eq:b-c}) leave only the MSSM gauge 
and matter fields below the compactification scale $1/R$.  Specifically, 
the matter supermultiplets yield three generations of quarks and leptons 
as the zero modes, while their super partners obtain the common soft mass 
of $\alpha/R$.  (Here, we have assumed that there are no 5D bulk mass 
terms for the matter multiplets.%
\footnote{This assumption can be justified by a local parity in the bulk; 
 see, e.g.,~\cite{Barbieri:2002sw}.})
The gauge supermultiplets give massless standard model gauge fields and 
gauginos of mass $\alpha/R$.  We therefore obtain the first two expressions 
in Eq.~(\ref{eq:tree}).  (The Kaluza--Klein excitations form ${\cal N}=2$ 
supermultiplets and have masses $\approx n/R$ ($n = 1,2,\cdots$), with 
supersymmetry-breaking mass splitting of order $\alpha/R$.)
We see these results explicitly in Sec.~\ref{sec:softmass}.

\subsection{Soft masses of scalar and gaugino}\label{sec:softmass}
Now let us investigate how the Scherk-Schwarz mechanism works. 
It can be seen that the supersymmetry breaking effect, particularly soft term, appears through extra dimensional derivative $\delfive$.
The matter scalars, $\phi_1$ and $\phi_2^*$, given in Eqs.(\ref{defH1},\ref{defH2}) form a $SU(2)_R$ doublet and transform under $\cal P$ and  $\cal T$,
	\begin{eqnarray}
	\left( \begin{array}{c}
      	\phi_1(-y)  \\ \phi_2^* (-y)
    	\end{array} \right)&=&
	\left( \begin{array}{c}
      	\phi_1(y)  \\ -\phi_2^* (y)
    	\end{array} \right), \label{boundary1}
\\\no\\
	\left( \begin{array}{c}
      	\phi_1(y+2\pi R)  \\ \phi_2^* (y+2\pi R)
    	\end{array} \right)&=&e^{i\sigma_2 (2\pi \alpha)}
	\left( \begin{array}{c}
      	\phi_1(y)  \\ \phi_2^* (y)
    	\end{array} \right).\label{boundary2}
	\end{eqnarray}
These conditions are satisfied by extracting an exponential factor of twist,
 	\begin{eqnarray}
	\left( \begin{array}{c}
      	\phi_1(y)  \\ \phi_2^* (y)
    	\end{array} \right)
	\to
	e^{i\sigma_2 \alpha y/R}
	\left( \begin{array}{c}
      	\phi_1(y)  \\ \phi_2^* (y)
    	\end{array} \right)\ . \label{eq:redef}
	\end{eqnarray}
Here the scalars $\phi_{1,2}$ on the right-hand side have simple boundary conditions, 
	\begin{eqnarray}
	\begin{array}{c c}
	\phi_1(y+2\pi R)=\phi_1(y), & \phi_1(-y)=\phi_1(y),
	\\ 
	\phi_2(y+2\pi R)=\phi_2(y), & \ \ \phi_2(-y)=-\phi_2(y),
    	\end{array} 
	\label{simplecondition}
	\end{eqnarray}
and their KK expansions are given by, 
	\begin{eqnarray}
	\phi_1=\frac{\phi_{1,0}}{\sqrt{2\pi R}}
	+\sum_{n=1}^\infty \frac{ \phi_{1,n}}{\sqrt{\pi R}}\cos\frac{ny}{R} \ , \quad
	\phi_2=
	\sum_{n=1}^\infty \frac{ \phi_{2,n}}{\sqrt{\pi R}}\sin\frac{ny}{R}\ .
	\end{eqnarray}
When the Lagrangian is written in terms of fields with simple boundary conditions of Eq.~(\ref{simplecondition}), the zero mode has wave function in the extra dimension for the exponential factor with non-zero $\alpha$. Having wave function in the extra dimension means that the mode has momentum in the extra dimensional direction which is seen as ``mass'' in the 4D picture. On the other hand, $SU(2)_R$ singlets cannot have such a wave function, and the zero mode of $SU(2)_R$ singlet is kept massless. 
One may worry that breaking supersymmetry with non-trivial boundary conditions is hard breaking, but the effect of breaking  appears only through derivative of fifth dimension $\delfive$, and thus this is always soft breaking.  

The derivative $\delfive$ interacts with the exponential factor, and extra terms (soft terms) are generated,
	\begin{eqnarray}
	{\cal L}_{5}^{\rm Hyper}
	&\supset& -({\cal D}_5 \phi_1)^\dag({\cal D}_5 \phi_1)
	-({\cal D}_5 \phi_2^*)^\dag({\cal D}_5 \phi_2^*)
	=-({\cal D}_5 \Phi_i)^\dag({\cal D}_5 \Phi_i)
	\\
	&\to& -\left({\cal D}_5 (e^{i\sigma_2 \alpha y/R})_{ij}\Phi_j\right)^\dag
	\left({\cal D}_5 (e^{i\sigma_2 \alpha y/R})_{ij}\Phi_j\right)
	\no\\
	&=&-\left(  (e^{i\sigma_2 \alpha y/R})_{ij}{\cal D}_5\Phi_j
	+(e^{i\sigma_2 \alpha y/R} i\sigma_2)_{ij}   \frac{\alpha}{R}\Phi_j
	\right)^\dag
	\no\\
	&&\quad \times
	\left(  (e^{i\sigma_2 \alpha y/R})_{il}{\cal D}_5\Phi_l
	+(e^{i\sigma_2 \alpha y/R} i\sigma_2)_{ik}   \frac{\alpha}{R}\Phi_k
	\right)
	\no\\
	&=&
	-({\cal D}_5 \Phi_i)^\dag({\cal D}_5 \Phi_i)-\left(\frac{\alpha}{R}\right)^2 \Phi_i^\dag\Phi_i
	\no\\&&\quad
	-({\cal D}_5 \Phi_i)^\dag  \frac{\alpha}{R} (i\sigma_2\Phi)_i
	+\Phi_i^\dag  \frac{\alpha}{R} (i\sigma_2{\cal D}_5\Phi)_i
	\no\\
	&=&-({\cal D}_5 \phi_1)^\dag({\cal D}_5 \phi_1)
	-({\cal D}_5 \phi_2^*)^\dag({\cal D}_5 \phi_2^*)
	-\left(\frac{\alpha}{R}\right)^2|\phi_1|^2-\left(\frac{\alpha}{R}\right)^2|\phi_2^*|^2
	\no\\
	&&+\frac{\alpha}{R}\left(
	\phi_1^\dag{\cal D}_5 \phi_2^*
	-(\phi_2^*)^\dag{\cal D}_5 \phi_1
	-({\cal D}_5\phi_1)^\dag \phi_2^*
	+({\cal D}_5\phi_2^*)^\dag \phi_1
	\right)
	\end{eqnarray}
where ``$\to$'' means extracting the exponential factor of Eq.~(\ref{boundary2}).
These terms form a mass matrix of scalars after the KK expansion, 
	\begin{eqnarray}
	&&\hspace{-20pt}\int \!\! dy \
	(\phi_1^*(y), \phi_2(y))
	 \left(
	 \begin{array}{c c }
     	\delfive^2 -(\alpha/R)^2   &2(\alpha/R)\delfive \\
     	-2(\alpha/R)\delfive  &\delfive^2 -(\alpha/R)^2  \\
    	\end{array} \ 
	\right)
	\left( \begin{array}{c } 
	 \phi_1(y) \\  \phi_2^* (y)
	 \end{array}\right)
	 \\
	&&\hspace{-20pt}=-\frac{\alpha^2}{R^2} |\phi_{1,0}|^2
	-\frac{1}{R^2} \sum_{n=1}^{\infty}
	(\phi_{1,n}^*, \phi_{2,n})	 
	 \left(
	 \begin{array}{c c }
     	{n}^2 +\alpha^2   &-2\alpha{n} \\
     	-2\alpha{n}  &{n}^2 +\alpha^2  \\
    	\end{array} \ 
	\right)
	\left( \begin{array}{c } 
	 \phi_{1,n} \\  \phi_{2,n}^* 
	 \end{array}\right). 
	\end{eqnarray}
The zero mode has soft mass of $\alpha/R$, and the non-zero mode have mass of $(\alpha\pm n)/R$ ($n$=1,2,$\cdots$). Using the mass eigenstates, the KK expansion is written in a more convenient way, 
	\begin{eqnarray}
	\phi_1(y)
	&=&\sum_{n=-\infty}^\infty \frac{\phi_{n}}{\sqrt{2\pi R}}\cos\frac{ny}{R},
	\\
	\phi_2^*(y)&=&-\sum_{n=-\infty}^\infty \frac{\phi_{n}}{\sqrt{2\pi R}}\sin\frac{ny}{R}.
	\end{eqnarray}
where the mass of $\phi_n$ is $(\alpha+{n})/R$. The detail is given in Appendix.~\ref{app:KKsquark}.

Conditions for gauginos are similar to those for the scalars. We simply replace $\phi_1$, $\phi_2^*$ with $\lambda_1$, $-\lambda_2$, respectively. Note that under $\cal P$ supermultiplets for gauge field transform as, 
	\begin{eqnarray}
	V(x,-y)=V(x,y), \quad \chi(x,-y)=-\chi(x,y),
	\end{eqnarray}
because $V$ contains $A_\mu$ while $\chi$ contains $A_5$.
 The gaugino $SU(2)_R$ doublet must  satisfy the boundary conditions for the matter scalars of Eqs.~(\ref{boundary1}, \ref{boundary2}) since $SU(2)_R$ is a common global symmetry. Thus the gaugino multiplet has also the exponential factor of twist, 
 	\begin{eqnarray}
	\left( \begin{array}{c}
      	\lambda_1(y)  \\ -\lambda_2 (y)
	\end{array} \right)	
	\to
	e^{i\sigma_2 \alpha y/R}
	\left( \begin{array}{c}
      	\lambda_1(y)  \\ -\lambda_2 (y)
    	\end{array} \right)
	\end{eqnarray}
Fields on the right-hand side are similar to those in Eq.(\ref{simplecondition}).  As in the case of scalars, there will be extra terms through $\delfive$ and we expect the same mass spectrum is generated. The kinetic term of gauginos is
	\begin{eqnarray}
	{\cal L}_5^{\rm YM} &\supset& 
	\frac{1}{2} \lambda_i^a (i\sigma_2)_{ij}{\cal D}_5 \lambda_j 
	\no\\
	&\to&
	\frac{1}{2} (e^{-i\sigma_2\alpha y/R})_{ij}\lambda^a_j (i\sigma_2)_{ik}{\cal D}_5
	(e^{-i\sigma_2\alpha y/R}\lambda^a )_k
	\no\\
	&=&
	\frac{1}{2} \lambda^a_i (e^{i\sigma_2\alpha y/R} i\sigma_2)_{ij}
	 \left(
	(e^{-i\sigma_2\alpha y/R})_{jk}{\cal D}_5\lambda^a_k
	+(-e^{-i\sigma_2\alpha y/R} i\sigma_2)_{jk}
	\frac{\alpha}{R}\lambda^a_k
	\right)
	\no\\
	&=&	
	\frac{1}{2} \lambda_i^a (i\sigma_2)_{ij}{\cal D}_5 \lambda_j 
	+\frac{\alpha}{2R}\lambda_i^a\lambda_i^a
	\end{eqnarray}
These terms form a mass matrix  after the KK expansion as follows,
	\begin{eqnarray}
	&&\int_0^{2\pi R}\!\!\!\!\!dy\ \frac{1}{2}
	\left(\lambda_1^a(y), \lambda_2^a(y)	\right)
	\left( \begin{array}{c c}
      	\alpha/R & -\delfive
    	\\ 
	\delfive & \alpha/R  
	\end{array} \right)	
	\left( \begin{array}{c}
      	\lambda_1^a(y)  \\ \lambda_2^a (y)
    	\end{array} \right)	
	\no\\&&=
	\frac{\alpha}{2R}\lambda_{1,0}^{a}\lambda_{1,0}^{a}
	+\sum_{n=1}^\infty 
	\left(\lambda_{1,n}^a, \lambda_{2,n}^a	\right)
	\left( \begin{array}{c c}
      	\alpha/R & n/R
    	\\ 
	n/R & \alpha/R  
	\end{array} \right)	
	\left( \begin{array}{c}
      	\lambda_{1,n}^a  \\ \lambda_{2,n}^a 
    	\end{array} \right)	\ ,
	\end{eqnarray}
In fact, these gauginos have the same mass spectrum of the scalars such that $m_{\rm gaugino}=\alpha/R, (\alpha\pm 1)/R, (\alpha\pm2)/R, \cdots$. Similarly it is understood that KK modes of gravitinos, another $SU(2)_R$ doublet,  should have the same mass spectrum. 

Let us summarize the generated soft breaking terms in the bulk,
	\begin{eqnarray}
	{\cal L}_{5,\rm soft}&=&
		-\left(\frac{\alpha}{R}\right)^2|\phi_1|^2-\left(\frac{\alpha}{R}\right)^2|\phi_2^*|^2
	\no\\
	&&+\frac{\alpha}{R}\left(
	\phi_1^\dag{\cal D}_5 \phi_2^*
	-(\phi_2^*)^\dag{\cal D}_5 \phi_1
	-({\cal D}_5\phi_1)^\dag \phi_2^*
	+({\cal D}_5\phi_2^*)^\dag \phi_1
	\right)
	\no\\&&
	+\frac{1}{2}\frac{\alpha}{R}\lambda_1^a\lambda_1^a +\frac{1}{2}\frac{\alpha}{R}\lambda_2^a\lambda_2^a +\hc \  \label{SSSBsoft}
	\end{eqnarray}
\subsection{Soft terms on brane}\label{subset:brane}
For a theory with only bulk fields, soft breaking terms  are only terms of Eq.~(\ref{SSSBsoft}), but once we consider Yukawa interactions on branes which only respects ${\cal N}=1,\ D=4$ supersymmetry, additional soft breaking terms appear. 
 Suppose a bulk filed $H_1(x,y)$ interacts with chiral superfields $S_{1,2}(x)$ on $y=0$ brane,
	\begin{eqnarray}
	W_{\rm brane1}=\delta(y)\lambda_1 S_1(x)S_2(x)H_1(x,y). 
	\end{eqnarray}
where $S_{1,2}=\phi_{S_{1,2}}+\psi_{S_{1,2}}\theta+F_{S_{1,2}}\theta^2.$
If some field of brane interactions lives in the bulk and hence receives SUSY breaking by the Scherk-Schwarz mechanism, soft term terms corresponding to superpotential appear through $\delfive$. It can be seen after  $\theta$ integral and field redefinition of $F$-term of bulk field, $H_1$, as in Eq.(\ref{redefineF1}),  
	\begin{eqnarray}
	\int \!d^2\theta\ W_{\rm brane1}&\supset& \delta(y)\lambda_1 \phi_{S_1}(x)\phi_{S_2}(x)F_1(x,y)
	\no\\&=&
	\delta(y)\lambda_1 \phi_{S_1}(x)\phi_{S_2}(x)\left(F'_1(x,y)+\big[ {\delfive} +g(\Sigma^a(x,y)-iA^a_5(x,y))T^a \big] \phi_2^*(x,y) \right)
	\no\\ \label{eq:brane0}
	\end{eqnarray}
Several terms vanishes because $\cal P$ odd field does not have wave function at $y=0$ such as $A_5,\Sigma\propto \sin(ny/R)$.  We rewrite Eq.~(\ref{eq:brane0}) by extracting $\alpha$ dependent factor from $\phi_{1,2}$ as Eq.~(\ref{eq:redef}),
	\begin{eqnarray}
	&&\!\!\!\!\!\!\!\!\!\!\!\! \delta(y)\lambda_1 \phi_{S_1}(x)\phi_{S_2}(x)\Big(F'_1(x,y)\no\\
	 &&\quad\quad
	 +\big[ {\delfive} +g(\Sigma^a(x,y)-iA^a_5(x,y))T^a \big]
	\left[ \cos\frac{\alpha y}{R}\phi_2^*(x,y)  - \sin\frac{\alpha y}{R}\phi_1(x,y)\right]\Big)
	\\
	=&&\delta(y)\lambda_1 \phi_{S_1}(x)\phi_{S_2}(x)\Big(F'_1(x,y)-\frac{\alpha}{R}\sin\frac{\alpha y}{R}\phi_2^*(x,y)   +\cos\frac{\alpha y}{R}\delfive\phi_2^*(x,y) 
	\no\\ &&\hspace{100pt}
	-\frac{\alpha}{R}\cos\frac{\alpha y}{R}\phi_1(x,y)    - \sin\frac{\alpha y}{R}\delfive\phi_1(x,y)\Big)
	\\
	=&&\delta(y)\lambda_1 \phi_{S_1}(x)\phi_{S_2}(x)\left(F'_1(x,y)
	+\delfive\phi_2^*(x,y)-\frac{\alpha}{R}\phi_1(x,y)	\right)
	\end{eqnarray}
The last term is soft SUSY breaking $A$-term. The size of soft terms on the brane is  uniquely determined by configuration of interaction. Let us consider other types of interaction, 
	\begin{eqnarray}
	W_{\rm brane2}=\delta(y)\frac{\lambda_2}{2} S_1(x)H_1^2(x,y), \quad 
	W_{\rm brane3}=\delta(y)\frac{\lambda_3}{3} H_1^3(x,y).
	\end{eqnarray}
As the previous procedure, we perform $\theta$ integral and redefine $F$-term of  $H_1$, and extract $\alpha$ dependent factor, 
	\begin{eqnarray}
	\int \!d^2\theta\ W_{\rm brane2}&\supset& \delta(y)\frac{\lambda_2}{2} \phi_{S_1}(x)\left(2\phi_{1}(x,y)F_1(x,y)\right)
	\no\\&=&
	\delta(y)\frac{\lambda_2}{2}2 \phi_{S_1}(x)\phi_{1}(x,y)\left(F'_1(x,y)+ {\delfive}  \phi_2^*(x,y) \right)
	\\&\to&
	\delta(y)\frac{\lambda_2}{2}2 \phi_{S_1}(x)\left[ \cos\frac{\alpha y}{R}\phi_1(x,y)  +\sin\frac{\alpha y}{R}\phi_2^*(x,y)\right]
	\no\\
	&&\quad\quad
	\times\left(F'_1(x,y)+ {\delfive} \left[ \cos\frac{\alpha y}{R}\phi_2^*(x,y)  -\sin\frac{\alpha y}{R}\phi_1(x,y)\right] \right)
	\\&=&
	\delta(y)\frac{\lambda_2}{2}2 \phi_{S_1}(x)\phi_1(x,y)
	\left(F'_1(x,y)+\delfive\phi_2^*(x,y)-\frac{\alpha}{R}\phi_1(x,y)	\right)
	 \label{eq:brane2}
	\end{eqnarray}
	\begin{eqnarray}
	\int \!d^2\theta\ W_{\rm brane3}&\supset& \delta(y)\frac{\lambda_3}{3} \left(3\phi_{1}^2(x,y)F_1(x,y)\right)
	\no\\&=&
	\delta(y)\frac{\lambda_3}{3}3 \phi_{1}^2(x,y)\left(F'_1(x,y)+ {\delfive}  \phi_2^*(x,y) \right)
	\\&\to&
	\delta(y)\frac{\lambda_3}{3}3 \left[ \cos\frac{\alpha y}{R}\phi_1(x,y)  +\sin\frac{\alpha y}{R}\phi_2^*(x,y)\right]^2
	\no\\
	&&\quad\quad
	\times\left(F'_1(x,y)+ {\delfive} \left[ \cos\frac{\alpha y}{R}\phi_2^*(x,y)  -\sin\frac{\alpha y}{R}\phi_1(x,y)\right] \right)
	\\&=&
	\delta(y)\frac{\lambda_3}{3}3 \phi_1^2(x,y)
	\left(F'_1(x,y)+\delfive\phi_2^*(x,y)-\frac{\alpha}{R}\phi_1(x,y)	\right)
	 \label{eq:brane3}
	\end{eqnarray}
Increasing number of bulk field on the brane Yukawa interactions, the size of $A$-term becomes larger. For a superpotential on brane 
	\begin{eqnarray}
	W_{\rm brane}&=&\delta(y)\left\{\lambda_1 S_1(x)S_2(x)H_1(x,y)	
	+\frac{\lambda_2}{2} S_1(x)H^2_1(x,y)	
	+\frac{\lambda_3}{3} H^3_1(x,y)	\right\},
	\end{eqnarray}
soft terms are generated through the Scherk-Schwarz mechanism as follows,  
	\begin{eqnarray}
	{\cal L}_{\rm brane,soft}  =\delta(y)\left\{\lambda_1 \frac{-\alpha}{R}  \phi_{S1}\phi_{S2}\phi_1
	+\frac{\lambda_2}{2} \frac{-2\alpha}{R} \phi_{S1}\phi_1^2 
	+\frac{\lambda_3}{3} \frac{-3\alpha}{R} \phi_1^3 \right\} .
	\label{eq:brsoft1}
	\end{eqnarray}

In the realistic situation, we will consider brane interactions which have bilinear of bulk fields. For example, $Q$ and $U$ are bulk chiral supefields that are $\cal P$ even part of hypermultiplets while $H_u$ is brane-localized chiral superfield, and their Yukawa interaction is localized on the brane, 
	\begin{eqnarray}
	W_{\rm brane}^{ y_U}=\delta(y)\ y_{U5}H_u(x)Q(x,y)U(x,y)\ .
	\end{eqnarray}
As before $F$-terms of $Q$ and $U$ are redefined and squarks have same $\alpha$ dependent wave function for the Scherk-Schwarz mechanism. Therefore, the soft term is generated from the Yukawa interaction, 
	\begin{eqnarray}
	{\cal L}_{{\rm brane, soft}}^{ y_U}=\delta(y)\ y_{U5}\left(\frac{-2\alpha}{R}\right)\phi_{H_u}(x)\phi_{Q}(x,y)\phi_{U}(x,y)\ .
	\end{eqnarray}
In this configuration, the large trilinear term, $A_t={-2\alpha}/{R}$, is naturally realized and it is close to  ``maximal mixing'' $(|A_t|=\sqrt{6}{\alpha}/{R})$.  The lightest Higgs mass is boosted by this correction without very heavy sparticles. 


\section{Radion Mediation }\label{sec:RM}
\subsection{Lagrangian}

In this section, we discuss the Radion mediation in the superfield formalism, and see that it actually  corresponds to the Scherk-Schwarz mechanism \cite{Marti:2001iw}. We see the dependence of the Radion at first,  and then na\"{i}vely construct the Lagrangian by promoting the radion field to a superfield. 

 In order to  extract an explicit radius dependence of Eqs.~(\ref{213628_19Apr12},\ref{matter-insuperspace}), the coordinate of the fifth dimension is parametrized by $y= R \varphi$  where $R$ is radius and $0\leq \varphi <2\pi$,  and we count powers of $R$. 
 The adjoint chilal field is redefined as $\chi \to \chi/R$ for convenience. The 4D Lagrangian is given by, 
	\begin{eqnarray}
	{\cal L}_4
	=\!\!    \int d\varphi \!\!\! &\Bigg\{&
          \frac{1}{16kg^2}{\rm Tr}\left[ \int d^2\theta \ R{\cal W^\alpha W_\alpha} \ 
				   +{\rm h.c.} \right]
	  \nonumber \\
	  &&+\frac{1}{4kg^2}   \int d^4\theta \
	   \frac{1}{R} {\rm Tr} 
	   \Big[ (\sqrt{2}\partial_\varphi+2g{\chi}^{\dag})e^{-2gV} 
	  (-\sqrt{2}\partial_\varphi+2g{\chi})
	  e^{2gV}
	  \nonumber \\    &&\hspace{3cm}
	  +\partial_\varphi e^{-2gV}\partial_\varphi e^{2gV}
	    +2g^2(\chi \chi+{\chi}^{\dag}{\chi}^{\dag})
	    \Big]
          \nonumber\\&&
	  +\int d^4\theta \ R\left(
			  {H_2} e^{2gV}{H_2^\dag} +{H_1^\dag} e^{-2gV}{H_1}
	\right)
	  \nonumber\\&&
	  +\int d^2\theta \  H_2 (\partial_\varphi  -\sqrt{2} g \chi) H_1 \ +\mbox{h.c.} 
	  \Bigg\}  .
	\end{eqnarray}
We incorporate $R$ into a chiral superfield,
	\begin{eqnarray}
 	T=R+iB_5 + \theta \Psi_R^5 +\theta^2 {F}_T \ . 
	\end{eqnarray}
The mass dimension of $T$ in this normalization is $-1$, and hence $F_T$ is dimensionless. 
Using this radion chiral superfield, we promote the radius $R$ to the superfield $T$ in such a way that 
	\begin{eqnarray}
 	\int d^2\theta  R^n &\to& \int d^2\theta \ T^n  \ , 
 	\nonumber\\
 	\int d^4\theta R^n &\to& \int d^4\theta \ \left(\frac{T+T^\dag}{2} \right)^n  \ . 
	\end{eqnarray}
Then the Lagrangian  becomes
\begin{eqnarray}
{\cal L}_4\!\!\!&=&\!\!\!\!\!\! \int  d\varphi \ \ \Bigg\{
          \frac{1}{16kg^2}{\rm Tr}\left[ \int d^2\theta \ T \ {\cal W^\alpha W_\alpha} \ 
				   +{\rm h.c.} \right]
	  \nonumber \\
	  &&+\frac{1}{4kg^2}  \int d^4\theta \
	   \frac{2}{T+{T^\dag} }  {\rm Tr} 
	   \Big[ (\sqrt{2}\partial_\varphi+2g{\chi}^{\dag})e^{-2gV} 
	  (-\sqrt{2}\partial_\varphi+2g{\chi})
	  e^{2gV}
	  \nonumber \\ 
           &&\hspace{100pt} +\partial_\varphi e^{-2gV}\partial_\varphi e^{2gV}
	    +2g^2(\chi \chi+{\chi}^{\dag}{\chi}^{\dag})
	    \Big]
          \nonumber\\&&
	  +\int d^4\theta \frac{T+{T^\dag} }{2} \left(
			  {H_2} e^{2gV}{H_2^\dag} +{H_1^\dag} e^{-2gV}{H_1}
	\right)
	  \nonumber\\&&
	  +\int d^2\theta \  H_2 (\partial_\varphi  -\sqrt{2} g \chi) H_1 \ +\hc  \Bigg\} \  .
\end{eqnarray}

\subsection{Radion mediation}
Now let us assume the Radion has a VEV of its $F$-term breaking supersymmetry when it stabilizes the radius,  such that $\langle T\rangle =R+\theta^2 {F}_T$. This is called Radion mediation, and we will see it is equivalent to the previous supersymmetry breaking by the Scherk-Schwarz mechanism. To be more explicit, we will show $F_T/2=\alpha$. 
Again we rescale the adjoint chiral superfiled as $\chi \to \chi R$, and then terms relevant to the supersymmetry breaking from the radion superfield are given by 
\begin{eqnarray}
	\!\!\!\!\! {\cal L}_4 \supset  \int  Rd\varphi \!\!\!\!\!&\Bigg\{&
	\frac{1}{16kg^2}{\rm Tr}\left[ \int d^2\theta \ \frac{T}{R} 
	\ {\cal W^\alpha W_\alpha} \ 	  +{\rm h.c.} \right]
       	\nonumber \\ &&
      	+ \frac{1}{k} \int d^4\theta
	\frac{2R}{T+{T^\dag}}  {\rm Tr} 
	\Big[
	 \chi{\chi}^{\dag} 
	 +\frac{\chi \chi+{\chi}^{\dag}{\chi}^{\dag}}{2}  \Big]
  	\no\\&&
          +\int d^4\theta 
	  \ \frac{T+T^\dag }{2R}
	  \Big\{  {H_2} {H_2^\dag} +{H_1^\dag} {H_1}
	  \Big\}
	  \no\\&&
	  +\int d^2\theta \ H_2 (R^{-1}\partial_\varphi  -\sqrt{2} g \chi) H_1 \ +\hc
	  \Bigg\} .  \label{RadionMed1}
\end{eqnarray}
One gaugino mass is generated from the first term, 
     \begin{eqnarray}
      &&\!\!\!\!\!\!\!\!\!\!\!\!
      \frac{1}{16kg^2}{\rm Tr}\left[ \int d^2\theta \ (1+2f\theta^2)
			       \ {\cal W^\alpha W_\alpha} \ 	  +{\rm h.c.} \right]
      \\\nonumber
     && \to
      \frac{1}{16kg^2}{\rm Tr}\left[ \int d^2\theta \
			       \ {\cal W^\alpha W_\alpha} \ 	  +{\rm h.c.} \right]      
      +\frac{f}{2}\lambda_1^a\lambda_1^a  
       +\frac{f^*}{2}\bar\lambda_1^a \bar\lambda_1^a  \ .
  \end{eqnarray}
  where $f\equiv F_T/(2R)$.
For $T$ in the denominator, we can expand in $\theta$ and $\bar\theta$,
\begin{eqnarray}
  	\frac{2R}{T+T^\dag}&=&1 
  	-\left(\frac{\theta^2 F_T+\bar{\theta}^2 F_T^* }{2R} \right)
  	+\left(\frac{\theta^2 F_T+\bar{\theta}^2 F_T^* }{2R} \right)^2
  	\nonumber\\
  	&=& 1- \theta^2\frac{ F_T }{2R}  -\bar{\theta}^2 \frac{F_T^* }{2R}
  	+ \theta^4 \frac{ |F_T|^2 }{2R}
	\\\nonumber
	\\&\equiv& 
	1 - \theta^2 f   -\bar{\theta}^2 f^* + 2|f|^2\theta^4 \ .
\end{eqnarray} 
Then $\chi$ is not  canonically normalized.  We redefine the adjoint chiral superfield,
     \begin{eqnarray}
      \chi \to (1+f \theta^2) \chi \ .
     \end{eqnarray}
Hence the first term in the second line of Eq.~(\ref{RadionMed1}) changes with this field redefinition, 
       \begin{eqnarray}
	&&\!\!\!\!\!\!\!\!  \frac{1}{k} \int d^4\theta
	\frac{2R}{T+{T^\dag}}  {\rm Tr} 
	 \Big[
	 \chi{\chi}^{\dag} 
	 \Big]
	\no\\&&\to
        \frac{1}{k} \int d^4\theta
	{\rm Tr} 
	\Big[
	(1 -f\theta^2  -f^* \bar\theta^2 +2|f|^2 \theta^4 )
	 (1+f \theta^2) (1+f^* \bar\theta^2)
	 \chi{\chi}^{\dag} 
	 \Big]
	 \nonumber\\ &&=
	 \frac{1}{k} \int d^4\theta
	 {\rm Tr} 
	 \Big[
       	 \chi{\chi}^{\dag} 
	 \Big] 
	 + |f|^2 |\phi_\chi^a|^2  \ ,
       \end{eqnarray}
and the second term changes as,
       \begin{eqnarray}
        &&\hspace{-20pt} \frac{1}{k} \int \!\!d^4\theta
	\frac{2R}{T+{T^\dag}}  {\rm Tr} 
	\Big[
	 \frac{\chi{\chi} +{\chi}^{\dag}{\chi}^{\dag} }{2}
	 \Big]
	\no \\
     &&\to  \frac{1}{k} \int d^4\theta
	\left(1 -f\theta^2  -f^* \bar\theta^2 +2|f|^2 \theta^4 \right)
	{\rm Tr} 
	\left[
	       (1+2f \theta^2)\frac{\chi{\chi} }{2}
	       +\hc
	      \right]
	 \nonumber\\&&=
        -\frac{1}{k} \int d^4\theta
	{\rm Tr} 
	\Big[
	       (f^* \bar\theta^2)\frac{\chi{\chi} }{2}
	       +\hc
	      \Big] 
	      \nonumber\\&&=
	      -f^* F^a_\chi \phi^a_\chi +\frac{f^*}{2} \lambda^a_2\lambda^a_2
	      \ +\mbox{h.c.}\ ,
       \end{eqnarray}
where we use a complex scalar in the adjoint representation,  
	\begin{eqnarray}
	 \phi^a_\chi\equiv \frac{\Sigma^a +iA_5^a }{\sqrt{2}} \ .
	\end{eqnarray}
The term of $|f|^2||\phi_\chi^a|^2$ will vanish after integration out of the auxiliary field $F_\chi$. 

For hypermultiplet, similarly we redefine the chiral superfields for canonical normalization, 
      \begin{eqnarray}
       H_1\to (1-f\theta^2)H_1, \ H_2\to (1-f\theta^2)H_2 \ .
       \label{172608_1May12}
      \end{eqnarray}
The third and fourth lines of Eq.~(\ref{RadionMed1}) are, 
     \begin{eqnarray}
        &&    \hspace{-30pt} \int d^4\theta 
	  \ \frac{T+T^\dag }{2R}
	  \Big\{  {H_2} {H_2^\dag} +{H_1^\dag} {H_1}
	  \Big\}
      \\&&\to
	   \int d^4\theta 
	   (1+f \theta^2 +f^* \bar\theta^2) 
	   (1-f \theta^2) 
	   (1-f^* \bar\theta^2)
	   \left\{ {H_2} {H_2^\dag} +{H_1^\dag} {H_1}        \right\}
	  \nonumber\\&&=
	  	   \int d^4\theta 
	   (1-|f|^2 \theta^4) \left\{ {H_2} {H_2^\dag} +{H_1^\dag} {H_1}        \right\}
	 \nonumber\\&&=
	  \int d^4\theta  \left\{ {H_2} {H_2^\dag} +{H_1^\dag} {H_1}        \right\}
	 -|f|^2 \left(  |\phi_1|^2   +|\phi_2|^2  \right) \ ,
     \end{eqnarray}    
and,
      \begin{eqnarray}
      &&\hspace{-30pt} \int d^2\theta 
	\ H_2 (R^{-1}\partial_\varphi  -\sqrt{2} g \chi) H_1   
       \no\\&&\to \int d^2\theta (1-f\theta^2)\ H_2 
	\left(\partial_5  -\sqrt{2} g (1+f\theta^2)\chi  \right) (1-f\theta^2)H_1   
      	\nonumber\\
       &&= \int d^2\theta (1-2f\theta^2)\ H_2 
	(\partial_5  -\sqrt{2} g \chi  ) H_1 
	-\theta^2 (\sqrt{2} g f  H_2 \chi H_1)
      \nonumber\\
       &&=  \int d^2\theta \left\{
	H_2 (\partial_5  -\sqrt{2} g \chi  ) H_1
	+\theta^2  f H_2 ( -2\delfive + \sqrt{2} g \chi )H_1
	\right\}
      \nonumber\\
       &&=  \int d^2\theta 
	H_2 (\partial_5  -\sqrt{2} g \chi  ) H_1
	+ f \phi_2 ( -2\delfive + \sqrt{2} g \phi^a_\chi T^a)\phi_1  \ . 
       \end{eqnarray}
So far, the extra terms generated by the Radion VEV of  $F_T(f)$  are
       \begin{eqnarray}
	\left( \frac{f}{2}\lambda_1^a\lambda_1^a  
	 +\frac{f^*}{2} \lambda^a_2\lambda^a_2
	 \ +\mbox{ h.c.}
      \right)
       	 	 -|f|^2 \left(  |\phi_1|^2   +|\phi_2|^2  \right) 
       \nonumber\label{135645_23Apr12}\\ 
	+|f|^2 |\phi^a_\chi|^2
	 + \left\{  f \phi_2 ( -2\delfive + \sqrt{2} g \phi^a_\chi T^a)\phi_1   \ +\mbox{ h.c.} \right\}
	 \nonumber\\
		 -f^* F^a_\chi \phi^a_\chi  -f F^{a*}_\chi \phi^{a*}_\chi \ . 
		  \label{160515_23Apr12}
       \end{eqnarray}
By integrating out $F_\chi$, the second and third lines of Eq.~(\ref{135645_23Apr12}) are changed into a different form which we can compare to the Scherk-Schwarz mechanism.  Terms involving $F_\chi$ in Eqs.~(\ref{141250_23Apr12}, \ref{140825_23Apr12}, \ref{135645_23Apr12}) are,
      \begin{eqnarray}
       |F_\chi|^2 
	-\sqrt{2} g F_\chi^a (\phi_2  T^a \phi_1)  
	-\sqrt{2} g F_\chi^{a*} (\phi_1^*  T^a \phi_2^*)  
	-f^* F^a_\chi \phi^a_\chi  -f F^{a*}_\chi \phi^{a*}_\chi \  ,
	\end{eqnarray}
and after integrating out $F_\chi$ we have
	\begin{eqnarray}
       && -2g^2 (\phi_2  T^a \phi_1)  (\phi_1^*  T^a \phi_2^*)  
		\nonumber\\
       &&-|f|^2|\phi^a_\chi|^2
	-\sqrt{2} g f \phi_\chi^{a*} (\phi_2  T^a \phi_1)
	-\sqrt{2} g f^* \phi_\chi^{a} (\phi_1^*  T^a \phi_2^*)  \ .\label{160438_23Apr12}  
      \end{eqnarray}
We combine the above soft terms with terms not involving $F_\chi$ from Eq.(\ref{160515_23Apr12}). The $|f|^2 |\phi_\chi^a|^2$ terms is cancelled and the rest terms are  
       \begin{eqnarray}
	{\cal L}_{5,\rm soft}&=&\Big(  \frac{f}{2}\lambda_1^a\lambda_1^a  
	 +\frac{f^*}{2} \lambda^a_2\lambda^a_2
	 + \mbox{ h.c.}
      \Big)
       	 	 -|f|^2 \left(  |\phi_1|^2   +|\phi_2|^2  \right) 
       \nonumber\\ 
	  &&-2 \left\{ f \phi_2 
	    \left( \delfive - \sqrt{2} g \frac{\phi^a_\chi -\phi^{a*}_\chi}{2} T^a \right)\phi_1  
	    \ +\mbox{ h.c.} \right\} 
	\nonumber\\ 
	&=&\left( \frac{f}{2}\lambda_1^a\lambda_1^a  
	 +\frac{f^*}{2} \lambda^a_2\lambda^a_2
	 \ +\mbox{ h.c.}
      \right)
       	 	 -|f|^2 \left(  |\phi_1|^2   +|\phi_2|^2  \right) 
       \nonumber\\ 
	 && -2 \left\{  f\phi_2 
	    \left( \delfive - i gA_5^a \ T^a \right)\phi_1  
	    \ +\mbox{ h.c.} \right\} 
	 \nonumber\\
	&=&\left( \frac{f}{2}\lambda_1^a\lambda_1^a  
	 +\frac{f^*}{2} \lambda^a_2\lambda^a_2
	 \ +\mbox{ h.c.}
      \right)
       	 	 -|f|^2 \left(  |\phi_1|^2   +|\phi_2|^2  \right) 
       \nonumber\\ 
	 &&
	  +f^*  \phi_1^*   ({\cal D}_5\phi_2^* )
	   -f  \phi_2  ( {\cal D}_5\phi_1  )
   	  -f^* ( {\cal D}_5\phi_1  )^\dag \phi_2^* 
	   +f  ({\cal D}_5\phi_2^*)^\dag   \phi_1  \ .
	  \label{Radionmedsoft}
       \end{eqnarray}
Since the phase of $f$ can be removed by $SU(2)_R$ rotation, it always becomes real. 
Therefore, soft terms in Eq.~(\ref{SSSBsoft}) and those generated by the Scherk-Schwarz mechanism in Eq.~(\ref{Radionmedsoft}) are same by a relation of
	\begin{eqnarray}
	f=\frac{F_T}{2R}=\frac{\alpha}{R}
	\end{eqnarray}

We can view this type of supersymmetry breaking in two different ways. The Scherk-Schwarz mechanism tells us that  this is non-local supersymmetry breaking in the 5D theory. This leads to an important consequence of radiative corrections that the radiative corrections which appear due to the SUSY breaking are UV finite because the UV behavior (divergence) is local effect while the SUSY breaking of this kind is not local (for more detail, see Ref.\cite{Barbieri:2001yz,ArkaniHamed:2001mi}). We see this in Sec.~\ref{subsec:threshold}.
The Radion mediation is formulated with ${\cal N}=1$ superfields yielding SUSY breaking in the wave function of hypermultiplet.  
\subsection{Soft terms on brane}
The Radion mediation requires the field redefinition, which generates soft terms once we consider  interactions on branes. 
As in Sec.\ref{subset:brane}, let us consider brane interactions, 
	\begin{eqnarray}
	W_{\rm brane}&=&\delta(y)\left\{\lambda_1 S_1(x)S_2(x)H_1(x,y)	
	+\frac{\lambda_2}{2} S_1(x)H^2_1(x,y)	
	+\frac{\lambda_3}{3} H^3_1(x,y) 	\right\}. \no
	\end{eqnarray}
 The bulk field $H_1$ are canonically normalized as Eq.~(\ref{172608_1May12}), which results in soft terms on the brane, 
	\begin{eqnarray}
	\int d^2\theta\ W_{\rm brane}&\to&\int d^2\theta\ \delta(y)\bigg\{\lambda_1 S_1(x)S_2(x)(1-f\theta^2)H_1(x,y)
	\no\\	&&\quad
	+\frac{\lambda_2}{2} S_1(x)(1-f\theta^2)^2H^2_1(x,y)	
	+\frac{\lambda_3}{3} (1-f\theta^2)^3H^3_1(x,y) 	\bigg\}
	\\	&=&
	\int d^2\theta\ \delta(y)\bigg\{\lambda_1 S_1(x)S_2(x)(1-f\theta^2)H_1(x,y)
	\no\\	&&\quad
	+\frac{\lambda_2}{2} S_1(x)(1-2f\theta^2)H^2_1(x,y)	
	+\frac{\lambda_3}{3} (1-3f\theta^2)H^3_1(x,y) 	\bigg\} , 
	\end{eqnarray}
and then,
	\begin{eqnarray}
	{\cal L}_{\rm brane,soft}  =\delta(y)\left\{\lambda_1 (-f) \phi_{S1}\phi_{S2}\phi_1
	+\frac{\lambda_2}{2} (-2f) \phi_{S1}\phi_1^2 
	+\frac{\lambda_3}{3} (-3f) \phi_1^3 \right\} .
	\end{eqnarray}
With $f=\alpha/R$, these soft terms are same as  those derived by the Scherk-Schwarz mechanism, Eq.~(\ref{eq:brsoft1}). From this point, we confirm the equivalence between Scherk-Schwarz mechanism and Radion mediation. 

\section{Model Setup}
Here a model setup of Compact Supersymmetry is described. 
\subsection{Bulk gauge and matter fields}
We consider a supersymmetric $SU(3)_C \times SU(2)_L \times U(1)_Y$ 
gauge theory in this spacetime, with the gauge and three generations 
of matter supermultiplets propagating in the bulk 
where fields must form ${\cal N}=2, D=4$ multiplet. Then, for instance, $SU(3)_C$ gauge supermultiplet  is composed  of a vector superfield $V_C$ (in ${\cal N}=1$) and a chiral superfield $\chi_C$, and similarly $SU(2)_L$ and $U(1)_Y$ gauge supermultiplets are given by $(V_L, \chi_L)$ and $(V_Y, \chi_Y)$, respectively.

A matter field form a hypermultiplet which requires a chiral superfield with the SM charge and another one with the opposite charge. For example $SU(2)_L$ doublet quark is a hypermultiplet of $(Q, Q^c)$, whose superscript $c$ means the opposite charge. The other fields are $(U,U^c), (D,D^c), (L,L^c), (E,E^c)$. 
The bulk Lagrangian of matters is given by Eq.~(\ref{matter-insuperspace}) replacing, 
	\begin{eqnarray}
	(H_1,H_2)\rightarrow (Q,Q^c),\ (U,U^c),\ (D,D^c),\ (L,L^c),\ (E,E^c) \ .	
	\end{eqnarray}
Under ${\cal P}:y\to -y$, fields with superscript $c$ is odd, and thus their zero modes are projected out. The soft masses of squark, slepton, and guagino are therefore 
\begin{equation}
\begin{array}{c}
  M_{1/2} = \frac{\alpha}{R},
\quad
  m_{\tilde{Q},\tilde{U},\tilde{D},\tilde{L},\tilde{E}}^2 
  = \left( \frac{\alpha}{R} \right)^2.
\quad \label{eq:softmass}
\end{array}
\end{equation}
in the language of MSSM. 

\subsection{Higgs fields and Yukawa interactions}
The Higgs chiral superfields $H_u$ and $H_d$ are located on one of the 
branes at $y=0$.  The Yukawa couplings, $\mu$ term, and kinetic terms can then be written 
on that brane:
\begin{figure}[t]
\begin{center}
  \includegraphics[clip,width=.65\textwidth]{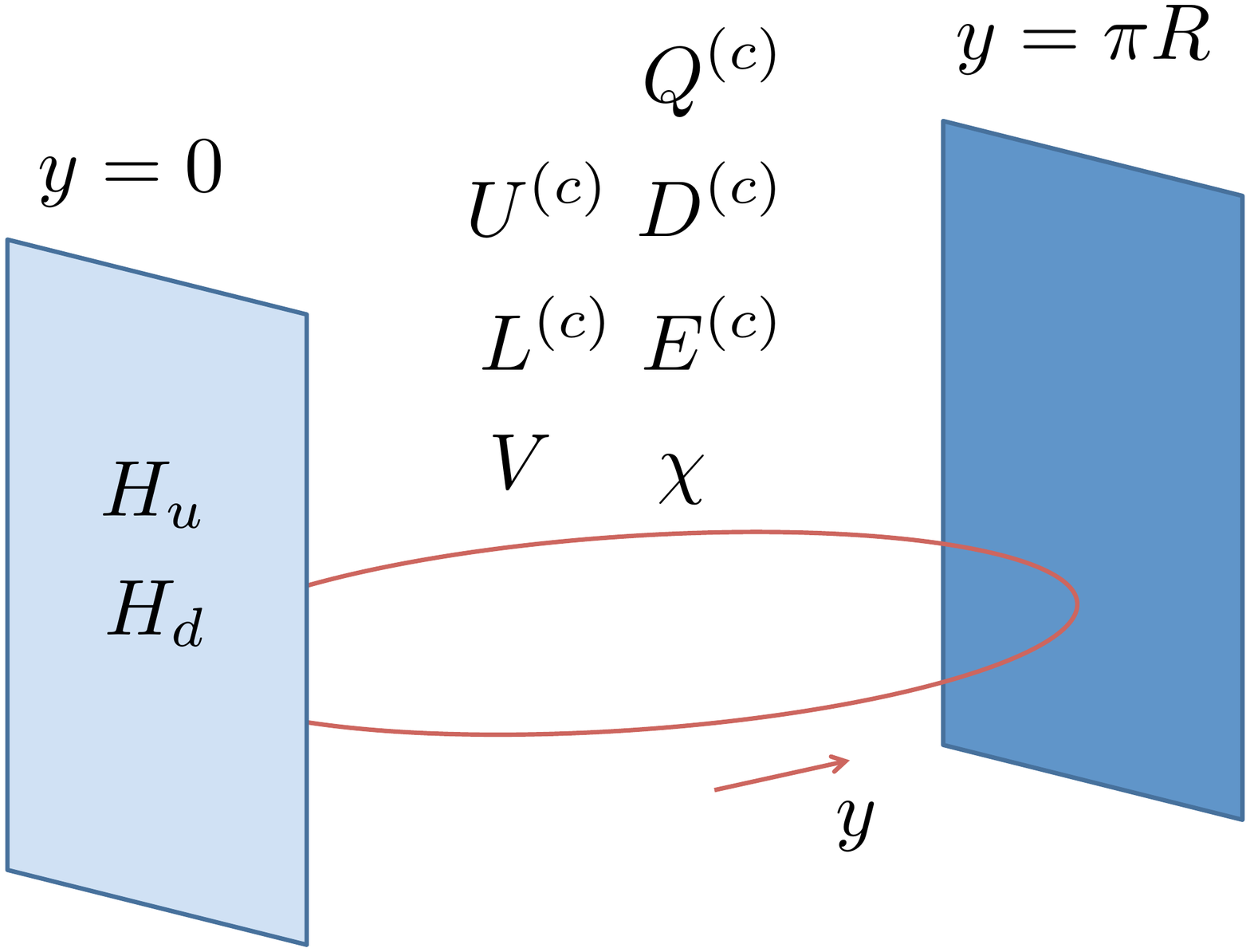}
\end{center}
\caption{Field configuration of Compact Supersymmetry Model}
\end{figure}
\begin{eqnarray}
  {\cal L}_{\rm brane} &=& \delta(y) \int\!d^2\theta \bigl\{ 
    y_{U5}^{ij} H_u Q_i U_j  + y_{D5}^{ij} H_dQ_i D_j 
   {} + y_{E5}^{ij} H_dL_i E_j  + \mu H_u H_d \bigr\}
\nonumber\\
  && +\delta(y) \int\!d^4\theta 
  \{H_u^\dag e^{-2gV} H_u +H_d^\dag e^{-2gV} H_d \}, 
\label{eq:brane-W}
\end{eqnarray}
where Yukawa couplings are proportional to those of the SM, and $i$ and $j$ are flavor indices.  The $\cal P$ odd fields ($Q^c, U^c, ...$) cannot have these Yukawa interactions since their wave functions are zero at $y=0$.
 (Here, 
we have simply assumed the existence of the $\mu$ term on the brane.  We 
leave discussions of its origin to future work.) 
In the Radion mediation, the fields living in the bulk have field redefinition in Eq.~(\ref{172608_1May12}) for the canonical normalization such that 
	\begin{eqnarray}
	\{Q,U,D, L, E\} \rightarrow \left(1-\frac{\alpha}{R}\theta^2 \right)\{Q,U,D, L, E\}.
	\end{eqnarray}
In general, $(1-F\theta^2)^n$  becomes ($1 -n F\theta^2$), and thus,
\begin{eqnarray}
  	{\cal L}_{\rm brane,soft} &=& \delta(y)  \biggl\{ 
   	 y_{U5}^{ij} \left(\frac{-2\alpha}{R}\right)\phi_{H_u} \phi_{Q_i} \phi_{U_j} + y_{D5}^{ij}\left(\frac{-2\alpha}{R}\right)\phi_{H_d}\phi_{Q_i} \phi_{D_j} 
	\no\\
	&&\quad\quad+ y_{E5}^{ij}\left(\frac{-2\alpha}{R}\right) \phi_{H_d} \phi_{L_i} \phi_{E_j} \biggr\} +\hc
\label{eq:brane-soft}
\end{eqnarray}
There are no soft masses of Higgs fields. 
Together with Eq.~(\ref{eq:softmass}), this leads to the expressions of (MSSM) soft terms we gave in Eq.~(\ref{eq:tree}) in the introduction,  
\begin{equation}
\begin{array}{c}
  M_{1/2} = \frac{\alpha}{R},
\quad
  m_{\tilde{Q},\tilde{U},\tilde{D},\tilde{L},\tilde{E}}^2 
  = \left( \frac{\alpha}{R} \right)^2,
\quad
  m_{H_u,H_d}^2 = 0,
\\[10pt]
  A_0 = -\frac{2\alpha}{R},
\quad
  B = 0, 
\end{array} \label{eq:summary}
\end{equation}
at tree-level. 
Note that the degeneracy among the three generations of scalars and gauginos is 
automatic because of the geometric nature of the SUSY breaking. 
In addition, since $\alpha$ and $\mu$ can always be taken real by phase 
redefinitions of fields associated with $R$ and Peccei-Quinn rotations, 
there is no physical phase in $M_{1/2}$, $A_0$, $\mu$, or $B$.  Therefore, 
the flavor problem as well as the $CP$ problem are automatically solved 
in this model.

 \begin{table}[htdp]
 \begin{center}\begin{tabular}{|c c | c c c c  c |}
 	\hline
	 && $SU(3)_C$ & $SU(2)_L$ & $U(1)_Y$ & $SU(2)_R$ & ${\cal P}:y\to -y$   \\
     	\hline
     	\hline
	$V_C \ \chi_C$ & $G_\mu$ & \bf 8 & \bf 1 & 0  & \bf 1 & $+$\\
	& $(G_{5}, \Sigma_G)$  & \bf 8 & \bf 1 & 0  & \bf 1 & $(-,-)$\\
	& $(\lambda_{G1}, -\lambda_{G2})$  & \bf 8 & \bf 1 & 0  & \bf 2 & $(+,-)$\\
	\hline
	$V_L \ \chi_L$ & $W_\mu$ & \bf 1 & \bf 3 & 0  & \bf 1 & $+$\\
	& $(W_{5}, \Sigma_W)$  & \bf 1 & \bf 3 & 0  & \bf 1 & $(-,-)$\\
	& $(\lambda_{W1}, -\lambda_{W2})$  & \bf 1 & \bf 3 & 0  & \bf 2 & $(+,-)$\\
	\hline
	$V_Y \ \chi_Y$ & $B_\mu$ & \bf 1 & \bf 1 & 0  & \bf 1 & $+$\\
	& $(B_{5}, \Sigma_B)$  & \bf 1 & \bf 1 & 0  & \bf 1 & $(-,-)$\\
	& $(\lambda_{B1}, -\lambda_{B2})$ & \bf 1 & \bf 1 & 0  & \bf 2 & $(+,-)$\\
	\hline
 \end{tabular} \caption{Representation of fields in gauge supermultipet.}
\end{center}
\label{tab:gauge}
\end{table}
 \begin{table}[htdp]
 \begin{center}\begin{tabular}{|c c | c c c c  c |}
 	\hline
	 && $SU(3)_C$ & $SU(2)_L$ & $U(1)_Y$ & $SU(2)_R$ & ${\cal P}:y\to -y$   \\
     	\hline
     	\hline
	$Q \ Q^{c\dag}$&$(\phi_Q, \phi_{Q^c}^*) $& \bf 3 & \bf 2 & 1/6  & \bf 2 & $(+,-)$\\
	&$(\psi_Q, \psi_{Q^c}^*)$ & \bf 3 & \bf 2 & 1/6  & \bf 1 & $(+,-)$\\
     	\hline
	$U \ U^{c\dag}$&$(\phi_U, \phi_{U^c}^*)$ & $\bf \bar{3}$ & \bf 1 & $-2/3$  & \rm \bf 2 & $(+,-)$\\
	&$(\psi_U, \psi_{U^c}^*)$ & $\bf \bar{3}$ & \bf 1 & $-2/3$  & \rm \bf 1 & $(+,-)$\\
     	\hline
	$D \ D^{c\dag}$&$(\phi_D, \phi_{D^c}^*)$ & $\bf \bar{3}$ & \bf 1 & 1/3  & \bf 2 & $(+,-)$\\
	&$(\psi_D, \psi_{D^c}^*)$ & $\bf \bar{3}$ & \bf 1 & 1/3  & \bf 1 & $(+,-)$\\
     	\hline
	$L \ L^{c\dag}$&$(\phi_L, \phi_{L^c}^*)$ & \bf 1 & \bf 2 & $-1/2$  & \rm \bf 2 & $(+,-)$\\
	&$(\psi_L, \psi_{L^c}^*)$ & \bf 1 & \bf 2 & $-1/2$  & \rm \bf 1 & $(+,-)$\\
   	\hline
	$E \ E^{c\dag}$&$(\phi_E, \phi_{E^c}^*) $& \bf 1 & \bf 1 & 1  & \rm \bf 2 & $(+,-)$\\
	&$(\psi_E, \psi_{E^c}^*)$ & \bf 1 & \bf 1 & 1  & \rm \bf 1 & $(+,-)$\\
     	\hline
	$H_u$& $\phi_{H_u}$ & \bf 1 & \bf 2 & 1/2  & \bf 1 & $+$\\
	& $\psi_{H_u}$ & \bf 1 & \bf 2 & 1/2  & \bf 1 & $+$\\
     	\hline
	$H_d$& $\phi_{H_d}$ & \bf 1 & \bf 2 & $-1/2$  & \bf 1 & $+$\\
	& $\psi_{H_d}$ & \bf 1 & \bf 2 & $-1/2$  & \bf 1 & $+$\\
     	\hline
 \end{tabular} \caption{Representation of fields in hypermultipet and Higgs chiral supermultiplet. }
\end{center}
\label{contents}
\end{table}

\subsection{Lagrangian on brane with Yukawa coupling}
\begin{figure}[t]
\begin{center}
  \includegraphics[clip,width=0.8\textwidth]{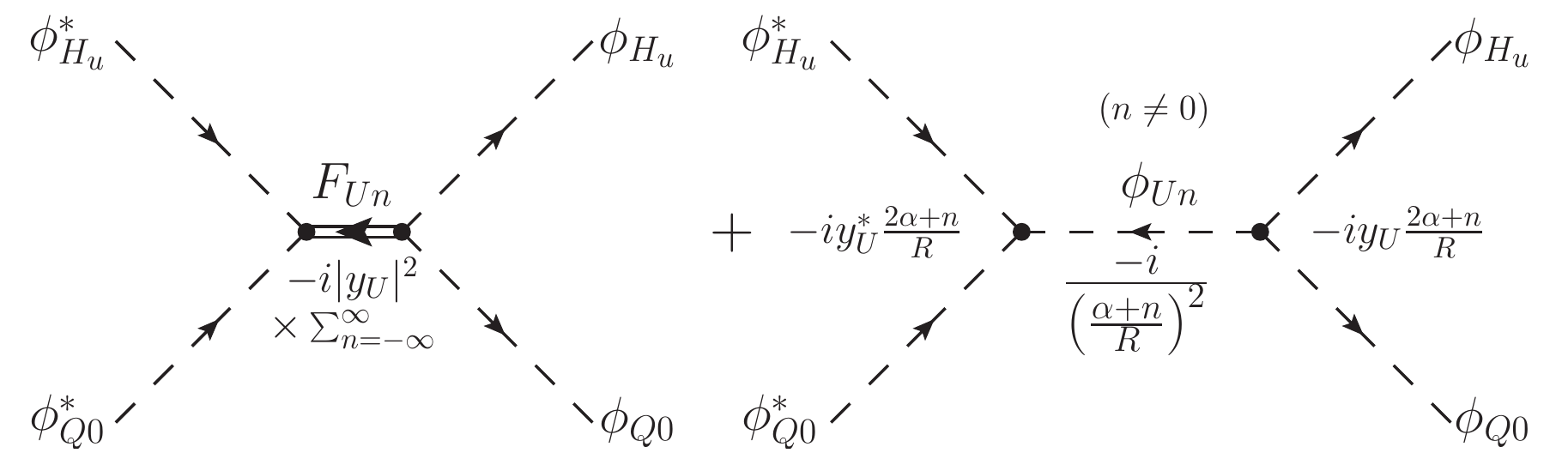}
\end{center}
\caption{Cancellation of infinite sum for $|\phi_{H_u}\phi_{Q_0}|^2$ vertex. }
\label{fig:fourpointKK}
\end{figure}

Let us investigate brane interactions involving up-type quark in KK modes. We use them in the later calculation of radiative corrections through top-Yukawa coupling in Sec.~\ref{subsec:threshold}. We omit flavor indices for simplicity. 
Component interactions from Eq.~(\ref{eq:brane-W}) are
\begin{eqnarray}
  	{\cal L}_{y_U5,\phi}
	&=& 
	\delta(y) \biggl[~y_{U5}  \bigl\{ 
   	F_{H_u}\phi_{Q}(y) \phi_{U}(y) 
	 +\phi_{H_u}F_{Q}(y) \phi_{U}(y) 
	 +\phi_{H_u}\phi_{Q}(y) F_{U}(y) 	\bigr\}
	\no\\
	&&\quad\quad+~\mu  \bigl\{  F_{H_u}\phi_{H_d}+\phi_{H_u}F_{H_d}
	\bigr\}  +~y_{U5} \left(\frac{-2\alpha}{R}\right)\phi_{H_u} \phi_{Q} \phi_{U}\biggr]+\hc
	\no\\
	&&+\delta(y)\{|F_{H_u}|^2+|F_{H_d}|^2\}+|F'_{Q}(y)|^2+|F'_{U}(y)|^2 
	\\
	&=&\delta(y)  \biggl[y_{U5} \Bigl\{ 
   	  F_{H_u}\phi_{Q}(y) \phi_{U}(y) 
	 + \phi_{H_u}\left[F'_{Q}(y) +\delfive\phi^*_{Q^c}(y)\right]\phi_{U}(y) 
	\no\\
	&&\quad\quad\quad\quad
	 + \phi_{H_u}\phi_{Q}(y) \left[F'_{U}(y) +\delfive\phi^*_{U^c}(y)\right]\Bigr\}
	\no\\
	 &&\quad\quad+~\mu\left\{F_{H_u}\phi_{H_d}+\phi_{H_u}F_{H_d}
	\right\} +~y_{U5} \left(\frac{-2\alpha}{R}\right)\phi_{H_u} \phi_{Q} \phi_{U}\biggr]+\hc
	\no\\
	&&+\delta(y)\{|F_{H_u}|^2+|F_{H_d}|^2\}+|F'_{Q}(y)|^2+|F'_{U}(y)|^2 \ ,
\end{eqnarray}
where $F$-terms in the bulk are already redefined as 
	\begin{eqnarray}
	F'_Q=F_Q+[-\delfive-g(\Sigma-iA_5)]\phi_{Q^c}^* , \label{FQredef}\\
	F'_U=F_U+[-\delfive-g(\Sigma-iA_5)]\phi_{U^c}^* ,\label{FUredef}
	\end{eqnarray}
and $\Sigma$ and $A_5$ vanishe since their wave functions are zero at $y=0$ brane. 
If we integrate out $F'$-term as 5D field, double delta function $\delta(y)\delta(y)\sim \delta(y)\sum_{n=-\infty}^\infty(2\pi R)^{-1}$ is led, and then the result seems meaningless. We perform the KK expansion first and integrate out the KK modes of $F'$-terms so that we understand  the meaning. We have the following KK expansions, 
	\begin{eqnarray}
	F'_{Q(U)}(y)=\sum_{n=-\infty}^\infty \frac{1}{\sqrt{2\pi R}}F'_{Q(U),n} \cos\frac{ny}{R}, \\
	\phi_{Q(U)}(y)=\sum_{n=-\infty}^\infty \frac{1}{\sqrt{2\pi R}}\phi_{Q(U)n} \cos\frac{ny}{R},\\
	\phi^*_{Q^c(U^c)}(y)=\sum_{n=-\infty}^\infty \frac{-1}{\sqrt{2\pi R}}\phi_{Q(U)n} \sin\frac{ny}{R},
	\end{eqnarray}
where $\phi_{Q(U)n}$ is a mass eigenstate with mass of $(\alpha+n)/R$. Perform $y$ integral, 
\begin{eqnarray}
	\int_0^{2\pi R}\!\!\!dy\ {\cal L}_{y_U5,\phi} \!\!\!\!
	&=&
	y_{U5} \Bigl\{ 
   	  F_{H_u}\sum_{n=-\infty}^\infty \frac{\phi_{Qn}}{\sqrt{2\pi R}} \sum_{m=-\infty}^\infty \frac{\phi_{Um}}{\sqrt{2\pi R}} 
	 \no\\&&\quad\quad
	 + \phi_{H_u}\sum_{n=-\infty}^\infty \frac{F'_{Qn}-\frac{n}{R}\phi_{Qn}}{\sqrt{2\pi R}} \sum_{m=-\infty}^\infty \frac{\phi_{Um}}{\sqrt{2\pi R}}  
	 \no\\&&\quad\quad
	 + \phi_{H_u}\sum_{n=-\infty}^\infty \frac{\phi_{Qn}}{\sqrt{2\pi R}} \sum_{m=-\infty}^\infty \frac{F'_{Um}-\frac{m}{R}\phi_{Um}}{\sqrt{2\pi R}} 
	\Bigr\}+\hc
	\no\\
	 &&+\mu\left\{F_{H_u}\phi_{H_d}+\phi_{H_u}F_{H_d}
	\right\}+\hc
	 \no\\&&
	 +y_{U5} \left(\frac{-2\alpha}{R}\right) \phi_{H_u}\sum_{n=-\infty}^\infty \frac{\phi_{Qn}}{\sqrt{2\pi R}} \sum_{m=-\infty}^\infty \frac{\phi_{Um}}{\sqrt{2\pi R}} +\hc
	\no\\
	&&+|F_{H_u}|^2+|F_{H_d}|^2+
	\sum_{n=-\infty}^\infty |F'_{Qn}|^2+	\sum_{m=-\infty}^\infty |F'_{Um}|^2 \ ,
\end{eqnarray}
and the equations of motion for $F$-terms, 
	\begin{eqnarray}
	&&F_{H_u}^*
	+y_U \sum_{n=-\infty}^\infty{\phi_{Qn}} \sum_{m=-\infty}^\infty {\phi_{Um}}
	+\mu\phi_{H_d}=0,
	\\
	&&F'^*_{Qn}+y_U\phi_{H_u}\sum_{m=-\infty}^\infty {\phi_{Um}} =0,
	\\
	&&F'^*_{Um}+y_U\phi_{H_u}\sum_{n=-\infty}^\infty {\phi_{Qn}} =0,
	\end{eqnarray}
where $y_U\equiv y_{U5}/(2\pi R)$. These conditions are substituted to the Lagrangian, 
	\begin{eqnarray}
		\int_0^{2\pi R}\!\!\!dy\ {\cal L}_{y_U5,\phi} \!\!\!\!
	&=&-\Big|	 y_U \sum_{n=-\infty}^\infty\phi_{Qn}\sum_{m=-\infty}^\infty \phi_{Um}
	+\mu\phi_{H_d}	\Big|^2
	\no\\&&
	-\Big| y_U\phi_{H_u}\sum_{m=-\infty}^\infty \phi_{Um} \Big|^2 \!\!\!\times\!\!\!\sum_{n=-\infty}^\infty
	-\Big| y_U\phi_{H_u}\sum_{m=-\infty}^\infty \phi_{Qm} \Big|^2 \!\!\!\times\!\!\!\sum_{n=-\infty}^\infty
	 \no\\
	 &&-\sum_{m,n=-\infty}^\infty y_{U} 
	 \frac{(\alpha+m)+(\alpha+n)}{R} \phi_{H_u}\phi_{Qm}\phi_{Un} +\hc
	 \label{eq:yukk}
	\end{eqnarray}
In the second line, there is infinite sum left which is originally from the sum of $F'_n$-terms. This itself looks to lead infinitely large coupling of $|\phi_{H_u} \phi_{U(Q)m}|^2$ interaction, and one may worry that the low-energy limit of this theory cannot be the MSSM.  However, the amplitude is certainly finite because the other bosonic interactions cancel the infinite sum as in Fig.~\ref{fig:fourpointKK}. For this cancellation, in the low-energy limit $E\ll 1/R$, vertices which are consistent with the MSSM actually remain, namely,
	\begin{eqnarray}
	&&-\left|	y_U \phi_{Q0} \phi_{U0}+\mu\phi_{H_d}	\right|^2
	-\left| y_U\phi_{H_u} \phi_{U0} \right|^2
	-\left| y_U\phi_{H_u} \phi_{Q0} \right|^2
	 \no\\
	 &&- y_{U} 
	 \frac{2\alpha}{R} \phi_{H_u}\phi_{Q0}\phi_{U0} +\hc \label{eq:yukawaKK}
	\end{eqnarray}
The same cancellation is important when we compute the radiative correction; it look dangerous on first sight, but finally gives a regularized result. 

Let us give a comment on the trilinear terms. In the Radion mediation, this dimensionful couplings come from the fifth derivative $\delfive$ due to the field redefinition of $F$ as in Eqs.~(\ref{FQredef},\ref{FUredef})  and from the soft supersymmetry breaking of Eq.~(\ref{eq:brane-soft}), and they are (luckily) combined to an organized form in Eq.~(\ref{eq:yukk}). This result is necessary because in the picture of the Scherk-Schwarz mechanism the soft terms come along with the fifth derivative $\delfive$, and therefore the dimensionful coupling is written in terms of KK mode mass such as $\frac{\alpha+n}{R}$.  If we consider a field configuration where $H_u$ is also living in the bulk as well as quark fields, analogous to Eq.~(\ref{eq:yukk}) the trilinear terms are expected to be 
	\begin{eqnarray}
	-\sum_{k,m,n=-\infty}^\infty y_{U} 
	 \frac{(\alpha+k)+(\alpha+m)+(\alpha+n)}{R} \phi_{H_uk}\phi_{Qm}\phi_{Un} .
	 \end{eqnarray}
This is certainly true. 

Next, we  give the fermionic interactions, which are much simpler compared to those of scalars. The fermionic interactions with $y_U$ are
	\begin{eqnarray}
	{\cal  L}_{y_U5,\psi}=\delta(y)~y_{U5}\left\{-\phi_{H_u}\psi_{Q}\psi_{U}
	-\psi_{H_u}\phi_{Q}\psi_{U}-\psi_{H_u}\psi_{Q}\phi_{U}\right\}+\hc
	\end{eqnarray}
and we perform the KK expansion and $y$ integral,
	\begin{eqnarray}
	 \int_0^{2\pi R}\!\!dy~{\cal  L}_{y_U5,\psi}\!\!\!
	&=&-\frac{y_{U5}}{2\pi R}\sum_{m,n=0}^\infty 2\eta_m\eta_n \phi_{H_u}\psi_{Q,m}\psi_{U,n}+\hc
	\no\\
	&&-\frac{y_{U5}}{2\pi R}\sum_{m=-\infty}^\infty\sum_{n=0}^\infty \sqrt{2}\eta_m \psi_{H_u}(\phi_{Qm}\psi_{U,n}+\psi_{Q,n}\phi_{Um})+\hc
	\end{eqnarray}
	where the KK expansion for fermions are
	\begin{eqnarray}
	\psi_{Q(U)}(y)=\sum_{n=0}^\infty \frac{\eta_n~\psi_{Q(U),n}}{\sqrt{\pi R}}\cos \frac{ny}{R} .
	\end{eqnarray}
Here, $\eta_n=(1/\sqrt{2})^{\delta_{n,0}}$. The non-zero modes are Dirac fermions, so we write the Lagrangian in four component spinors, 
	\begin{eqnarray}
	\!\!\!\!\!\!\int_0^{2\pi R}\!\!dy~{\cal  L}_{y_U5,\psi}\!\!\!
	&=&
	\sum_{m,n=0}^\infty (-2\eta_m\eta_n)
	\left( y_U\phi_{H_u}\overline{\Psi^c}_{Qm}P_L\Psi_{Un}+y_U^*\overline{\Psi}_{Un}P_R\Psi_{Qm}^c \phi_{H_u}^*\right)
	\no\\
	&&\!\!\!\!\!\!+\sum_{m=-\infty}^\infty\sum_{n=0}^\infty (-\sqrt{2}\eta_n) 
	\left(y_U\overline\Psi_{H_u}\phi_{Qm}P_L\Psi_{Un}
	+y_U^*\overline{\Psi}_{Un}\phi_{Qm}^*P_R\Psi_{H_u} 	\right)
	\no\\
	&&\!\!\!\!\!\!+\sum_{m=0}^\infty\sum_{n=-\infty}^\infty (-\sqrt{2}\eta_m) 
	\left(y_U\overline\Psi_{H_u}P_L\Psi_{Qm}\phi_{Un}
	+y_U^*\phi_{Un}^*\overline\Psi_{Qm}P_R\Psi_{H_u} 	\right)
	\end{eqnarray} 
where the four component spinors are
	\begin{eqnarray}
	\Psi_{H_u}\equiv \left( \begin{array}{c}
      	\psi_{H_u}  \\ \bar\psi_{H_u} \end{array} \right),\ 
	 \Psi_{Q(U)0}\equiv \left( \begin{array}{c}
      	\psi_{Q(U),0}  \\ \bar\psi_{Q(U),0} \end{array} \right)	,\ 
	\Psi_{Q(U)n}\equiv \left( \begin{array}{c}
      	\psi_{Q(U),n}  \\ \bar\psi_{Q^c(U^c),n} \end{array} \right).
	\no\\
	\end{eqnarray}
Note that we can rewrite the above interactions according to the situation using relations below,
	\begin{eqnarray}
	&&\overline{\Psi^c}_{Qm}P_L\Psi_{Un}=\overline{\Psi^c}_{Un}P_L\Psi_{Qm},
	\quad \overline\Psi_{Un}P_R\Psi^c_{Qm}=\overline\Psi_{Qm}P_R\Psi^c_{Un},
	\no\\
	&&\overline\Psi_{H_u}P_L\Psi_{Un}=\overline{\Psi^c}_{Un}P_L\Psi_{H_u}, 
	\quad  \overline\Psi_{Un}P_L\Psi_{H_u}=\overline\Psi_{H_u}P_R\Psi_{Un}^c,
	\\
	&&\overline\Psi_{H_u}P_L\Psi_{Qn}=-\overline{\Psi^c}_{Qn}P_L\Psi_{H_u}, 
	\quad  \overline\Psi_{Qn}P_L\Psi_{H_u}=-\overline\Psi_{H_u}P_R\Psi_{Qn}^c .
	\no
	\end{eqnarray}
The minus sign in the last line is due to $SU(2)_L$ contraction. 

\subsection{Lagrangian on brane with gauge coupling}
Now we investigate gauge interactions on the brane. Here we consider gauge fields of non-abelian group. 
These gauge interactions are basically as same as those in ordinary ${{\cal N}=1, D=4}$, except for $\delta(y)$ factor. The interactions together with bulk $D$-term are 
	\begin{eqnarray}
	{\cal L}_{{\rm gauge5},\phi_H}\!\!\!\!&=&
	\frac{1}{2}D'^a(y) D'^a(y) 
	-\delta(y)\left\{g_5 D^a(y)  \phi_{H_u}^* T^a \phi_{H_u}\right\}
	\no\\&&
	+\delta(y)\left(ig_5 A^a_\mu(y)\right)\left\{\phi_{H_u}^{*}T^a(\partial^\mu \phi_{H_u})
 	-(\partial^\mu \phi_{H_u}^{*})T^a \phi_{H_u}	\right\}
	\no\\&&
	+\delta(y)\left\{g_5^2 A_\mu^a(y) A^{b\mu}(y)\phi_{H_u}^{*}T^a T^b \phi_{H_u}  \right\}
	\no\\&&
	+\delta(y)\left\{ -\sqrt{2}g_5\phi_{H_u}^*T^a \psi_{H_u} \lambda^a_1(y)
	-\sqrt{2}g_5 \bar\lambda^a_1(y)\bar\psi_{H_u}T^a \phi_{H_u} \right\}
	\no\\&& \quad \quad+({H_u} \leftrightarrow {H_d})
	\no\\
	&=&\frac{1}{2}D'^a(y) D'^a(y) 
	-\delta(y)\left\{ g_5 \left(D'^a(y)-{\cal D}_5\Sigma^a(y) \right) 
	 \phi_{H_u}^* T^a \phi_{H_u}\right\}\no\\
	&&
	+\delta(y)\left(ig_5 A^a_\mu(y)\right)\left\{\phi_{H_u}^{*}T^a(\partial^\mu \phi_{H_u})
 	-(\partial^\mu \phi_{H_u}^{*})T^a \phi_{H_u}	\right\}
	\no\\&&
	+\delta(y)\left\{g_5^2 A_\mu^a(y) A^{b\mu}(y)\phi_{H_u}^{*}T^a T^b \phi_{H_u}  \right\}
	\no\\&&
	+\delta(y)\left\{ -\sqrt{2}g_5\phi_{H_u}^*T^a \psi_{H_u} \lambda^a_1(y)
	-\sqrt{2}g_5 \bar\lambda^a_1(y)\bar\psi_{H_u}T^a \phi_{H_u} \right\}
	\no\\&&  \quad \quad+({H_u} \leftrightarrow {H_d})
	\end{eqnarray}
where $D'^a=D^a+{\cal D}_5 \Sigma^a$.   As for Yukawa interaction, we perform the KK expansion and $y$ integral,
	\begin{eqnarray}
	\hspace{-10pt}\int_{0}^{2\pi R}\!\!\!dy~{\cal L}_{{\rm gauge5},\phi_H}
	\hspace{-20pt}&&=
	\frac{1}{2}\sum_{n=-\infty}^\infty (D'^{a}_n)^2
	-\sum_{n=-\infty}^\infty \!\! g~D'^{a}_n
	( \phi_{H_u}^* T^a \phi_{H_u})\no\\
	&&+\sum_{n=1}^\infty \sqrt{2}g\frac{n}{R}
	~\Sigma^a_n \left\{ \phi_{H_u}^* T^a \phi_{H_u}+\phi_{H_d}^* T^a \phi_{H_d} \right\}
	\no\\
	&&
	+ \sum_{n=0}^\infty (i\sqrt{2}\eta_n )gA^a_{(n)\mu}
	\left\{\phi_{H_u}^{*}T^a(\partial^\mu \phi_{H_u})
 	-(\partial^\mu \phi_{H_u}^{*})T^a \phi_{H_u}	\right\}
	\no\\&&
	+\sum_{m,n=0}^\infty (2\eta_m\eta_n)gA_{(m)\mu}^a A^{b\mu}_{(n)}\phi_{H_u}^{*}T^a T^b \phi_{H_u}  
	\no\\&&
	-\sum_{n=-\infty}^\infty \sqrt{2}g \left\{ \phi_{H_u}^*T^a \overline\Psi_{H_u} P_L \lambda^a_n
	+\bar\lambda^a_nP_R \Psi_{H_u}T^a \phi_{H_u} \right\}
	\no\\&&  \quad \quad+({H_u} \leftrightarrow {H_d})
	\end{eqnarray}
where $g\equiv g_5/\sqrt{2\pi R}$,  and the KK expansions of $D', A_\mu,\Sigma$, and $\lambda_1$, are,
	\begin{eqnarray}
	D'(y)=\sum_{n=-\infty}^\infty\frac{D'_n}{\sqrt{2\pi R}} \cos\frac{ny}{R}, \quad
	A_\mu(y) =\sum_{n=0}^\infty\frac{\eta_n~A_{(n)\mu}}{\sqrt{\pi R}} \cos\frac{ny}{R},
	\no\\
	\Sigma(y) =\sum_{n=0}^\infty\frac{ \Sigma_n}{\sqrt{\pi R}} \sin\frac{ny}{R}, \quad 
	\lambda_1(y)=\sum_{n=-\infty}^\infty P_L\frac{\lambda_n}{\sqrt{2\pi R}} \cos\frac{ny}{R}.
	\end{eqnarray}
The equation of motion for $D'$-terms is given by, 
	\begin{eqnarray}
	D'^a_n= g~( \phi_{H_u}^* T^a \phi_{H_u}) 
	+g~( \phi_{H_d}^* T^a \phi_{H_d})+\cdots .
	\end{eqnarray}
Here, $\cdots$ represents terms from other matters, but they are irrelevant for the later calculation of $\phi_{H_{u,d}}$ vacuum polarizations since the generator is traceless ${\rm Tr}(T^a)=0$. 
\footnote{For the $U(1)_Y$ case, the generator is not traceless, but the terms from matters such as $|\phi_{H_u}|^2|\phi_{Q}|^2$ are again irrelevant. This comes from the fact that $U(1)_Y$ gauge symmetry is anomaly free but not from consequence of supersymmetry. This is understood by the renormalization of tadpole term $\int d^4\theta V\supset D$ which is allowed only for abelian gauge field. Since this is in the K\"ahlar potential and has a coefficient of (mass)$^2$, quadratic divergences proportional to $U(1)_Y$ charge actually appear for this $D$-term. The cancellation condition of this term is Tr$Y$=0 which corresponds to one of the conditions of anomaly cancellation. See Ref.~\cite{Fischler:1981zk} for derivation. }
The equation of motion for $D$-term leads to, 
	\begin{eqnarray}
	&&\!\!\!\!\!\!\frac{1}{2}\sum_{n=-\infty}^\infty (D'^{a}_n)^2
	-\sum_{n=-\infty}^\infty \!\! 
	g~D'^{a}_n( \phi_{H_u}^* T^a \phi_{H_u}+ \phi_{H_d}^* T^a \phi_{H_d}+\cdots)\no\\	
	&&\rightarrow
	-\frac{g^2}{2}\sum_{n=-\infty}^\infty \left( \phi_{H_u}^* T^a \phi_{H_u}+ \phi_{H_d}^* T^a \phi_{H_d}+\cdots\right)^2 .
	\end{eqnarray}
Similar to the case of Yukawa interaction, these quartic couplings have infinite sum, but they are also cancelled by the effective quartic couplings from $\Sigma$ exchange. 
 
\section{Radiative Corrections}
\subsection{Dimensional analysis for radiative corrections}\label{sec:dimanalysis}
The expressions in Eq.~(\ref{eq:summary}) receive corrections from 
physics above and at $1/R$.  In the 5D picture, corrections above 
$1/R$ come from brane-localized kinetic terms for the gauge and matter 
supermultiplets, and affect $M_{1/2}$, $m_{\tilde{f}}^2 \equiv 
m_{\tilde{Q},\tilde{U},\tilde{D},\tilde{L},\tilde{E}}^2$, and $A_0$. 
These terms have tree-level (threshold) contributions at $\Lambda$ and radiative 
ones between $1/R$ and $\Lambda$.  From dimensional analysis, the 
size of the radiative contributions is
\begin{equation}
  \frac{\delta M_{1/2}}{M_{1/2}},\, 
    \frac{\delta m_{\tilde{f}}^2}{m_{\tilde{f}}^2},\, 
    \frac{\delta A_0}{A_0}
  \approx {\cal O}\left(\frac{g^2,\, y^2}{16\pi^2} \log(\Lambda R)\right).
\label{eq:corr-1}
\end{equation}

Then we discuss what scale is appropriate for the cutoff, to be more explicit,  up to what scale of the cutoff the prediction from the Scherk-Schwarz mechanism is valid. This is estimated by {\it Na\"ive Dimensional Analysis} in higher dimension \cite{Chacko:1999hg}. 
Higher dimensional theories are not renormalizable and considered as effective theories.  
The effective theory picture is not valid at certain energy scale where the the radiative correction exceeds tree-level term. 

The size of radiative corrections is related to the phase-space factors in loop integrals which is dimension-dependent. In $D$-dimensional theories, the typical loop integral is
	\begin{eqnarray}
	\int \frac{d^Dp}{(2\pi)^D}&\sim& \frac{\Omega_{D-1}}{(2\pi)^2}\int \frac{dp^2}{2}p^{D-2}f(p^2)
	=
	\frac{\pi^{D/2}}{(2\pi)^D\Gamma(D/2)}\int {dp^2}p^{D-2}f(p^2) 
	\end{eqnarray}
Then each loop integral is suppressed by the loop factor $L_D$, 
	\begin{eqnarray}
	{L_D}=(4\pi)^{D/2}\Gamma(D/2) \ .
	\end{eqnarray}
The specific values of $L_D$ are $16\pi^2  \ (D=4)$, $24\pi^3  \ (D=5)$, $128\pi^3  \ (D=6),\cdots$. 
When one computes radiative corrections of gauge and Yukawa couplings, there appear quadratic- and linear-divergence rather than log-divergence,
\footnote{Power of divergences can be understood by dimensional analysis. For instance, the gauge coupling in the 5D has mass dimension of $-1/2$, and the radiative corrections  along with $g_5^2$ are linearly divergent. 
} 
	\begin{eqnarray}
	{\cal L}_5&\sim&
	\int d^2\theta\ \delta(y)\left( 
	1+ \frac{{\cal O}(g_4^2)}{16\pi^2}\Lambda R
	+ \frac{{\cal O}(y_4^2)}{16\pi^2}(\Lambda R)^2 \right )H^\dag H
	\no\\&&\quad
	+\int d^2\theta\ \frac{1}{g_5^2}\left( 
	1+ \frac{{\cal O}(g_5^2)}{24\pi^3}\Lambda \right )W^\alpha W_\alpha .
	\end{eqnarray}
The loop factors are different because the Yukawa interactions are localized on the brane at $y=0$ while the gauge fields are living in the bulk. 
Thus our effective higher dimensional field theory is valid for $\Lambda R \lesssim 4\pi$, and 
accoding to Eqs.~(\ref{eq:corr-1},\ref{eq:corr-thresh})  the corrections 
to Eq.~(\ref{eq:summary}) from physics above $1/R$ are always {subdominant}. 
%
%
(The same can also be seen in 
the 4D picture.  In this picture, ${\cal N}=2$ supersymmetry existing for 
the $n>0$ modes leads to nontrivial cancellations of the corrections 
to $M_{1/2}$, $m_{\tilde{f}}^2$ and $A_0$ from these modes.  In order 
to see the cancellations for the gauge multiplets, the effect of anomalies 
must be taken into account correctly.  The explicit demonstration 
of these nontrivial cancellations will be given elsewhere.)

\subsection{Finite threshold corrections}\label{subsec:threshold}

The corrections from physics at $1/R$ arise from non-local operators in 
5D.  They affect all the supersymmetry-breaking masses, and are of order 
$1/16\pi^2$.  Here we calculate only the contributions to the Higgs mass 
parameters, which could potentially affect the analysis of electroweak 
symmetry breaking. 

It is known that these radiative corrections from all the KK modes are surprisingly finite. The reason is because SUSY breaking of the Scherk-Schwartz mechanism is non-local in 5D while UV divergence is local effect. More discussion is found in Ref.~\cite{ArkaniHamed:2001mi} and also even two-loop level calculation is performed in Ref.~\cite{Delgado:2001xr}. 

In order to {\it match} and {\it run}, we need to know the threshold corrections at $E\approx 1/R$ after integrating out the non-zero KK modes.  Since the low-energy theory is matched onto the MSSM, we can use MSSM renormalization group running.  While the radiative corrections from all the KK modes are finite, they include the MSSM (zero mode) contributions, and then we subtract them in $\overline{DR}$ scheme. 
Schematically, the procedure is as follows. For some mass parameter $M$, the corrections in full theory and the MSSM are
	\begin{eqnarray}
	&&\delta M|_{\rm KK}= {\rm finite}\\
	&&\delta M|_{\rm MSSM}(Q)=\mbox{log-div}\ 
	\end{eqnarray}
where $Q$ is the renormalization scale in the $\overline{DR}$ scheme. Then the threshold correction at $E\approx 1/R$ is given by
	\begin{eqnarray}
	&&\delta M_{\rm}(Q)=\delta M|_{\rm KK}-\delta M|_{\rm MSSM}(Q). \label{eq:match}
	\end{eqnarray}
This value can be used as initial condition of RG running at scale $Q$. For the check of this calculation, we must observe the cancelation of $\log \alpha$ because this term comes from IR-physics and is not the effect of integrating out massive particles.

\begin{figure*}[ht]
\begin{center}
  \includegraphics[clip,width=.34\textwidth]{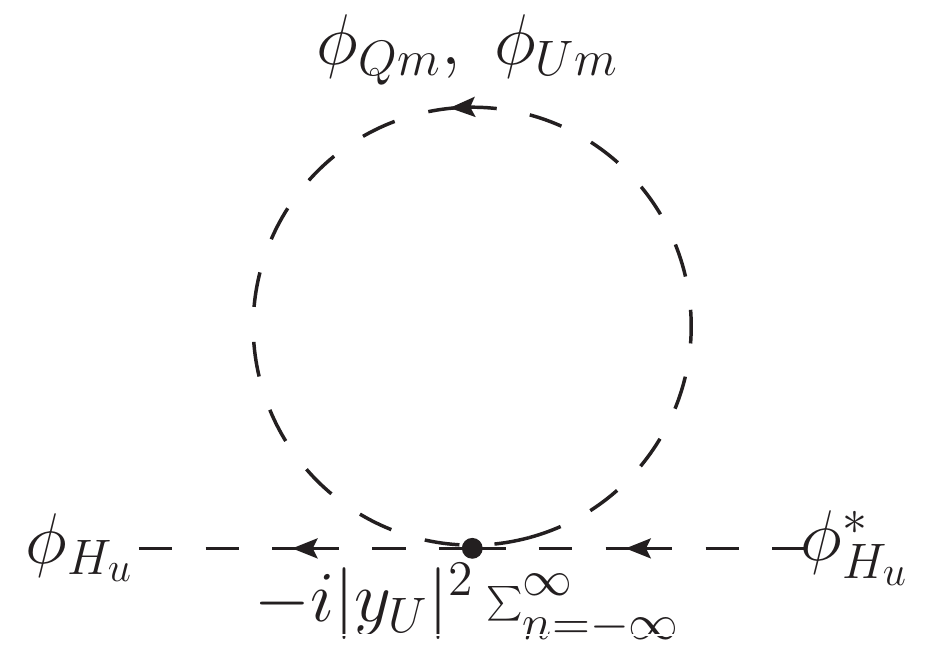}\hspace{10pt}
  \includegraphics[clip,width=.40\textwidth]{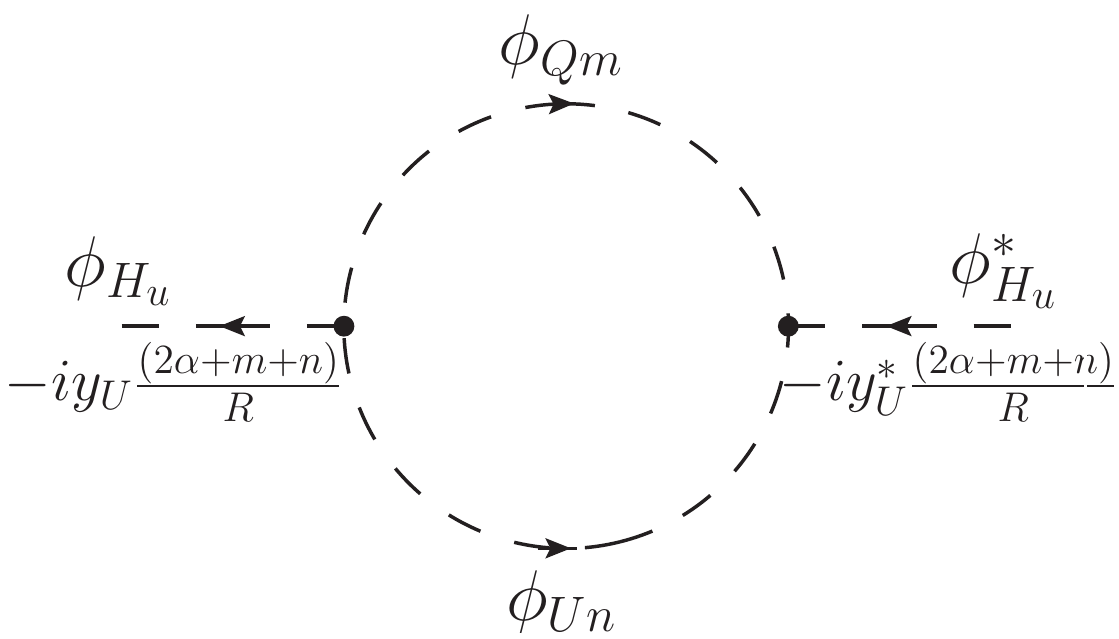}
  \includegraphics[clip,width=.40\textwidth]{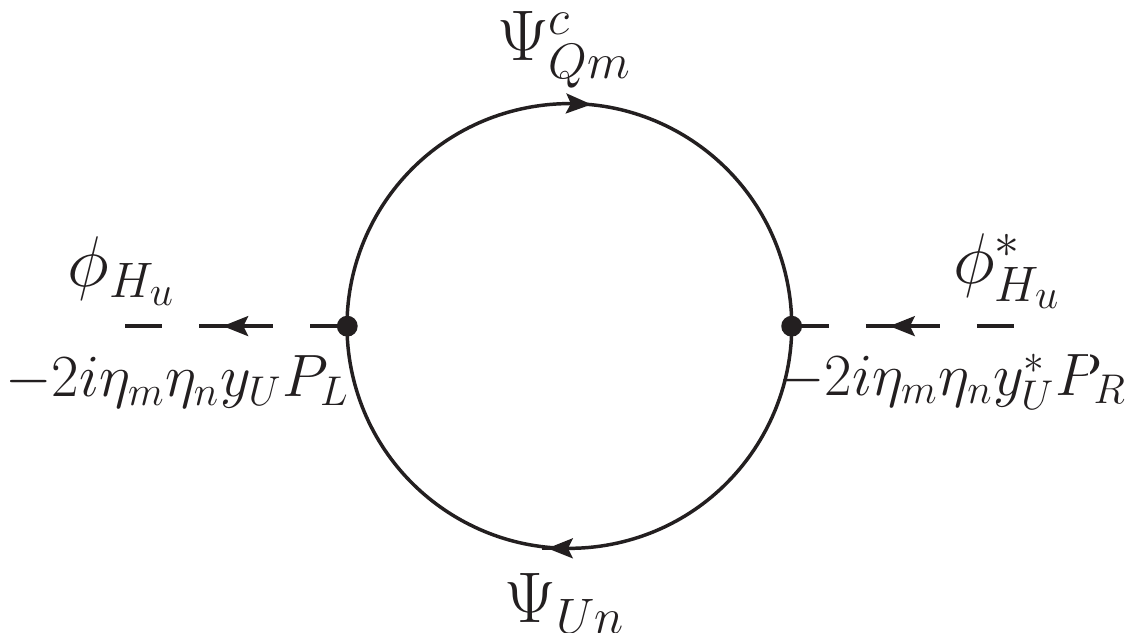}
\caption{Bosonic loops of radiative corrections to $m_{H_u}^2$ with Yukawa coupling are on the top, and fermionic loops are on the bottom. }
\label{mhu2ytB}
\end{center}
\end{figure*}
We first calculate radiative corrections from all the KK modes to $m_{H_{u}}^2$ with Yukawa interaction for up-type quark  in Fig.~\ref{mhu2ytB} . The contributions of bosonic loops in the limit of zero external momentum are
	\begin{eqnarray}
	-i\Pi_{\rm Bosonic}^{y_U}&&=N_c\int \frac{d^4k}{(2\pi)^4} \sum_{m,n=-\infty}^\infty\bigg\{
	 -i|y_U|^2\frac{i}{k^2-(\hat{m}+\hata)^2}\times2
	\no\\
	&&  -iy_U\left(2\hata+\hat{m}+\hat{n}\right)\frac{i}{k^2-(\hata+\hat{n})^2}(-iy_U^*)\left(2\hata+\hat{m}+\hat{n}\right)\frac{i}{k^2-(\hata+\hat{m})^2}
	\bigg\}
	\no\\\\
	&&=N_c\int \frac{d^4k}{(2\pi)^4} \sum_{m,n=-\infty}^\infty\bigg\{
	 2|y_U|^2 k^2~\frac{1}{k^2-(\hat{m}+\hata)^2}\frac{1}{k^2-(\hat{n}+\hata)^2}
	\no\\
	&&   \hspace{120pt}
	+2|y_U|^2\frac{\hat{m}+\hata}{k^2-(\hat{m}+\hata)^2}\frac{\hat{n}+\hata}{k^2-(\hat{n}+\hata)^2}
	\bigg\},
	\end{eqnarray}
where $\hata=\frac{\alpha}{R}, \hat{m}=\frac{m}{R}$ and $\hat{n}=\frac{n}{R}$. The contributions of fermionic loops are
	\begin{eqnarray}
	-i\Pi_{\rm Fermionic}^{y_U}&&=-N_c\int \frac{d^4k}{(2\pi)^4} \sum_{m,n=0}^\infty
	\eta_n^2\eta_m^2~{\rm Tr}\bigg[
	 (-2 i y_U) P_L\frac{i}{\slashed{k}+\hat{m}}
	 (-2 i y_U^*)P_R\frac{i}{\slashed{k}+\hat{n}}
	 \bigg]
 	\no\\
	&&=N_c\int \frac{d^4k}{(2\pi)^4} \sum_{m,n=0}^\infty
	(-4\times2)\eta_n^2\eta_m^2~|y_U|^2 k^2~
	  \frac{1}{k^2-\hat{m}^2}
	 \frac{1}{k^2-\hat{n}^2}
	\no\\
	&&=N_c\int \frac{d^4k}{(2\pi)^4} \sum_{m,n=-\infty}^\infty
	(-2)~|y_U|^2 k^2~
	  \frac{1}{k^2-\hat{m}^2}
	 \frac{1}{k^2-\hat{n}^2}.
	\end{eqnarray}
We perform summation and integral after combining them, 
	\begin{eqnarray}
	\delta m_{H_u}^2|^{y_U}_{\rm KK}\!\!\!\!&=&\Pi_{\rm Bosonic+Fermionic}^{y_t}\no\\
	&=&2i N_c |y_U|^2 \int \frac{d^4k}{(2\pi)^4} \sum_{m,n=-\infty}^\infty
	\bigg\{  \frac{k^2}{(k^2-(\hata+\hat{m})^2)(k^2-(\hata+\hat{n})^2)}
	\no\\  
	&&\quad\quad
	\frac{\hat{m}+\hata}{k^2-(\hat{m}+\hata)^2}\frac{\hat{n}+\hata}{k^2-(\hat{n}+\hata)^2}
	  -\frac{k^2}{(k^2-\hat{m}^2)(k^2-\hat{n}^2)} \bigg\}
	\\
	&=&\frac{N_c|y_U|^2}{16\pi^2}\frac{3}{\pi^2 R^2}
	\left[ {\rm Li}_3(e^{2\pi i \alpha}) +{\rm Li}_3(e^{-2\pi i \alpha})-2\zeta(3)\right]
	\\
	&\approx&
	\frac{N_c|y_U|^2}{16\pi^2}\left(\frac{\alpha}{R}\right)^2
	[6\log(2\pi \alpha)^2	-18].  \label{eq:kk1}
	\end{eqnarray}
We leave the detail calculations in the {Appendix}~\ref{app:thresh}. A key for analytic calculations is that sums of $m$ and $n$ are factorized. Divergences appear from bosonic and fermionic diagrams with the same amount and are cancelled exactly leaving the finite results. In the last line, we expand with respect to $\alpha$.

\begin{figure*}[t]
\begin{center}
  \includegraphics[clip,width=.35\textwidth]{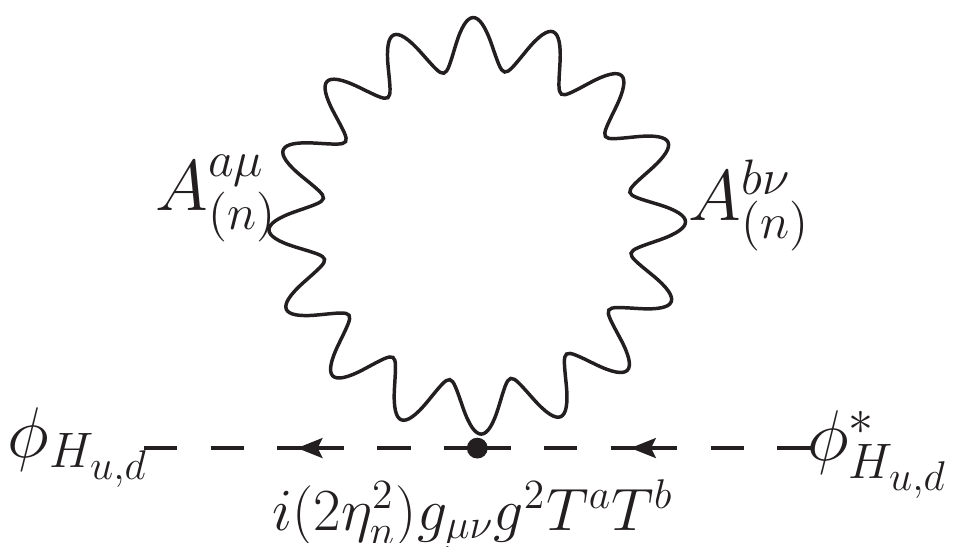}\hspace{10pt}
  \includegraphics[clip,width=.35\textwidth]{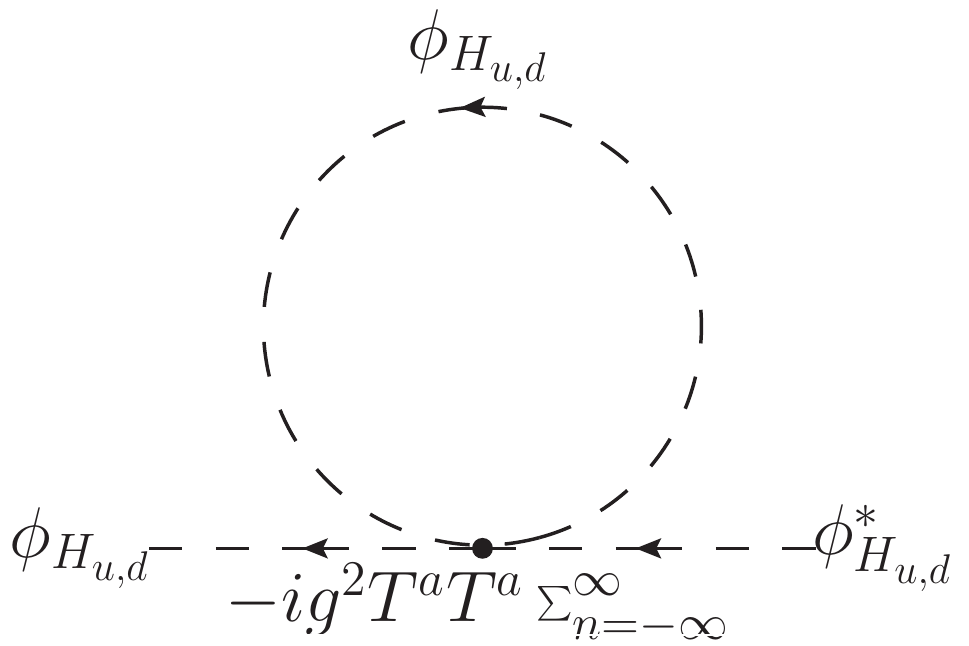}
\end{center}
\begin{center}
  \includegraphics[clip,width=.35\textwidth]{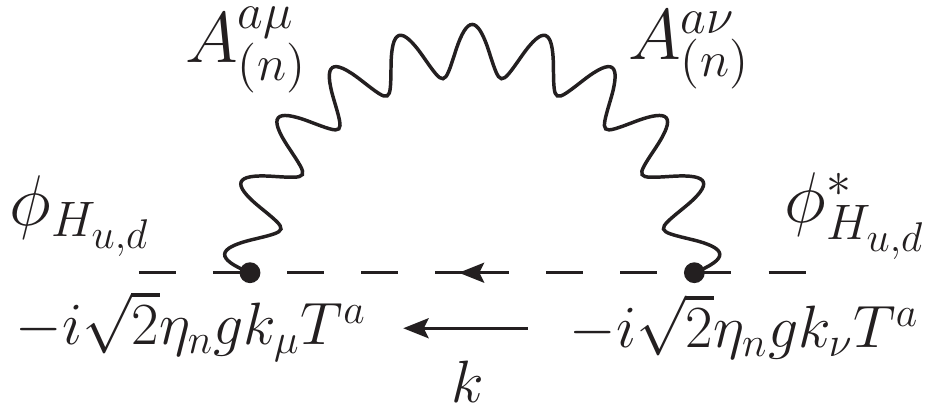}\hspace{10pt}
  \includegraphics[clip,width=.35\textwidth]{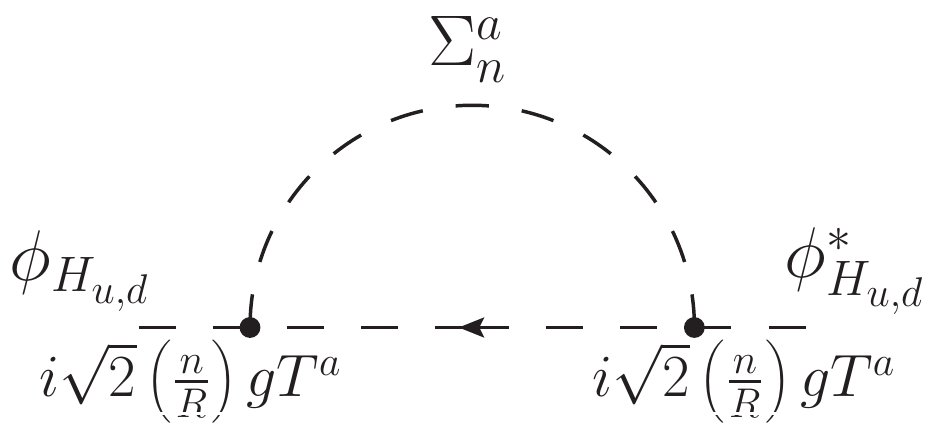}
\end{center}
\caption{Bosonic loops of radiative corrections to  $m_{H_{u,d}}^2$ with the gauge coupling. }
\label{mhu2gB}
\begin{center}
  \includegraphics[clip,width=.38\textwidth]{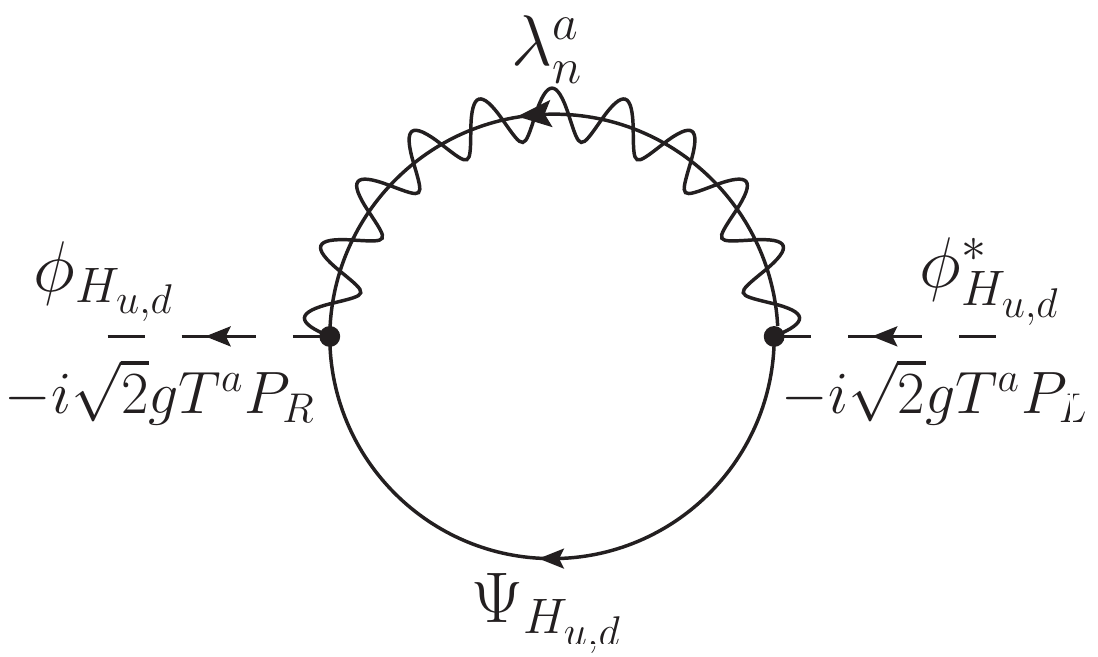}
\caption{Fermionic loops of radiative corrections to  $m_{H_{u,d}}^2$with the gauge coupling. }
\label{mhu2gF}
\end{center}
\end{figure*}

Radiative corrections from all the KK modes with gauge interaction to $m_{H_{u,d}}^2$ are calculated.   The contributions of bosonic loops in Fig.~\ref{mhu2gB}
 are
	\begin{eqnarray}
	\!\!\!\!-i\Pi_{\rm Bosonic}^{g}
	\!\!\!&=&C_2(r)\int \frac{d^4k}{(2\pi)^4} \bigg\{\sum_{n=0}^\infty
	 i(2\eta_n^2) g^2g_{\mu\nu}\frac{-ig^{\mu\nu}}{k^2-\hat{n}^2} 
	+ \sum_{n=-\infty}^\infty  -ig^2 \frac{i}{k^2}
	\no\\
	&&+ \sum_{n=0}^\infty(-i\sqrt{2}\eta_n  g) k_\mu \frac{-ig_{\mu\nu}}{k^2-\hat{n}^2}
	(-i\sqrt{2}\eta_n  g) k_\nu \frac{i}{k^2}
	\no\\&&
	+\sum_{n=0}^\infty\left(i\sqrt{2}g\hat{n}\right)\frac{i}{k^2-\hat{n}^2}\left(i\sqrt{2}g\hat{n}\right)\frac{i}{k^2}
	\bigg\}
	\\
	&=&C_2(r)g^2\int \frac{d^4k}{(2\pi)^4} \sum_{n=-\infty}^\infty
	\bigg\{\frac{4}{k^2-\hat{n}^2}+\frac{1}{k^2}
	-\frac{1}{k^2-\hat{n}^2}
	+\frac{\hat{n}^2}{k^2-\hat{n}^2}\frac{1}{k^2}
	\bigg\}
	\\
	&=&C_2(r)g^2\int \frac{d^4k}{(2\pi)^4} \sum_{n=-\infty}^\infty
	\frac{4}{k^2-\hat{n}^2},
	\end{eqnarray}
where $C_2(r)$ is Casimir invariant. Contributions of fermionic loops in Fig.~\ref{mhu2gF}
 are
	\begin{eqnarray}
	-i\Pi_{\rm Fermionic}^{g}\!\!\!\!
	&=&\!\!\!-C_2(r)\int \frac{d^4k}{(2\pi)^4} 
	\sum_{n=-\infty}^\infty\!\!\!{\rm Tr}\left[-i\sqrt{2} g P_R  \frac{i}{\slashed{k}+\hata+\hat{n}}
	(-i\sqrt{2} g P_L) \frac{i}{\slashed{k}} \right]
	\no\\&=&
	-C_2(r)g^2\int \frac{d^4k}{(2\pi)^4} 
	\sum_{n=-\infty}^\infty \frac{4}{k^2-(\hata+\hat{n})^2}.
	\end{eqnarray}
We combine them and perform summation and integral, 
	\begin{eqnarray}
	\delta m_{H}^2|^{g}_{\rm KK}\!\!&=&
	\Pi_{\rm Bosonic+Fermionic}^{g}\no\\
	&=&
	\!\! iC_2(r)g^2
		\int \frac{d^4k}{(2\pi)^4} 
	\!\sum_{n=-\infty}^\infty \!\!
	\bigg\{\frac{4}{k^2-\hat{n}^2} -\frac{4}{k^2-(\hata+\hat{n})^2} \bigg\}
	\\
	&=&\frac{C_2(r)g^2}{16\pi^2}\frac{-2}{\pi^2 R^2}
	\left[ {\rm Li}_3(e^{2\pi i \alpha}) +{\rm Li}_3(e^{-2\pi i \alpha})-2\zeta(3)\right]
	\\
	&\approx&
	\frac{C_2(r)g^2}{16\pi^2}\left(\frac{\alpha}{R}\right)^2
	[-4\log(2\pi \alpha)^2	+12] \label{eq:kk2}
	\end{eqnarray}

\begin{figure*}[t]
\begin{center}
  \includegraphics[clip,width=.38\textwidth]{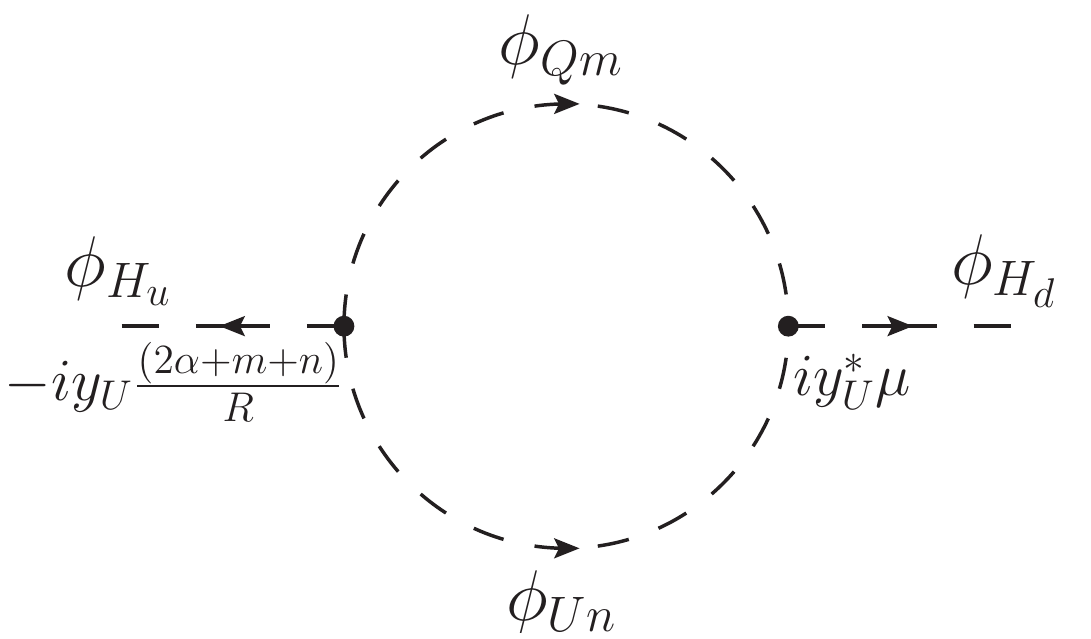}
  \includegraphics[clip,width=.38\textwidth]{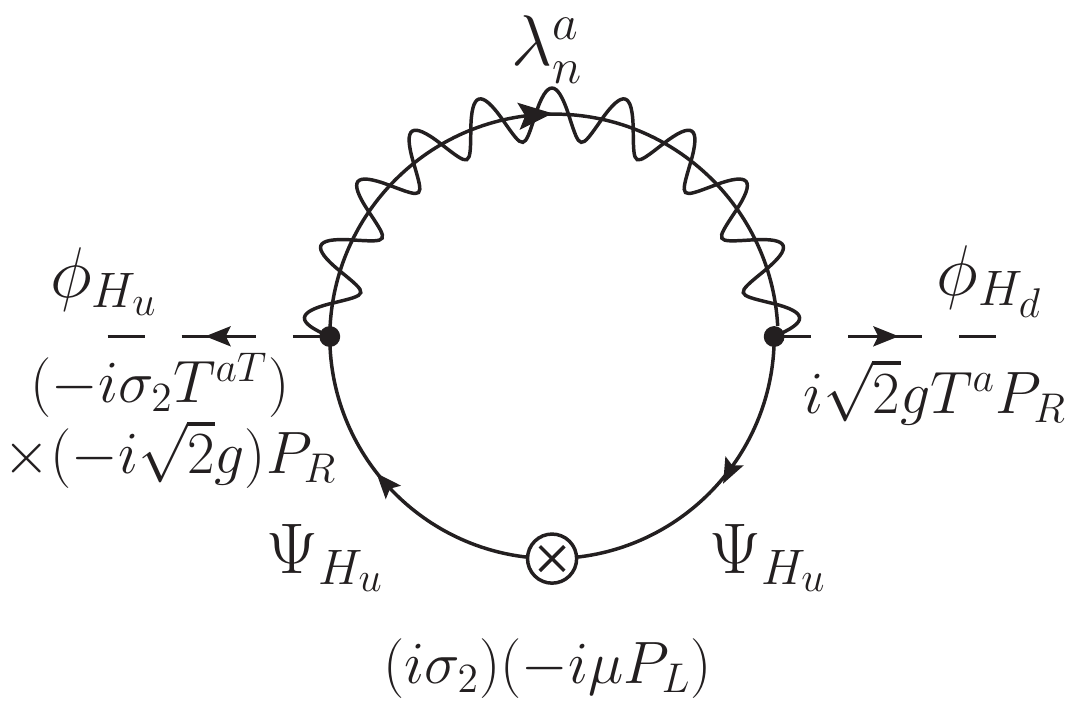}
  \caption{Loops of radiative corrections to  $b$ through Yukawa coupling  ({\it left}) and through gauge coupling  ({\it right}). }
\label{fig:b}
\end{center}
\end{figure*}

Also another Higgs soft mass, $b$, is radiatively generated. The contributions from all the KK modes with Yukawa coupling  in left of Fig.~\ref{fig:b} are
	\begin{eqnarray}
	-i\left(\delta b|^{y_U}_{\rm KK}\right)&&=N_c\int \frac{d^4k}{(2\pi)^4}\!\!\!\! \sum_{m,n=-\infty}^\infty
	\!\!\!\!-iy_U(2\hata+\hat{m}+\hat{n})
	\frac{i}{k^2-(\hata+\hat{m})^2}\frac{i}{k^2-(\hata+\hat{n})^2}
	iy_U^* \mu
	\no\\
	&&=-2N_c|y_U|^2 \mu  \int \frac{d^4k}{(2\pi)^4} \sum_{m,n=-\infty}^\infty
	\frac{(\hata+\hat{m})}{k^2-(\hata+\hat{m})^2}\frac{1}{k^2-(\hata+\hat{n})^2}
	\\
	&&=
	\frac{N_c |y_U|^2}{16\pi^2}\mu\frac{(-1)}{\pi R}
	\left[ {\rm Li}_2(e^{2\pi i \alpha}) -{\rm Li}_2(e^{-2\pi i \alpha})\right]
	\\
	&&\approx-i
	\frac{N_c |y_U|^2}{16\pi^2}\mu\left(\frac{\alpha}{R}\right)
	[-2\log(2\pi \alpha)^2	+4]. \label{eq:kk3}
	\end{eqnarray}
The contributions from all the KK modes with gauge coupling in right graph of Fig.~\ref{fig:b} are
	\begin{eqnarray}
	-i\left(\delta b|^{g}_{\rm KK}\right)&&=C_2(r)\int \frac{d^4k}{(2\pi)^4}\!\!\!\! \sum_{n=-\infty}^\infty
	\!\!{\rm Tr}\left[(-i\sqrt{2}gP_R)
	\frac{i}{\slashed{k}}(-i\mu P_L)\frac{i}{\slashed{k}}(i\sqrt{2}g P_R)\frac{i}{\slashed{k}+\hata+\hat{n}}\right]
	\no\\
	&&=4C_2(r) g^2\mu\int \frac{d^4k}{(2\pi)^4}\sum_{n=-\infty}^\infty
	\frac{(\hata+\hat{n})}{k^2-(\hata+\hat{n})^2}\frac{1}{k^2}
	\\
	&&=\frac{C_2(r) g^2}{16\pi^2}
	\mu\frac{2}{\pi R}
	\left[ {\rm Li}_2(e^{2\pi i \alpha}) -{\rm Li}_2(e^{-2\pi i \alpha})\right]
	\\
	&&\approx
	-i\frac{C_2(r) g^2}{16\pi^2}\mu\left(\frac{\alpha}{R}\right)
	[4\log(2\pi \alpha)^2	-8]	\label{eq:kk4}
	\end{eqnarray}
where mass insertion is used for $\mu$.
The mass insertion of $\mu$ and gaugino-Higgs-Higgsino interaction are given by
	\begin{eqnarray}
	 -\mu\psi_{H_u}\psi_{H_d}&=&-\mu\psi_{H_u}^\alpha(i\sigma_2)_{\alpha\beta}\psi_{H_d}^\beta
	\no\\&=&-\mu\overline\Psi_{H_u}^\alpha(i\sigma_2)_{\alpha\beta}P_L\Psi_{H_d}^\beta
	 \\
	\bar\lambda^a_nP_R \Psi_{H_u}T^a \phi_{H_u}
	&=&\phi_{H_u}T^{aT}\bar\lambda^a_nP_R \Psi_{H_u} 
	\no\\&=&(\phi_{H_u}i\sigma_2)(-i\sigma_2T^{aT})\bar\lambda^a_nP_R \Psi_{H_u} 
	\end{eqnarray}
Here we explicitly  show $(i\sigma_2)$ for contraction of $SU(2)_L$ doublets. And we use a relation,
	\begin{eqnarray}
	(-i\sigma_2)T^{aT}(i\sigma_2)=-T^a \ .
	\end{eqnarray}

On the other hand, contributions from the MSSM to the same parameters  are computed subtracting logarithmic divergences in $\overline{DR}$ scheme. The results   are given by 
	\begin{eqnarray}
	\delta{m_{H_u}^2}|_{{\rm MSSM}}^{y_U}&=&\frac{N_c|y_U|^2}{16\pi^2}
		\left(\frac{\alpha}{R}\right)^2
		\left[	-6\log\left(\frac{Q^2}{\alpha^2/R^2}\right) -2	\right], \label{eq:mssm1}
		\\
	\delta{m_{H_{u,d}}^2}|_{\rm MSSM}^{g}&=&
	\frac{C_2(r)}{16\pi^2}\left(\frac{\alpha}{R}\right)^2
	\left[	4\log\left(\frac{Q^2}{\alpha^2/R^2}\right) +4	\right],	\label{eq:mssm2}
	\\
	\delta{b}|_{{\rm MSSM}}^{y_U}&=&\frac{N_c |y_U|^2}{16\pi^2}
	\mu\left(\frac{\alpha}{R}\right)
	\left[	2\log\left(\frac{Q^2}{\alpha^2/R^2}\right) 	\right], \label{eq:mssm3}
	\\
	\delta{b}|_{\rm MSSM}^{g}&=&
	\frac{C_2(r)}{16\pi^2}\mu\left(\frac{\alpha}{R}\right)
	\left[	-4\log\left(\frac{Q^2}{\alpha^2/R^2}\right) -4	\right],	\label{eq:mssm4}
	\end{eqnarray}
We the corrections of $m_{H_u}^2, m_{H_u}^2$ and $b$ from all the KK modes and from only the MSSM, respectively, and apply the results for $y_t, g_1$, and $g_2$ couplings following Eq.~(\ref{eq:match}).  The Casimir invariants for $SU(2)_L$ and $U(1)_Y$ in $SU(5)$ normalization are $C_{2,SU(2)_L}^h=3/4$ and $C_{2,U(1)_Y}^h=3/20$.
We combine Eqs.~(\ref{eq:kk1}, \ref{eq:kk2}, \ref{eq:kk3}, \ref{eq:kk4}) with Eqs.~(\ref{eq:mssm1}, \ref{eq:mssm2}, \ref{eq:mssm3}, \ref{eq:mssm4}), 
	\begin{eqnarray}
	\delta{m_{H_u}^2}(Q)&=&\frac{N_cy_t^2}{16\pi^2}\left(\frac{\alpha}{R}\right)^2
	\left\{6\log\left[\frac{Q^2}{(2\pi R )^{-2}}\right]	-16\right\}
	\nonumber\\&&
	+\frac{\sum_{A=1,2} C_A^h g_A^2}{16\pi^2}\left(\frac{\alpha}{R}\right)^2
	\left\{-4\log\left[\frac{Q^2}{(2\pi R )^{-2}}\right]	+8\right\}+{\cal O}(\alpha^4)
	\\
	\delta{m_{H_d}^2}(Q)&=&
	\frac{\sum_{A=1,2} C_A^h g_A^2}{16\pi^2}\left(\frac{\alpha}{R}\right)^2
	\left\{-4\log\left[\frac{Q^2}{(2\pi R )^{-2}}\right]	+8\right\}+{\cal O}(\alpha^4)
	\\
	\delta{b}(Q)&=&
	\frac{N_cy_t^2}{16\pi^2}\mu \left(\frac{\alpha}{R}\right)
	\left\{-2\log\left[\frac{Q^2}{(2\pi R )^{-2}}\right]	+4\right\}
	\nonumber\\&&
	+\frac{C_2^h g_A^2}{16\pi^2}\mu\left(\frac{\alpha}{R}\right)
	\left\{4\log\left[\frac{Q^2}{(2\pi R )^{-2}}\right]	-4\right\}+{\cal O}(\alpha^3)
	\end{eqnarray}
We check  $\log\alpha$ are certainly cancelled. And we choose the renormalization scale, 
	\begin{eqnarray}
	Q_{\rm RG}=\frac{1}{2\pi R}. 
	\end{eqnarray}
and then we find 
\begin{eqnarray}
  \delta m_{H_u}^2 
  &=& \left( -\frac{3 y_t^2}{\pi^2} 
    + \frac{3 (g_2^2 + g_1^2/5)}{8\pi^2} \right) 
    \left( \frac{\alpha}{R} \right)^2,
\\[5pt]
  \delta m_{H_d}^2 
  &=& \frac{3 (g_2^2 + g_1^2/5)}{8\pi^2} 
    \left( \frac{\alpha}{R} \right)^2,
\\[5pt]
  \delta b 
  &=& \left( \frac{3 y_t^2}{4\pi^2} 
    - \frac{3 (g_2^2 + g_1^2/5)}{16\pi^2} \right) 
    \mu\frac{\alpha}{R}.
\label{eq:corr-2}
\end{eqnarray}
Using the above results, we run MSSM parameters down to low-energy to discuss phenomenology in the next chapter. \footnote{The factor difference from the original result in Ref.~\cite{Murayama:2012jh} is due to scheme difference. In Ref.~\cite{Murayama:2012jh} cutoff regularization rather than $\overline{DR}$ scheme is used when zero mode contributions are subtracted. } 

%

\section{Brane Kinetic Terms at Tree-Level}\label{sec:branekin}
In addition to brane-kinetic terms which are radiatively generated, those at tree-level  which can be considered as threshold corrections at the cutoff scale  are potentially dangerous because they may change the prediction of Eq.~(\ref{eq:summary}) or may be strongly constrainted by low-energy measurements. Therefore, basically we assume the tree-level contributions do not exceed the radiative ones of Eq.~(\ref{eq:corr-1}) with $\log(\Lambda R) 
\rightarrow {\cal O}(1)$, that is, 
\begin{equation}
{  \frac{\delta M_{1/2}}{M_{1/2}},\, 
    \frac{\delta m_{\tilde{f}}^2}{m_{\tilde{f}}^2},\, 
    \frac{\delta A_0}{A_0}\Bigg|_{{\rm at }\ \Lambda}
  \approx {\cal O}\left(\frac{g^2,\, y^2}{16\pi^2} \right).}
\label{eq:corr-thresh}
\end{equation}

This assumption can be reasonable because effects from brane kinetic terms are reduced by the volume of the extra dimension \cite{Hall:2002rk}.  Let us  focus on $\cal P$ even chiral superfields which have zero modes and hence are relevant to low-energy physics. Then kinetic terms for them are, for example, 
	\begin{eqnarray}\label{eq:brane01}
	\int d^4x\ dy \int d^4\theta \left[Z_{{\rm bulk},Q}\delta^{ij}+Z_{{\rm brane},Q}^{ij} \delta(y)\right]Q_{i}^\dag Q_{j}, 
	\end{eqnarray}
where $i,j$ are flavor indices. Now suppose the bulk and brane kinetic terms have ``comparable'' sizes at the cutoff scale $\Lambda$.  It is natural to rescale $Z_{\rm brane}$ by $\Lambda$ such as $Z_{\rm brane}={\tilde Z}_{\rm brane}/\Lambda$, and the comparable sizes of kinetic terms imply
	\begin{eqnarray}
	Z_{\rm bulk}\approx {\tilde Z}_{\rm brane}\ .
	\end{eqnarray}
We normalize $Z_{\rm bulk}$ to be 1, and Eq.~(\ref{eq:brane01}) is rewritten as
	\begin{eqnarray}\label{eq:brane1}
	\int d^4x\ dy \int d^4\theta \left[1+\frac{c^{ij}_Q}{\Lambda} \delta(y)\right]Q_{i}^\dag Q_{j}. 
	\end{eqnarray}
where $c^{ij}_Q=Z_{{\rm bulk},Q}/Z_{{\rm bulk},Q}\approx {\cal O}(1)$. If we perform $y$ integral and KK-expansion of the chiral superfield, the kinetic term for zero mode is
	\begin{eqnarray}\label{eq:brane1}
	\int d^4x \int d^4\theta \left[1+\frac{c^{ij}_Q}{(2\pi R)\Lambda} \right]Q_{i}^{(0)\dag} Q_{j}^{(0)}. 
	\end{eqnarray}
When we take the maximum value of the cutoff which we discussed in Sec.~\ref{sec:dimanalysis}, the effect of brane terms is suppressed, 
	\begin{eqnarray}
	\frac{c_Q}{2\pi\Lambda R}\sim \frac{c_Q}{8\pi^2}\sim {\cal O}(0.01).
	\end{eqnarray}
Thus we expect the effect of brane kinetic terms is small. Although unlikely, it is possible to have  anomalously large coefficients for brane terms. We briefly discuss this impact to the phenomenology in Sec.~\ref{sec:comment-branekin}. 


\chapter{Phenomenology of Compact Supersymmetry}\label{compact:pheno}
The Compact Supersymmetry leads to  viable phenomenology despite the fact that it essentially has two less free parameters than the conventional CMSSM. In the low-energy, we can treat the model as the MSSM. Using the common notation in the MSSM, parameters at the scale $\frac{1}{2\pi R}$ are given by 
\begin{equation}
\begin{array}{c}
  M_{1/2} = \frac{\alpha}{R},
\quad
  m_{\tilde{Q},\tilde{U},\tilde{D},\tilde{L},\tilde{E}}^2 
  = \left( \frac{\alpha}{R} \right)^2,
\quad
  A_0 = -\frac{2\alpha}{R},
\vspace{5pt}\\
  \mu \neq 0,
\quad
  b =  \left( \frac{3 y_t^2}{4\pi^2} 
    - \frac{3 (g_2^2 + g_1^2/5)}{16\pi^2} \right) 
    \mu\frac{\alpha}{R},
\\[10pt]
  m_{H_u}^2 =\left( -\frac{3 y_t^2}{\pi^2} 
    + \frac{3 (g_2^2 + g_1^2/5)}{8\pi^2} \right) 
    \left( \frac{\alpha}{R} \right)^2,
\quad
  m_{H_d}^2 = \frac{3 (g_2^2 + g_1^2/5)}{8\pi^2} 
    \left( \frac{\alpha}{R} \right)^2.
\quad
\end{array}
\end{equation}
We discuss the Higgs sector, the spectrum, experimental limits and dark matter phenomenology. 

\section{Higgs Sector}

\subsection{Electroweak symmetry breaking}\label{sec:EWSB}
\begin{figure}[ht]
\begin{center}
  \includegraphics[clip,width=.7\textwidth]{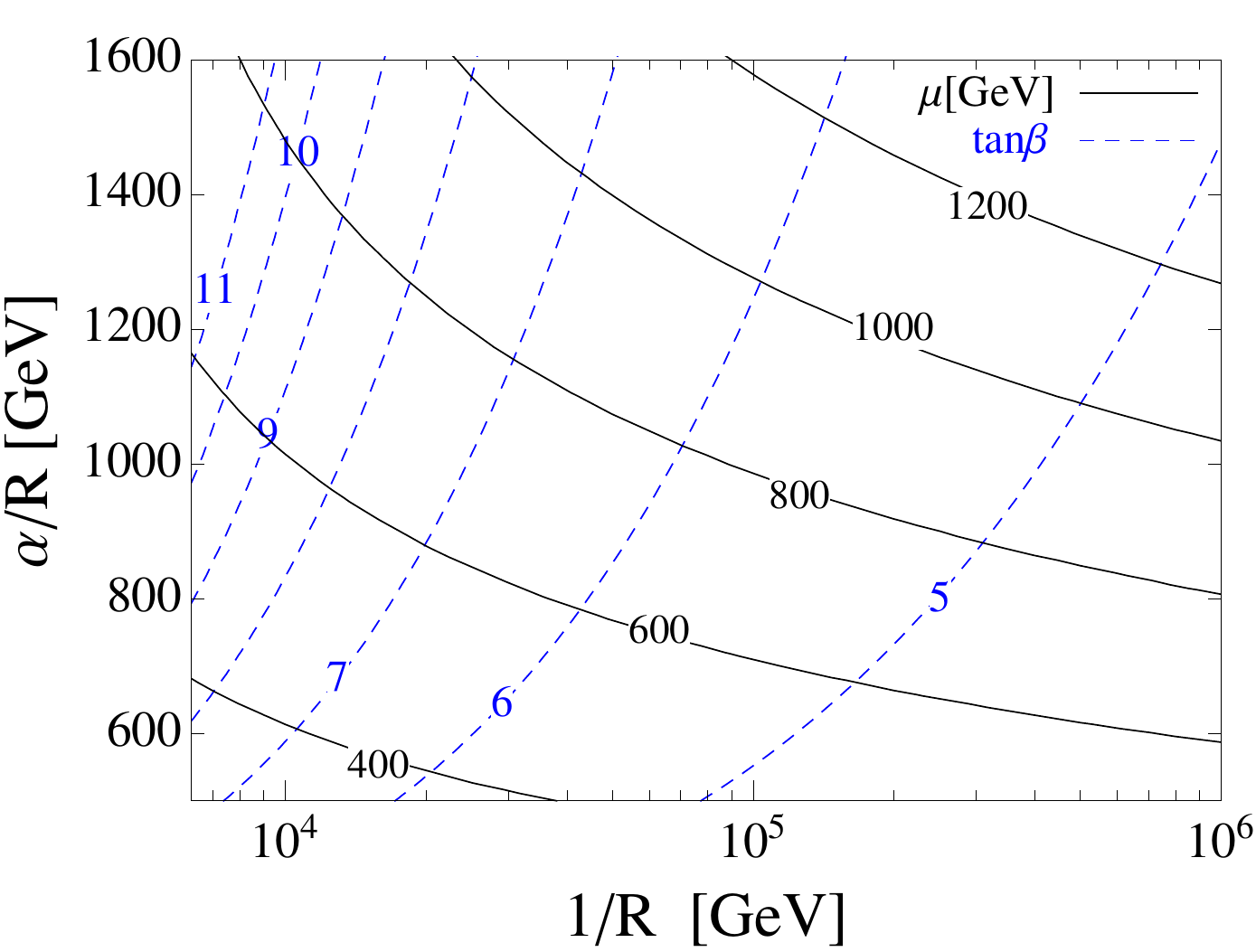}
\end{center}
\caption{$\mu$ and $\tan\beta$ at SUSY scale ($Q=\alpha/R$) as a function of $\alpha/R$ and $1/R$.}
\label{fig:mutanb}
\end{figure}

This  model has essentially three parameters: $\mu, 1/R, \alpha/R$. Since the measured VEV can be as a condition of EWSB, there are only two free parameters. We treat $\alpha/R$ and $1/R$ as free parameters and the other electroweak parameters ($\mu, \tan\beta$) are determined by the EWSB conditions. In Fig.~\ref{fig:mutanb},  we show $\mu$ and $\tan\beta$ at the scale of $\alpha/R$ in terms of $\alpha/R$ and $1/R$. For this calculation {\tt SOFTSUSY~3.3.1} \cite{Allanach:2001kg}  is used. 
 
The result is basically understood by tree-level EWSB conditions. The vacuum conditions for $H_u$ and $H_d$ are
	\begin{eqnarray}
	&&m^2_{H_u} + |\mu|^2 - b\cot \beta - (m^2_Z /2) \cos(2\beta) = 0, \label{Humin}
	\\
	&&m^2_{H_d} + |\mu|^2 - b\tan \beta + (m^2_Z /2) \cos(2\beta) = 0.\label{Hdmin}
	\end{eqnarray}
Let us estimate $\mu$ analytically. Eq.~(\ref{Humin}) is approximated to $-|\mu|^2 \approx m_{H_u}^2$ neglecting $m_Z$ and $b$, and $m_{H_u}^2$ is estimated by the threshold correction and one-loop RGE, 
	\begin{eqnarray}
	-|\mu|^2\approx m_{H_u}^2 &\approx&\delta m_{H_u}^2 +\frac{d m_{H_u}^2}{d \log Q} L
	\no\\ &\sim&-\frac{3y_t^2}{\pi^2}\left(\frac{\alpha}{R}\right)^2 
	-\frac{9y_t^2}{4\pi^2}\left(\frac{\alpha}{R}\right)^2 L
	 . \label{eq:mu}
	\end{eqnarray}
where $L=\log \frac{(2\pi R)^{-1}}{\alpha/R}$ is RGE leading-logarithm. The size of  $\mu$ scales with $\alpha/R$,  and $\mu$ becomes larger for larger $1/R$ as shown in Fig.~\ref{fig:mutanb}. For better estimation, we should consider next-to-leading logs by RGEs of $m_{\tilde{Q}_3}$, $m_{\tilde{u}_3}$, and $A_t$.

It is rather complicated for $\tan\beta$ behavior.  We approximate Eq.(\ref{Hdmin}) and consider RGE of $b$, 
	\begin{eqnarray}
	&&\tan\beta\approx \frac{|\mu|^2}{b}, \label{eq:tanb}\\ 
	&&\frac{db}{d\log Q}L\approx \frac{3y_t^2}{4\pi^2} \mu \frac{\alpha}{R}L . \label{eq:b}
	\end{eqnarray}
From Eqs.~(\ref{eq:mu},\ref{eq:tanb},\ref{eq:b}), roughly speaking, $b$ grows with $L^{3/2}$ whereas $|\mu|^2$ grows with $L$, and hence $\tan\beta$ decreases for large $1/R$. The numerical result is shown by dashed line of Fig.~\ref{fig:mutanb}.

\subsection{Higgs mass and naturalness}
\begin{figure}[ht]
\begin{center}
  \includegraphics[clip,width=.7\textwidth]{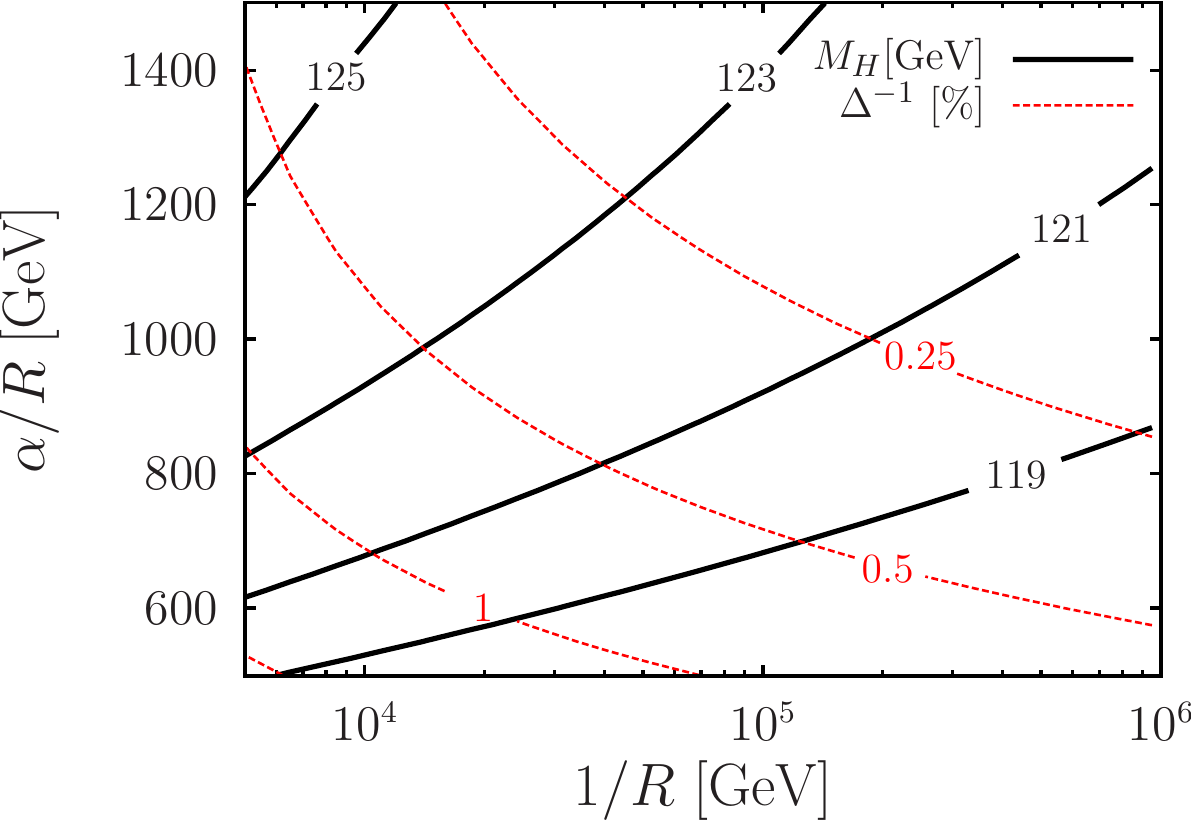}
\end{center}
\caption{The lightest Higgs boson mass $M_H$ (in GeV) and the fine-tune 
 parameter $\Delta^{-1}$.  Note that there is an approximately $3~{\rm GeV}$ 
 systematic error in theoretical computation of $M_H$.}
\label{fig:higgs}
\end{figure}

We calculate the MSSM mass spectrum using 
{\tt SOFTSUSY~3.3.1}~\cite{Allanach:2001kg} and the lightest Higgs 
boson mass using {\tt FeynHiggs~2.8.6}~\cite{Heinemeyer:1998yj}.  In 
Fig.~\ref{fig:higgs}, we plot the contours of the mass of the lightest 
Higgs boson, $M_H$, and the fine-tune parameter, defined by 
$\Delta^{-1}$ where
	\begin{eqnarray}
	\Delta \equiv \max\left|\frac{\partial \log m_Z^2}{\partial \log x} \right|,
	 \ \ x=\alpha, \mu, 1/R, g_3, y_t   . 
	\end{eqnarray}
%
in the $1/R$-$\alpha/R$ plane.  The fine-tuning parameter is determined mostly by $x = \mu$. It is understood from Eq.(\ref{Humin}) some cancellation between $m_{H_u}^2$ and $|\mu|^2$ is needed to give observed $m_Z$. It turns out that  the tuning level is $\cal O$(1)\%. 

The lightest Higgs boson mass is enhanced because the radiative correction from stop left-right mixing $A_t$ is nearly maximized. The overall mass is larger with large $\tan\beta$ (small $1/R$). 
In the calculation, we have used the top-quark mass of $m_t = 
173.2~{\rm GeV}$~\cite{Lancaster:2011wr}.  Varying it by $1\sigma$, 
$\varDelta m_t = \pm 0.9~{\rm GeV}$, affects the Higgs boson mass 
by $\varDelta M_H \approx \pm 1~{\rm GeV}$.  Also, theoretical 
errors in $M_H$ are large with $|\varDelta M_H| \approx 
2~\mbox{--}~3~{\rm GeV}$ in {\tt FeynHiggs~2.8.6}, which is also implied by Ref.~\cite{Allanach:2004rh}, so the regions 
with $M_H \gtrsim 121~\mbox{--}~123~{\rm GeV}$ in the plot are not 
necessarily incompatible with the $125~{\rm GeV}$ Higgs boson at the LHC~\cite{:2012gk, ATLAS:2012ae}. 
Indeed, using the recently-released 
program {\tt H3m}~\cite{Kant:2010tf}, which includes a partial three-loop 
effect, we find that the corrections to $M_H$ from higher order effects 
are positive and of order a few GeV in most of the parameter region in 
the plot.


\section{Sparticle Spectrum}
\begin{figure}[ht]
\begin{center}
  \includegraphics[clip,width=.7\textwidth]{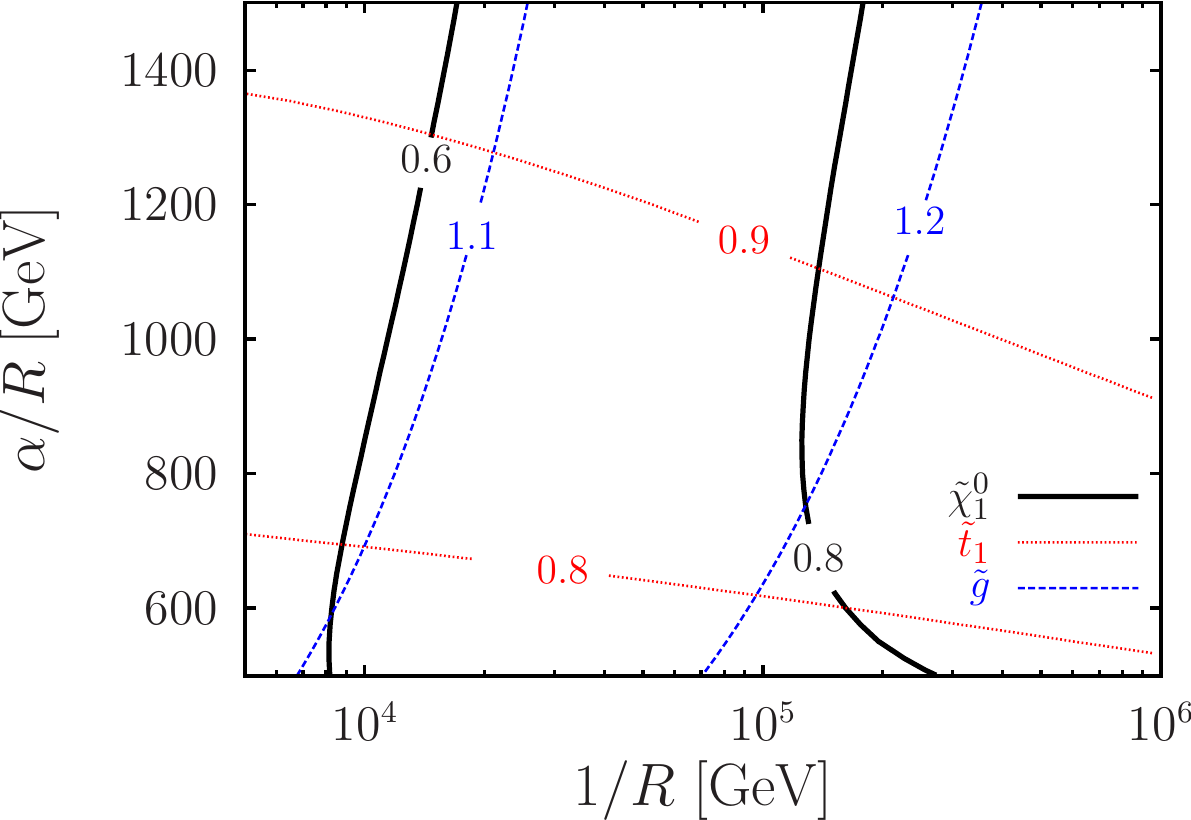}
\end{center}
\caption{Masses of the lightest neutralino $\tilde{\chi}^0_1$, the lighter 
 top squark $\tilde{t}_1$, and the gluino $\tilde{g}$ normalized to 
 $\alpha/R$.}
\label{fig:mass}
\end{figure}

In Fig.~\ref{fig:mass}, the masses of selected sparticles (the 
lightest neutralino $\tilde{\chi}^0_1$, the lighter top squark 
$\tilde{t}_1$, and the gluino $\tilde{g}$) are shown.  
The gluino is generically is the heaviest for the RGE effect.  For the stops, the left-right mixing term, $y_t A_0 v \sin\beta \approx -2y_t \frac{\alpha}{R}v \sin\beta$,  leads to a mass splitting. Since the stop mass-squared is 
	\begin{eqnarray}
	m^2_{\tilde{t}_{1,2}}\approx \left(\frac{\alpha}{R}\right)^2 \pm 2y_t \frac{\alpha}{R}v \sin\beta,
	\end{eqnarray}
the relative mass splitting becomes small as $\alpha/R$ increases. 
The lightest neutralino is mostly Higgsino-like and the mass scale is mostly determined by $\mu$. 
As discussed in Sec.~\ref{sec:EWSB}, $\mu$ basically scales with $\alpha/R$.    
When $\mu$ approaches $\alpha/R$ in larger $1/R$, the LSP has a significant composition of EW-inos. 

\begin{table}[ht!]
\begin{center}
\begin{tabular}{lcc|lcc}
\hline
Particle &  Point1 & Point2 & Particle & Point1 & Point2 
\\ \hline
$\tilde{g}$          & $1494$ & $1174$ &                 --   &     -- &    -- \\
$\tilde{u}_L$        & $1467$ & $1162$ & $\tilde{u}_R$        & $1459$ & $1144$ \\
$\tilde{d}_L$        & $1469$ & $1165$ & $\tilde{d}_R$        & $1458$ & $1143$ \\
$\tilde{b}_2$        & $1460$ & $1142$ & $\tilde{b}_1$        & $1430$ & $1086$ \\
$\tilde{t}_2$        & $1557$ & $1192$ & $\tilde{t}_1$        & $1267$ & $885$ \\
$\tilde{\nu}$        & $1411$ & $1027$ & $\tilde{\nu}_\tau$   & $1410$ & $1027$ \\
$\tilde{e}_L$        & $1413$ & $1030$ & $\tilde{e}_R$        & $1406$ & $1013$ \\
$\tilde{\tau}_2$     & $1417$ & $1032$ & $\tilde{\tau}_1$     & $1402$ & $1011$ \\
$\tilde{\chi}^0_1$   &  $767$ & $783$ & $\tilde{\chi}^0_2$   &  $777$ & $815$ \\
$\tilde{\chi}^0_3$   & $1384$ & $946$ & $\tilde{\chi}^0_4$   & $1410$ & $1010$ \\
$\tilde{\chi}^\pm_1$ &  $771$ & $793$ & $\tilde{\chi}^\pm_2$ & $1409$ & $1008$ \\
$h^0$                &  $125$ & $122$ & $H^0$                &  $819$ & $872$ \\
$A^0$                &  $819$ & $872$ & $H^\pm$              &  $822$ & $875$ \\
\hline
\end{tabular}
\end{center}
\caption{Phenomenologically viable mass spectrum of the benchmark points 
 (in GeV).  Point1: $1/R = 10^4~{\rm GeV}$, $\alpha/R = 1400~{\rm GeV}$ 
 and Point2: $1/R = 10^5~{\rm GeV}$, $\alpha/R = 1000~{\rm GeV}$.}
\label{tab:masses}
\end{table}

The masses of the first and second generation squarks are almost the 
same as  gluino mass.  The masses of the electroweak sparticles 
are close to $\alpha/R$, except for the lightest two neutralinos 
$\tilde{\chi}^0_{1,2}$ and the lighter chargino $\tilde{\chi}^+_1$, 
which are Higgsino-like (and thus close in mass) in most of the parameter 
space. The spectrum is more compressed for larger $1/R$ where $\mu$ approaches $\alpha/R$. We find that the masses of the sparticles are compressed 
at a $20\%$ level, exceptfor the Higgsinos which can be 
significantly lighter (up to a factor of $\approx 2$). 
We give full mass spectra for two points in Table.~\ref{tab:masses}.

\section{Experimental Limits}
\subsection{LHC Run I}
As we have seen, the model naturally 
predicts a compressed mass spectrum for sparticles.  This has 
strong implications on supersymmetry searches at the LHC.  Because 
of the mass degeneracy, production of high $p_{\rm T}$ jets and large 
missing energy is suppressed.  Therefore, typical searches, based on 
high $p_{\rm T}$ jets and large missing energy, are less effective for 
the present model. We first show limits presented in our work of Ref.~\cite{Murayama:2012jh} based 7 TeV run, and next show an updated limit based on 8 TeV run. 

\begin{figure}[ht]
\begin{center}
  \includegraphics[clip,width=.7\textwidth]{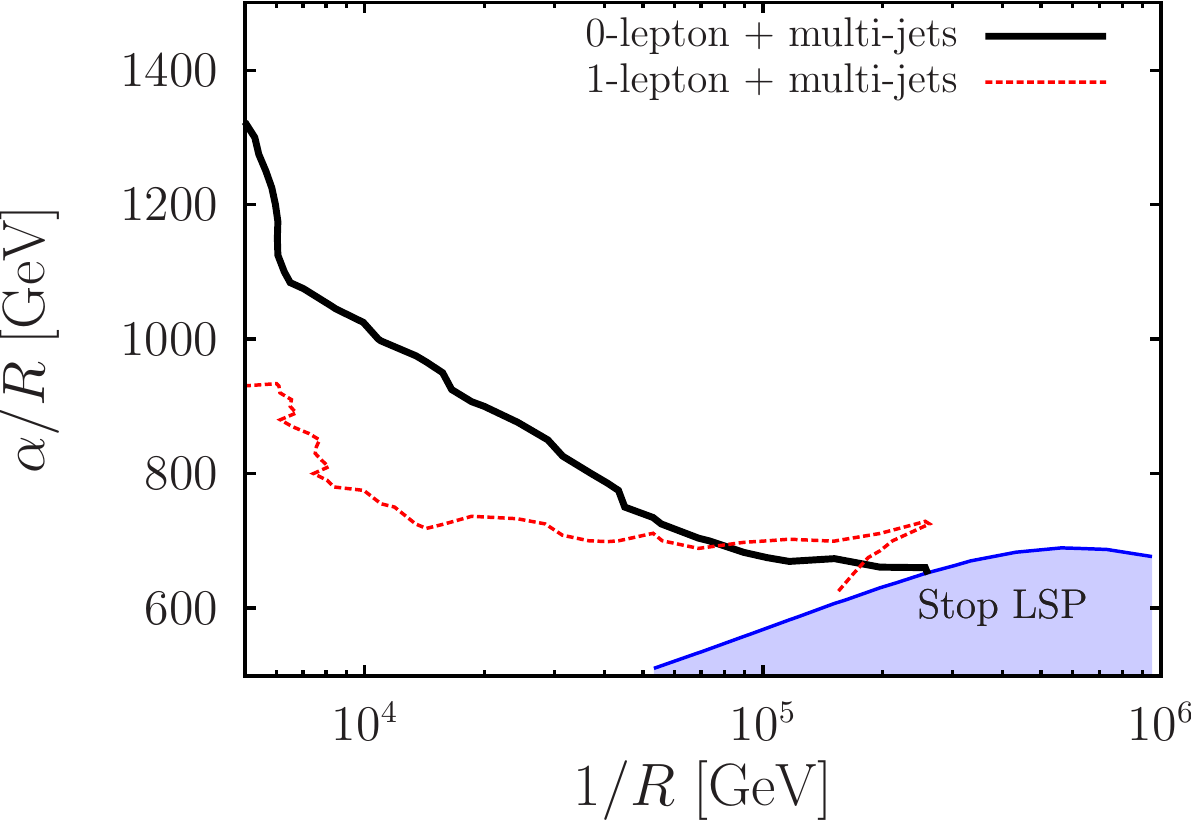}
\end{center}
\caption{The LHC constraint on the model using 7 TeV run data.}
\label{fig:LHC}
\end{figure}

In 7 TeV study, in order to estimate the number of supersymmetric events, we have used 
{\tt ISAJET~7.72}~\cite{Paige:2003mg} for the decay table of sparticles, 
{\tt Herwig~6.520}~\cite{Corcella:2000bw} for the generation of 
supersymmetric events, {\tt AcerDET~1.0}~\cite{RichterWas:2002ch} 
for the detector simulation, and {\tt NLL-fast}~\cite{NLLFAST} for 
estimation of the production cross section including next-to-leading
order QCD corrections and the resummation at next-to-
leading-logarithmic accuracy.  To constrain the 
parameter space, we compare the obtained event numbers with the results 
of ATLAS searches for multi-jets plus large missing energy with and 
without a lepton at $L = 4.7~{\rm fb}^{-1}$ at $\sqrt{s} = 7~{\rm 
TeV}$~\cite{ATLAS1,ATLAS2}.  In Fig.~\ref{fig:LHC}, we show the resulting 
LHC constraint on the model.
\begin{figure}[ht!]
\begin{center}
\hspace{25pt}  \includegraphics[clip,width=.63\textwidth]{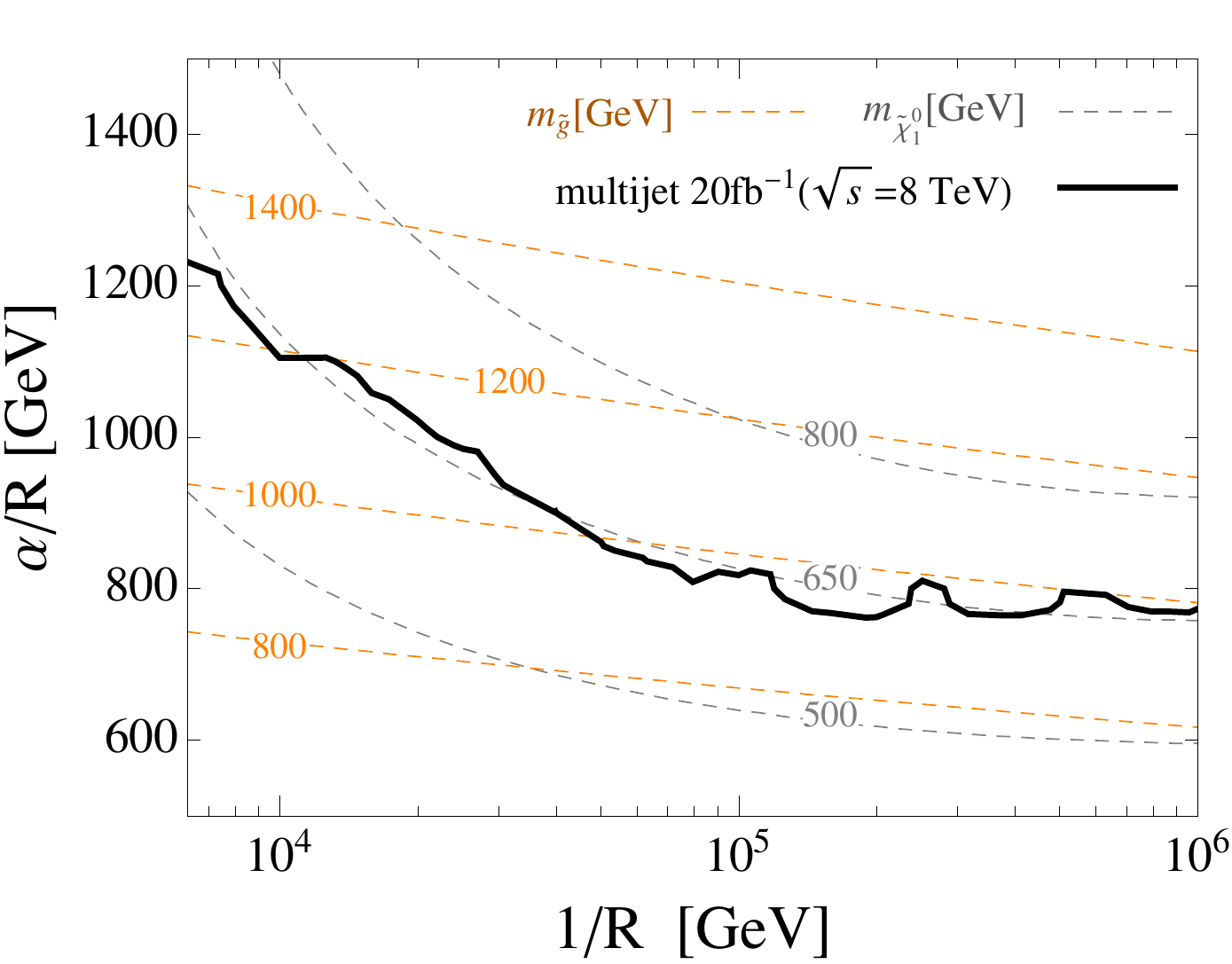}
\end{center}
\caption{The LHC constraint on the model using 8 TeV run data.}
\label{fig:LHC8}
\end{figure}

Other searches such as those for $b$-jets and/or multi-leptons are less 
effective.  We find that for $1/R \gtrsim 10^5~{\rm GeV}$, the case 
that $m_{\tilde{g}} \simeq m_{\tilde{q}} \lesssim 1~{\rm TeV}$ is 
allowed for this integrated luminosity.  This constraint is significantly weaker than that 
on the CMSSM, which excludes $m_{\tilde{g}} \lesssim 1.5~{\rm TeV}$ 
for $m_{\tilde{g}} \simeq m_{\tilde{q}}$~\cite{ATLAS1}.  (We have 
checked that our naive method of estimating the LHC constraints adopted 
here reproduces this bound for the CMSSM spectra.)

We also study a limit based on the result of $\sqrt{s}=8\TeV$ run. Here, {\tt PYTHIA 6.4}~\cite{Sjostrand:2006za} is adopted for supersymmetric event generation and showering, and  {\tt PGS 4}$~\cite{PGS}$ is used for the detector simulation. We again use {\tt NLL-fast}~\cite{NLLFAST} for estimation of the production cross section. 
We compare the obtained event numbers with the results 
of ATLAS searches for multi-jets plus large missing energy without lepton at $L = 20.3~{\rm fb}^{-1}$ at $\sqrt{s} = 8~{\rm 
TeV}$~\cite{ATLAS:multijet2013}, and the result is shown in Fig.~\ref{fig:LHC8}. 
We find the exclusion limit is extended up to $m_{\tilde{q}}\simeq 1\TeV$, and for a region at $1/R\sim 10^4 \GeV$ the limit is stronger as $m_{\tilde{q}}\gtrsim 1.2\TeV$. In contrast, the CMSSM and simplified model are more constrained as the limit is $m_{\tilde{q}}\gtrsim 1.7\TeV$. 

Hence, in the Compact Supersymmetry, the bounds from the current searches at the LHC are certainly ameliorated. We discuss potential improvement of the search in Appendix~\ref{ch:mt2} based on Ref.~\cite{Murayama:2011hj}.
This possibility is that a kinematic variable, $M_{T2}$, can be useful since the Standard Model background is systematically removed by requiring $M_{T2}>m_{\rm top}$. On the other hand, the signal along with a compressed spectrum is still extracted because  $M_{T2}$ for the signal becomes larger with energetic initial state radiation. We demonstrate this technique for a typical compressed model, the Minimal Universal Extra Dimension (MUED).

\subsection{Other measurements}
First of all, for the geometric nature of supersymmetry breaking there are no extra flavor-violation and CP-violation.  
We typically find $\tan\beta \sim 4~\mbox{--}~10$, which allows for the 
model to avoid the constraint from $b \rightarrow s\gamma$, despite the 
large $A$ terms.

The contribution of the non-zero KK states to the electroweak precision 
parameters bounds $1/R \gtrsim \mbox{a few}~{\rm TeV}$~\cite{Delgado:2001si}. 
Since we consider the region $1/R \gtrsim 10~{\rm TeV}$, 
however, the model is not constrained by the electroweak precision data.

\subsection{Comment on brane kinetic terms}\label{sec:comment-branekin}
As we mentioned in Sec.~\ref{sec:branekin}, if the brane kinetic terms introduce additional wave functions to fields, they affect tree-level prediction of the model. Here, we discuss brane kinetic terms for quarks, parameterized by
	\begin{eqnarray}\label{eq:brane02}
	\int d^4x \int d^4\theta \ \left[1+\frac{c^{ij}_Q}{8\pi^2 } \right]Q_{i}^{(0)\dag} Q_{j}^{(0)}
	+\left[1+\frac{c^{ij}_U}{8\pi^2 } \right]U_{i}^{(0)\dag} U_{j}^{(0)} 
	+\left[1+\frac{c^{ij}_D}{8\pi^2 } \right]D_{i}^{(0)\dag} D_{j}^{(0)} .
	\end{eqnarray}
When $c_{Q,U,D}\approx {\cal O}(1)$, the bulk and brane kinetic terms have comparable sizes. First, let us consider a case of no additional flavor violation through the kinetic terms, that is, $c^{ij}_{Q,U}\propto \delta^{ij}$. Then, the brane kinetic terms still change mass of each squark, and an important effect is from $c^{33}_{Q,U}$ because one stop is light due to the large mixing and can be the LSP with some amount of $c^{33}_{Q,U}$. Fig.~\ref{fig:mass} tells that 30\% reduction of the lighter stop mass leads the stop to be the LSP and LHC phenomenology is completely changed. So if we have large brane terms such as
	\begin{eqnarray}
	c^{33}_{Q}=c^{33}_{U}\gtrsim 0.3\times 8\pi^2\sim24, 
	\end{eqnarray}
the LSP is the lighter stop rather than neutral sparticle. 

Next, we consider the flavor violations in brane kinetic terms. There many constraints on additional flavor violations from low-energy measurements, and here let us consider only a stringent constraint from $K^0-{\bar K}^0$ mixing. Assuming no CP violation, analysis based on mass insertion method \cite{Altmannshofer:2009ne} together with a constraint from Ref.~\cite{Bona:2007vi} gives limits for soft mass mixing between first and second generations. If $c^{12}_{D}$ is an unique source of flavor violation, $K^0-{\bar K}^0$ mixing excludes
	\begin{eqnarray}
	c^{12}_{D} \gtrsim {\cal O}(0.1)\times8\pi^2 \sim {\cal O}(10), 
	\end{eqnarray}
for MSSM particles at 1 TeV. 
If we introduce an additional source, $c^{12}_{Q}$,  such that $c^{12}_{D}=c^{12}_{Q}$, the exclusion limit gets severer, 
	\begin{eqnarray}
	c^{12}_{D}=c^{12}_{Q} \gtrsim {\cal O}(0.01)\times8\pi^2 \sim {\cal O}(1). 
	\end{eqnarray}
If there is additionally CP violation, the limit could be stronger by one order of magnitude. 
Thus brane kinetic terms have to be adequately small in presence of these flavor violations.   
 
\section{Dark Matter}
\begin{figure}[t]
\begin{center}
  \includegraphics[clip,width=.7\textwidth]{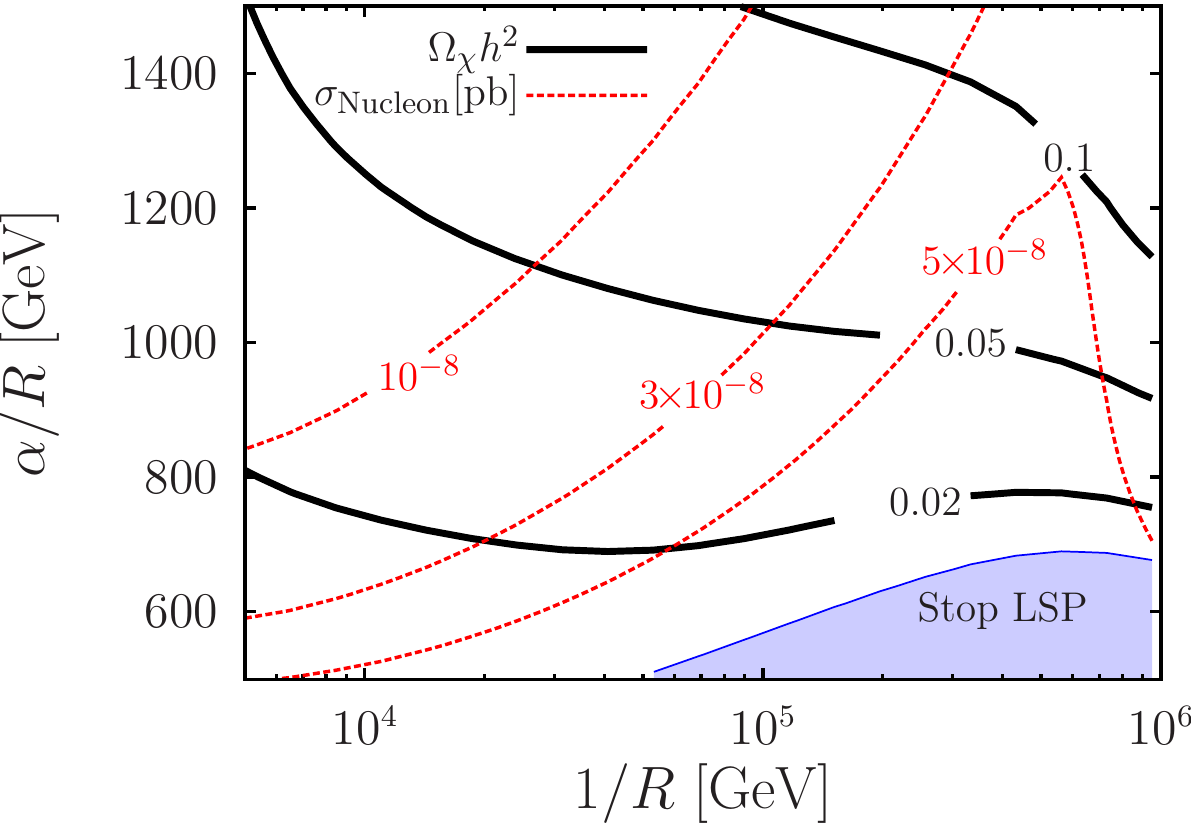}
\end{center}
\caption{The thermal relic abundance, $\Omega_\chi h^2$, and the 
 spin-independent  cross section with a nucleon, $\sigma_{\rm Nucleon}$, 
 of $\tilde{\chi}^0_1$.  The solid (black) lines are the contours of 
 $\Omega_\chi h^2$, while the dotted (red) lines are those of 
 $\sigma_{\rm Nucleon}$.}
\label{fig:DM}
\end{figure}

In the present model, the dark matter candidate is 
the lightest neutralino $\tilde{\chi}^0_1$, whose dominant component is 
the Higgsino.  
In Fig.~\ref{fig:DM}, we show the thermal relic abundance, 
$\Omega_\chi h^2$, and the spin-independent cross section with a nucleon, 
$\sigma_{\rm Nucleon}$, of $\tilde{\chi}^0_1$, assuming $R$-parity 
conservation.
To estimate these, we have used {\tt micrOMEGAs~2.4}~\cite{Belanger:2010gh}. 
For the strange quark form factor we have adopted $f_s = 0.02$, suggested 
by lattice calculations~\cite{Ohki:2008ff}, instead of the default value 
of {\tt micrOMEGAs} ($f_s = 0.26$).

  As seen in Fig.~\ref{fig:DM}, the 
thermal relic abundance of $\tilde{\chi}^0_1$ is much smaller than the 
observed dark matter density $\Omega_{\rm DM} h^2 \simeq 0.1$, unless 
$\tilde{\chi}^0_1$ is rather heavy $\sim {\rm TeV}$ (in the upper-right 
corner of the plot). This is due to the Higgsino-like nature of $\tilde{\chi}^0_1$. Therefore, in most parameter regions, 
$\tilde{\chi}^0_1$ cannot be the dominant component of dark matter 
if only the thermal relic abundance is assumed.  It must be produced 
nonthermally to saturate $\Omega_{\rm DM} h^2$, or some other 
particle(s), e.g.\ the axion/axino, must make up the rest.

The spin-independent scattering with nuclear of $\tilde{\chi}^0_1$ is mostly mediated  by the lightest Higgs boson exchange between the lightest neutralino and quarks inside the nucleon. The relevant coupling at tree-level originate from gaugino-Higgsino-Higgs interactions, $H\widetilde{H}\widetilde{B}$ and $H\widetilde{H}\widetilde{W}$. Hence, the scattering does not occur for pure Higgsino or pure EW-ino. When the lightest neutralino is well-tempered mixture of Higgsinos and pure EW-inos, the scattering is enhanced, and the enhancement in $1/R\sim 10^{5.7}\GeV$ and  $\alpha/R \lesssim 1.2\TeV$ is for this reason. Beyond $1/R\sim 10^{5.7}\GeV$, the scattering cross section is damped for the purity of Bino. 

\begin{figure}[ht!]
\begin{center}
  \includegraphics[clip,width=.7\textwidth]{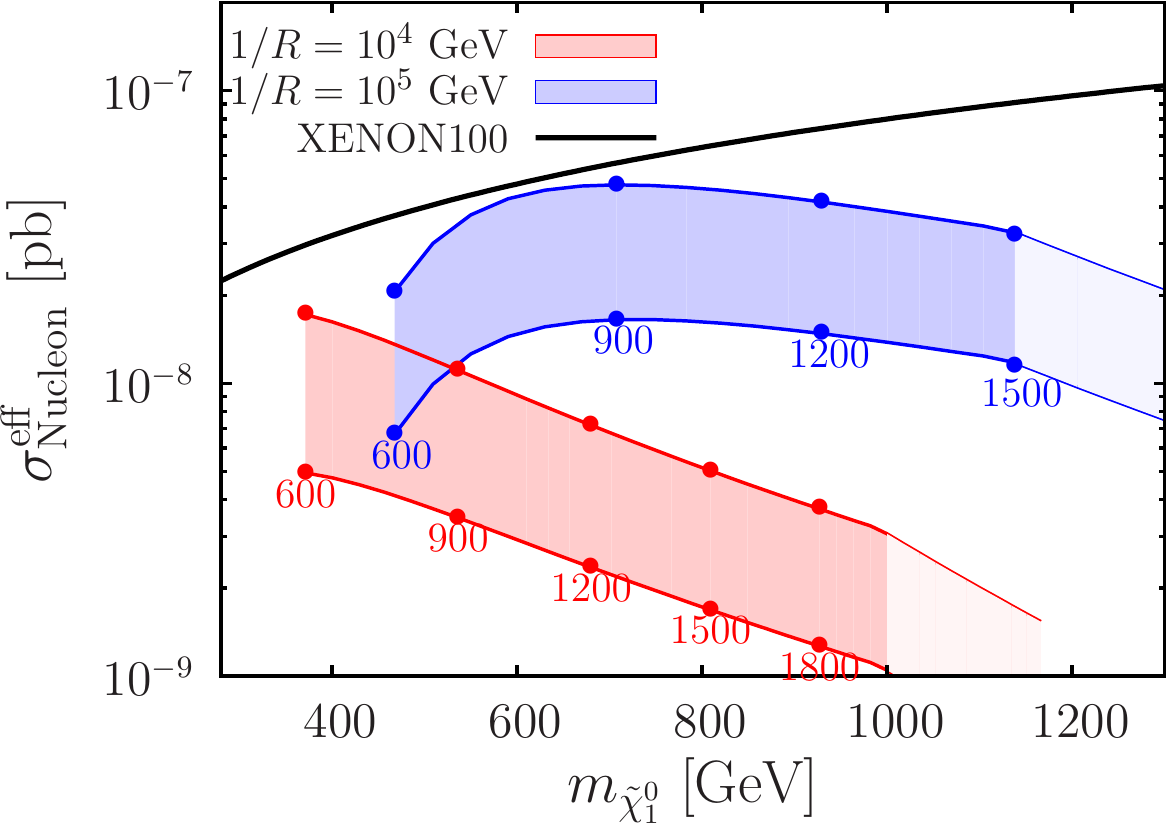}
\end{center}
\caption{Effective dark matter-nucleon cross section for $1/R = 10^4~{\rm 
 GeV}$ (lower, red shaded) and $10^5~{\rm GeV}$ (upper, blue shaded). 
 In each region, the upper and lower boarders correspond to $f_s=0.26$ 
 and $0.02$, respectively, and the dots represent the corresponding 
 values of $\alpha/R$. (The very light shaded regions are those in which 
 the thermal abundance exceeds $\Omega_{\rm DM}$.)  The solid (black) 
 line shows the current upper bound from XENON100.}
\label{fig:eff}
\end{figure}
It is natural to expect that at least the thermal abundance 
of $\tilde{\chi}^0_1$ remains as a (sub-)component of dark matter.  In 
this case, direct and indirect signatures of the relic neutralino are 
expected.  To discuss the direct-detection signal, is is useful to define
\begin{equation}
  \sigma_{\rm Nucleon}^{\rm eff} \equiv \sigma_{\rm Nucleon} 
    \frac{{\rm min}\{\Omega_\chi, \Omega_{\rm DM}\}}{\Omega_{\rm DM}},
\label{eq:sigma_eff}
\end{equation}
which is the quantity to be compared with the dark matter-nucleon cross 
section in the usual direct-detection exclusion plots (which assume 
$\Omega_\chi = \Omega_{\rm DM}$).  In Fig.~\ref{fig:eff}, we plot 
$\sigma_{\rm Nucleon}^{\rm eff}$ as a function of $m_{\tilde{\chi}^0_1}$ 
for $1/R = 10^4~{\rm GeV}$ and $10^5~{\rm GeV}$.
To represent the uncertainty from the nucleon matrix element, we show both 
the $f_s=0.02$ and $0.26$ cases.  We also present the current upper bound 
on $\sigma_{\rm Nucleon}^{\rm eff}$ from XENON100~\cite{Aprile:2011hi}. 
We find that improving the bound by one or two orders of magnitude will 
cover a significant portion of the parameter space of the model.
\footnote{Recently the LUX experiment reported a new result of spin-independent WIMP-nucleon elastic scattering \cite{Akerib:2013tjd}. The upper limit was improved by a factor compared to XENON100 limit. }

\section{Summary and Discussion}
In Ch.~\ref{compact:model} and Ch.~\ref{compact:pheno}, we pointed out that supersymmetry 
broken by boundary conditions in extra dimensions, the Scherk-Scwartz mechanism, leads naturally to a compressed sparticle spectrum, ameliorating the limits from experimental 
searches.  We demonstrated the Scherk-Schwarz mechanism is equivalent to the Radion mediation. 
We presented the simplest model in the $S^1/{\mathbb Z}_2$ 
orbifold, the Compact Supersymmetry. Despite the fact 
that it essentially has two less free parameters than the CMSSM: $1/R$, 
$\alpha/R$, and $\mu$, the model can accommodate the Higgs mass and is less fine-tuned than many models. However, the  theoretical error of the Higgs mass is still large, $\Delta M_H \approx 2\sim3 \GeV$.

The LHC limit is weaker by about 500 GeV compared to the CMSSM-like models. 
We find the LSP is mostly Higgsino-like and can be a component of dark matter. The direct detection experiments for dark matter particle have potential to search for the model. 
In Table~\ref{tab:masses} we give two representative 
points in the parameter space, which can serve benchmark points for 
further phenomenological studies.
%

The theory presented here can be extended in several different ways. 
An interesting one is to introduce a singlet field $S$ in the bulk together with 
superpotential interactions on the $y=0$ brane: $\lambda S H_u H_d + f(S)$, 
where $f(S)$ is a polynomial of $S$ with the simplest possibility being 
$f(S) = -\kappa S^3/3$.  This allows for an extra contribution to the 
Higgs boson mass from $\lambda$, and can make the lightest neutralino 
(which would now contain a singlino component as well) saturate the 
observed dark matter abundance without resorting to nonthermal production.
Furthermore,  $\mu$ term could be dynamically generated by $\langle S\rangle$ in the same scale of $\alpha/R$, and therefore the mass degeneracy is theoretically more reasonable. 
Detailed studies of this possibility will be presented elsewhere.

\part{Natural Higgs Mass in Supersymmetry with Two Singlets}

\chapter{Dirac NMSSM}\label{dnmssm:model}
\section{Introduction}
The discovery of a new resonance at 125~GeV~\cite{:2012gk}, that appears to be the
long-sought Higgs boson, marks a great triumph of experimental and
theoretical physics.  On the other hand,  the presence of this light scalar forces us to face the naturalness problem of its mass.
Arguably, the best known mechanism
to ease the naturalness problem is weak-scale supersymmetry (SUSY), but the lack of experimental signatures is pushing 
supersymmetry into a tight corner.  In
addition, the observed mass of the Higgs boson 
is higher than what was expected in the Minimal Supersymmetric Standard
Model (MSSM), requiring fine-tuning of parameters at the 1\% level or worse~\cite{Hall:2011aa}.
This comes from a fact that it is necessary for the observed Higgs mass to have the large radiative corrections by large stop mass, $m^2_{\tilde{t}}$, or by large left-right mixing of stops, $X_t$, where the Higgs mass formula is
	\begin{eqnarray}
	m_{h,\rm MSSM}^2 \simeq m_Z^2 \cos 2\beta+ \frac{3m_t^4}{4\pi^2 v^2}\left(\log\frac{m_{\tilde{t}}^2}{m_t^2} +m_{\tilde{t}}^2\left(1-\frac{X_t^2}{12m_{\tilde{t}}^2} \right) \right),
	\end{eqnarray}
while these large $m_{\tilde{t}}$ and $A_t$ make the theory unnatural because the Higgs soft mass can be very different from the weak scale by a fast Renormalization Group (RG) evolution,
	\begin{eqnarray}
	\frac{d}{d\log Q}m_{H_u}^2\simeq \frac{3y_t^2}{8\pi^2}(2m_{\tilde{t}}^2+|A_t|^2),
	\end{eqnarray}
where $A_t$ is a soft breaking part of $X_t$. 


If supersymmetry is realized in nature, one possibility is to give up on naturalness~\cite{Wells:2003tf, ArkaniHamed:2004fb}.  Alternatively, theories that retain naturalness must address two problems, (I) the missing superpartners and (II) the Higgs mass.  The collider limits on superpartners are highly model-dependent and can be relaxed when superpartners unnecessary for naturalness are taken to be heavy~\cite{Dimopoulos:1995mi}, when less missing energy is produced due to a compressed mass spectrum~\cite{Murayama:2012jh,LeCompte:2011fh} or due to decays to new states~\cite{Fan:2011yu}, and 
 when $R$-parity is violated~\cite{Barbier:2004ez}.  Even if superpartners have evaded detection for one of these reasons, we must address the surprisingly heavy Higgs mass.

\subsection{Beyond the MSSM}
There have been many attempts to extend the MSSM to accommodate the Higgs mass.  In such extensions, new states interact with the Higgs, raising its mass by increasing the strength of the quartic interaction of the scalar potential.  If the new states are integrated out supersymmetrically, their effects decouple and the Higgs mass is not increased.  On the other hand, SUSY breaking can lead to non-decoupling effects that increase the Higgs mass.  One possibility is a non-decoupling $F$-term, as in the  NMSSM (MSSM plus a singlet)~\cite{Espinosa:1991gr, Nomura:2005rk} or $\lambda$SUSY (allowing for a Landau
pole)~\cite{Harnik:2003rs}.  A second possibility is a non-decoupling $D$-term that results if the Higgs is charged under a new gauge group~\cite{Batra:2003nj}.   In general, these extensions require new states at the few hundred GeV scale, so that the new sources of SUSY breaking do not spoil naturalness.

\begin{figure}[h!]
  \centering
  \includegraphics[width=3cm]{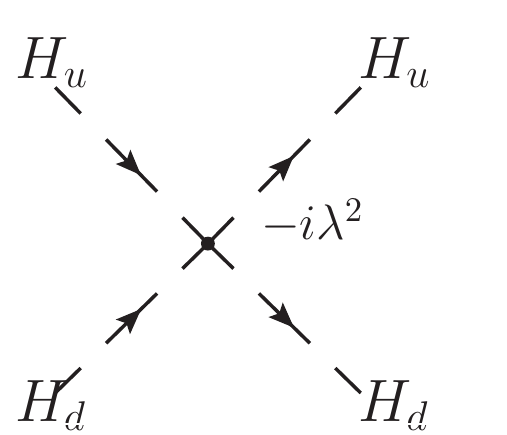}\hspace{1cm}
  \includegraphics[width=4.3cm]{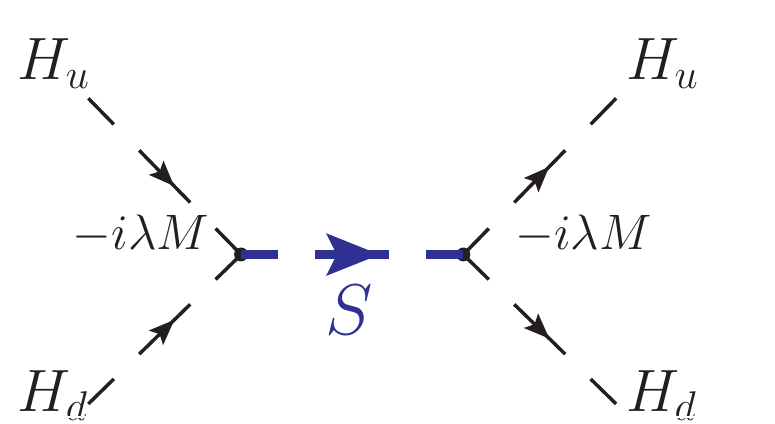}
\caption{Additional Higgs quartic terms}
\label{fig:NMSSM1}
\end{figure}

For example, consider the NMSSM, where a singlet superfield, $S$, interacts with the MSSM Higgses, $H_{u,d}$, through the superpotential,
\footnote{When $\lambda \langle S\rangle$ is small, explicit $\mu$ term is necessary to avoid LEP chargino bounds.}
\begin{equation}
  W \supset \lambda \, S H_u H_d + \frac{M}{2} S^2 +\mu \, H_u H_d.
  \label{eq:NMSSM}
\end{equation}
The $F$-term of $S$ gives 
	\begin{eqnarray}
	V\supset|F_S|^2=|\lambda H_u H_d +MS|^2 \ .
	\end{eqnarray}
It generates additional Higgs quartic terms which potentially increase the Higgs mass. The potential is 
	\begin{eqnarray}
	\Delta V=\lambda^2|H_uH_d|^2 -\frac{(\lambda M)^2 }{M^2+m_S^2}|H_uH_d|^2 ,
	\end{eqnarray}
and we can understand it diagrammatically by Fig.~\ref{fig:NMSSM1}.

\begin{figure}[h]
  \centering
  \includegraphics[width=5cm]{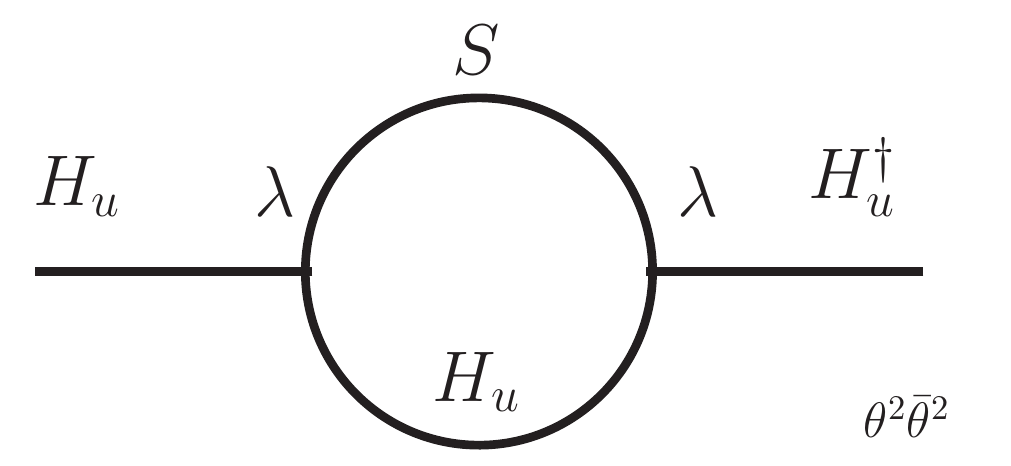}
\caption{Supergraph}
\label{fig:NMSSM2}
\end{figure}

The Higgs mass is increased by,
\begin{equation} \label{eq:nmssm_mass}
\Delta m_h^2 = \lambda^2 v^2 \sin^2 2 \beta \left( \frac{m_S^2}{M^2 + m_S^2} \right),
\end{equation}
where $m_S^2$ is the SUSY breaking soft mass $m_S^2 |S|^2$, $\tan \beta = v_u / v_d$ is the ratio of the VEVs of the up and down-type Higgses,  and $v=\sqrt{v_u^2+v_d^2} = 174$~GeV\@.  Notice that this term decouples in the supersymmetric limit, $M \gg m_S$, which means $m_S$ should not be too small.  On the other hand, $m_S$ feeds into the Higgs soft masses, $m_{H_{u,d}}^2$ at one-loop,
	\begin{eqnarray}
	 \frac{d m_{H_{u,d}^2}}{d\log Q}\supset\frac{\lambda^2m_S^{2} }{8\pi^2} 
	\label{eq:NMSSM1}
	\end{eqnarray}
 requiring fine-tuning if $m_S \gg m_h$. This can be easily understood since the singlet directly couples to the Higgs superfields which leads one-loop Supergraph of Fig.~\ref{fig:NMSSM2}. The soft mass of $S$, that is $m_S^2\theta^2\bar{\theta}^2 S^\dagger S$, gives $m_{H_u}^2$ along with a logarithmic divergence, and we have Eq.(\ref{eq:NMSSM1}) as a consequence. Therefore, there is tension between raising the Higgs mass, which requires large $m_S$, and naturalness, which demands small $m_S$. Of course, in the limit of $m_S\to0$, similar tension which is discussed in the beginning exists with respect to $m_{\tilde{t}}$.

In this chapter, we point out that, contrary to the above example,  a {\it lack of light
  scalars}\/ can help raise the Higgs mass without a cost to
naturalness, if the singlet has a Dirac mass.

\section{Model Setup}

We consider a modification of Eq.~\ref{eq:NMSSM} where $S$ receives a Dirac mass with another singlet, $\bar S$,
\begin{equation}
  W = \lambda \, S  H_u H_d + M S \bar{S} +\mu \, H_u H_d.
\end{equation}
We call this model the {\it Dirac NMSSM}.  

\subsection{Spurion analysis}
The absence of various dangerous operators (such as large tadpoles $M^2_{pl}S$ for the singlets) follows from a $U(1)_{PQ}\times U(1)_{\bar{S}}$ Peccei-Quinn-like symmetry,
\begin{table}[h!]
\centering
  \begin{tabular}{|c| c  c  c c c c c | c c |} \hline
      &Matter&$H_u$& $H_d$ & $S$ &$\bar{S}$ & $\mu$ & $M$  & $\epsilon_\mu$ & $\epsilon_M$  \\\hline
     $U(1)_{PQ}$ & $-1/2$& $1$ & $1$ & $-2$ & $-2$ & $-2$ & $4$ & $-2$ & $4$ 
    \\
     $U(1)_{\bar{S}}$& $0$& $0$& $0$& $0$ & $1$ &$0$  & $-1$ &$0$  & $-1$
\\ \hline
  \end{tabular}
\end{table}\\
Here, $U(1)_{\bar{S}}$ has the effect of differentiating $S$ and $\bar
S$ and forbidding the operator $\bar S H_u H_d$.  Because $\mu$
  and $M$ explicitly break the $U(1)_{PQ}\times U(1)_{\bar{S}}$ symmetry, we
  regard them to be spurions originating from chiral superfields (``flavons'' \cite{ArkaniHamed:1996xm})
 so that
  the superpotential should not depend on their complex conjugates to
  avoid certain unwanted terms (``SUSY zeros'' \cite{Leurer:1993gy}).
  Namely, 
	\begin{eqnarray}
	\epsilon_\mu(-2,0)\sim\frac{\mu}{M_{pl}}, \  \  \  \  \ 
	\epsilon_M(4,-1)\sim\frac{M}{M_{pl}}	
	\end{eqnarray}
are small breaking parameters. 
  
  By classifying all possible operators induced by these spurions, we
  see that a 
tadpole for $\bar{S}$ is suppressed until the weak scale,
\begin{eqnarray}
	W \supset c_{\bar{S}} \mu M \bar{S} \sim \epsilon_\mu\epsilon_M M_{pl}^2\bar{S} \ ,
\end{eqnarray}
where $c_{\bar{S}}$ is a ${\cal O}(1)$ coefficient. 
Some terms involving only singlets
	\begin{eqnarray}
	S, S^2, S^3, S^2\bar{S}
	\end{eqnarray}
 are forbidden by the symmetries.  The other terms are suppressed by higher order of $\epsilon_{\mu,M}$,
	\begin{eqnarray}
	(\epsilon_\mu \epsilon_M )^2 M_{pl} \bar{S}^2, 
	(\epsilon_\mu \epsilon_M )^3\bar{S}^3, 
	\epsilon_\mu \epsilon_M^2  S\bar{S}^2.
	\end{eqnarray}

Potentially dangerous tadpoles can appear from K\"alher potential with SUSY breaking. Using SUSY breaking spurion ${\cal Z}$ 
\footnote{{One may worry that a similar SUSY breaking term, ${\bar{\theta}}^2\mu^\dag S$, in K\"alher potential behaves as tadpole of $S$ in superpotential, but it is removed by filed redefinition of $\bar{S}$. }}
	\begin{eqnarray}
	{\cal Z}=m_{SUSY}\theta^2, \ \  m_{SUSY}\sim{\cal O}(1 \rm TeV),
	\end{eqnarray}
we can classify such terms and see the size turns out to be safe, 
	\begin{eqnarray}
	\int d^4\theta {\cal Z}^\dagger {\cal Z}\mu^\dagger S =m_{SUSY}^2\mu^\dagger S.
	\end{eqnarray}

\subsection{Dirac NMSSM}
\begin{figure}[h!]
  \centering
  \includegraphics[width=6cm]{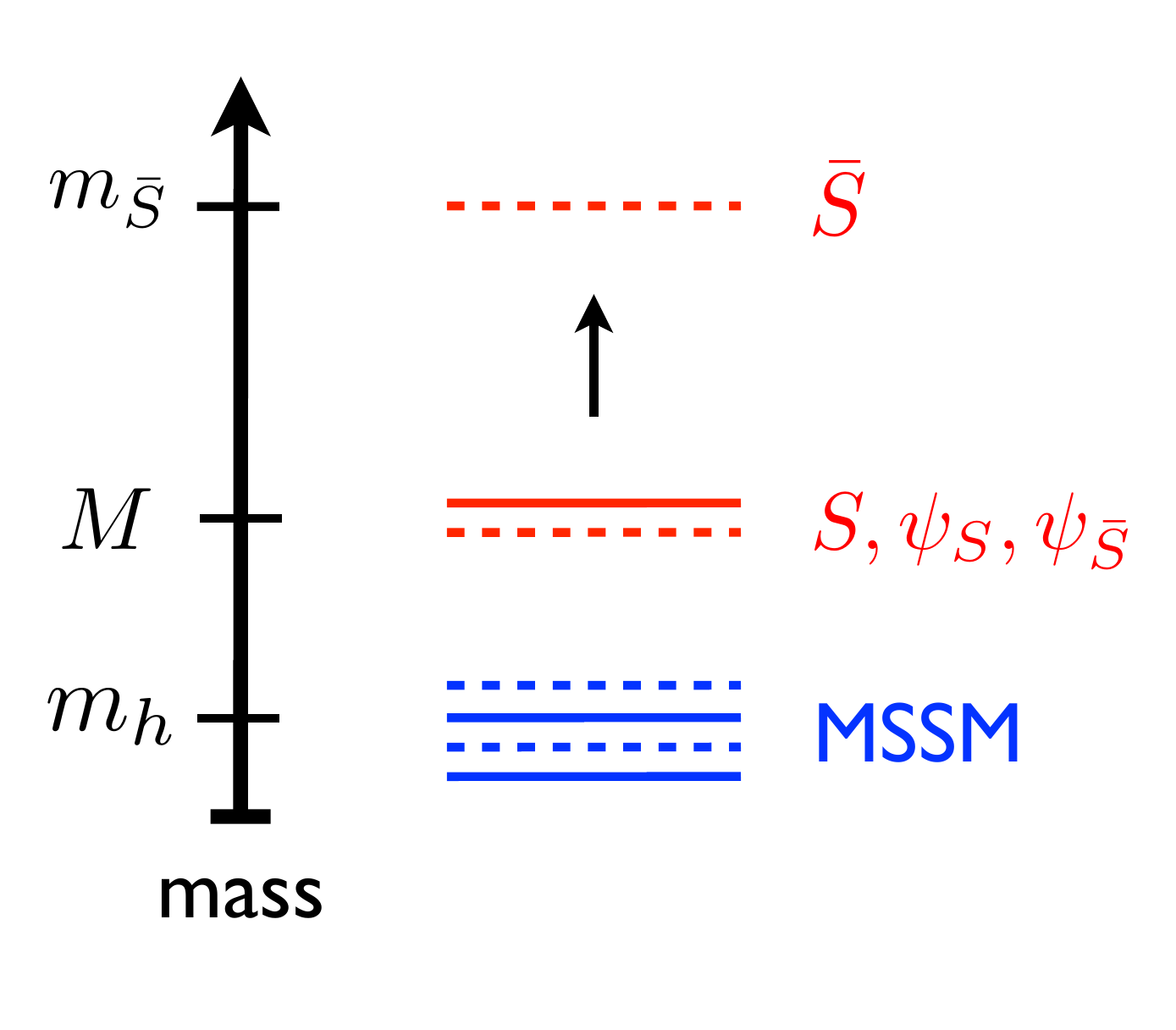}
  \caption{Schematic mass spectrum we consider in the Dirac NMSSM.}
  \label{fig:schema}
\end{figure}

From the spurion analysis the superpotential of Higgs sector in the Dirac NMSSM is given by
	\begin{eqnarray}
	  W_{\rm Dirac} = \lambda \, S  H_u H_d + M S \bar{S} +\mu \, H_u H_d
	  +c_{\bar{S}}\mu M\bar{S},
	\end{eqnarray}
and the corresponding potential is
	\begin{eqnarray}
	V_{\rm Dirac}&=&|F_{\bar{S}}|^2+|F_S|^2+|F_{H_u}|^2+|F_{H_d}|^2
	\nonumber\\
	&=&|\lambda H_u H_d+MS|^2+|MS+c_{\bar{S}}\mu M|^2
	\nonumber\\
	&&+|(\lambda S+\mu) H_d|^2+|(\lambda S+\mu) H_u|^2
	\end{eqnarray}
where terms involving quark and lepton are omitted. 
The  following soft supersymmetry breaking terms are allowed by the symmetries,
\begin{eqnarray}
   V_{\rm Dirac}^{\it soft} &=& m_{H_u}^2 |H_u|^2+m_{H_d}^2 |H_d|^2+
  m_S^2 |S|^2 + m_{\bar{S}}^2 |\bar{S}|^2 
  \nonumber\\ 
  &&+  \lambda A_\lambda {S} H_u H_d +  M B_S S\bar{S} +  \mu B H_u H_d  +c.c.
  \nonumber\\ 
  &&+ t_{\bar{S}} \bar{S}+ t_S S+c.c.
\end{eqnarray}
The last tadpole arises from a non-holomorphic term $\mu^\dagger S$. Both soft tadpoles naturally have weak-scale sizes due to the symmetry and spurion structure.
As described later, the spectrum we consider is one shown in Fig.~\ref{fig:schema}

\subsection{NMSSM}
In the following discussion we consider the NMSSM for comparison. Here $U(1)_{\bar{S}}$ is absent. The non-negligible superpotential classified by spurion analysis of $U(1)_{PQ}$ is
	\begin{eqnarray}
	W_{\rm NMSSM}= \lambda \, S  H_u H_d + \frac{M}{2} S^2 +\mu \, H_u H_d
	  +c_{{S}}\mu M{S} \ ,
	\end{eqnarray}
and the corresponding potential is
	\begin{eqnarray}
V_{\rm NMSSM}&=&|F_S|^2+|F_{H_u}|^2+|F_{H_d}|^2
	\nonumber\\
	&=&|\lambda H_u H_d+MS+c_{{S}}\mu M|^2+
	|(\lambda S+\mu) H_d|^2+|(\lambda S+\mu) H_u| ^2
	\end{eqnarray}
where terms involving quark and lepton are omitted.
The soft SUSY breaking terms are given by
	\begin{eqnarray}
	   V_{\rm NMSSM}^{\it soft} &=& m_{H_u}^2 |H_u|^2+m_{H_d}^2 |H_d|^2+
  m_S^2 |S|^2  
  \nonumber\\ 
  &&+  \lambda A_\lambda {S} H_u H_d +  \frac{M}{2} B_S S^2 +  \mu B H_u H_d  + t_S S+c.c.
	\end{eqnarray}
\section{Raising the Higgs Mass without Fine-tuning}
Non-decoupling effects to boost the Higgs mass remain when $\mSbar$ is extremely large in the Dirac NMSSM. We see the Higgs mass parameters are still stable against the radiative corrections unlike the NMSSM. 

\subsection{Non-decoupling effects}
\begin{figure}[t!]
  \centering
  \includegraphics[width=3cm]{figs/Dirac-1.pdf}\hspace{1cm}
  \includegraphics[width=4.3cm]{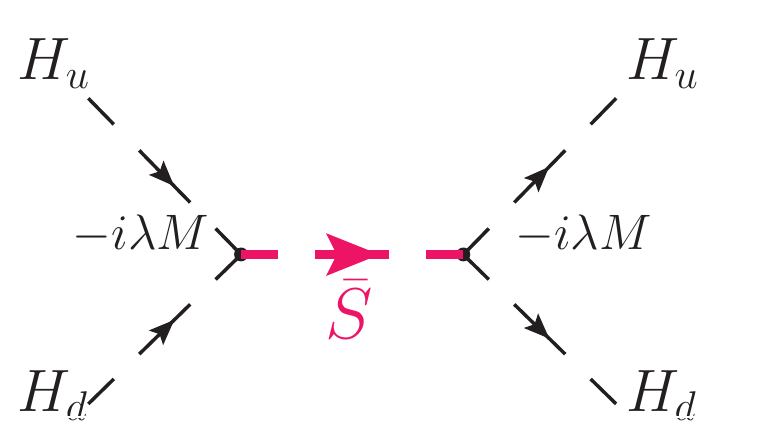}
\caption{Non-decoupling effects  in Dirac NMSSM}\vspace{20pt}
\label{fig:Non-decouplingDNMSSM}
  \includegraphics[width=12.5cm]{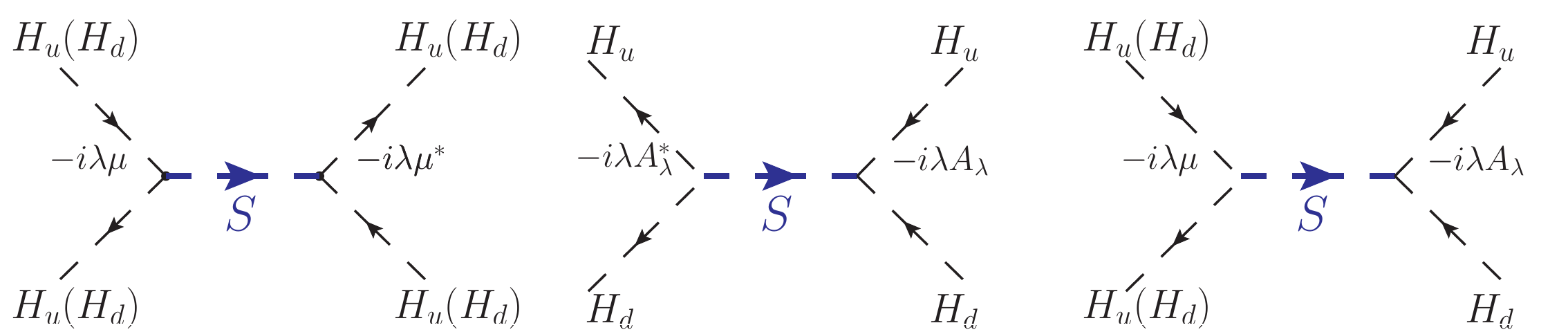}
\caption{Other Higgs quartic terms in Dirac NMSSM}
\label{fig:Dirac1}
\end{figure}
We would like to understand whether the new quartic term, $|\lambda H_u H_d|^2$,  can naturally 
raise the Higgs mass. 
When we integrate out the $S$ and $\bar{S}$ chiral multiplets, normally we expect that the quartic potential decouples in the limit of heavy singlets.  However, we find the $S$ and $\bar{S}$ exchanges do not cancel the quartic term,
\begin{eqnarray}
  V_{\it eff}
  &=& |\lambda H_u H_d|^2 \left( 1 - \frac{M^2}{M^2 + m_{\bar{S}}^2}  \right)
	- \frac{\lambda^2 }
    {M^2 + m_{S}^2 }\left| A_\lambda H_u H_d+{\mu^*}(|H_u|^2+|H_u|^2) \right|^2,
  \no\\\label{eq:Veff}   
\end{eqnarray}
where we keep leading $(M^2 + m_{S, \bar{S}}^2)^{-1}$  terms and neglect the tadpole terms for simplicity.
The new contribution to the Higgs quartic does not decouple when $m^2_{\bar{S}}$ is large. 
The SM-like Higgs mass becomes
\begin{eqnarray}
\!\!\!\!\!  m_h^2 \!\!\!&=& m_{h,{\rm MSSM}}^2(m_{\tilde{t}}) + \lambda^2 v^2 \sin^2 2\beta \left( \frac{m_{\bar{S}}^2}{M^2 + m_{\bar{S}}^2} \right)
  \nonumber \\
  && 
  -\frac{ \lambda^2 v^2 }{M^2 + m_S^2}\left|
    A_{\lambda} \sin 2\beta-2\mu^*
     \right|^2, 
     \label{eq:higgsmass}
\end{eqnarray}
in the limit where the VEVs and mass-eigenstates are aligned, $H_u\to v_u +h \sin\beta$ and $H_d\to v_d +h \cos\beta$. The second term, coming from diagrams of Fig. \ref{fig:Non-decouplingDNMSSM}, shows so-called non-decoupling effect which is maximized by large $m_{\bar{S}}$. 
The second line of Eq.(\ref{eq:higgsmass}) can be understood by diagrams of Fig.~\ref{fig:Dirac1}, and it always reduces the size of quartic coupling.

By the way, in a limit of $M\gg m_{\bar{S}}$, the non-decoupling effect is easily derived by integrating out of $S$ and $\bar{S}$ using equations of motions of superpotential,
	\begin{eqnarray}
	\int d^4\theta \ (1-m_{\bar{S}}^2 \theta^4)\bar{S}^\dag\bar{S}+S^\dagger S
	+\left(\int d^2\theta \ \lambda S H_u H_d +M\bar{S}S +h.c.\right)
	\\
	\rightarrow \int {{d^4}\theta } \,\frac{\lambda^2 (1-m_{\bar{S}}^2 \theta^4)}{M^2}{({H_u}{H_d})^\dag }({H_u}{H_d})\,\,\,.
	\end{eqnarray}
The SUSY breaking term leads to a new Higgs quartic coupling. 
Similar analysis for various extensions of MSSM is found in Ref.~\cite{Dine:2007xi}.

\subsection{Renormalization group equations}
\begin{figure}[ht!]
  \centering
  \includegraphics[width=6cm]{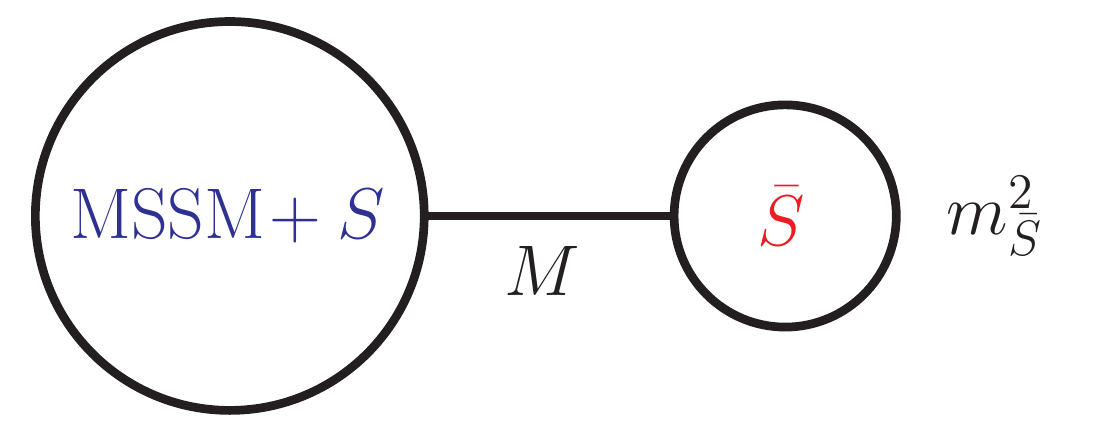}
\caption{Schematic structure of Dirac NMSSM.}
\label{fig:structure}
\end{figure}

The Higgs sector is natural when there are no large radiative corrections to $m_{H_{u,d}}^2$.
The renormalization group equations (RGEs) for $m_{H_{u,d}}^2$ are
\begin{eqnarray}
   \frac{d}{d\log Q} m_{H_u}^2 &=& \frac{1}{8\pi^2} 
   \Big\{3 y_t^2 \left(m_{\tilde{Q}_3}^2+m_{\tilde{t}_R}^2 +m_{H_u}^2 +|A_t|^2 \right)
   \no\\
   &&\ \ +\lambda^2\left(m_{S}^2 +m_{H_u}^2+m_{H_d}^2 +|A_\lambda|^2\right)
   \no\\
   &&\ \ -3g_2^2 M_2^2-g_1^2 M_1^2
   \Big\} ,\\
      \frac{d}{d \log Q} m_{H_d}^2 &=& \frac{1}{8\pi^2} 
   \Big\{\lambda^2\left(m_{S}^2 +m_{H_u}^2+m_{H_d}^2 +|A_\lambda|^2\right)
   \no\\
   &&\ \ -3g_2^2 M_2^2-g_1^2 M_1^2
   \Big\} \ .
  \label{eq:RGE}
\end{eqnarray}
While heavy stops or $m_{S}^2$ lead to fine-tuning, we find that $m_{\bar S}^2$ does not appear.
In fact, the RGEs for  $m_{H_{u,d}}^2$ are independent of $m_{\bar S}^2$ to all orders in mass-independent schemes such as $\overline{MS}$ and $\overline{DR}$ schemes. 
This is clarified by dimensional analysis. 

First of all, because $\bar{S}$ couples to the MSSM+$S$ sector only through the  dimensionful coupling $M$ as in Fig.~\ref{fig:structure}, their interaction vanishes in $M\to 0$ limit, and then terms involving $m_{\bar{S}}^2$ must proportional to $M$. Next, for the $U(1)_{\bar{S}}$ conservation, a combination of lowest mass dimension is $|M|^2 m_{\bar{S}}^2$, which has too high mass dimension to enter RGEs of the Higgs parameters, $m_{H_{u}}^2$, $m_{H_{d}}^2$ and $\mu B$, 
whose mass dimension is only two. Hence, the large $m_{\bar{S}}$ does not upset naturalness through  RGEs. 

\subsection{Threshold corrections}
We consider threshold corrections in the effective theory where the scalar component of $\bar{S}$ is integrated for $m_{\bar{S}}\gg M, \mu, m_{S}, m_{H_{u,d}}, A_\lambda ...,$ and see if important corrections appear for $m_{H_{u,d}}^2$. One can see that the double insertion of $M$ is needed for $\bar{S}$ to involve $m_{H_{u,d}}^2$ as shown in Fig.~\ref{eq:threshold1} and consequently there is no quadratic sensitivity to $m_{\bar{S}}$.  
In the diagram shown in  right-bottom of Fig.~\ref{fig:threshold}, $m_{\bar{S}}$ changes only finite piece of log-divergence. 
 \begin{figure}[t!]
  \centering
  \includegraphics[width=5.8cm]{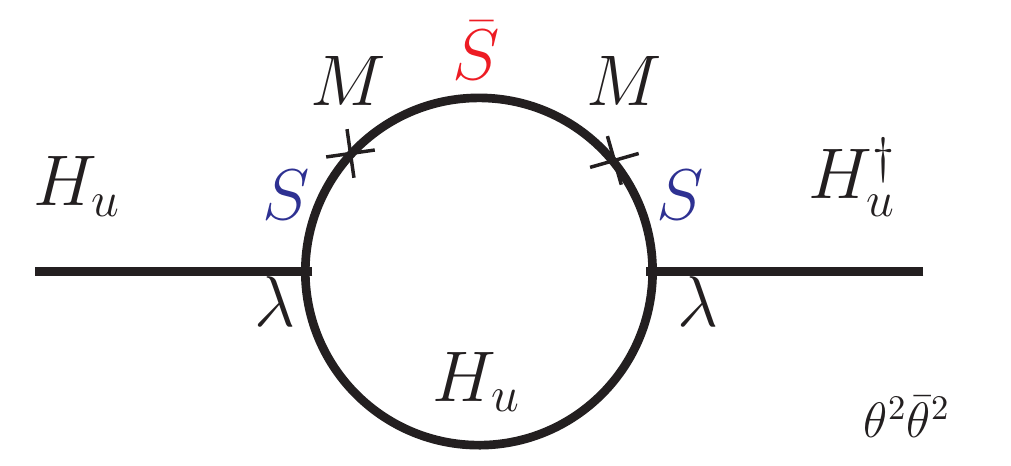}
\caption{Supergraph of $m_{H_u}^2$ correction}
\end{figure}
Hence there is only logarithmic sensitivity to $m_{\bar{S}}^2$ from the one-loop finite
threshold correction in Fig.~\ref{fig:threshold},
\begin{eqnarray}
  \delta m_H^2 \equiv \delta m_{H_{u,d}}^2 
  = \frac{(\lambda M)^2}{(4\pi)^2} \log \frac{M^2 + m_{\bar{S}}^2}{M^2}\ ,
  \label{eq:threshold1}
\end{eqnarray}
which still allows for very heavy $m_{\bar{S}}^2$ without fine-tuning.

\begin{figure}[h!]
  \centering
  \includegraphics[width=0.55\linewidth]{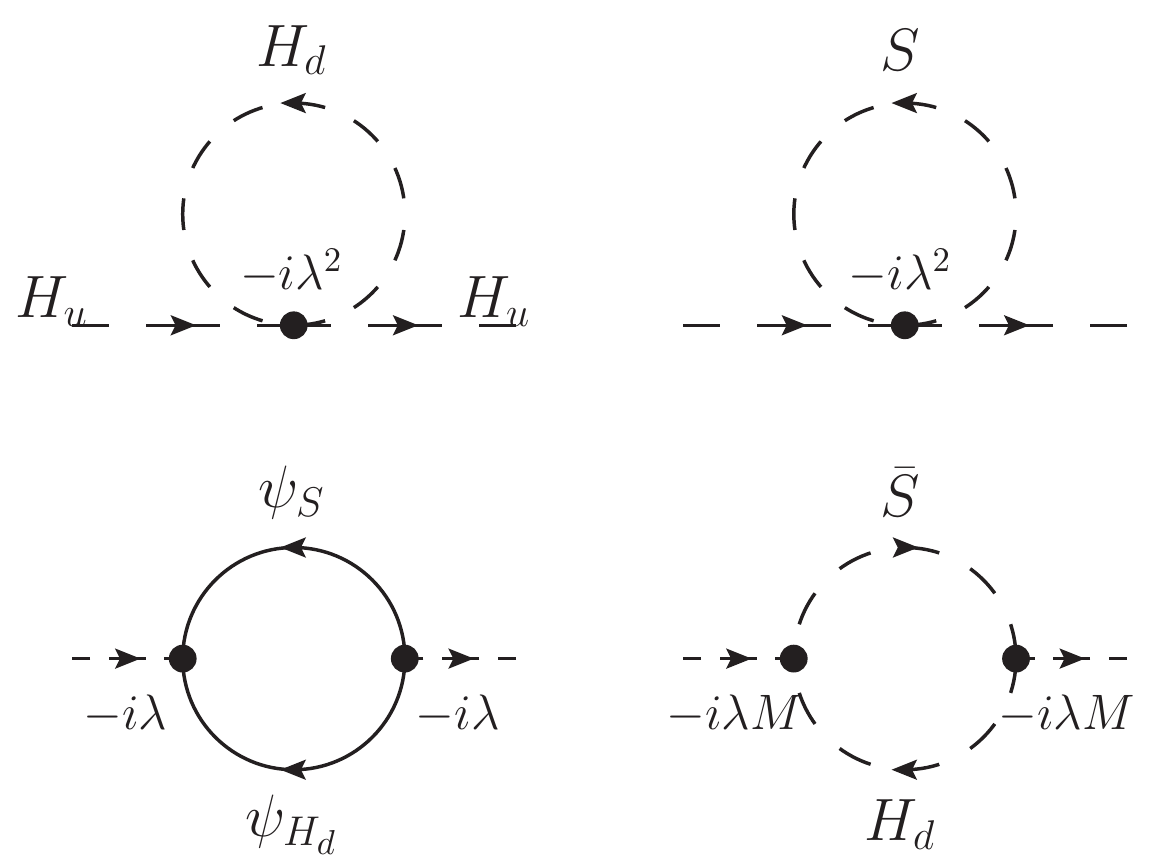}
\caption{Threshold corrections to $m_{H_{u}}^2$}
\label{fig:threshold}
\end{figure}

One may wonder if there are dangerous finite threshold corrections to $m_{H_u}^2$ at higher order, after integrating out $\bar S$.  In fact, there is no quadratic sensitivity to $m_{\bar S}^2$ to all orders again.  This follows because any dependence on $m_{\bar S}^2$ must be proportional to $|M|^2$ (since $\bar S$ decouples when $M \rightarrow 0$ and by conservation of $U(1)_{\bar S}$), but $ |M|^2 m_{\bar S}^2$ has too high mass dimension.  The mass dimension cannot be reduced from other mass parameters appearing in the denominator because threshold corrections are always analytic functions of IR mass parameters~\cite{Georgi:1994qn}.
This fact is from the matching of IR singularities. The full (UV) theory and the effective theory where the high energy dynamics down to a certain energy scale, e.g., $m_{\bar{S}}$, is integrated out must have same IR structure, namely, IR singularities are same in the both theories. If there are threshold corrections which are not analytic for IR mass parameters, such as, 
	\begin{eqnarray}
	\delta m^2 \sim \frac{|M|^2 m_{\bar S}^2 }{m_{\rm IR}^2},
	\end{eqnarray}
they will be new IR singularities when $m_{\rm IR}^2\to0$, which contradicts the IR matching.

\subsection{Semi-soft SUSY breaking}
It sounds contradictory that naturalness is maintained in the limit of very heavy $m_{\bar S}$, since removing the $\bar S$ scalar from the spectrum constitutes a hard breaking of supersymmetry\@.  
In the effective theory with $m_{\bar{S}}\to \infty$, there exist the $\bar S$ fermion with a Dirac mass and  several terms originating from $F_{\bar{S}}$,
	\begin{eqnarray}
	{\cal L}\supset i \bar{\psi}_{\bar{S}}\bar{\sigma}^\mu \partial_\mu \psi_{\bar{S}}
	-M \psi_S \psi_{\bar{S}}
	-M^2|S+c_{\bar{S}}\mu|^2 \,\, ,
	\label{eq:eff1}
	\end{eqnarray}
as well as terms derived by 
	\begin{eqnarray}
	{\cal K}&\supset&H_u^\dag H_u +H_d^\dag H_d +S^\dag S \ ,
	\no\\  
	W_{\rm} &\supset&\lambda S H_u H_d +\mu H_u H_d \ .
	\end{eqnarray}

This theory can be written in superfields and soft breakings if we reintroduce the scalar of $\phi_{\bar{S}'}$ and $F$-term of $F_{\bar{S}'}$ to form a chiral supermultiplet of 
	\begin{eqnarray}
	\bar{S}'=\phi_{\bar{S}'}+\theta \psi_{\bar{S}}+\theta^2 F_{\bar{S}'}
	\end{eqnarray}
Eq.(\ref{eq:eff1}) is rewritten in superspace as
	\begin{eqnarray}
	{\cal K}_{\it eff}&=& \bar{S}'^\dag \bar{S}' 
	-\theta^2\bar{\theta}^2 \left(M^2 | S + c_{\bar{S}} \mu|^2 \right)	\ , \no\\
	W_{\it eff} &=& -\theta^2 \left( M\, {\cal D}^\alpha  {S} \ {\cal D}_\alpha  \bar{S}' \right)
	\end{eqnarray}
where $\phi_{\bar{S}'}$  and $F_{\bar{S}'}$ are completely decoupled from the other states. 
Therefore, the low energy effective theory even in absence of scalar  $\bar{S}$ is equivalent to a theory with only softly broken supersymmetry. 
%
We call this mechanism {\it semi-soft supersymmetry breaking}. It is crucial that $\bar{S}$ couples to the other fields only through dimensionful couplings. 
Note that Dirac gauginos are a different example where adding new fields can lead to improved naturalness properties \cite{Fox:2002bu}.

\section{Higgs Mass at Tree-Level}
The most natural region of parameter space, summarized in Fig.~\ref{fig:schema}, has $m_S$ and $M$ at the hundreds of GeV scale, to avoid large corrections to $m_{H_u}$, and large $m_{\bar S} \gtrsim 10$~TeV, to maximize the second term of Eq.~(\ref{eq:higgsmass}).  The tree-level contribution to the Higgs mass can be large enough such that $m_{\tilde{t}}$ takes a natural value at the hundreds of GeV scale.

\subsection{CP-even neutral scalars}
For simplicity we assume there are no CP violations in the following studies.  
The minimization conditions for the Higgs and Singlet scalars are
	\begin{eqnarray}
	&&\mu_{\it eff}^2 +m_{H_d}^2-b_{\it eff}\tan\beta +\frac{m_Z^2}{2}\cos 2\beta +\lambda^2 v^2 \sin^2\beta=0, \label{EOMvu}
	\\
	&&\mu_{\it eff}^2 +m_{H_u}^2-b_{\it eff}\cot\beta -\frac{m_Z^2}{2}\cos 2\beta +\lambda^2 v^2 \cos^2\beta=0, \label{EOMvd}
	\\
	&&M^2+m_{S}^2+ \frac{\lambda v^2}{v_S} (\mu +\lambda v_S)
	+\frac{1}{v_{S}}\left(M B_S v_{\bar{S}}-\frac{\lambda^2 A_\lambda v^2}{2} \sin 2\beta 
	+c_{\bar{S}} M^2\mu+t_s \right)=0,\quad\quad\quad
	\\
	&&M^2+m_{\bar{S}}^2+\frac{1}{v_{\bar{S}}}
	\left( M B_S v_S  -\frac{\lambda M v^2}{2} \sin 2\beta +t_{\bar S}\right) =0 .
	\end{eqnarray}
where $v_S=\langle S\rangle$ and $v_{\bar{S}}=\langle \bar{S}\rangle$, and it is convenient to use
	\begin{eqnarray}
	&&b_{\it eff}\equiv \mu B+\lambda(A_\lambda v_S+Mv_{\bar{S}}),\label{beff}\\
	&&\mu_{\it eff}\equiv \mu+\lambda v_S, \label{mueff}\\
	&&\bar{m}_{A}^2 \equiv b_{\it eff}/(s_\beta c_\beta).	
	\end{eqnarray}
The mass matrix of CP-even neutral scalars at tree level is given by ${\cal M}^2_{H^0,ij}(=\M^2_{H^0,ji})$ in the base of Re$(H_d^0,H_u^0, S, \bar{S})^T/\sqrt{2}$ where
	\begin{eqnarray}
&&\M_{H^0}^2 =\no\\
&&\left(
	\begin{array}{cc cc}
\M_{H^0,11}^2 & &  & \\ \\
	\M_{H^0,21}^2& \M_{H^0,22}^2&  &\\ \\
	 \lambda v(2\mu_{\it eff}  c_\beta-A_\lambda  s_\beta)&\!\!\! \lambda v(2\mu_{\it eff}  c_\beta-A_\lambda  s_\beta)  & M^2+m_S^2+\lambda^2 v^2\\ \\
	 -\lambda v M s_\beta & -\lambda v M c_\beta & MB_S &M^2+m_{\bar{S}}^2
	\end{array}
\right)
\label{CPevenmatrix} 	\no
\\
	\end{eqnarray}
and where 
	\begin{eqnarray}
	\M^2_{H^0, 11}&=&\bar{m}_{A}^2 s^2_\beta+m_Z^2 c^2_\beta,\nonumber\\
	\M^2_{H^0, 22}&=&\bar{m}_{A}^2 c^2_\beta+m_Z^2 s^2_\beta, \nonumber\\
	\M^2_{H^0, 21}&=&-(\bar{m}_{A}^2+m_Z^2-2\lambda^2 v^2)s_\beta c_\beta . 
	\end{eqnarray}
Here $s_\beta (c_\beta)$ denotes $\sin\beta(\cos\beta)$, and the vacuum conditions of Eqs.(\ref{EOMvu}, \ref{EOMvd}) are used for $\M^2_{H^0, 11}$ and $ \M^2_{H^0, 22}$. Note that non-decoupling effect can be seen in $\M^2_{H^0,12}$, which completely remains when $m_{\bar{S}}\to \infty$.

\section{Radiative Corrections}
We take into account radiative corrections to the mass matrix. It is particularly important to estimate  stop mass. The one-loop level calculation of $y_t$ is easy to perform, which raises the SM-like Higgs mass, but it is well-known that the two-loop contributions tend to reduce the Higgs mass. Since only one-loop calculation leads optimistic result about naturalness, we adopt RG-improved one-loop calculation which is consistent enough with two-loop calculation of the MSSM. The similar result is given by Ref.\cite{Carena:1995bx}.

\subsection{One-loop corrections}
The corrections are included by two parts. One is at ${\cal O}(y_t^4)$ calculated by effective potential, and the other is at ${\cal O}(y_t^2 g_1^2, y_t^2 g_2^2)$ involving ${\cal O}(y_t^2)$ corrections to wave function. The effective potential at one-loop is given by
	\begin{eqnarray}
	\Delta V_{\it}^{(1)}(H, Q)&=&\frac{3}{32\pi^2}\Bigg\{
	m_{\tilde{t}_1}^4(H) \left(\log \frac{m_{\tilde{t}_1}^2(H)}{Q^2}  -\frac{3}{2}\right) 
	+m_{\tilde{t}_2}^4(H) \left(\log \frac{m_{\tilde{t}_2}^2(H)}{Q^2}  -\frac{3}{2}\right) 
	\no\\
	&&\ \ \ \ \ \ \
	 -2m_t^4(H) \left(\log \frac{m_t^2(H)}{Q^2}  -\frac{3}{2}\right)
	\Bigg\} \ \ ,
	\end{eqnarray}
where the mass parameters are $H_{u,d}$ fields dependent,
	\begin{eqnarray}
	&&m_t^2(H)=|y_t H_u|^2 \ , \\ 
	&&m_{\tilde{t}_1}^2(H)=m_{\tilde{t}}^2+|y_t H_u^0|^2- |y_t \mu H_d^0|  \ , \\
	&&m_{\tilde{t}_2}^2(H)=m_{\tilde{t}}^2+|y_t H_u^0|^2+ |y_t \mu H_d^0|  \ .
	\end{eqnarray}
Here we assume $A_t=0$ and $m_{\tilde{t}_R}^2=m_{\tilde{Q}_3(\tilde{t}_L)}^2=m_{\tilde{t}}^2$ for simplicity. 

\begin{figure*}[h!]
  \centering
  \includegraphics[width=9.cm]{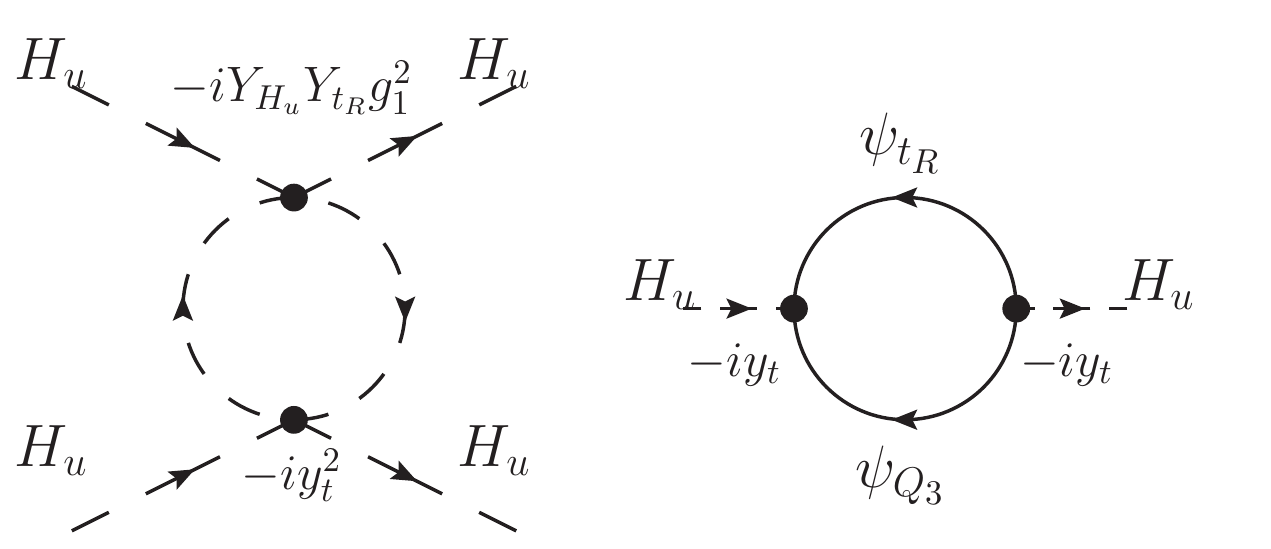}\hspace{20pt}
  \includegraphics[width=4.cm]{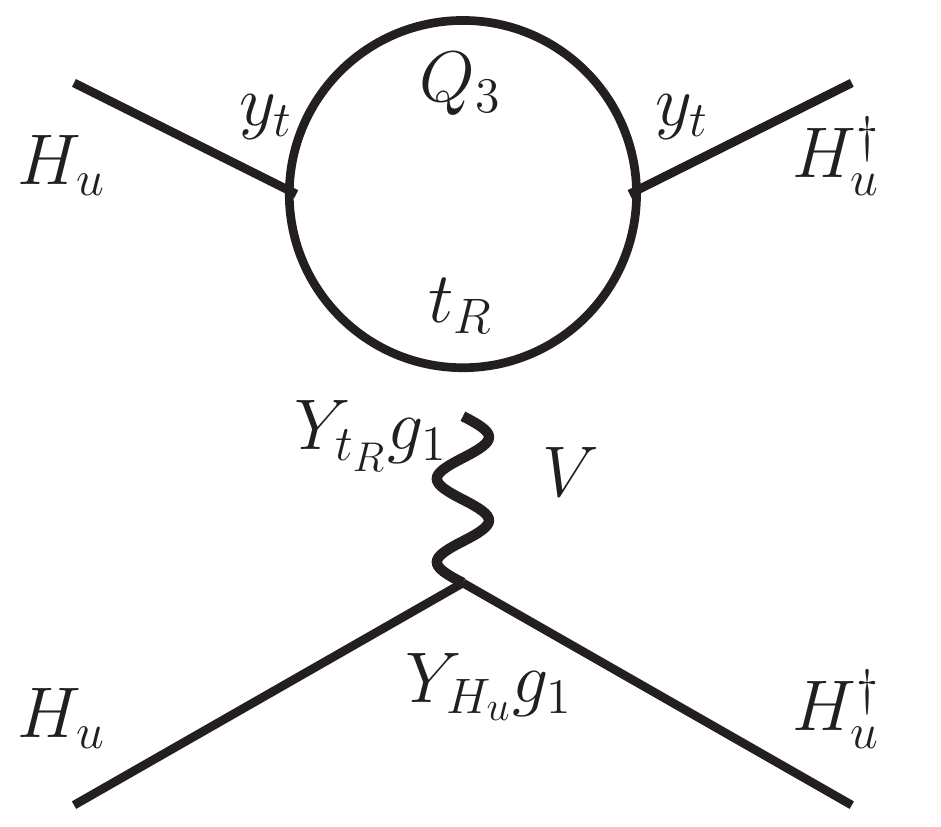}
\caption{{\it Left:}  ${\cal O}(y_t^2 g_1^2)$ vertext correction, {\it center:} wave function renormalization leading corrections of  ${\cal O}(y_t^2 g_1^2)$. {\it Right:} One-loop supergraph. The supergraph without attachment of gauge field leads wavefunction renormalization of {\it center}, and that with gauge field leads vertex corrections of {\it left}. }
\label{fig:Correction2}
\end{figure*}

In addition, the corrections of ${\cal O}(y_t^2 g_1^2, y_t^2 g_2^2)$ require wave function renormalization as well as vertex corrections. Since both corrections are from the same one-loop supergraph that the fermionic part enters wave function renormalization and the bosonic part enters vertex correction as seen Fig.~\ref{fig:Correction2}, their divergences are canceled leaving finite corrections. We match them onto the Higgs potential,
	\begin{eqnarray}
	\Delta V^{(2)}_{\it} (H)
	=-\frac{g_1^2+g_2^2}{4}\frac{3y_t^2}{8\pi^2} \ln \frac{m_{\tilde{t}}^2 +m_t^2}{m_t^2} 
	\left( |H_u^0|^4 -\frac{1}{2} |H_u^0|^2|H_d^0|^2 \right) \ ,
	\end{eqnarray}
where we neglect the mixing between $\tilde{t}_R$ and $\tilde{t}_L$.

For the new potential at one-loop level, 
the vacuum conditions for  $H_{u,d}$ are shifted,
	\begin{eqnarray}
	&&\mu_{\it eff}^2 +m_{H_d}^2-b_{\it eff}\tan\beta +\frac{m_Z^2}{2}\cos 2\beta +\lambda^2 v^2 \sin^2\beta+\Big\langle\frac{1}{2H_d^0}\frac{\partial  \Delta V}{\partial H_d^0}\Big\rangle=0 \ , \label{EOMvu2}
	\\
	&&\mu_{\it eff}^2 +m_{H_u}^2-b_{\it eff}\cot\beta -\frac{m_Z^2}{2}\cos 2\beta +\lambda^2 v^2 \cos^2\beta
	+\Big\langle\frac{1}{2H_u^0}\frac{\partial \Delta V}{\partial H_u^0}\Big\rangle=0 \ . \label{EOMvd2}
	\end{eqnarray}
Hence, the mass matrix of Eq.(\ref{CPevenmatrix}) is modified by 
	\begin{eqnarray}
	\Delta\M^2_{H^0, 11}&=&
	\frac{1}{2}\Big\langle\frac{\partial^2  \Delta V}{\partial (H_d^0)^{2}}\Big\rangle
	-\Big\langle\frac{1}{2H_d^0}\frac{\partial \Delta V}{\partial H_d^0}\Big\rangle 
	\no\\
	&=& \frac{3}{32\pi^2}\left(
	(y_t\mu)^2-\frac{y_t\mu(m_{\tilde{t}}^2+m_t^2)}{v c_\beta}\log\frac{m_{\tilde{t}_1}^2}{m_{\tilde{t}_2}^2}
	\right) \label{deltaM11}
	\ ,\\
	\Delta\M^2_{H^0, 22}&=&
	\frac{1}{2}\Big\langle\frac{\partial^2  \Delta V}{\partial (H_u^0)^2}\Big\rangle
	-\Big\langle\frac{1}{2H_u^0}\frac{\partial \Delta V}{\partial H_u^0}\Big\rangle 
	\no\\
	&=&\frac{3y_t^2 m_t^2}{8\pi^2} \ln \frac{m_{\tilde{t}_1}^2 m_{\tilde{t}_2}^2}{m_t^4}
	-m_Z^2s_\beta^2 \left(\frac{3y_t^2}{8\pi^2} \ln \frac{m_{\tilde{t}}^2 +m_t^2}{m_t^2} \right)
	\label{deltaM22}
	\ ,\\
	\Delta\M^2_{H^0, 21}&=&\frac{1}{2}\Big\langle\frac{\partial^2  \Delta V}{\partial H_u^0\partial H_d^0}\Big\rangle 
	\no\\
	&=&\frac{3y_t^2 m_t \mu}{16\pi^2}\log\frac{m_{\tilde{t}_1}^2}{m_{\tilde{t}_2}^2}
	+m_Z^2s_\beta c_\beta \left(\frac{3y_t^2}{16\pi^2} \ln \frac{m_{\tilde{t}}^2 +m_t^2}{m_t^2} \right)
	\label{deltaM21}  . \ 
	\end{eqnarray}
In Eqs.(\ref{deltaM11},\ref{deltaM22}), the vacuum shifts must be considered since the vacuum conditions are already used in Eq.(\ref{CPevenmatrix}). This result is consistent with Refs.\cite{Carena:1995bx}.
 
\subsection{RG-improved calculation}
In the MSSM, it is known that two-loop corrections are not negligible for the Higgs mass estimation. We have taken into account ${\cal O}(y_t^4)$ and ${\cal O}(y_t^2 g_{1,2}^2)$ corrections above. As two-loop contributions the dominant corrections should be at ${\cal O}(y_t^4 g_s^2)$ and ${\cal O}(y_t^6)$. 

Although we do not perform explicit two-loop calculation,  RG-improved one-loop calculation developed in Ref.~\cite{Carena:1995bx} deals with some part of corrections of ${\cal O}(y_t^4 g_s^2)$ and ${\cal O}(y_t^6)$. In the one-loop correction with $y_t$, top quark and squark are always involved and we use $y_t$ value at weak scale. However, the two particles have two different scales, weak scale and $m_{SUSY}$, and then the scale of $y_t$ we use is not the best one and more appropriately it should be a geometric mean of the two scales, 
	\begin{eqnarray}
	Q\sim \sqrt{m_t(m_{\tilde{t}}^2+m_t^2)^{1/2}}. 
	\end{eqnarray}
When we change the scale of $m_t$, beta functions of $y_t^2$ and $g_s^2$ are dominant,
	\begin{eqnarray}
	m_t\to m_t\left( 1+\left( \beta_t  +\beta_s \right)\log\frac{Q}{m_t} \right)
	\end{eqnarray}
 Thus original correction at ${\cal O}(y_t^4)$ leads to ${\cal O}(y_t^4 g_s^2)$ and ${\cal O}(y_t^6)$ through RGE. We expect this scale discrepancy is the main source of two-loop corrections. 
 We adopt beta functions of low energy effective theory that is Type-II two Higgs doublet model,
	\begin{eqnarray}
	\beta_t=\frac{3y_t^2}{32\pi^2}\ , \ \  \beta_s=-\frac{g_s^2}{2\pi^2} \ .
	\end{eqnarray}

\begin{figure*}[h!]
  \centering
  \includegraphics[width=8.cm]{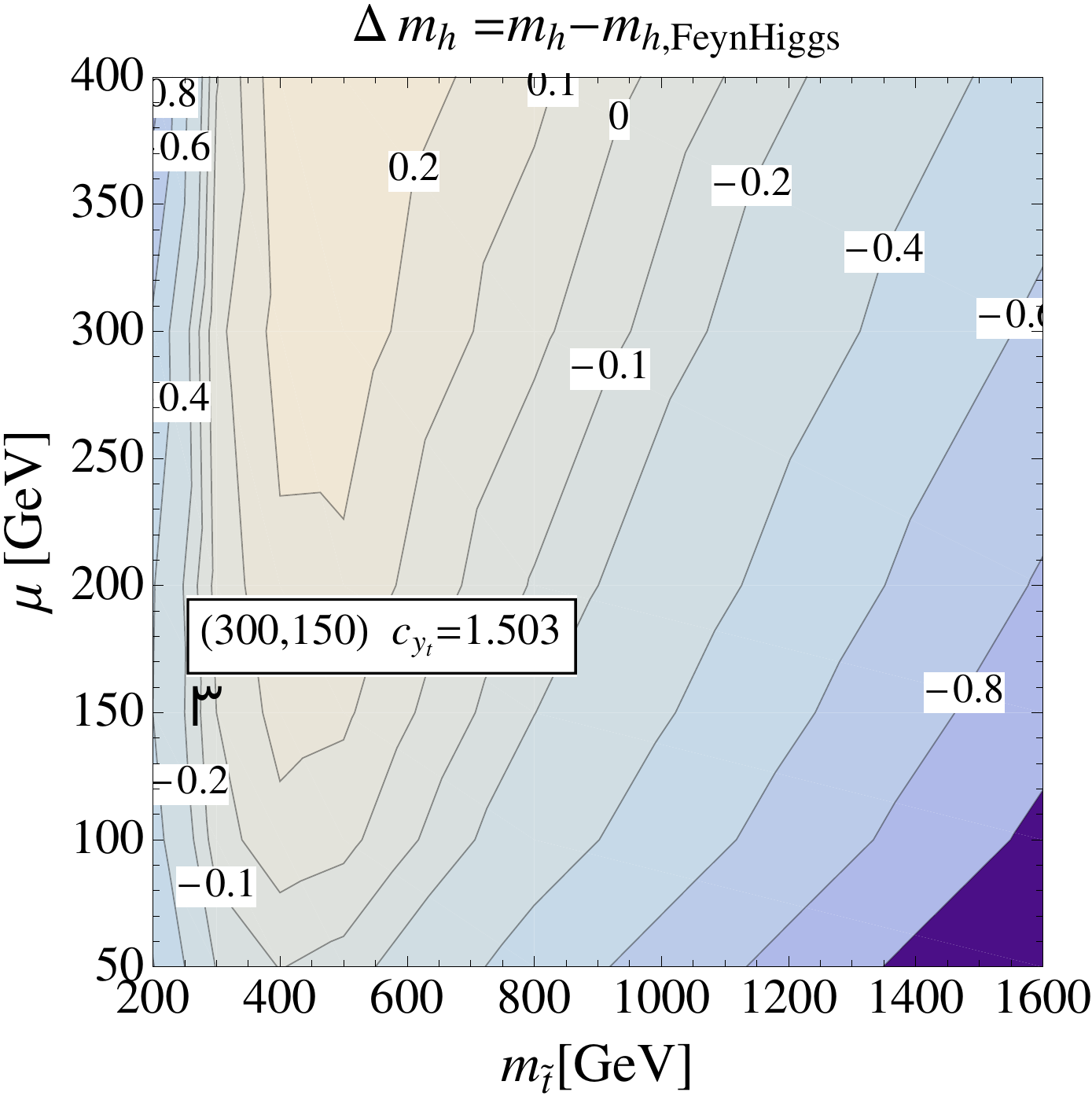}
\caption{The mass difference between the lightest Higgs mass calculated by RG-improved method and that calculated by {\tt FeynHiggs} (in GeV). Our calculation is matched with a result of {\tt FeynHiggs} at $\mu=150 \GeV$ and $m_{\tilde{t}}=300 \GeV$. Other parameters in {\tt FeynHiggs} are set to be $\tan\beta=2, \ A_t=0, \ m_{A_0}=1 \TeV, \ (M_1, M_1, M_1)=(300, 400, 1000) \GeV$.}
  \label{matchingc}
\end{figure*}

Since the two-loop calculations in the MSSM are available, we can make further improvement. We basically use the renormalization scale, 
	\begin{eqnarray}
	Q=c_{\rm y_t} \sqrt{m_t(m_{\tilde{t}}^2+m_t^2)^{1/2}} \ \ ,
	\end{eqnarray}
and $c_{\rm y_t}$ should be an $\order(1)$ coefficient but  is not strictly determined. We choose $c_{\rm y_t}$ so that our MSSM result ($\lambda \to0$) matches with the full two-loop MSSM result given by a software {\tt FeynHiggs} \cite{Frank:2006yh}. When we choose the matching point in the MSSM of $\mu=150 \GeV$ and $m_{\tilde{t}}=300 \GeV$, the coefficient $c_{\rm y_t}$ turns out to be $\simeq1.5$ as shown in Fig.~\ref{matchingc}. For other values of $\mu$ and $m_{\tilde{t}}$ the difference of the lightest Higgs mass is quite small, $|\Delta m_h|<1 \GeV$. Then we reliably apply this calculation for the upper-left $2\times2$ mass matrix of CP-even scalars in Eq.~(\ref{CPevenmatrix}).


\section{Benchmark Parameters and Stop Mass}
When we estimate the stop mass in the Dirac NMSSM, we take $M$ and $m_{\bar{S}}$ to be free parameters since $M$ is common in both models and $m_{\bar{S}}$ is essential to discuss the non-decoupling effect. Other parameters are fixed according to the table of benchmark parameters (Table.\ref{tab:bench}). We have chosen $\lambda$ to saturate the upper-limit such that it does not reach a Landau pole below the unification scale~\cite{Barbieri:2007tu}.
For the NMSSM, we treat $m_{{S}}$ as a free parameter instead of $m_{\bar{S}}$ and correspondingly   $b_{\it eff}$ has a different definition,
	\begin{eqnarray}
	b_{\it eff}\equiv \mu B+\lambda v_S(A_\lambda +M).
	  \label{beff2}
	\end{eqnarray}
The other parameters are as same as in the Dirac NMSSM. 

\begin{table}[h!]
\centering
  \begin{tabular}{|c   c  c |}
\hline
 $\lambda = 0.74$ & $\tan \beta = 2$ &\quad $\mu_{\it eff} = 150$~GeV \\
 $b_{\it eff}$ =(190 GeV)$^2$ &$A_\lambda = 0$ & $B_s = 100$~GeV\\ 
 \multicolumn{3}{|c|}{$m_S=800$~GeV}\\
 \hline
  \end{tabular}
  \caption{Benchmark parameters}
  \label{tab:bench}
\end{table}
One may worry about the size of singlet VEVs especially from the tadpoles because we fixed $\mu_{\it eff}$ and $b_{\it eff}$ defined in Eqs.(\ref{beff},\ref{mueff}). However we can see the VEVs  adequately small, 
	\begin{eqnarray}
	&&v_S \simeq \frac{(c^*_{\bar{S}}\mu^*|M|^2 +t_{S}^*)(|M|^2+m^2_{\bar{S}}) }
	{(|M|^2+m^2_{{S}})(|M|^2+m^2_{\bar{S}})} \sim \mu^* 
	\\
	&&v_{\bar{S}} \simeq \frac{t_{\bar{S}}^*(|M|^2+m^2_{S})}{(|M|^2+m^2_{{S}})(|M|^2+m^2_{\bar{S}})}
	\sim \frac{(\mu M)^* m_{\bar{S}}}{|M|^2+m^2_{\bar{S}}}
	\end{eqnarray}
where the tadpoles scale as $t_S\sim \mu^* m_{S}^2$ and $t_{\bar{S}}\sim \mu M m_{\bar{S}}$.

\begin{figure*}[t!]
  \centering 
  \includegraphics[width=\linewidth]{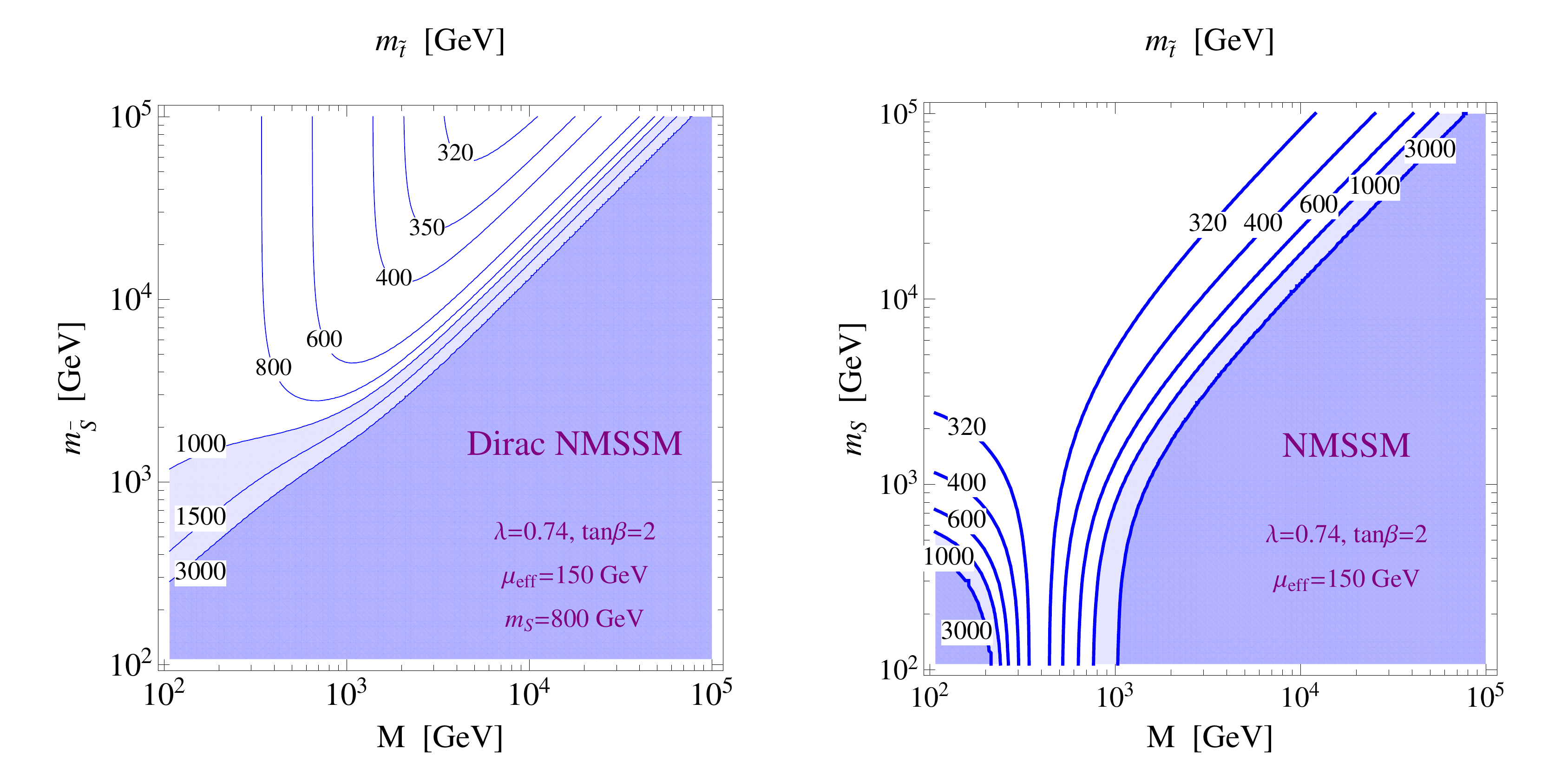}
  \caption{Stop soft mass in the Dirac NMMSM ({\it left}) and NMMSM ({\it right}). 
  }
  \label{fig:stopmass}
 \end{figure*}

We now estimate the stop mass. For each value
of $(M, \mSbar) $, the stop soft masses, $m_{\tilde{t}} = m_{\tilde{t_R}} = m_{\tilde{Q_3}}$, are
chosen to maintain the lightest scalar mass at 125 GeV. As results are shown in Fig.~\ref{fig:stopmass}, 
basically the stop mass becomes small as $\mSbar$ increases for the fixed $M$ because more non-decoupling effect remains.  Hence, the natural region should spread in large $\mSbar$ where naturalness does not suffer from fine-tuning thanks to the semi-soft breaking.  
However, this argument does not apply for the  low $M$ region. This is because the $S$ exchange effect, which is given by the second line of Eq.(\ref{eq:higgsmass}), is enhanced and decreases the tree-level Higgs mass when $M$ is small.  

\section{Raising Higgs Mass for Explicit $\mu$ in the NMSSM}\label{mudriven}
It is worth mentioning the behavior of stop mass in the NMSSM. Like in the Dirac NMSSM, the larger $m_S$ leads to the more non-decoupling effect, that is the more Higgs mass at tree level. Hence, the stop mass becomes lighter as $m_S$ increases for the fixed $M$. In this limit, naturalness is worsened by $m_S$ rather than $m_{\tilde{t}}$.

Apart from the non-decoupling effect along with the SUSY breaking, there is a supersymmetric effect to enhance the tree-level Higgs mass in presence of explicit $\mu$ term while keeping natural EWSB \cite{Nomura:2005rk}. After the singlet field is integrated out, new Higgs potential is generated,
	\begin{eqnarray}
	V_{\it eff}=\frac{\mu^2 M^2-\lambda^2 \mu M(H_u H_d+h.c.)+\lambda^4|H_uH_d|^2}{M^2+\lambda^2(|H_u|^2+|H_d|^2)}(|H_u|^2+|H_d|^2)
	\end{eqnarray}
where $m_S\to0$. It leads to additional Higgs mass at tree level,
	\begin{eqnarray}
	\Delta m_h^2 = \frac{4\lambda^2(M\mu \sin2\beta -\mu^2)}{M^2}
	\end{eqnarray}
where the VEVs and mass-eigenstates are aligned, $H_u\to v_u +h \sin\beta$ and $H_d\to v_d +h \cos\beta$. This term positively remains only if $M$ and $\mu$ have the same sign and the same size. As seen in Fig.~\ref{fig:stopmass} {\it right}, the mass correction vanishes when $M$ is a few times larger or smaller than $\mu$. 

\section{Naturalness}
In this section, naturalness is evaluated. 
There are several measures, and we take a bottom-up approach described in Ref.~\cite{Kitano:2006gv} 
since we do not specify any UV completions and discuss based on low-energy parameters. 

\begin{figure}[t!]
\centering
\includegraphics[width=1.0\linewidth]{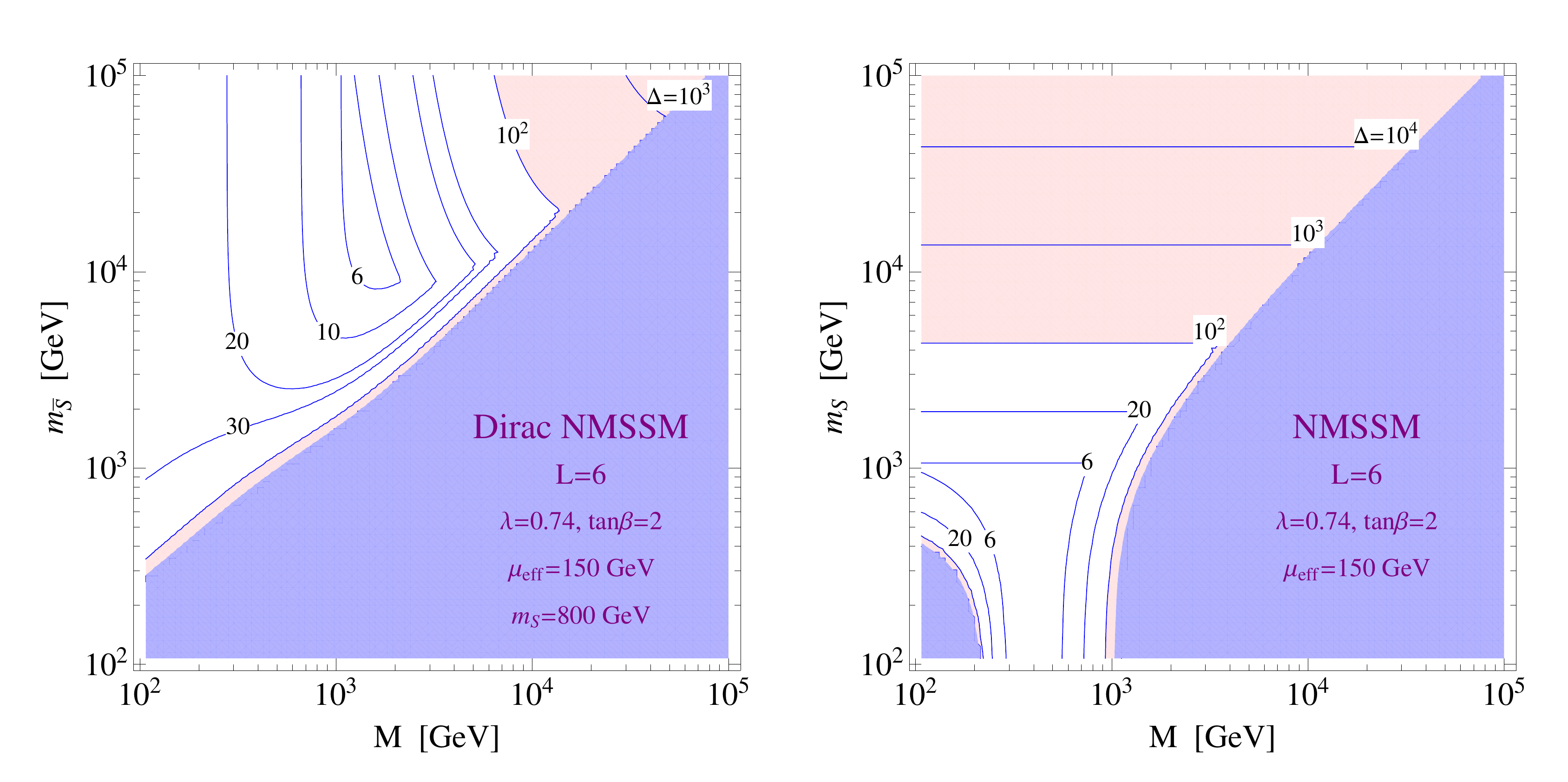}
\includegraphics[width=1.0\linewidth]{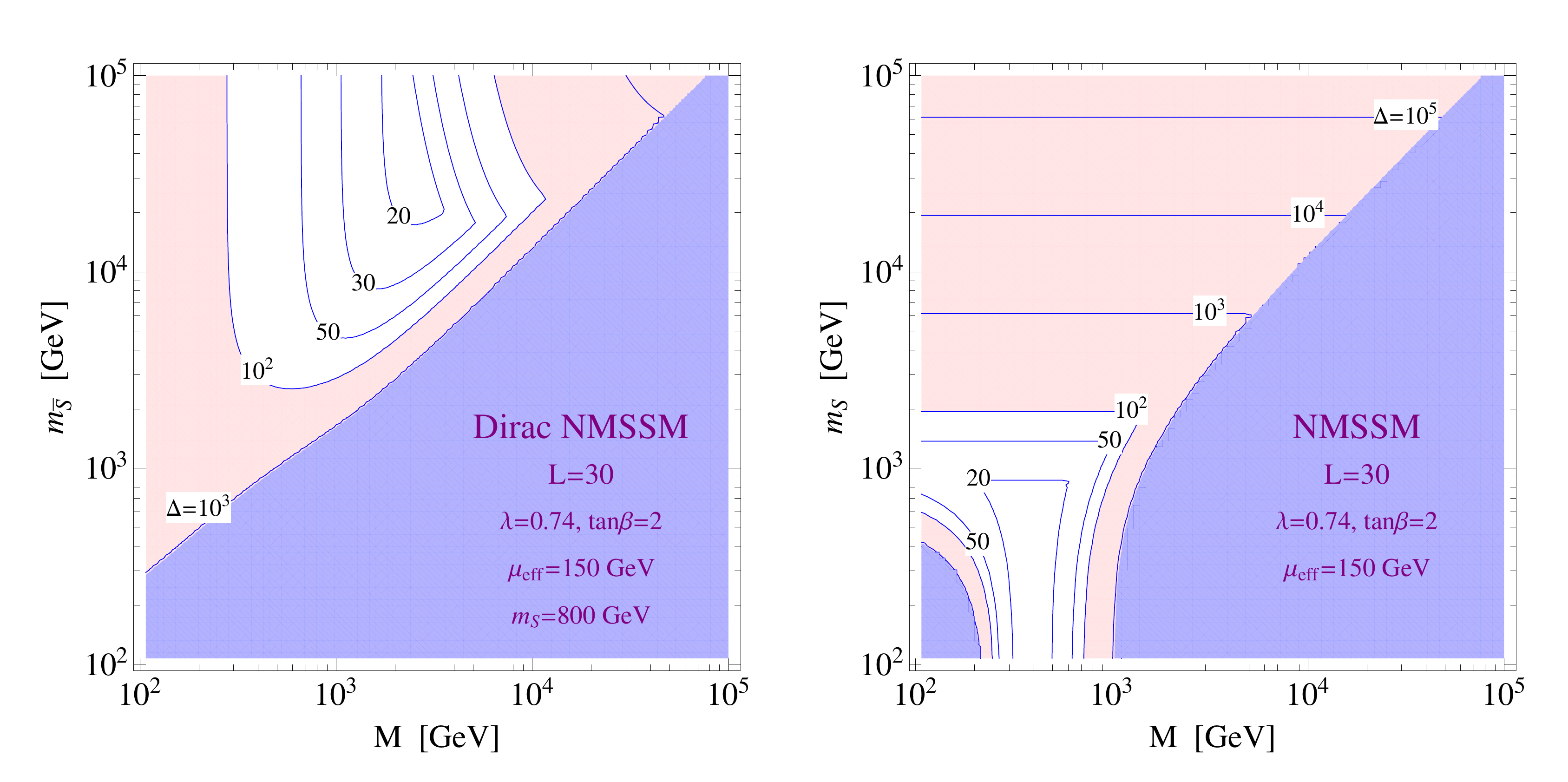}
\caption{\label{fig:tuning}
 The tuning $\Delta$, defined in Eq.~{eq:tune}, for the Dirac
 NMSSM is shown on the {\it left}\/ as a function of $M$ and $m_{\bar
   S}$.   For comparison, the tuning of the NMSSM is shown
 on the {\it right}\/, as a function of $M$ and
 $m_S$.  
 The red region has high fine-tuning, $\Delta >
 100$, and the purple region requires $m_{\tilde t} > 2$~TeV, signaling severe fine-tuning $\gtrsim \mathcal{O}(10^3)$. The tuning is evaluated with $L=6$ ($30$) in the {\it upper} ({\it lower}) plots.}
\vspace{-5pt}
\end{figure}

The degree of fine-tuning is
estimated by 
\begin{equation} \label{eq:tune}
  \Delta = \frac{2}{m_h^2} {\rm max} \left( m_{H_u}^2, m_{H_d}^2, b_{\it eff}, \frac{d m_{H_u}^2 }{d \ln \mu}
    L, \frac{d m_{H_d}^2}{d \ln \mu}  L, \delta m_H^2 \right),
\end{equation}
where $L \equiv \log ( \Lambda_{\rm mess} / m_{\tilde t}) $, and $\Lambda_{\rm mess}$ is a scale at which particle masses  are generated. 
The first three terms of Eq.~(\ref{eq:tune}) represent magnitude of cancellation between parameters at the electroweak scale, and here $m_{H_u}^2$ and $m_{H_d}^2$ are determined by Eqs.~(\ref{EOMvu}, \ref{EOMvd}) with benchmark parameters. If any of them is much larger than the electroweak scale, tuning is required to give the right value of VEV. The rest terms of Eq.~(\ref{eq:tune}) shows fine-tuning from high-scale. Higgs soft masses are radiatively corrected by, for instance, top-stop sector, and if stop mass is extremely large, cancellation among contributions to $m_{H_u}^2$ is needed so that $m_{H_u}^2$ is finally at the electroweak scale. 
We take two values of $\Lambda_{\rm mess}$ corresponding to low-scale ($L=6$) and high-scale ($L=30$) SUSY breaking.  
We assume that contributions through gauge couplings to the RGEs for $m_{H_{u,d}}^2$ are subdominant. To be more concrete about Eq.~(\ref{eq:tune}), the RGE effects to $m_{H_u}$, for example,  are separated by each parameter, 
	\begin{eqnarray}
	\frac{d m_{H_u}^2}{d \ln \mu}  L  
	=\left(\frac{3y_t^2 m_{\tilde{Q}_3}^2  }{8\pi^2}L, \ 
	\frac{3y_t^2 m_{\tilde{t}_R}^2  }{8\pi^2}L, \ 
	\frac{\lambda^2 m_{S}^2  }{8\pi^2}L 
	\ \ldots
		\right) .
	\end{eqnarray}
The results are shown in Fig.~\ref{fig:tuning}.

In the Drac NMSSM, the degree of tuning $\Delta$ is mostly determined by the RGE effect from $m_{\tilde{t}}$. When $M$ is quite large, $M>1\TeV$, $\Delta$ is determined by the threshold correction given in Eq.~(\ref{eq:threshold1}). Actually as shown in Figs.~\ref{fig:stopmass} and \ref{fig:tuning}, the shape of curves almost corresponds to that of stop mass, and only upper-right region is dominated by the threshold correction. The difference between low- and high-scale SUSY breaking cases is basically the difference of absolute value by a factor of 5.  The tuning level is as low as $\Delta^{-1} \approx  20\%\ (5\%)$ for low-scale (high-scale) SUSY breaking.

We see that the least-tuned region of the Dirac NMSSM corresponds to $M \sim 2$~TeV and $m_{\bar S} \gtrsim 10$~TeV, where the tree-level correction to the Higgs mass is maximized. The fact that the large SUSY breaking, $m_{\bar S}$, leads to more natural theory is counter-intuitive, but, thanks to the mechanism introduced as semi-soft SUSY breaking, this particular SUSY breaking is irrelevant to naturalness. 

  On the other-hand, the tuning in the NMSSM is determined by the RGE effects not only through stop mass ($m_{\tilde{t}_R},m_{\tilde{Q}_3}$) but also $m_{S}$. 
  The NMSSM becomes highly tuned when $m_S$ is large, 
and then $m_S\lesssim 500~$GeV is favored. 
Note that region of low-tuning in the NMSSM extends to the supersymmetric limit, $m_{S} \rightarrow 0$.  In this region the Higgs mass is increased by a new contribution to the quartic coupling proportional to 
$\lambda^2(M\mu\sin 2\beta -\mu^2)/M^2$ as described in Sec.~\ref{mudriven}~(see Ref.\cite{Nomura:2005rk} for more details).

In order to estimate the SM-like Higgs mass at one-loop level, we only need to consider Higgs and singlet sector plus corrections from stop-top sector, and do not specify gaugino masses.  However, when the mediation scale is high or gluino mass is large, the main source of tuning can be gluino mass scale because stop masses are corrected by gluino mass with a large coefficient through RGE, 
	\begin{eqnarray}
	m_{\tilde{Q}_3, \tilde{t}_R}=\frac{g_s^2}{4\pi^2}\frac{N_C^2-1}{N_C}M_3^2+...
	\end{eqnarray}
This point is studied with the latest LHC constraints in Ref.\cite{Arvanitaki:2013yja}, and a recent study discusses it in the context of Dirac NMSSM \cite{Kaminska:2014wia} . The gluino mass contribution to tuning is roughly estimated by
	\begin{eqnarray}
	\Delta_{\tilde g} =\frac{2}{m_h^2}\left[\frac{3y_t^2}{8\pi^2}\frac{8\alpha_s}{3\pi}M_3^2 \right]\frac{L^2}{2}\sim 30 \left(\frac{M_3}{400\GeV}\right)^2 \left(\frac{L}{30}\right)^2. 
	\end{eqnarray}
This is independent of $M$, $m_S$, and $m_{\bar S}$. For high-scale mediation where $L=30$, gluino mass more than a few hundreds GeV becomes dominant in tuning and affects the lowest tuned parameter region of Fig.~\ref{fig:tuning},
regardless of $m_{\tilde{t}}$, $m_S$, and $m_{\bar S}$.


\chapter{Phenomenology of Dirac NMSSM}\label{dnmssm:pheno}
We now discuss the experimental signatures of the Dirac NMSSM\@.  The phenomenology of the NMSSM is well-studied~\cite{Hall:2011aa, Ellwanger:2011aa}.  In the natural region of the Dirac NMSSM, the singlet states are too heavy to be produced at the LHC\@.  The low-energy Higgs phenomenology is that of a two Higgs doublet model, and we focus here on the nature of the lightest SM-like Higgs, $h$, and the heavier doublet-like Higgs, $H$ \cite{Craig:2012pu}.
The properties of the two doublets differ from the MSSM due to the presence of the non-decoupling quartic
coupling $|\lambda H_u H_d|^2$ in Eq.~(\ref{CPevenmatrix}), which raises the Higgs mass by the semi-soft SUSY breaking, described above.  

For the fixed $\lambda$ and $\tan\beta$, the essential parameter is only $b_{\it eff}$ in upper-left $2\times 2$ mass matrix of Eq.~(\ref{CPevenmatrix}). Therefore, we treat $b_{\it eff}$ as a free parameter, and all the results are given in terms of the heavier Higgs mass, $m_H$, which is roughly proportional to $b_{\it eff}$.  
The other parameters are fixed as follows.
\begin{table}[h!]
\centering 
  \begin{tabular}{|c   c  c  c|}
\hline
 \multicolumn{4}{|c|}{benchmark parameters} \\ \hline
 $\lambda = 0.74$ & $\tan \beta = 2$ &\quad $\mu_{\it eff} = 150$~GeV &$A_\lambda = 0$\\
 $M = 1$~TeV   & $m_{\bar S} = 10$~TeV& $B_s = 100$~GeV&\\ 
 \hline
  \end{tabular}
\end{table}

\section{Current Constraints}

\subsection{SM-like Higgs}
\begin{figure}[t!]
  \centering
  \includegraphics[width=8.cm]{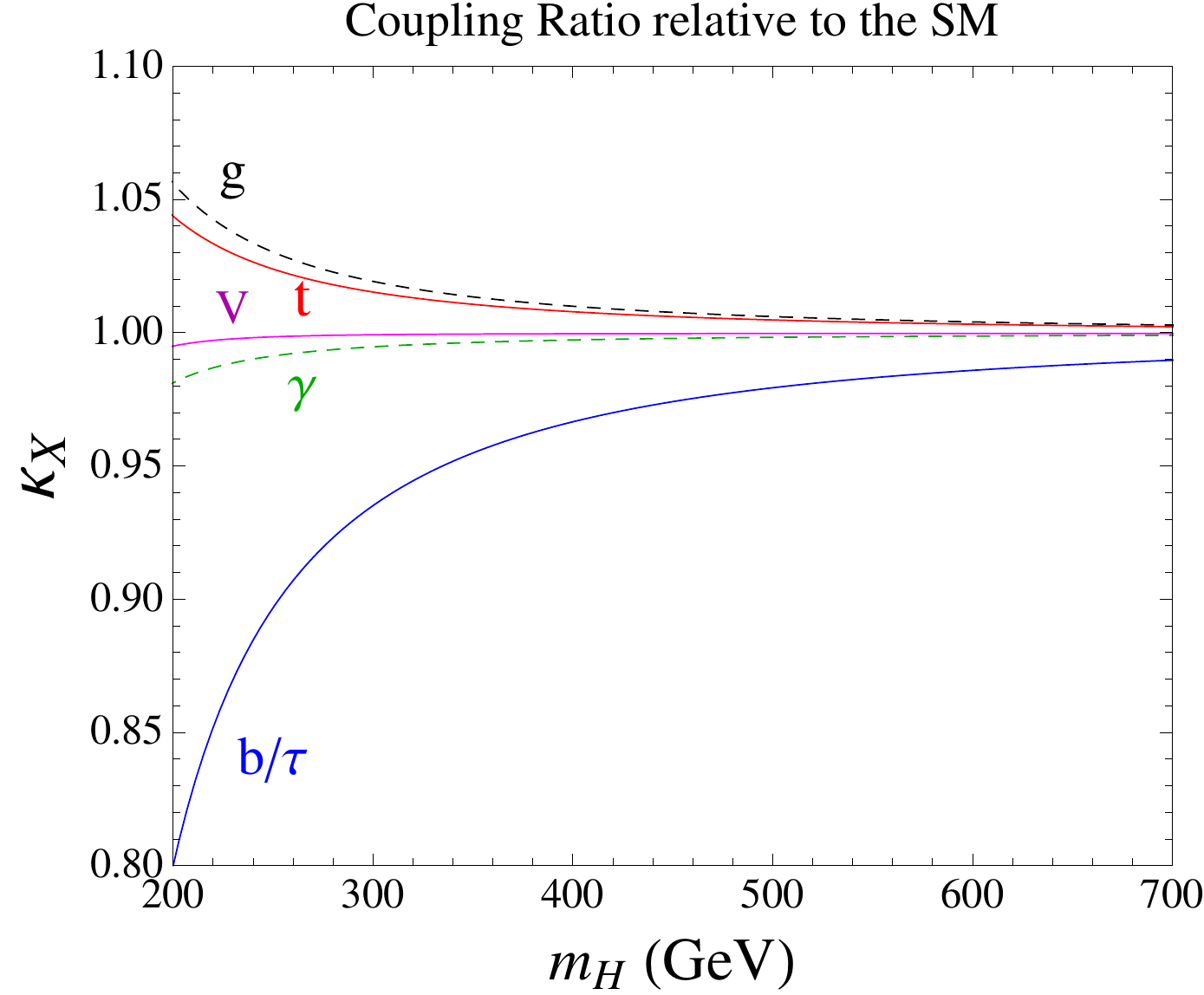}\hspace{0.cm}
\caption{Deviations of SM-like Higgs couplings normalized to the SM values on the left as a function of $m_H$. Here $t,\ b,$ and $\tau$ represent all up-type quarks, down-type quarks, and charged leptons, respectively.}
\label{fig:ratio1}
\vspace{0.3cm}
  \centering
  \includegraphics[width=8.cm]{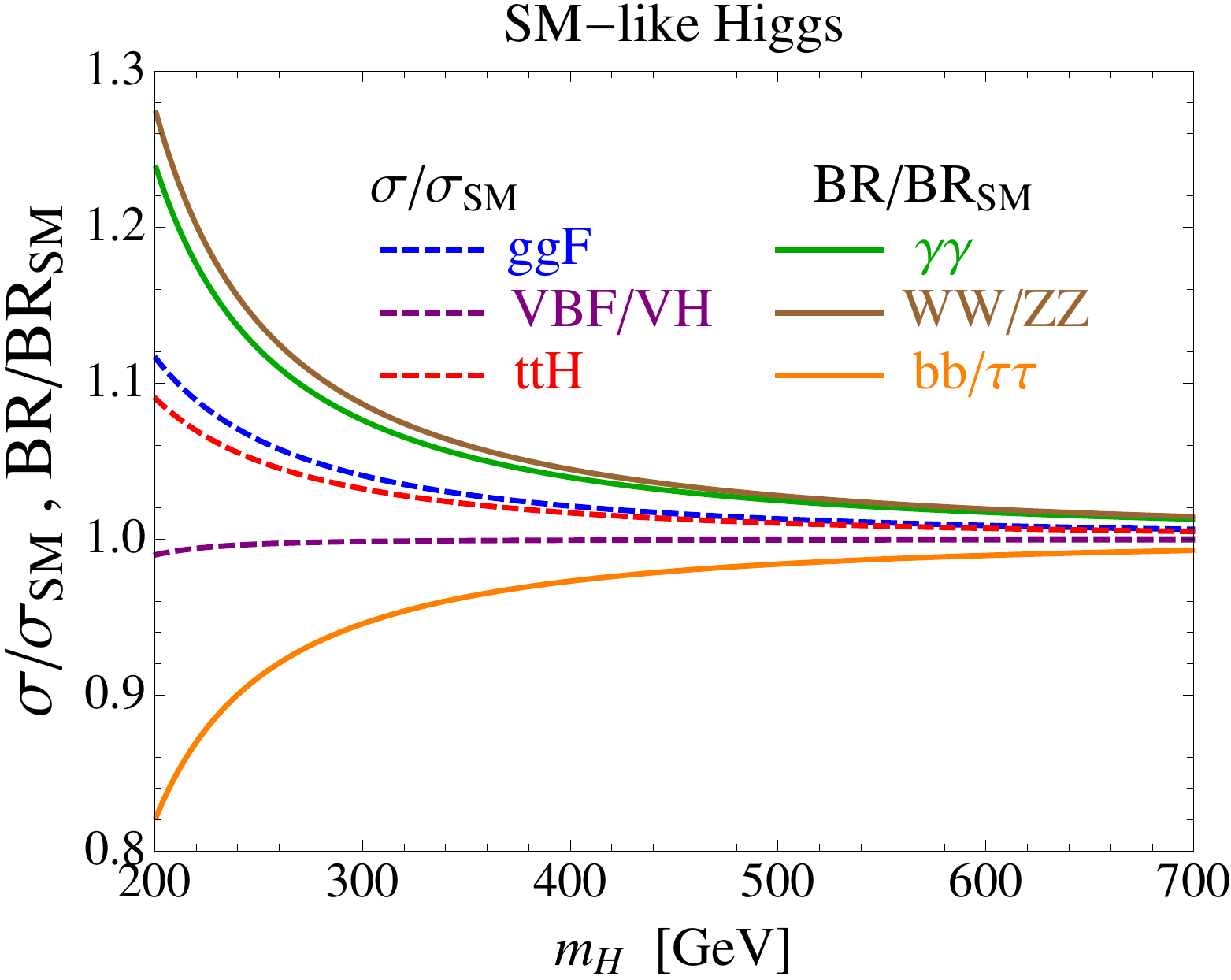}
\caption{The branching rations and production cross sections of the SM-like Higgs are shown normalized to the SM values on the right  as a function of $m_H$ }
\label{fig:ratio2}
\end{figure}

We consider the potential with radiative corrections from top-stop sector, and we find that the couplings of the SM-like Higgs to leptons and down-type quarks are lowered due to mixing between two Higgs fields, while couplings to up-type quarks are slightly increased compared to those in the SM for relatively low $\tan\beta$. In Fig.~\ref{fig:ratio1}, we plot deviations of couplings relative to values in the SM  using 
	\begin{eqnarray}
	\kappa_X \equiv \frac{{XXh \ \rm coupling}|_{\rm Dirac\ NMSSM}}{{XXh \ \rm coupling}|_{\rm SM}} \ .
	\end{eqnarray}
The couplings of Higgs to gluons and photons occur at loop levels, and then the deviation of them comes from the other couplings. For example, the $h\gamma\gamma$ effective coupling has mainly two contributions that are from the $W$ boson loop and top quark loop, and we simply multiply $\kappa_V^2$ and $\kappa_t^2$ in each loop diagram. 

\begin{figure}[th!]
  \centering
  \includegraphics[width=0.45\linewidth]{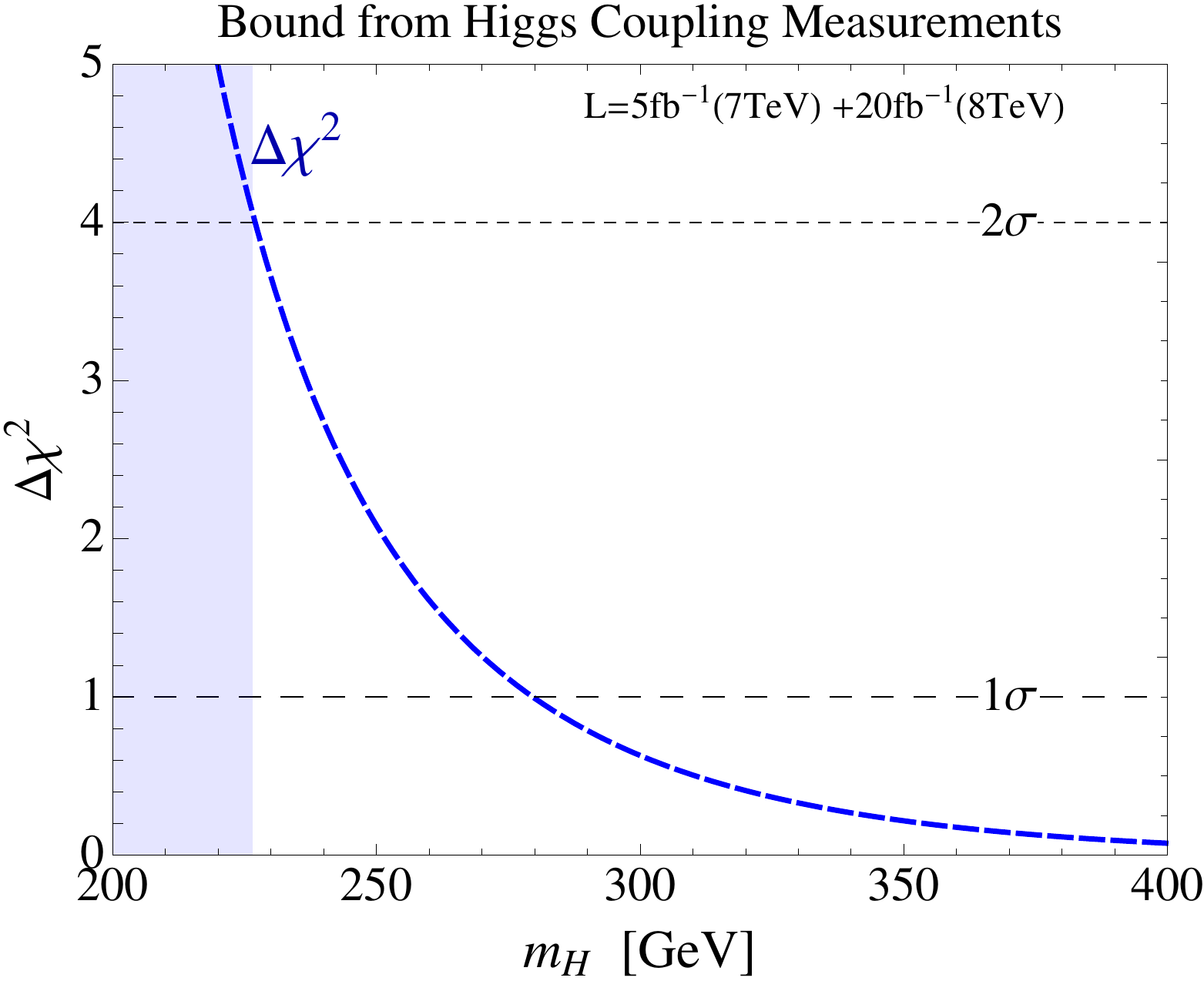}
\caption{$\Delta \chi^2$ for the SM-like Higgs couplings with the 7 and 8 TeV datasets from the ATLAS and CMS \cite{ATLAS:2013sla, ATLASheavyHiggs, CMS:ril, CMSheavyHiggs}. The shaded region is excluded at 95\%C.L.}
\end{figure}

 The smaller coupling of $hb\bar{b}$ results in the deviations to the cross sections and decay patterns shown in Fig.~\ref{fig:ratio2}. 
 These effects decouple in a limit of $m_H \gg m_h$, which corresponds to large $b_{\it eff}$.\footnote{Corrections from sparticles to the Higgs decay and production are not included. 
} 
 For the branching fraction calculation, we utilize the Higgs decay widths given by the LHC Higgs Cross Section Working Group \cite{Dittmaier:2011ti} and multiply each width by appropriate factors of $\kappa$ to get our width.

The first constraint on the Higgs sector of the Dirac NMSSM\@ comes from measurements of the couplings of the SM-like Higgs from ATLAS~\cite{ATLAS:2013sla, ATLASheavyHiggs} and CMS~\cite{CMS:ril, CMSheavyHiggs} with integrated luminosity of 5 fb$^{-1}$ at 7 TeV and 20 fb$^{-1}$ at 8 TeV.  We perform $\Delta\chi^2$ test on the Dirac NMSSM, which excludes $m_H \lesssim 220$~GeV at $95\%$C.L.

\subsection{Heavier doublet-like Higgs}
\begin{figure}[ht!]
  \centering
  \includegraphics[width=0.6\linewidth]{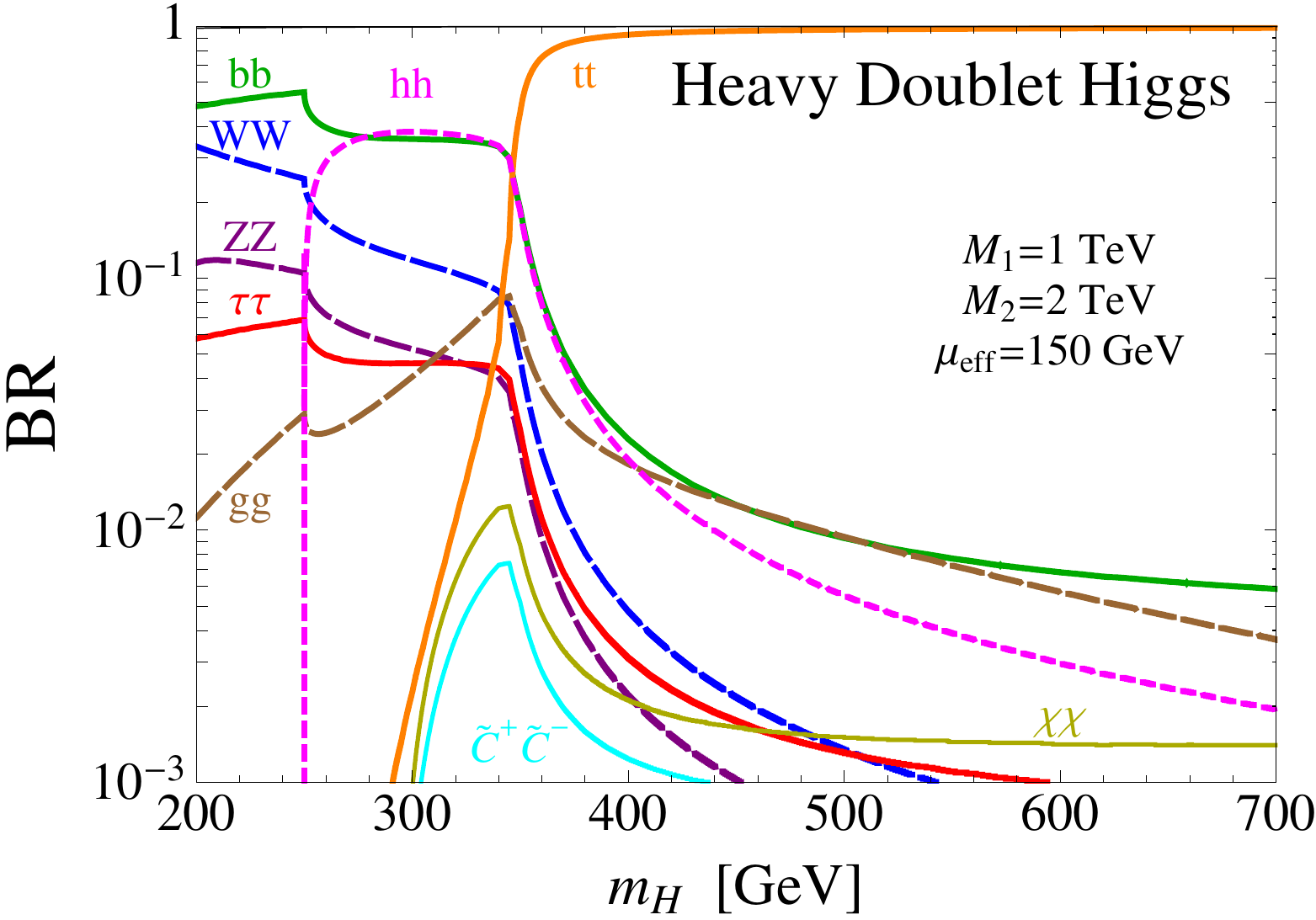}
  \caption{Several branching ratios of the heavy doublet-like Higgs as a function of its mass.  Note that the location of the chargino/neutralino thresholds depend on the -ino spectrum.  Here we take heavy gauginos and $\mu_{\it eff} = 150$~GeV.}
  \label{fig:Hdecay}
\vspace{20pt}
  \includegraphics[width=0.6\linewidth]{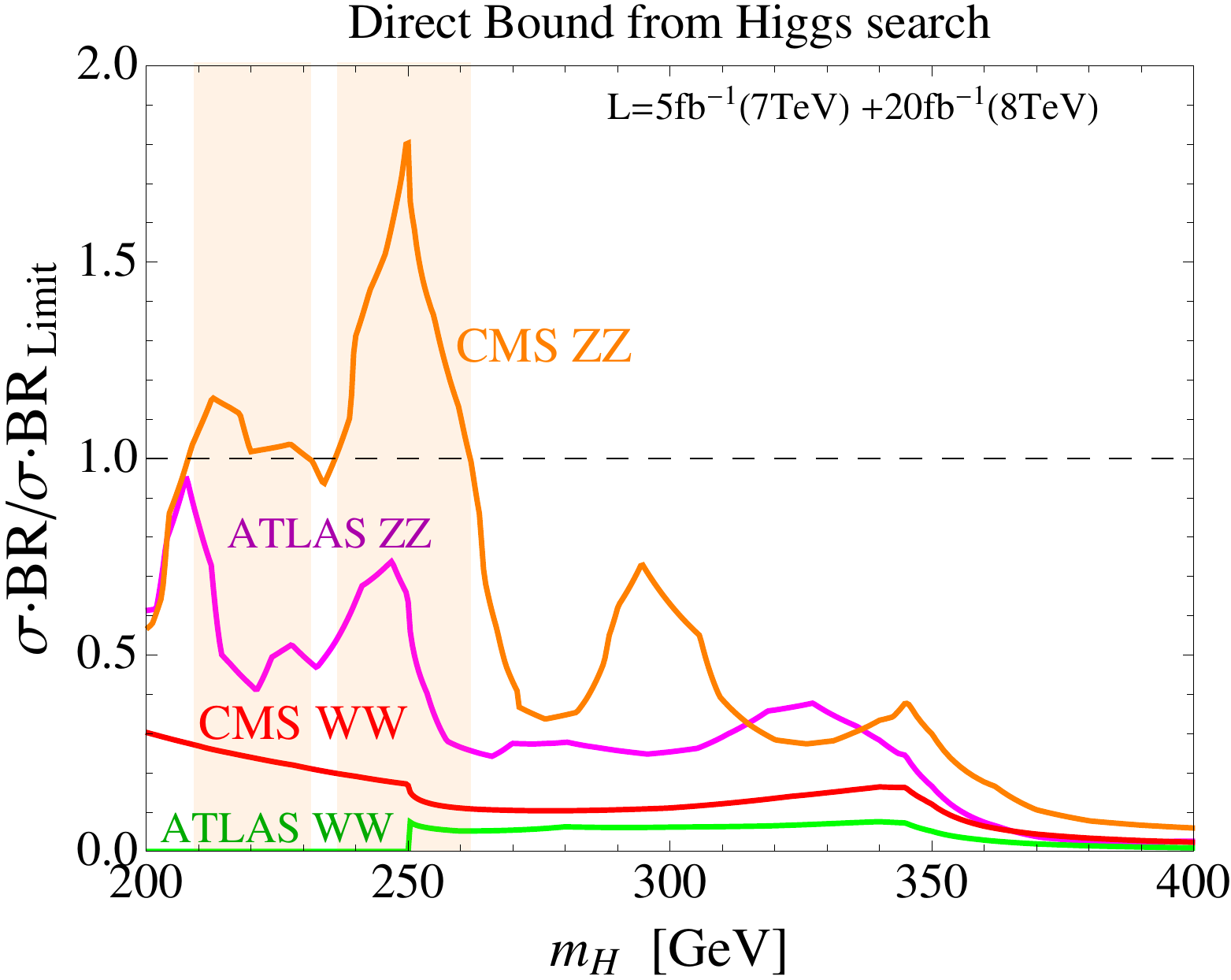}
\caption{$\sigma / \sigma_{lim}$ for direct searches, $H \rightarrow ZZ,WW$~\cite{ATLASheavyHiggs, CMSheavyHiggs}. The shaded region is excluded at 95\%C.L.}
\label{fig:Hdirect}
\end{figure}

 We also show, in Fig.~\ref{fig:Hdecay}, the decay branching ratios of $H$. Due to the non-decoupling term, di-Higgs decay, $H \rightarrow 2h$, becomes the dominant decay once its threshold is opened, $m_H\gtrsim 250$~GeV.  For the calculation of decay widths to SM particles ($t,b,Z,W,\tau, g$), the widths given by the LHC Higgs Cross Section Working Group \cite{Dittmaier:2011ti} are utilized again. 
 
 Bellow this threshold, direct searchers of the heavier state decaying to dibosons, $H \rightarrow ZZ,WW$~\cite{ATLASheavyHiggs, CMSheavyHiggs}, can constraint the model. It extends the limit to $m_H \sim 260$~GeV as  in Fig.~\ref{fig:Hdirect} (except for a small gap near $m_H \approx 235$~GeV).

\section{Future Reach}
We also estimate the future reach to probe $m_H$ with future
Higgs coupling (or width) measurements
\cite{ATLAS-collaboration:2012iza, Peskin:2012we, Klute:2013cx, Snowmass}. Those precisions are summarized in  Tables~\ref{tab:HLHC} and \ref{tab:ILC}.

Two different studies of ILC \cite{Peskin:2012we, Klute:2013cx} have different estimates of theoretical errors while the same experimental precisions are used in both. And recently (2013 August) Snowmass Higgs working group gave an estimate of coupling measurements with the highest precision based on integration of high-luminosity ILC runs.  This is not only experimental challenge but also theoretical challenge because the theoretical error of $b$ quark mass  is expected to decrease up to 0.1\% level. 

\begin{figure}[h!]
  \centering
  \includegraphics[width=10cm]{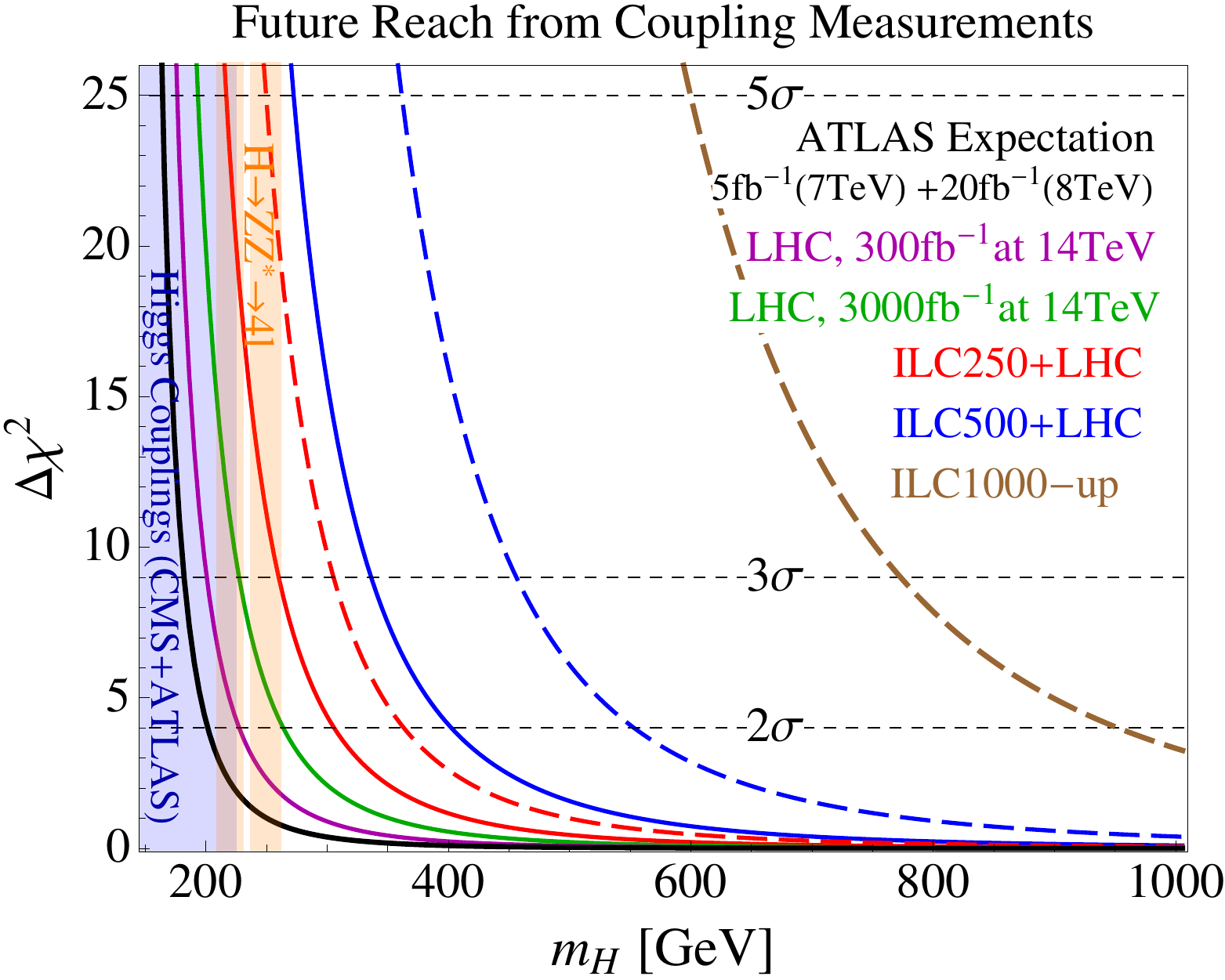}
\caption{
The curves show the expected $\Delta \chi^2$ from combined measurements of the Higgs-like couplings at the high-luminosity LHC at $\sqrt s = 14$~TeV~\cite{ATLAS-collaboration:2012iza} and the ILC at $\sqrt s = 250, 500, 1000$~GeV\@.  The optimistic~\cite{Peskin:2012we,Snowmass} (conservative~\cite{Klute:2013cx}) ILC reach curves are solid (dashed) and  neglect (include) theoretical uncertainties in the Higgs branching ratios. The ILC analyses include the expected LHC measurements.  For comparison we show the present limits and also the expected limit of the current ATLAS measurements (solid, black).
The shaded regions show current constraints on our model by the SM-like Higgs coupling measurements and direct searches of $H$.}
\label{fig:LimitReach}
\end{figure}

\begin{table}[h!]
\centering
  \begin{tabular}{|c |   c  c|}
\hline
 &HL-LHC &HL-LHC  
\\
 & 300 fb$^{-1}$& 3000 fb$^{-1}$ \\
 \hline
 $\Gamma_Z /\Gamma_g $ 	&0.523	& 0.284   \\
 $\Gamma_t /\Gamma_g $ 	&0.519	& 0.230   \\
 $\Gamma_\mu /\Gamma_Z $ 	&0.448	& 0.142   \\
 $\Gamma_\tau /\Gamma_Z $ 	&0.417	& 0.206   \\
 $\Gamma_\gamma /\Gamma_Z $ &0.110& 0.029   \\
 $\Gamma_g \cdot \Gamma_Z /\Gamma_H $ &0.159& 0.029   \\
 \hline
  \end{tabular}
  \caption{Expected relative uncertainties on the ratios of Higgs boson partial widths with theory systematic errors (Tables 6, 7 in Ref.~\cite{ATLAS-collaboration:2012iza}).
  $\Gamma_H $ is the total width, and $\sqrt{s}=14\TeV$. 
   }\label{tab:HLHC}
\end{table}

\begin{table}[h!]
\centering
  \begin{tabular}{|c |  c  c c c  c |}
\hline
 &(a) ILC250 &(b)ILC250 &(a)ILC500 
 &(b)ILC500 & ILC1000-up \\
 in \% &+LHC&   +LHC&  +LHC & +LHC &\\
 \hline
 $\Delta\kappa_W$ &$\pm$1.9 &$\pm$4 & $\pm$0.24 &$+1.0\atop -1.1$ & $\pm$0.13\\
 \hline
 $\Delta\kappa_Z$ &$\pm$0.44 &$+0.9\atop -1.0$& $\pm$0.30 &$\pm$0.8  & $\pm$0.22\\ 
 \hline
$\Delta\kappa_b$ &$\pm$2.7 & ${+5\atop -4}$& $\pm$0.94&$+3\atop-2$ &$\pm$0.31\\ 
\hline
 $\Delta\kappa_\tau$ &$\pm$3.3& $\pm$4 & $\pm$1.9& $\pm$3& $\pm$0.72\\ 
\hline
 $\Delta\kappa_{t,c}$ &$\pm$4.7 &$\pm$4  & $\pm$2.5&$\pm$3&$\pm$0.76\\ 
 \hline
  \end{tabular}
  \caption{Expected precisions on the Higgs couplings. 
  ILC250(ILC500) represents an analysis with ILC data of 250(500) fb$^{-1}$ at $\sqrt{s}=250(500)\GeV$ plus LHC data of 300 fb$^{-1}$ at at $\sqrt{s}=14\TeV$. 
  The values in (a) are given by M.~Peskin \cite{Peskin:2012we} and used in ILC Technical Design Report(TDR), and the different group includes more theoretical errors giving conservative values of (b) \cite{Klute:2013cx}.
  ILC1000-up assumes upgraded ILC and integrates all the data of 
  $1150(\sqrt{s}=250\GeV)+1600(500\GeV)+2500(1\TeV) \ {\rm fb}^{-1}$ \cite{Snowmass}.
 }\label{tab:ILC}
\end{table}

The expected accuracies of coupling measurement give the future reach  in Fig.~\ref{fig:LimitReach}.
 The $2 \sigma$ exclusion reach is
$m_H\simeq 280$~GeV at the high-luminosity LHC \cite{ATLAS-collaboration:2012iza},
$m_H\simeq 400$~GeV with theoretical uncertainty at ILC500 \cite{Klute:2013cx},
and $m_H\simeq 550$~GeV without theoretical uncertainty. 
The reach extends to $m_H\simeq 950$~GeV at upgraded ILC1000 \cite{Snowmass}. 
The increased sensitivity at the ILC is dominated by the improved measurements projected for the $b\bar{b}$ and $\tau^+\tau^- $ couplings~\cite{Peskin:2012we,Snowmass} since those have larger deviation compared to SM values as in Fig.~\ref{fig:ratio1}. 

\pagebreak
\section{Summary and Discussion}
The LHC has discovered a new particle, consistent with the Higgs boson, with a mass near 125~GeV\@.  Weak-scale supersymmetry must be reevaluated in light of this discovery.
Naturalness demands new dynamics beyond the minimal theory, such as a non-decoupling $F$-term, but this implies new sources of SUSY breaking that themselves threaten naturalness.  In Ch.~\ref{dnmssm:model}, we have identified a new model where the Higgs couples to a singlet field with a Dirac mass.  The non-decoupling $F$-term is naturally realized through semi-soft SUSY breaking, because large $m_{\bar S}$ helps raise the Higgs mass but does not threaten naturalness.  

The key feature of semi-soft SUSY breaking in the Dirac NMSSM is that $\bar S$ couples to the MSSM only through the dimensionful Dirac mass, $M$.   We noted that interactions between $\bar S$ and other new states are not constrained by naturalness, even if these states experience SUSY breaking.  Therefore, the Dirac NMSSM represents a new type of portal, whereby our sector can interact with new sectors, with large SUSY breaking, without spoiling naturalness in our sector.

The first collider signatures of the Dirac NMSSM are expected to be those of the MSSM fields, with the singlet sector naturally heavier than 1 TeV.
In Ch.~\ref{dnmssm:pheno}, we discussed the phenomenology focusing on the Higgs sector, which is a two Higgs doublet model with low $\tan \beta$. We obtained constraints from direct searches for heavier Higgs boson and coupling measurements for the lightest Higgs boson at the LHC. We also studied the future reach based on prospects of high-luminosity LHC and ILC, and showed large parameter space can be probed.

\chapter{Overall Summary}\label{ch:last}

 The main accomplishment of LHC Run I is discovery of a Higgs boson, and it is a momentous step towards understanding electroweak symmetry breaking. 
 However, there are still many experimental evidences that cannot be explained in the Standard Model and require some new physics, as represented by dark matter. 
Furthermore the SM Higgs sector is not theoretically satisfactory for the naturalness problem. Supersymmetry cures this problem and is regarded as a prime candidate for physics beyond the Standard Model. 

However there are tensions between low-energy supersymmetry and the LHC results. We discussed them in Ch.~1. First of all, searches at the LHC for sparticles have not found any signal and give strong bounds on the conventional  CMSSM. Next, the observed Higgs mass of 125 GeV is not easily accommodated in the MSSM, where one has to rely on the radiative corrections to boost the Higgs mass beyond the tree-level upper bound of $m_Z\simeq91\GeV$. This requirement push sparticles scale well beyond the TeV within CMSSM, leading to fine-tuning of $\Delta^{-1}\lesssim 1\%$.  
In this thesis we particularly have investigated two scenarios of supersymmetry originating the tensions. 
\\

A  compressed  spectrum ameliorates the bounds from the current searches at the LHC whereas the CMSSM typically generates a widely spread spectrum. For the lack of SUSY signal, the scenario with a compressed spectrum recently has more attentions but it has not been  theoretically justified  based on simple and explicit models of SUSY breaking.

Supersymmetry broken geometrically in extra dimensions, by the Scherk-Schwartz mechanism, naturally leads to a compressed spectrum. 
 In Ch.~\ref{compact:model} we have built a minimal such model with a single extra dimension of  $S^1/{\mathbb Z}_2$, ``Compact Supersymmetry.'' After reviewing construction of  5D SUSY, we demonstrated the Scherk-Schwarz mechanism is equivalent to the Radion Mediation.
 In the model,  gauge, quark and lepton superfields are living in the full 5D while Higgs fields are localized on a brane.
 The model has only three parameters, a size of extra dimension, SUSY breaking scale, and supersymmetric Higgs mass,  and hence it is explicitly testable. For the nature of geometrical SUSY breaking, the universality of gaugino masses and fermion masses is present.  
 We presented radiative corrections generating Higgs parameters which are absent at tree-level.  

We matched the full theory onto the MSSM and studied phenomenology of the model in Ch.~\ref{compact:pheno}. The conditions for the EWSB were studied at first to determine all the parameters. The spectrum certainly tends to be compressed. 
We found predicted near-maximal mixing in the scalar top sector with $|A_t|\approx 2m_{\tilde{t}}$ boosts the lightest Higgs boson mass. The  Higgs mass in large parameter space is about $121 \sim 125 \GeV$, but this is not necessarily incompatible with 125 GeV for the large theory uncertainty. 
Despite the rather constrained structure, the theory is less fine-tuned than many supersymmetric models. 
 The LSP is Higgsino-like and can be a component of  dark matter. We found direct detection experiments will cover a large portion of parameter space. 
 The theory does not suffer from the supersymmetric flavor or CP problem because of universality of geometric breaking.
The collider constraint on the Compact Supersymmetry is certainly weaker than that on the CMSSM such that gluino and squark mass bound is relaxed from $m_{\tilde{g},\tilde{q}}\lesssim 1.7\TeV$ down to $m_{\tilde{g},\tilde{q}}\lesssim 1\TeV$.  
%
%
\\

Naturalness implies new dynamics beyond the minimal theory. 
There have been many attempts to extend the MSSM to accommodate the Higgs mass. 
If the new states are integrated out supersymmetrically, their effects decouple and the Higgs mass is not increased. On the other hand, 
SUSY breaking can lead to non-decoupling effects that increase the Higgs mass.  However, in general, these extensions require new states at the few hundred GeV scale, so that the new sources of SUSY breaking do not spoil naturalness.

In Ch.~\ref{dnmssm:model}, we have identified a new model where the Higgs couples to two singlet fields with a Dirac mass, which we call Dirac NMSSM,
	\begin{eqnarray}
	W\supset \lambda S H_u H_d +M S \bar{S}. \nonumber
	\end{eqnarray}
The non-decoupling $F$-term increases the Higgs mass while maintaining  naturalness even in the presence of large SUSY breaking in the singlet sector, namely, $m_{\bar S}\gtrsim 10\TeV$.
The key feature  in the Dirac NMSSM is that $\bar S$ couples to the MSSM only through the dimensionful Dirac mass, $M$. We pointed out that interactions between $\bar S$  and other new states are not constrained by naturalness, even if $\Sbar$ states experience SUSY breaking. We call this mechanism semi-soft SUSY breaking. Therefore, the Dirac NMSSM represents a new type of portal, whereby our sector can interact with new sectors, with large SUSY breaking, without spoiling naturalness in our sector.

Collider signatures of the Dirac NMSSM are  discussed in Ch.~\ref{dnmssm:pheno}. The low-energy  phenomenology is that of a two Higgs doublet model.  We obtained constraints from direct searches for heavier Higgs boson and coupling measurements for the lightest Higgs boson at the LHC. We also studied the future reach based on prospects of high-luminosity LHC and future international linear collider, and show large parameter space can be probed. 


\chapter*{Acknowledgments}
\addcontentsline{toc}{chapter}{Acknowledgments}
\thispagestyle{empty}

First of all, I am deeply grateful to Prof. Hitoshi Murayama. He always encourages me for  the research progress, and he leads me to the interesting fields in particle physics with his knowledge, experience, and foresight. 
I also appreciate that he let me stay at University of California, Berkeley for about a year. 
During the stay, I had the best lectures ever by him about Quantum Field Theory and particle phenomenology. 
His international connections enabled me to know various researchers all over the world even when I was at Kavli IPMU. These extraordinary experience brought by him stimulates my research and makes me enjoy the research. 

I wish to acknowledge Yuuko Enomoto for her great help. For her support and corporation, I could find discussion time with my advisor and conduct research at Kavli IPMU.  
I would like to express my gratitude to Mihoko M. Nojiri for patiently teaching me the basics of collider phenomenology. Without her guidance, my first research project would not be possible. 

Many results obtained in this dissertation were never obtained without collaborations with Yasunori Nomura and Satoshi Shirai, and Xiaochuan Lu and Joshua T. Ruderman at University of California, Berkeley. I have learned a lot of things about supersymmety and phenomenology from them. I also thank Brian Henning for a lot of discussion. 
The collaborations and communication made my stay at Berkeley very fruitful.  I thank collaborators of the other projects, Tomohiro Abe, Keisuke Harigaya, Junji Hisano, Teppei Kitahara, Shigeki Matsumoto, Ryosuke Sato and Norimi Yokozaki. I would particularly thank Junji Hisano for taking care of informal reading course of supersymmetry in the beginning of graduate program.

I spent most of my graduate school days at Kavli IPMU.
I enjoyed talking with people in various research fields at the tea time
of the institute. I would like to thank all researchers, visitors, and
administrative stuff in the institute for making my daily life enjoyable
and supporting my research activities.
At Kavli IPMU, I had fruitful discussion with researchers,  Won Sang Cho, Masahiro Ibe, Ryoichi Nishio, Seong Chan Park, Ryosuke Sato, Taizan Watari, Tsutomu Yanagida, Kazuya Yonekura and many others.  
I would like to especially thank students with whom
I shared my office, namely,
Takeshi Kobayashi, Hironao Miyatake, Ryoichi Nishio, Sourav Mandal, Takashi Moriya,
William L. Klemm, Xu-Feng Wang, Kimihiko Nakajima,
Tomonori Ugajin, Ayuki Kamada,
Tomohiro Fujita, Keisuke Harigaya, Masato Shirasaki, Gen Chiaki,
Wen Yin, Hidemasa Oda, Yuuki Nakaguchi, Yusuke Ono,
Noriaki Watanabe, Ryo Matsuda
and many visitors from all over the world.

I enjoyed a lot of informal discussion with friends in the same grade, Hiraku Fukushima, Manami Hashi, Yuji Hirono, Kenta Hotokezaka, Yohei Kikuta and Natsumi Nagata. Communication with them always raises my motivation of research. 

I would like to express my sincere gratitude to my parents,
Hisaya Tobioka and Shizuko Tobioka.
My interest in nature was originated from my young days
with my parents in Aso, Kumamoto.
I appreciate they allowed me to study at Tohoku University, Sendai, and even supported my studying abroad at University of California, San Diego during undergraduate. And without their special supports I would not have accomplished my research. 
\\\\\

\vfill

{\small
This research is supported by the Japan Society for the Promotion of
Science Research Fellowship for Young Scientists, 
World Premier International Research Center Initiative (WPI Initiative),
MEXT, Japan.
}

\thispagestyle{empty}

\appendix

\chapter{Structure of 5D Supersymmetry}
\section{Gauge Invariance of 5D Supersymmetry Action}\label{gaugeinv5D}
We check the gauge invariance of 5D gauge field strength. The non-trivial part in Eq.~(\ref{213628_19Apr12}) is
	\begin{eqnarray}
	\int d^4\theta \ {\rm Tr}\left[ (\sqrt{2}\partial_5+\chi^\dagger)e^{-V} (-\sqrt{2}\partial_5+\chi)e^{V}+(\partial_5 e^{-V})(\partial_5 e^{V})+
	\frac{\chi^\dagger \chi^\dagger+\chi\chi}{2} \right].
	\hspace{-20pt}
	\label{Eq:5DYM}
	\end{eqnarray} 
We redefine gauge fields to absorb gauge coupling as $2gV\to V,\ 2g\chi\to \chi$.
Note that the last term vanishes after performing $\theta$ integrals, but it is essential for gauge invariance. In addition, in the Radion mediation, we have supersymmetry breaking in front this term, which means it does not vanish by $\theta$ integrals.

Under the non-abelian gauge transformation, the vector and adjoint chiral superfields transform as
	\begin{eqnarray}
	e^{-V}\to U^\dagger e^{-V} U, & \ \ \ & e^{V}\to U^{-1\dagger} e^{-V} U^{-1}
	\nonumber\\
	\chi \to U^{-1}(\chi -\sqrt{2}\partial_5)U, &\ \ \ &
	\chi^\dagger \to U^\dagger(\chi^\dagger+\sqrt{2}\partial_5)U^{-1\dagger}].
		\nonumber\\
	H_1\to U^{-1} H_1, &\ \ \ & H_2 \to H_2 U^{\dag}
	\end{eqnarray}
First let us see the first term, 
	\begin{eqnarray}
	\hspace{-20pt}
	(\sqrt{2}\partial_5+\chi^\dagger)e^{-V} 
	\hspace{-10pt}&\to& \sqrt{2}\partial_5 (U^\dagger e^{-V} U ) 
	+[U^\dagger(\chi^\dagger+\sqrt{2}\partial_5)U^{-1\dagger}]U^\dagger e^{-V} U 
	\\
	&=& \sqrt{2}\left\{
	(\partial_5 U^\dagger)e^{-V}U +U^\dagger(\partial_5 e^{-V})U+U^\dagger e^{-V}\partial_5 U
	 \right\}
	\nonumber
	 \\&&+U^\dagger \chi^\dagger e^{-V}U+\sqrt{2}U^\dagger(\partial_5U^{-1\dagger})U^\dagger e^{-V} U
	\nonumber\\
	&=& \sqrt{2}\left\{
	(\partial_5 U^\dagger)e^{-V}U +U^\dagger(\partial_5 e^{-V})U+U^\dagger e^{-V}\partial_5 U
	 \right\}
	\nonumber
	 \\&&+U^\dagger \chi^\dagger e^{-V}U  -\sqrt{2}U^\dagger U^{-\dagger}(\partial_5U^{\dagger}) e^{-V} U
	\nonumber\\
	&=& \sqrt{2}\left\{
	U^\dagger(\partial_5 e^{-V})U+U^\dagger  e^{-V}\partial_5U
	 \right\} +U^\dagger \chi^\dagger e^{-V}U
	\nonumber\\
	&=&U^\dagger \left\{
	 \sqrt{2}\partial_5 e^{-V}+\chi^\dagger e^{-V}
	 \right\}U +\sqrt{2}U^\dagger e^{-V}\partial_5 U
	 \label{Eq:inv1}
	\end{eqnarray}
Here we used an identity 
	\begin{eqnarray}
	\partial_5(U^{-1}U)=(\partial_5 U^{-1})U+U^{-1}(\partial_5 U)=0\ .
	\end{eqnarray}
Similarly,
	\begin{eqnarray}
	(-\sqrt{2}\partial_5+\chi)e^{V}
		&\to& \sqrt{2}\partial_5 (U^{-1} e^{V} U^{-1\dagger} ) 
	+[U^{-1}(\chi -\sqrt{2}\partial_5)U]U^{-1} e^{V} U^{-1\dagger} 
	\\
	&=& \sqrt{2}\left\{
	(\partial_5 U^{-1})e^{V}U^{-1\dagger} +U^{-1}(\partial_5 e^{V})U^{-1\dagger}
	+U^{-1} e^{V}\partial_5 U^{-1\dagger}
	 \right\}
	\nonumber
	 \\&&+U^{-1} \chi e^{V}U^{-1\dagger}-\sqrt{2}U^{-1}[-U(\partial_5 U^{-1})] e^{V} U^{-1\dagger}
 	\nonumber\\
	&=&U^{-1} \left\{
	 \sqrt{2}\partial_5 e^{V}+\chi e^{V}
	 \right\}U^{-1\dagger} +\sqrt{2}U^{-1} e^{V}\partial_5 U^{-1\dagger}.
	 \label{Eq:inv2}
	\end{eqnarray}
We combine Eq.(\ref{Eq:inv1},\ref{Eq:inv2}) and take trace of it,
	\begin{eqnarray}
	&&\hspace{-20pt}{\rm Tr[(\ref{Eq:inv1})\times (\ref{Eq:inv2})]}\\
	&=\hspace{-5pt}&
	{\rm Tr}\Big[( \sqrt{2}\partial_5 e^{-V}+\chi^\dagger e^{-V}) (\sqrt{2}\partial_5 e^{V}+\chi e^{V})
	-2 U^\dagger e^{-V}(\partial_5 U) U^{-1} e^{V}(\partial_5 U^{-1\dagger})
	\nonumber
	\\&& \ \ \
	-\sqrt{2}U^\dagger( \sqrt{2}\partial_5 e^{-V}+\chi^\dagger e^{-V})e^V(\partial_5 U^{-1\dagger}) 
	\nonumber
	\\&& \ \ \
	+\sqrt{2}e^{-V}(\partial_5 U)U^{-1}(\sqrt{2}\partial_5 e^{V}+\chi e^{V})
	\Big]
	\nonumber\\
	&=\hspace{-5pt}&
	{\rm Tr}\Big[( \sqrt{2}\partial_5 e^{-V}+\chi^\dagger e^{-V}) (\sqrt{2}\partial_5 e^{V}+\chi e^{V})
	-2 ^\dagger e^{-V}(\partial_5 U) U^{-1} e^{V}(\partial_5 U^{-1\dagger})U	
	\nonumber
	\\&& 	\ \ \ 
	-2 (\partial_5 e^{-V})e^V(\partial_5 U^{-1\dagger})U^{\dagger} 
	-\sqrt{2}\chi^\dagger(\partial_5U^{-1\dagger})U^\dagger
	\no\\&&\ \ \ 
	-2 (\partial_5 e^{V})e^{-V}(\partial_5 U)U^{-1} 
	+\sqrt{2}\chi(\partial_5U)U^{-1}  
	\Big]
 	\label{Eq:inv3}
	\end{eqnarray}
The second term of Eq.(\ref{Eq:5DYM}) transforms as
	\begin{eqnarray}
	&&\hspace{-10pt}
	{\rm Tr}\left[ (\partial_5 e^{-V})(\partial_5 e^{V})
	\right]
	\\\ &&\to
	{\rm Tr}\Big[ 
	\big\{(\partial_5 U^\dagger)e^{-V}U+U^\dagger(\partial_5 e^{-V})U+U^\dagger e^{-V}\partial_5 U\big\}
	\nonumber \\&&\hspace{30pt} \times 
	\big\{(\partial_5 U^{-1})e^{V}U^{-1\dagger}+U^{-1}(\partial_5 e^{V})U^{-1\dagger}+U^{-1} e^{V}\partial_5 U^{-1\dagger} \big\}
	\Big]	
	\\&&=
	{\rm Tr}\Big[ 
	(\partial_5 e^{-V})(\partial_5 e^{V})+2(\partial_5 e^{V})e^{-V}(\partial_5 U)U^{-1}
	+2(\partial_5 e^{-V})e^{V}(\partial_5 U^{-1\dagger})U^{\dagger}	
	\nonumber\\
	&&\hspace{30pt}
	+2e^{-V}(\partial_5 U)U^{-1}e^{V}(\partial_5 U^{-1\dagger})U^{\dagger}
	+\partial_5 U^{\dagger}\partial_5 U^{-1\dagger}
	+\partial_5 U^{}\partial_5 U^{-1}\Big], \label{Eq:inv4}
	\no\\
	\end{eqnarray}
and the last terms of Eq.(\ref{Eq:5DYM}) transform as
	\begin{eqnarray}
	{\rm Tr}\left[\frac{\chi\chi}{2}\right]
	&\to& \frac{1}{2}{\rm Tr}\left[
	U^{-1}(\chi U-\sqrt{2}\partial_5 U)U^{-1}(\chi U-\sqrt{2}\partial_5U)
	\right]\\
	&=& 
	\frac{1}{2}{\rm Tr}\left[
	\chi\chi+2U^{-1}(\partial_5 U)U^{-1}(\partial_5 U)-2\sqrt{2}(\partial_5 U)U^{-1}\chi
	\right]
	\nonumber\\&=&
	{\rm Tr}\left[
	\frac{\chi\chi}{2}-(\partial_5 U^{-1})(\partial_5 U)-\sqrt{2}\chi(\partial_5 U)U^{-1}
	\right]\label{Eq:inv5}
	\\
	{\rm Tr}\Big[\frac{\chi^\dagger \chi^\dagger}{2}\Big]
	&\to&
	{\rm Tr}\Big[\frac{\chi^\dagger \chi^\dagger}{2}
	-(\partial_5 U^{\dagger})(\partial_5 U^{-1\dagger})-\sqrt{2}\chi^\dagger U^{-1\dagger}(\partial_5 U^{\dagger})
	\Big].\label{Eq:inv6}
	\end{eqnarray}
As a result, we prove the gauge invariance of Eq.(\ref{Eq:5DYM}),
	\begin{eqnarray}
	&&\hspace{-20pt}{\rm (\ref{Eq:inv3})+(\ref{Eq:inv4})+(\ref{Eq:inv5})+(\ref{Eq:inv6}) }
	\nonumber\\
	&=&
	{\rm Tr}\left[ (\sqrt{2}\partial_5+\chi^\dagger)e^{-V} (-\sqrt{2}\partial_5+\chi)e^{V}+(\partial_5 e^{-V})(\partial_5 e^{V})+
	\frac{\chi^\dagger \chi^\dagger+\chi\chi}{2} \right].
	\end{eqnarray}
\section{$SU(2)_R$ invariance}\label{explicitSU2}
In this section, we prove the $SU(2)_R$ invariance for Eq.(\ref{LagrangianHyper}).
The terms with auxiliary fileds $F_\chi$ and $D^a$ in Eq.(\ref{LagrangianHyper}) do not seem  $SU(2)_R$ invariant. To see the invariance, we must combine them with $|F^a_\chi|^2$ and $(D^a)^2$ from the gauge sector. 
We obtain following four point interactions for squarks (sleptons) by completing squares:
\begin{eqnarray}
   |F_\chi^a|-(\sqrt{2}gF_\chi^a \phi_2 T^a \phi_1 +\mbox{h.c.})+ 
    \frac{1}{2}(D^a)^2  -gD^a(\phi^*_1 T^a \phi_1 - \phi_2 T^a \phi^*_2)
    \\
 \to -2g^2 (\phi_1^* T^a \phi_2^*)(\phi_2 T^a \phi_1) -\frac{1}{2}g^2 (\phi^*_1 T^a \phi_1 - \phi_2 T^a \phi^*_2)^2 \ .\label{211636_21Apr12}
\end{eqnarray}
For convenience, we use $\Phi_i$ as $(\Phi_1, \ \Phi_2 ) = (\phi_1, \ \phi_2^*)$ and we omit the gauge coupling.  Then eq.(\ref{211636_21Apr12}) is, 
\begin{eqnarray}
 -\frac{1}{2}(\Phi_1^* T^a \Phi_1)^2  -\frac{1}{2}(\Phi_2^* T^a \Phi_2)^2 + (\Phi_1^* T^a \Phi_1)(\Phi_2^* T^a \Phi_2)
  -2 (\Phi_1^* T^a \Phi_2)(\Phi_2^* T^a \Phi_1) \ ,  
\end{eqnarray}
but the invariance is still not clear.  

Now we pick up an explicit $SU(2)_R$ invariant form of the squark four-point interection. 
\begin{eqnarray}
 \sum_{m=1,2,3} (\Phi_i^*  \sigma^m_{ij} T^a \Phi_j)(\Phi_k^*  \sigma^m_{kl} T^a \Phi_l)\label{212455_21Apr12}
\end{eqnarray}
where $i,j,k,$ and $l$ are indices of the $SU(R)$ doublet. 
For this term, we can use a formula of pauli matrices (Fierz identity): 
\begin{eqnarray}
 \sum_{m} \sigma^m_{ij} \sigma^m_{kl} = 2\delta_{il} \delta_{jk} - \delta_{ij}\delta_{kl},
\end{eqnarray}
and then eq.(\ref{212455_21Apr12}) becomes, 
\begin{eqnarray}
&& \hspace{-30pt}
(\Phi_i^*   T^a \Phi_j)(\Phi_k^*  T^a \Phi_l)(2\delta_{il} \delta_{jk} - \delta_{ij}\delta_{kl})
\\
&=&2\Phi_1^* T^a (\Phi_1\Phi_1^* +\Phi_2\Phi_2^*)  T^a \Phi_1
 +2\Phi_2^* T^a (\Phi_1\Phi_1^* +\Phi_2\Phi_2^*)  T^a \Phi_2
\nonumber\\
&&-(\Phi_1^* T^a\Phi_1 +\Phi_2^* T^a \Phi_2)^2 
\\
&=&(\Phi_1^* T^a\Phi_1)^2 +(\Phi_2^* T^a\Phi_2)^2
-2 (\Phi_1^* T^a\Phi_1 )(\Phi_2^* T^a\Phi_2)\label{212547_21Apr12}
\nonumber\\
&&+4 (\Phi_1^* T^a\Phi_2 )(\Phi_2^* T^a\Phi_1)
\end{eqnarray}
As a result,  $(\ref{211636_21Apr12}) =-\frac{1}{2}\times(\ref{212547_21Apr12})$, and therefore the four-point interactions are $SU(2)_R$ invariant. 

\chapter{Kaluza-Klein Expansion}
Here is a summary of bulk Lagrangian:
\begin{eqnarray}
{\cal L}_5^{\rm YM}
&=& -\frac{1}{4}F^{a MN}F^a_{MN}
 +\frac{1}{2} \Big[  
  i\bar{\lambda}^a_i \sigmabar^\mu {\cal D}_\mu \lambda_i^a 
  + {\lambda}^a_i \epsilon_{ij}{\cal D}_5 {\lambda}_j^a 
  \ +\mbox{h.c.} \
  \Big]
\nonumber\\
&& +\frac{1}{2} {\cal D}_M \Sigma^a {\cal D}^M \Sigma^a 
 +\frac{g}{2}f^{abc} \Sigma^a(i\lambda_i^b \epsilon_{ij} \lambda^c_j  \ +\mbox{h.c.})
+\frac{1}{2}(D'^a)^2 
  +| F^a_\chi|^2
  \\\no\\
  {\cal L}^{\rm Hyper}_5 &=&
   |F'_1|^2+ |F'_2|^2
    +({\cal D}^M \phi_1)^\dag {\cal D}_M \phi_1 
    +({\cal D}^M \phi_2^*)^\dag {\cal D}_M \phi^*_2
\nonumber\\&& 
+i\left(\psi_2, \bar\psi \right)
\left(
 \begin{array}{c c }
  -i {\cal D}_5 & \sigma^\mu{\cal D}_\mu \\
      \sigmabar^\mu{\cal D}_\mu & i{\cal D}_5 \\
 \end{array}
\right)
\left(
 \begin{array}{c}
  \psi_{1} \\
  \bar\psi_2 \\
 \end{array}
\right) \ 
\nonumber\\ &&
   - g^2(\Sigma^aT^a \phi_1)^\dag 
     (\Sigma^aT^a \phi_1)
       - g^2(\Sigma^aT^a \phi_2^*)^\dag 
       (\Sigma^aT^a \phi_2^*)
\nonumber\\ &&
	   +g \Sigma^a (\psi_2   T^a   \psi_1 +\bar\psi_1    T^a   \bar\psi_2    ) \ 
\nonumber\\ &&
    +\sqrt{2}g
    (\phi_1^* , \phi_2)T^a 
    \left\{
        \psi_1 
	\left(
	 \begin{array}{c}
	-  \lambda^a_{1} \\
	  \lambda^a_2 \\
	 \end{array}
	\right)
	+
	\bar\psi_2 
	\left(
	 \begin{array}{c}
	  \bar\lambda^a_{2} \\
	  \bar\lambda^a_1 \\
	 \end{array}
	\right)
	       \right\}
    	   \ +\mbox{h.c.}  	   
     \nonumber\\ && 
     -gD'^a(\phi_1^* T^a \phi_1)+gD'^a(\phi_2 T^a \phi_2^*)
     \nonumber\\ &&
     +\phi_2 \big[ -\sqrt{2} g F_\chi^a T^a \big]\phi_1     \ +\mbox{h.c.}  \\\no\\
	{\cal L}_{5,\rm soft}&=&
		-\left(\frac{\alpha}{R}\right)^2|\phi_1|^2-\left(\frac{\alpha}{R}\right)^2|\phi_2^*|^2
	\no\\
	&&+\frac{\alpha}{R}\left(
	\phi_1^\dag{\cal D}_5 \phi_2^*
	-(\phi_2^*)^\dag{\cal D}_5 \phi_1
	-({\cal D}_5\phi_1)^\dag \phi_2^*
	+({\cal D}_5\phi_2^*)^\dag \phi_1
	\right)
	\no\\&&
	+\frac{1}{2}\frac{\alpha}{R}\lambda_1^a\lambda_1^a +\frac{1}{2}\frac{\alpha}{R}\lambda_2^a\lambda_2^a +\hc \ 
\end{eqnarray}

\section{KK Expansion in $S^1/{\mathbb Z}_2$}
In a flat extra dimension without orbifold, the field which has a condition, $\varphi(x,y+2\pi R)=\varphi(x,y)$,  is expanded to discrete Fourier modes, called Kaluza-Klein modes,
	\begin{eqnarray}
	\varphi(x,y) = \frac{1}{\sqrt{2\pi R}}\sum_{n=-\infty}^{\infty} \varphi_{n}(x) e^{iny/R}\ .
	\end{eqnarray}
From here we omit 4D position labels such as $x$ for simplicity. Under ${\cal P}: y\to -y$ (orbifold), the field can have two different conditions,
	\begin{eqnarray}
	\phi_{\rm even}(y)&=&+\phi_{\rm even}(-y) , 
	\\
	\phi_{\rm odd}(y)&=&-\phi_{\rm odd}(-y).
	\end{eqnarray}
We introduce SUSY breaking by Radion Mediation, which is equivalent to the Scherk-Schwarz mechanism but  is simpler in calculation than using the twist. The even and odd fields  are expanded as 
	\begin{eqnarray}
	\phi_{\rm even}(y)=\sum^{\infty}_{n=0}\!\! 	 \frac{\eta_n }{\sqrt{\pi R}}~\phi_{n}^{\rm even}\cos\frac{ny}{R} , \label{KKOdd}
	\\
	\phi_{\rm odd}(y)=\sum^{\infty}_{n=1} \!\!	 \frac{1}{\sqrt{\pi R}}~\phi_{n}^{\rm odd}\sin\frac{ny}{R} \ ,	\label{KKEven}
	\end{eqnarray}
where we have a wave function factor,
	\begin{eqnarray}
	\eta_n\equiv 
	 \Big\{
    \begin{array}{c c }
      \frac{1}{\sqrt{2}}  &(n=0) \\
      1&(n\neq 0) \\
    \end{array} \  .
	\end{eqnarray}
The wave function factor is necessary for the proper normalization of bilinear terms since $y$ integral gives, 
	\begin{eqnarray}
	\int_0^{2\pi R}\!\!\!\! dy \ \cos\frac{my}{R}\cos\frac{ny}{R}&=&
	  \ \delta_{m,n} \times\Big\{
    \begin{array}{c c }
     2\pi R   &(n=0) \\
     \ \pi R  &(n\neq 0)  \\
    \end{array} \  ,
    \\
    	\int_0^{2\pi R}\!\!\!\! dy \ \sin\frac{my}{R}\sin\frac{ny}{R}&=&
	  \ \delta_{m,n} \pi R \ .
	\end{eqnarray}

\section{Mass Matrix and Propagator}

\subsection{Squarks and sleptons} \label{app:KKsquark}
This is the easiest example. For the requirement of chiral fermion, one chiral superfield in Hypermultiplet must be odd under ${\cal P}: y\to -y$ while the other chiral superfield is even, and correspondingly one complex scalar, say $\phi_1$, is even and the other one, $\phi_2$, is even. The mass matrix of them is 
	\begin{eqnarray}
	&&\hspace{-20pt}{\cal L}_{\phi}(x)\no\\
	&&\hspace{-20pt}=\int \!\! dy \
	(\phi_1^*(y), \phi_2(y))
	 \left(
	 \begin{array}{c c }
     	-\partial^2+\delfive^2 -\hata^2   &2\hata\delfive \\
     	-2\hata\delfive  &-\partial^2+\delfive^2 -\hata^2  \\
    	\end{array} \ 
	\right)
	\left( \begin{array}{c } 
	 \phi_1(y) \\  \phi_2^* (y)
	 \end{array}\right)
	 \no\\\\
	&&\hspace{-20pt}=\phi_{1,0}^*(-\partial^2-\hata^2 )\phi_{1,0}
	\no\\&&\hspace{-10pt}
	+ \sum_{n=1}^{\infty}
	(\phi_{1,n}^*, \phi_{2,n})	 
	 \left(
	 \begin{array}{c c }
     	-\partial^2-\hat{n}^2 -\hata^2   &2\hata\hat{n} \\
     	2\hata\hat{n}  &-\partial^2-\hat{n}^2 -\hata^2  \\
    	\end{array} \ 
	\right)
	\left( \begin{array}{c } 
	 \phi_{1,n} \\  \phi_{2,n}^* 
	 \end{array}\right) 
	\end{eqnarray}
where $\hat\alpha=\alpha/R$ and $\hat n=n/R$. Here we used
	\begin{eqnarray}
	\int \!\! dy \ 2\hata~\phi_2(y) \delfive \phi_1(y)&=&
	2\hata\int \!\! dy \sum_{n,m}
	\Big[\frac{\phi_{2,m}}{\sqrt{\pi R}}\sin\frac{my}{R}\Big]
	\left(\frac{n}{R}\right)
	\Big[\frac{\eta_n \phi_{1,n}}{\sqrt{\pi R}}\cos\frac{ny}{R}\Big]
	\no\\
	&=&2\hata \sum_{n,m} \phi_{2,m}\phi_{1,n}(-\delta_{m,n})\left(\frac{n}{R}\right)
	\no\\
	&=&-2\hata \hat{n}\sum_{n=1} \phi_{2,n}\phi_{1,n}
	\end{eqnarray}
for off-diagonal elements. 
Mass eigenstates are given by
	\begin{eqnarray}
	\phi_{\pm,n}\equiv \frac{\phi_{1,n}\mp\phi_{2,n}^*}{\sqrt{2}} \ \ (n>0),
	\end{eqnarray}
with mass $(\hata\pm\hat{n})^2$. It is convenient that they are combined with zero mode, 
	\begin{eqnarray}
	\phi_n\equiv 	
	 \Bigg\{
	 \begin{array}{c c }
     	\phi_{+,n}   &(n>0)\\
     	 \phi_{1,0} &(n=0)  \\
     	\phi_{-,|n|}   &(n<0)\\
    	\end{array} ,
	\end{eqnarray}
Using these mass eigenstates the KK expansion is rewritten by, 
	\begin{eqnarray}
	\phi_1(y)&=&\frac{1}{\sqrt{2\pi R}}\phi_{1,0} +\sum_{n=1}^\infty \frac{\phi_{+,n}+\phi_{-,n}  }{\sqrt{2\pi R}}\cos\frac{ny}{R}
	\no\\
	&=&\sum_{n=-\infty}^\infty \frac{\phi_{n}}{\sqrt{2\pi R}}\cos\frac{ny}{R},
	\\
	\phi_2^*(y)&=&\sum_{n=1}^\infty \frac{-\phi_{+,n}+\phi_{-,n}  }{\sqrt{2\pi R}}\sin\frac{ny}{R}\no\\
	&=&-\sum_{n=-\infty}^\infty \frac{\phi_{n}}{\sqrt{2\pi R}}\sin\frac{ny}{R}.
	\end{eqnarray}
Then the mass matrix is simplified,
	\begin{eqnarray}
	{\cal L}_{\phi}=\sum_{n=-\infty}^{\infty}
	\phi_n^* \{\partial^2-(\hata+\hat{n})^2\}\phi_n \ ,
	\end{eqnarray}
and the propagator is
	\begin{eqnarray}
	\langle \phi_n(p)\phi^*_n(p) \rangle \sim \frac{i}{p^2 -(\hata+\hat{n}^2)} \ .
	\end{eqnarray}
Here we denote,
	\begin{eqnarray}
	\!\!\!\!\langle \phi_n(p)\phi^*_n(q) \rangle
	\!\!\!\!&&\equiv \int \!\!d^4x \int \!\!d^4y\ e^{ipx}e^{-iqy} \langle |{\rm T} \phi_n(x)\phi^*_n(y)| \rangle 
	\\
	&&=\int \!\!d^4x \int \!\!d^4y\ e^{ipx}e^{-iqy} 
	\int \!\!\frac{d^4k}{(2\pi)^4}\frac{i e^{-ik(x-y)}}{k^2-(\hata+\hat{n})^2+i\epsilon}
	\no\\
	&&=(2\pi)^4\delta^{4}(p-q) 
	\left(\frac{i }{p^2-(\hata+\hat{n})^2+i\epsilon}\right) \ .
	\end{eqnarray}
The prefactor of $(2\pi)^4\delta^{4}(p-q)$ is common, so we omit this. For the case of real field propagator, the delta function becomes $\delta^{4}(p+q)$.

\subsection{Quarks and leptons}
To obtain a chiral fermion, one fermion $\psi_1$ is even under ${\cal P}: y\to -y$, while the other fermion $\psi_2$ is odd. The kinetic term of quark is
	\begin{eqnarray}
	{\cal L}_{\psi}&=&	\int \!\! dy \
	i\left(\psi_2(y), \bar\psi_1(y) \right)
	\left(
	 \begin{array}{c c }
	  i \delfive & \sigma^\mu{\partial}_\mu \\
	      \sigmabar^\mu{\partial}_\mu & -i\delfive \\
	 \end{array}
	\right)
	\left(
	 \begin{array}{c}
	  \psi_{1}(y) \\  \bar\psi_2(y) \\
	 \end{array}
	\right) 
	\no\\
	&=&i{\bar\psi_{1,0}}\bar{\sigma}^\mu\partial_\mu\psi_{1,0}
	+\sum_{n=1}^\infty
		\left(\psi_{2,n}, \bar\psi_{1,n} \right)
	\left(
	 \begin{array}{c c }
	  \hat{n}& i\sigma^\mu{\partial}_\mu \\
	      i\sigmabar^\mu{\partial}_\mu & \hat{n}\\
	 \end{array}
	\right)
	\left(
	 \begin{array}{c}
	  \psi_{1,n} \\  \bar\psi_{2,n} \\
	 \end{array}
	\right) 
	\no\\
	&=&\frac{i}{2}\bar{\Psi}_0 \slashed{\partial}\Psi_0 
	+\sum_{n=1}^\infty 	\overline{\Psi}_n (i\slashed{\partial} +\hat{n})\Psi_n 
	=\sum_{n=0}^\infty (\eta_n^2)
	 \overline{\Psi}_n \left(i\slashed{\partial} +\hat{n}\right)\Psi_n 
	\end{eqnarray}
where 
	\begin{eqnarray}
	\Psi_0\equiv 
	\left(
	 \begin{array}{c}
	  \psi_{1,0} \\  \bar\psi_{1,0} \\
	 \end{array}
	\right), \ \
	 \Psi_n\equiv 
	\left(
	 \begin{array}{c}
	  \psi_{1,n} \\  \bar\psi_{2,n} \\
	 \end{array}
	\right)\ .
	\end{eqnarray}
Since we work on calculations in four components notation, it is convenient to mention charge conjugate of the fermions, that is,  
	\begin{eqnarray}
	 \Psi_n^c=C\overline{\Psi}_n^T\equiv 
	\left(
	 \begin{array}{c}
	  \psi_{2,n} \\  \bar\psi_{1,n} \\
	 \end{array}
	\right)\ ,
	\end{eqnarray}
where $C=i\gamma_0\gamma_2$.
The propagator is given by
	\begin{eqnarray}
	\langle \Psi_n(p)\overline\Psi_n(p) \rangle
	\sim
	\langle \Psi_n^c(p)\overline\Psi_n^c(p) \rangle
	 \sim \frac{i}{\slashed{p}+\hat{n}+i\epsilon}.
	\end{eqnarray}

\subsection{Gauginos}
Gaugino $\lambda_1$ is even under ${\cal P}: y\to -y$ because $V\supset A_\mu$ must be even for the gauge invariance while $\lambda_2$ which comes along with $A_5$ is odd. 
Their kinetic term with soft breaking is given by, 
	\begin{eqnarray}
		{\cal L}_{\lambda}&=&	\int \!\! dy \
	i\left(\lambda_2(y), \bar\lambda_1(y) \right)
	\left(
	 \begin{array}{c c }
	  i \delfive & \sigma^\mu{\partial}_\mu \\
	      \sigmabar^\mu{\partial}_\mu &- i\delfive \\
	 \end{array}
	\right)
	\left(
	 \begin{array}{c}
	  \lambda_{1}(y) \\  \bar\lambda_2(y) \\
	 \end{array}
	\right)
	\no\\
	&&\ \ \ +\frac{1}{2}\hata\left(\lambda_1(y)\lambda_1(y)+\lambda_2(y)\lambda_2(y)
	+{\rm h.c.}\right) 
	\no\\
	&=&\sum_{n=0}^\infty \Big\{
		\left(\lambda_{2,n}, \bar\lambda_{1,n} \right)
	\left(
	 \begin{array}{c c }
	  \hat{n} & i\sigma^\mu{\partial}_\mu \\
	    i  \sigmabar^\mu{\partial}_\mu & \hat{n} \\
	 \end{array}
	\right)
	\left(
	 \begin{array}{c}
	  \lambda_{1,n} \\  \bar\lambda_{2,n} \\
	 \end{array}
	\right)
	\no\\
	&&\ \ \ \ \ 
	+\frac{1}{2} \hata\left( \lambda_{1,n}\lambda_{1,n}
	+\lambda_{2,n}\lambda_{2,n}
	\right) +{\rm h.c.}
	\Big\}.
	\end{eqnarray}
where $\lambda_{2,0}=0$ is kept for convenience. For mass eigenstates, we use two majorana fermions, 
	\begin{eqnarray}
	\lambda_0\equiv 
	\left(
	 \begin{array}{c}
	  \lambda_{1,0} \\  \bar\lambda_{1,0} \\
	 \end{array}
	\right), \ \
	 \lambda_{\pm,n}\equiv 
	\frac{1}{\sqrt{2}}
	\left(
	 \begin{array}{c}
	  \lambda_{1,n}\pm\lambda_{2,n} \\  \bar\lambda_{1,n}\pm\bar\lambda_{2,n} \\
	 \end{array} 
	\right) \ (n>0),
	\end{eqnarray}
where the mass of $\lambda_0$ is $\hata$, and the masses of $\lambda_{\pm,n}$ are $(\hata+\hat{n})$. Similar to the squark and slepton, we combine these gauginos of mass eigenstate,
	\begin{eqnarray}
	\lambda_n\equiv 	
	 \Bigg\{
	 \begin{array}{c c }
     	\lambda_{+,n}   &(n>0)\\
     	 \lambda_{0} &(n=0)  \\
     	\lambda_{-,|n|}   &(n<0)\\
    	\end{array} ,
	\end{eqnarray}
and the KK expansion of gauginos are also given in the mass eigenstates, 
	\begin{eqnarray}
	\lambda_1(y)&=&\frac{1}{\sqrt{2\pi R}}P_L\lambda_{0} 
	+\sum_{n=1}^\infty P_L\frac{\lambda_{+,n}+\lambda_{-,n}}{\sqrt{2\pi R}}\cos\frac{ny}{R}
	\no\\
	&=&\sum_{n=-\infty}^\infty P_L\frac{\lambda_{n}}{\sqrt{2\pi R}}\cos\frac{ny}{R},
	\\
	\lambda_2(y)&=&\sum_{n=1}^\infty P_L\frac{\lambda_{+,n}-\lambda_{-,n}  }{\sqrt{2\pi R}}\sin\frac{ny}{R}\no\\
	&=&\sum_{n=-\infty}^\infty P_L\frac{\lambda_{n}}{\sqrt{2\pi R}}\sin\frac{ny}{R}.
	\end{eqnarray}
The kinetic terms are written by
	\begin{eqnarray}
	{\cal L}_{\lambda}&=&
	\frac{1}{2}\bar{\lambda}_0(i\slashed{\partial}+\hata)\lambda_0
	\no\\
	&&
	+	\sum_{n=1}^\infty \frac{1}{2}\bar{\lambda}_{-,n}(i\slashed{\partial}+(\hata-\hat{n}))\lambda_{-,n}
	+	\sum_{n=1}^\infty \frac{1}{2}\bar{\lambda}_{+,n}(i\slashed{\partial}+(\hata+\hat{n}))\lambda_{+,n}
	\no\\
	&=&	\sum_{n=-\infty}^\infty  \frac{1}{2}\bar{\lambda}_n(i\slashed{\partial}+(\hata+\hat{n}))\lambda_n, 
	\end{eqnarray}
and then the propagator is
	\begin{eqnarray}
	\langle \lambda_n(p) \bar\lambda_n(p)\rangle
	\sim \frac{i}{\slashed{p}+(\hata+\hat{n}) +i\epsilon}\ .
	\end{eqnarray}

\subsection{Gauge fields}
We calculate the propagator of $U(1)$ gauge fields, and the extension to non-abelian cases is straightforward. In the 4D picture, non-zero KK modes of $A_\mu$ are massive and their longitudinal degrees of freedom comes from non-zero KK modes of $A_5$, and therefore $A_5$ behaves as Nambu-Goldstone boson.  
So we take $R_\xi$ gauge, 
	\begin{eqnarray}
	{\cal L}_{5,R_\xi}= 
	-\frac{1}{2\xi}(\partial^\mu A_{\mu}+\xi\partial^5 A_5)^2 \ ,
	\end{eqnarray}
and the field strength is,
	\begin{eqnarray}
	{\cal L}_{5,\rm gauge}&=& -\frac{1}{4}F^{MN}F_{MN}
	=-\frac{1}{4}F^{\mu\nu}F_{\mu\nu}-\frac{1}{2}F^{\mu5}F_{\mu5} \ .
	\end{eqnarray}
Under $y\to-y$, clearly $A_\mu$ is even and $A_5$ is odd. Then KK expansion of $A_\mu$ and $A_5$ are given by Eq.~(\ref{KKEven}) and Eq.~(\ref{KKOdd}), respectively. The mixing term of them vanishes, 
	\begin{eqnarray}
	-\frac{1}{2}F^{\mu5}F_{\mu5}+{\cal L}_{5,R_\xi}
	\!\!\!&=&-\frac{1}{2}\left(\partial^\mu A^5 \partial_\mu A_5
	+\partial^5 A^\mu \partial_5 A_\mu
	-2 \partial^\mu A^5 \partial_5 A_\mu
	\right)\no
	\\
	&&-\frac{1}{2\xi}\left\{(\partial^\mu A_{\mu})^2+\xi^2(\partial^5 A_5)^2 \right\}
	-\partial^\mu A^5 \partial_5 A_\mu
	\\
	&=&\frac{1}{2}\left\{ \partial^\mu A_5 \partial_\mu A_5
	-\xi(\delfive A_5)^2
	\right\}
	\no\\
	&&+\frac{1}{2}\left\{
	\partial_5 A^\mu \partial_5 A_\mu
	-\frac{1}{\xi}(\partial^\mu A_{\mu})^2
	\right\}.
	\end{eqnarray}
For $A_5$ KK modes,
	\begin{eqnarray}
	\int \!\! dy\ \frac{1}{2}\left\{ \partial^\mu A_5(y) \partial_\mu A_5(y)
	-\xi(\delfive A_5(y))^2 \right\}
	=\sum_{n=1}^\infty \frac{1}{2}A_{5 n} \left(-\partial^2  -\xi \hat{n}^2 \right) A_{5 n},
	\end{eqnarray}
the propagator is, 
	\begin{eqnarray}
	\langle A_{5,n}(p)A_{5,n}(-p)\rangle \sim\frac{i}{p^2-\xi \hat{n}^2+i\epsilon}.
	\end{eqnarray}
For $A_\mu$ KK modes,
	\begin{eqnarray}
&&
	\int\!\! dy\ 
	-\frac{1}{4}F^{\mu\nu}F_{\mu\nu}
	+\frac{1}{2}
	\partial_5 A^\mu(y) \partial_5 A_\mu(y)
	-\frac{1}{2\xi}(\partial^\mu A_{\mu}(y))^2 
	\no\\&&=
	\int \!\! dy\ 
	\frac{1}{2}A_\mu(y)\left\{ 
	g^{\mu\nu}(\partial^2 -\delfive^2)-\partial^\mu \partial^\nu(1-\xi^{-1})
	\right\}A_\nu(y)
	\no\\&&=
	\sum_{n=0}^\infty \frac{1}{2}A_{(n)\mu} 
	\left\{g^{\mu\nu}\left(\partial^2 +\hat{n}^2\right) -\partial^\mu \partial^\nu(1-\xi^{-1})
	\right\} A_{(n) \nu},
	\end{eqnarray}
the propagator is, 
	\begin{eqnarray}
	\langle A_{(n)\mu}(p)A_{(n)\nu}(-p)\rangle \sim
	\frac{-i}{p^2-\hat{n}^2+i\epsilon}
	\left(g_{\mu\nu} -(1-\xi)\frac{p_\mu p_\nu}{p^2-\xi \hat{n}^2}\right).
	\end{eqnarray}

\subsection{Adjoint real scalars}
 The mass and propagator of adjoint real scalar $\Sigma$ is simple,
	\begin{eqnarray}
	{\cal L}_{\Sigma} &=&\int_0^{2\pi R}\!\! dy\ \frac{1}{2}\Sigma^a(y)(-\partial^2+\partial_5^2) \Sigma^a(y)
	\\&=&\sum_{n=1}^\infty\frac{1}{2}\Sigma^a_n(-\partial^2-\hat{n}^2) \Sigma^a_n \ ,
	\end{eqnarray}
and the propagator is
	\begin{eqnarray}
	\langle \Sigma^a_{n}(p)\Sigma^a_{n}(-p)\rangle \sim\frac{i}{p^2- \hat{n}^2+i\epsilon} \ .
	\end{eqnarray}



\section{Formulae}
\subsection{Formulae for KK Expansion}
In order to obtain interactions of KK modes, there appear many integrals of $\sin$ and $\cos$. The interactions are either four-point or three-point, the integrals are summarized in the following formulae.  
For  integers, $k,l,m,n$, four point interactions have integrals of 
	\begin{eqnarray}
	M_1^{klmn}&\equiv&\frac{2}{\pi}\int_0^{2\pi}\!\!\!\! dx \cos(kx)\cos(lx)\cos(mx)\cos(nx)
	\no\\
	&=&
		\delta_{k+l+m+n,0}+\delta_{k+l-m-n,0}+\delta_{k+l-m+n,0}+\delta_{k+l+m-n,0}
	\no\\
	&&+\delta_{k-l+m+n,0}+\delta_{k-l-m-n,0}+\delta_{k-l-m+n,0}+\delta_{k-l+m-n,0},
	\\
	M_2^{kl,mn}&\equiv&\frac{2}{\pi}\int_0^{2\pi}\!\!\!\! dx \cos(kx)\cos(lx)\sin(mx)\sin(nx)
	\no\\
	&=&
	-\delta_{k+l+m+n,0}-\delta_{k+l-m-n,0}+\delta_{k+l-m+n,0}+\delta_{k+l+m-n,0}
	\no\\
	&&-\delta_{k-l+m+n,0}-\delta_{k-l-m-n,0}+\delta_{k-l-m+n,0}+\delta_{k-l+m-n,0},
		\\
	M_3^{klmn}&\equiv&\frac{2}{\pi}\int_0^{2\pi}\!\!\!\! dx \sin(kx)\sin(lx)\sin(mx)\sin(nx)
	\no\\
	&=&
		\delta_{k+l+m+n,0}+\delta_{k+l-m-n,0}-\delta_{k+l-m+n,0}-\delta_{k+l+m-n,0}
	\no\\
	&&-\delta_{k-l+m+n,0}-\delta_{k-l-m-n,0}+\delta_{k-l-m+n,0}+\delta_{k-l+m-n,0}.
	\end{eqnarray}
Relative signs are easily understood by  changing a sign of argument in $\sin$ functions, for instance, 
$M_2^{kl,(-m)n}=-M_2^{kl,mn}$. This results in that the integral of odd number of sin functions vanishes. The useful combinations are
	\begin{eqnarray}
	M_4^{kl,mn}\!\!\!\!\!&\equiv& \frac{1}{2}(M_1^{klmn}+M_2^{kl,mn})
	\\
	&=&\delta_{k+l+m-n,0}+\delta_{k+l-m+n,0}+\delta_{k-l+m-n,0}+\delta_{k-l-m+n,0},
	\no\\
	\!\!\!M_5^{kl,mn}\!\!\!&\equiv& \frac{1}{2}(M_1^{klmn}-M_2^{kl,mn})
	\\
	&=&\delta_{k+l+m+n,0}+\delta_{k+l-m-n,0}+\delta_{k-l+m+n,0}+\delta_{k-l-m-n,0},
	\no\\
	\!\!\!M_6^{kl,mn}\!\!\!&\equiv& \frac{1}{2}(M_2^{kl,mn}+M_3^{kl,mn})
	\\
	&=&\!\!\!\!-\delta_{k-l+m+n,0}-\delta_{k-l-m-n,0}+\delta_{k-l+m-n,0}+\delta_{k-l-m+n,0},
	\no\\
	M_7^{kl,mn}\!\!\!&\equiv&\frac{1}{2} (M_2^{kl,mn}-M_3^{kl,mn})
	\\
	&=&\!\!\!\!-\delta_{k+l+m+n,0}-\delta_{k+l-m-n,0}+\delta_{k+l+m-n,0}+\delta_{k+l-m+n,0}.
	\no
	\end{eqnarray}
Formulae for three-point interactions are
	\begin{eqnarray}
	N_1^{kmn}&\equiv&
	\frac{2}{\pi}\int_0^{2\pi}\!\!\!\! dx \cos(kx)\cos(mx)\cos(nx)
	\no\\
	&=&
	\delta_{k+m+n,0}+\delta_{k-m-n,0}+\delta_{k-m+n,0}+\delta_{k+m-n,0},
	\\
	N_2^{k,mn}&\equiv&\frac{2}{\pi}\int_0^{2\pi}\!\!\!\! dx \cos(kx)\sin(mx)\sin(nx)
	\no\\
	&=&
	-\delta_{k+m+n,0}-\delta_{k-m-n,0}+\delta_{k-m+n,0}+\delta_{k+m-n,0},
	\\
	\no\\
	N_3^{k,mn}&\equiv&\frac{1}{2}(N_1^{kmn}+N_2^{k,mn})
	=\delta_{k+m-n,0}+\delta_{k-m+n,0},
	\\
	N_4^{k,mn}&\equiv&\frac{1}{2}(N_1^{kmn}-N_2^{k,mn})
	=\delta_{k+m+n,0}+\delta_{k-m-n,0}.
	\end{eqnarray}

\section{Interactions}
Using the above formulae, we can derive some Feynman rules. 
Note that some interactions with 5D derivative must be combined with soft terms since they are essentially from the same term in the 5D. Here is an example, squark-squark-$A_{\mu(5)}$ Interaction. 
\begin{figure}[h!]
\begin{center}
  \includegraphics[clip,width=1.\textwidth]{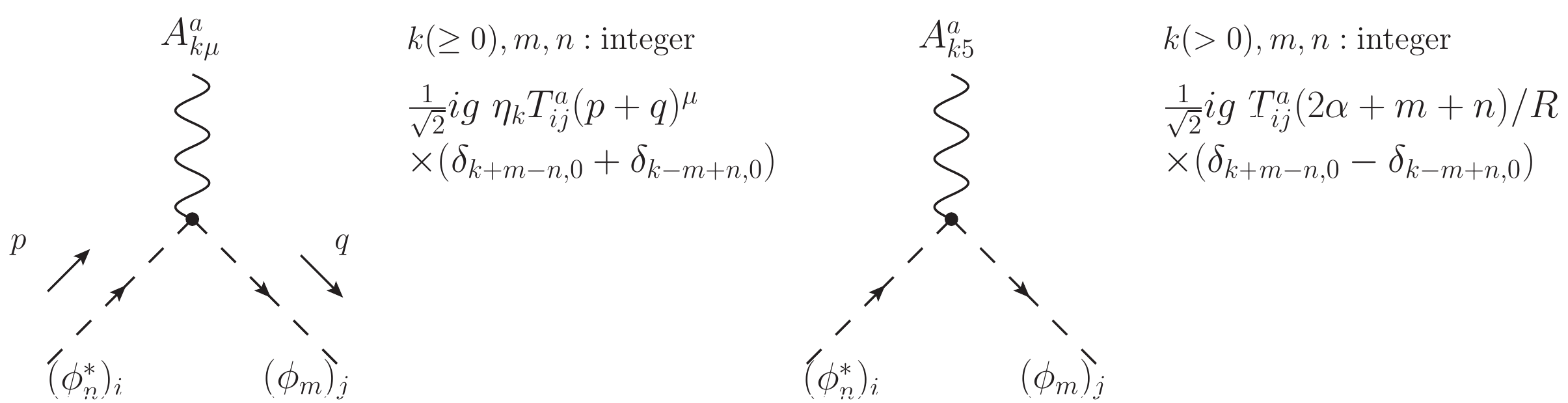}
\end{center}
\vspace{-10pt}
\caption{Squark-Squark-$A_{\mu}$ interaction and Squark-Squark-$A_{5}$ interaction.}
\vspace{-10pt}
\end{figure}




\chapter{Formulae for Summation and Integral}\label{app:thresh}
\section{Infinite Sum to Analytic Function}
For a function that has no singularities on the real $z$ axis, the useful relation is
	\begin{eqnarray}
	\sum_{n=-\infty}^{\infty} f(k, n+\alpha)= \sum_{n=-\infty}^{\infty} \oint_{C_n^\alpha}dz \ 
	f(k,z) \frac{\coth[i \pi(z-\alpha)]}{2}.
	\end{eqnarray}
Where the contour $C_n^\alpha$ is a path which rounds about $z=n+\alpha$ with a small radius. 
When ${z\to\alpha}$, 
	\begin{eqnarray}
	\frac{\coth[i \pi(z-\alpha)]}{2} = \frac{1}{2i \pi (z-\alpha)} +{\cal O}(z-\alpha) \ , 
	\end{eqnarray}
and the function above is periodic under a transformation of $z \to z+ n\pi$, so each $\oint_{C_n^\alpha}dz$ generates discrete point of $f(k,z)$. 
\begin{figure}[htb]
 \begin{center}
  \includegraphics[width=100mm]{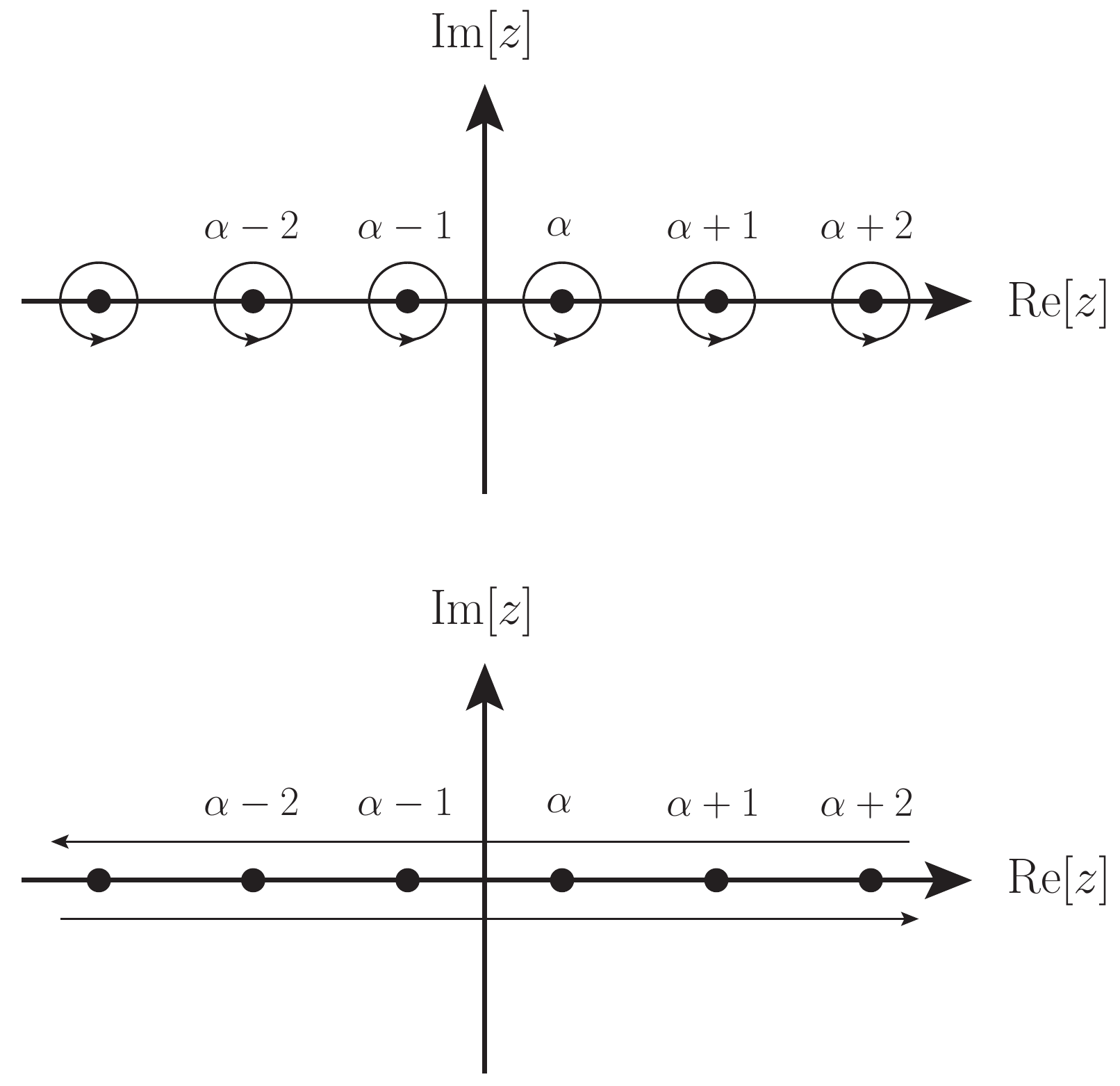}
 \end{center}
 \caption{Integral path of $C_n^\alpha$. They are combined as shown in the lower plot.}
 \label{fig:path}
\end{figure}
We change the path as in \ref{fig:path}, 
	\begin{eqnarray}
	&&\sum_{n=-\infty}^{\infty} \oint_{C_n^\alpha}dz \ f(k,z) \frac{\coth[i \pi(z-\alpha)]}{2}
	\no\\&&=
	\left(\int_{\infty+i\epsilon}^{-\infty+i\epsilon} dz +\int_{-\infty-i\epsilon}^{\infty-i\epsilon} dz	\right)
	  f(k,z) \frac{\coth[i \pi(z-\alpha)]}{2}
	\\&&=
	 \int_{-\infty-i\epsilon}^{\infty-i\epsilon} dz	\left\{
	  f(k,z) \frac{\coth[i \pi(z-\alpha)]}{2} -f(k,-z) \frac{\coth[i \pi(-z-\alpha)]}{2}
	 \right\} \no
	\\&&=
	 \int_{-\infty}^{\infty} dz	\left\{
	  \frac{f(k,z)+f(k,-z)}{2} \right\}
	\no \\ 
	&&\quad +\int_{-\infty-i\epsilon}^{\infty-i\epsilon} dz\left\{
	  \frac{f(k,z)}{e^{2i \pi(z-\alpha)} -1} +\frac{f(k,-z)}{e^{2i \pi(z+\alpha)} -1}
	 \right\}
	 \label{f1}	\ . 
	\end{eqnarray}
Here, we used
	\begin{eqnarray}
	\coth(x)=1+\frac{2}{e^{2x}-1}=-\left( 1+\frac{2}{e^{-2x}-1} \right)\ .
	\end{eqnarray}
We are interested in $f(k,z)=1/(k^2+z^2),\ z/(k^2+z^2) $, and hence, for such functions that damps 
for $z\to \pm\infty-i\epsilon$ and can be suppressed by $e^{2\pi|z|}$ for $z \to -i\infty$,  the second expression of Eq.(\ref{f1}) can enclose the path, referred as to $C_\bigtriangledown$,  in the negative imaginary $z$ plane,
	\begin{eqnarray}
	\sum_{n=-\infty}^{\infty} f(k, n+\alpha)
	&&=	 \int_{-\infty}^{\infty} dz	\left\{
	  \frac{f(k,z)+f(k,-z)}{2} \right\}
	\no\\&&
	+\oint_{C_\bigtriangledown} dz\left\{
	  \frac{f(k,z)}{e^{2i \pi(z-\alpha)} -1} +\frac{f(k,-z)}{e^{2i \pi(z+\alpha)} -1}
	 \right\}.  
	 \label{formula:sum}
	\end{eqnarray}
If $f(k,z)$ has poles inside the closed path $C_\bigtriangledown$, the second term on the right-hand side becomes a function of $k$, otherwise it vanishes. In the following section, we will see the cases of, 
	\begin{eqnarray}
	f(k,z)=\frac{1}{k^2+z^2}, \ \frac{z}{k^2+z^2} \ .
	\end{eqnarray}

\subsection{Formulae}
Using Eq.~(\ref{formula:sum}) formulae for infinite sum: 
	\begin{eqnarray}
	\bullet \ F_1^\alpha(k) &\equiv&\no	\sum_{n=-\infty}^{\infty} \frac{1}{k^2+(n+\alpha)^2}
	\\&=&	\int_{-\infty}^\infty \frac{dz}{k^2+z^2} 
	+\frac{\pi}{k}\left\{ \frac{1}{e^{2 \pi(k-i\alpha)} -1} +\frac{1}{e^{2 \pi(k+i\alpha)} -1} \right\}
	\\
	&=&\frac{\pi}{k}
	+\frac{\pi}{k}\left\{ \frac{1}{e^{2 \pi(k-i\alpha)} -1} +\frac{1}{e^{2 \pi(k+i\alpha)} -1} \right\}
	\ \ (\text{if } \ k\neq 0)
	\no\\\\
	\bullet \ F_2^\alpha(k) &\equiv&\no \sum_{n=-\infty}^{\infty} \frac{(n+\alpha)}{k^2+(n+\alpha)^2}\\
	&=&
	(-i\pi) \left\{ \frac{1}{e^{2 \pi(k-i\alpha)} -1} -\frac{1}{e^{2 \pi(k+i\alpha)} -1} \right\} \ 
	\end{eqnarray}
Formulae  for momentum integral:
	\begin{eqnarray}
	&\bullet& \ G^n_1(\pm\alpha) \equiv \int_{0}^\infty dx \frac{x^n}{e^{x\pm2\pi i \alpha}-1}
	\ =n! \ {\rm Li}_{n+1}[e^{\mp2\pi i \alpha}]
	\\
	&\bullet& \ G_2^n(\pm\alpha) \equiv \int_{0}^\infty dx \frac{x^n}{(e^{x\pm2\pi i \alpha}-1)^2 }
	=n! \left( {\rm Li}_{n}[e^{\mp2\pi i \alpha}]-{\rm Li}_{n+1}[e^{\mp2\pi i \alpha}] \right)
	\no\\\\
	&\bullet& \ G_3^n(\pm\alpha) \equiv \int_{0}^\infty dx 
	\frac{x^n}{(e^{x+2\pi i \alpha}-1)(e^{x-2\pi i \alpha}-1) }
	\no\\
	&& 	\hspace{45pt}
	=\frac{i }{2\sin(2\pi\alpha)}  
	n! \left( {\rm Li}_{n+1}[e^{\mp2\pi i \alpha}]-{\rm Li}_{n+1}[e^{\pm2\pi i \alpha}] \right)
	\\ 
	&& \hspace{55pt}
	\to n!\left[\zeta(n)-\zeta(n+1) \right] \ \ (\text{if }\ \alpha \to 0)
	\\\no
	\\
	&\bullet&  G^n_1(\pm\alpha)+G^n_2(\pm\alpha)
	=n! \ {\rm Li}_{n}[e^{\mp2\pi i \alpha}]
	\end{eqnarray}
\subsection{Loop integral of infinite sum }
We combine formulae for infinite sum ($F$) and those for momentum integral ($G$) for radiative corrections of KK tower. 
	\begin{eqnarray}
	\bullet &&\int_{0}^{\infty}\frac{dk }{8\pi^2} k^n\Big( F_1^\alpha(k)\Big) \no\\
	&=& \frac{1}{(2\pi)^{n+1} }\int_{0}^{\infty}\frac{dx }{8\pi^2} x^n\Big( F_1^\alpha\left(\frac{x}{2\pi} \right)\Big)
	\\
	&=& \frac{2\pi^2}{(2\pi)^{n+1}8\pi^2}\left[ \left( \int_{0}^{\infty}dx x^{n-1} \right)
	+G^{n-1}_1(+\alpha)+G^{n-1}_1(-\alpha)
	\right]
	\\
	&=& \frac{1}{4(2\pi)^{n+1}}\left[ \left( \int_{0}^{\infty}dx x^{n-1} \right)
	+(n-1)!\Big({\rm Li}_{n}[e^{2\pi i \alpha}]+{\rm Li}_{n}[e^{-2\pi i \alpha}] \Big)
	\right]\no
\\\no\\
	\end{eqnarray}
	\begin{eqnarray}
	\bullet &&\int_{0}^{\infty}\frac{dk }{8\pi^2} k^n\Big( F_1^\alpha(k)\Big)^2\no\\
	&=& \frac{1}{(2\pi)^{n+1} }\int_{0}^{\infty}\frac{dx }{8\pi^2} x^n\Big( F_1^\alpha\left(\frac{x}{2\pi} \right)\Big)^2
	\\
	&=& \frac{(2\pi)^2\pi^2}{(2\pi)^{n+1}8\pi^2}\Bigg[ \left( \int_{0}^{\infty}dx x^{n-2} \right)
	+2\Big(G^{n-2}_1(\alpha)+G^{n-2}_1(-\alpha)\Big)
	\no\\
	&&\hspace{80pt}\ +\Big(G^{n-2}_2(\alpha)+G^{n-2}_2(-\alpha)  +2G^{n-2}_3(\alpha) \Big)
	\Bigg]
	\\
	&=& \frac{1}{8(2\pi)^{n-1}}\Bigg[ \left( \int_{0}^{\infty}dx x^{n-2} \right)
	+2(n-2)!\Big({\rm Li}_{n-1}[e^{2\pi i \alpha}]+{\rm Li}_{n-1}[e^{-2\pi i \alpha}]\Big)
	\no\\
	&&\hspace{-10pt}\ +(n-2)!\Big(
	{\rm Li}_{n-2}[e^{2\pi i \alpha}]-{\rm Li}_{n-1}[e^{2\pi i \alpha}]
	+{\rm Li}_{n-2}[e^{-2\pi i \alpha}]-{\rm Li}_{n-1}[e^{-2\pi i \alpha}]\Big) 
	\no\\
	&&\hspace{10pt}\ +2(n-2)!\frac{i}{2\sin(2\pi \alpha)}\Big(
	e^{2\pi i \alpha}{\rm Li}_{n-1}[e^{-2\pi i \alpha}]-e^{-2\pi i \alpha}{\rm Li}_{n-1}[e^{2\pi i \alpha}]
	\Big) \Bigg] \no
	\\
	&=& \frac{1}{8(2\pi)^{n-1}}\Bigg[ \left( \int_{0}^{\infty}dx x^{n-2} \right)
	\no\\
	&&\hspace{-10pt}\ +(n-2)!\Big(
	{\rm Li}_{n-2}[e^{2\pi i \alpha}] +{\rm Li}_{n-1}[e^{2\pi i \alpha}]
	+{\rm Li}_{n-2}[e^{-2\pi i \alpha}] +{\rm Li}_{n-1}[e^{-2\pi i \alpha}]\Big) 
	\no\\
	&&\hspace{10pt}\ +2(n-2)!\frac{i}{2\sin(2\pi \alpha)}\Big(
	e^{2\pi i \alpha}{\rm Li}_{n-1}[e^{-2\pi i \alpha}]-e^{-2\pi i \alpha}{\rm Li}_{n-1}[e^{2\pi i \alpha}]
	\Big) \Bigg] \no
	\\
	\\
	\bullet &&\int_{0}^{\infty}\frac{dk }{8\pi^2} k^n\Big( F_2^\alpha(k)\Big)^2 \no\\
	&=& \frac{1}{(2\pi)^{n+1} }\int_{0}^{\infty}\frac{dx }{8\pi^2} x^n\Big( F_2^\alpha\left(\frac{x}{2\pi} \right)\Big)^2
	\\
	&=& \frac{-\pi^2}{(2\pi)^{n+1}8\pi^2}
	\big[ G^{n}_2(\alpha)+G^{n}_2(-\alpha)-2G^{n}_3(\alpha)\big]
	\\&=&
	\frac{-1}{8(2\pi)^{n+1}} \bigg[
	n!\Big({\rm Li}_{n}[e^{2\pi i \alpha}]-{\rm Li}_{n+1}[e^{2\pi i \alpha}]
	+{\rm Li}_{n}[e^{-2\pi i \alpha}]-{\rm Li}_{n+1}[e^{-2\pi i \alpha}]\Big) 
	\no\\
	&&\hspace{10pt}\ -2n!\frac{i}{2\sin(2\pi \alpha)}\Big(
	e^{2\pi i \alpha}{\rm Li}_{n-1}[e^{-2\pi i \alpha}]-e^{-2\pi i \alpha}{\rm Li}_{n-1}[e^{2\pi i \alpha}]
	\Big) \bigg]
	\no\\
	\end{eqnarray}

Useful combinations:
	\begin{eqnarray}
	\bullet&& \int_{0}^{\infty}\frac{dk }{8\pi^2}k^n 
	\left[  F_1^\alpha(k)- F_1^0(k)\right]	
	\no\\&&= \frac{(n-1)!}{4(2\pi)^{n+1}}
	\Big({\rm Li}_{n}[e^{2\pi i \alpha}]+{\rm Li}_{n}[e^{-2\pi i \alpha}] -2\zeta(n) \Big)
	\end{eqnarray}
	\begin{eqnarray}
	&\bullet& \int_{0}^{\infty}\frac{dk }{8\pi^2}k^n 
	\left[  k^2\Big(F_1^\alpha(k)\Big)^2-k^2\Big(F_1^0(k) \Big)^2 -\Big(F_2^\alpha(k)\Big)^2	\right]	
	\no\\
	&&= \frac{1}{8(2\pi)^{n+1}}
	\Bigg[ \left( \int_{0}^{\infty}dx x^{n} \right)
	\no
	\\
	&&\hspace{50pt} 
	+\Big(2G^{n}_1(\alpha)+2G^{n}_1(-\alpha) 
	+G^{n}_2(\alpha)+G^{n}_2(-\alpha)+2G^{n}_3(\alpha)\Big)
	\no
	\\
	&&\hspace{50pt} 
	-\left( \int_{0}^{\infty}dx x^{n} \right)
	-\Big(4G^{n}_1(0) +4G^{n}_2(0)\Big)
	\no
	\\
	&&\hspace{50pt} 
	+\Big(G^{n}_2(\alpha)+G^{n}_2(-\alpha) -2G^{n}_3(\alpha)\Big)\Bigg]
	\no\\
	&&= \frac{1}{4(2\pi)^{n+1}}
	\Big[G^{n}_1(\alpha)+G^{n}_1(-\alpha) 
	+G^{n}_2(\alpha)+G^{n}_2(-\alpha)	-2G^{n}_1(0)-2G^{n}_2(0)	\Big]
	\no\\
	\no\\
	&&= \frac{n!}{4(2\pi)^{n+1}}
	\left[{\rm Li}_{n}[e^{2\pi i \alpha}]+{\rm Li}_{n}[e^{-2\pi i \alpha}] -2\zeta(3)
	\right]\label{eq:usefulcombi}
	\end{eqnarray}

\section{Example:\ $m_{H_u}^2$ from all KK particles with top-Yukawa coupling}
We use Eq.~(\ref{eq:usefulcombi}) for threshold correction to  $m_{H_u}^2$. All the infinite sums are factorized in the following way, 
	\begin{eqnarray}
	-i m_{H_u}^2|_{\rm Yukawa} 
	&=& \frac{iN_c y_t^2}{R^2}  \int_{0}^{\infty}\frac{dk \ k^{3}}{8\pi^2}
	\Bigg[
	2k^2 \left(\sum_{n=-\infty}^{\infty}\frac{1}{k^2+n^2}	\right)\left(\sum_{m=-\infty}^{\infty}\frac{1}{k^2+m^2}	\right)
	\no\\
	&&\hspace{20pt}
	-2k^2 \left(\sum_{n=-\infty}^{\infty}\frac{1}{k^2+(n+\alpha)^2}	\right)\left(\sum_{m=-\infty}^{\infty}\frac{1}{k^2+(m+\alpha)^2}	\right)
	\no\\
	&&\hspace{20pt}
	+2 \left(\sum_{n=-\infty}^{\infty}\frac{n+\alpha}{k^2+(n+\alpha)^2}	\right)\left(\sum_{m=-\infty}^{\infty}\frac{m+\alpha}{k^2+(m+\alpha)^2}	\right)
	\Bigg]
	\\
	&=& \frac{iN_c y_t^2}{R^2}(-2)  \int_{0}^{\infty}\frac{dk \ k^3}{8\pi^2}
	\left[  k^2\Big(F_1^\alpha(k)\Big)^2-k^2\Big(F_1^0(k) \Big)^2 -\Big(F_2^\alpha(k)\Big)^2	\right]
	\no\\
	&=&\frac{iN_c y_t^2}{R^2}  \frac{-2\cdot 3!}{4(2\pi)^{4}}
	\left[{{\rm Li}_{3}[e^{2\pi i \alpha}]+{\rm Li}_{3}[e^{-2\pi i \alpha}]}-2\zeta(3)
	\right] \ .
	\end{eqnarray}
If we expand Polylog functions with respect to $\alpha$, 
	\begin{eqnarray}
	{\rm Li}_{3}[e^{2\pi i \alpha}]
	=\zeta(3)+\frac{i\pi^3}{3}\alpha
	+\pi^2\left(-3+2\ln(2\pi\alpha)-\frac{i\pi}{2} \right)\alpha^2 	+{\cal O}(\alpha^3)
	\\
	{\rm Li}_{3}[e^{2\pi i \alpha}]+{\rm Li}_{3}[e^{-2\pi i \alpha}]
	=2\zeta(3)
	+2\pi^2\left(-3+2\ln(2\pi\alpha) \right)\alpha^2 	+{\cal O}(\alpha^3).
	\end{eqnarray}
Hence, 
	\begin{eqnarray}
	-i m_{H_u}^2|_{\rm Yukawa} 
	=
	\frac{-iN_c y_t^2}{16\pi^2}\left(\frac{\alpha^2}{R^2}\right) 
	\Big( 12\ln(2\pi \alpha)-18 \Big)
	+{\cal O}(\alpha^3) \ .
	\end{eqnarray}

\chapter{Search for New Physics with Compressed Spectrum by $M_{T2}$}\label{ch:mt2}
\section{Introduction}

When the MSSM has a compressed (nearly degenerate) spectrum like the Compact Supersymmetry,  their limits from the LHC are much weaker than those on the CMSSM which  has a widely spread spectrum. Ameliorating limits is generally true for various new physics models with a compressed spectrum and a invisible stable particle like dark matter particle. We refer these (BSM) models to as compressed models. 
In this chapter we point out a kinematic variable, $M_{T2}$, is effective for a compressed model search. The effectiveness of $\mttwo$ \cite{Lester:1999tx, Barr:2003rg} to search for conventional MSSM was discussed in \cite{Barr:2009wu, Lester:2007fq}, and $\mttwo$ was
already applied in several searches for supersymmetry  \cite{daCosta:2011qk, CMS-PAS-SUS-11-005}.

New physics models, as represented by the MSSM, are basically searched based on the missing energy, $\slashed{E}_T$, because created colored particles like gluino and squark decay into invisible particles with large momentum  thanks to expected large mass gap between  them. Large $\slashed{E}_T$ is a distinct signature from the SM processes. 
The difficulty of the search for compressed models is that $\slashed{E}_T$ is significantly smaller in each event due to mass degeneracy even if their mass scale of the process is very high. 
In the SM processes missing energy is actually generated for the neutrino emission, and in particular $t\bar{t}$ pair production, with a subsequent  decay of $t\to bW\to bl\nu$, leads to $\slashed{E}_T$. This SM process is quite similar to the new physics process. 
The current searches by multijet, {\it e.g.} in Ref.~\cite{ATLAS1}, are optimized to search for  the CMSSM, and therefore the event elections imposed here is too strong to keep enough signal events of compressed models. On the other hand, if selection criterion are weakened, there could be a lot of contamination of background. 

Note that initial state radiation (ISR) is important in searches for compressed models. 
While energy emitted from the decay relies on the mass gap, energy scale of QCD radiation, particularly ISR, roughly increases with mass scale of colored particle, and therefore the QCD radiation scale of new physics is larger than that of the SM. 
In the search for compressed model, this is an important feature. 

Throughout this appendix we adopt the MUED as a  benchmark model \cite{Appelquist:2000nn} (see \cite{Hooper:2007qk} for 
review).
In the MUED all the SM fields propagate in a compactified flat extra dimension, $S^1/{\mathbb Z}_2$, and the model provides a good candidate of dark matter particle \cite{Servant:2002aq,  Belanger:2010yx}. 
The collider phenomenology is basically same as the MSSM: first excited KK states of SM field behave as superpartners of the SUSY, and KK parity instead of $R$-parity stabilize the lightest particle, called LKP.  The spectrum is highly compressed at masses of $1/R$ where $R$ is radius of the extra dimension. 
Experimental sensitivity to the MUED needs to be improved since the favored scale is pretty high as $1/R\sim 1.5\TeV$ \cite{Belanger:2010yx} while current LHC  limit is still  $1/R\lesssim 800\GeV$ \cite{ATLAS1}.

After briefly describing the MUED, we review the definition of $M_{T2}$ and next effectiveness of this variable as a event selection as discussed in \cite{Barr:2009wu, Lester:2007fq}. Then we discuss $M_{T2}$ cut is useful for compressed model search. Event selections with  $M_{T2}$ cut is applied for a typical compressed model, the MUED \cite{Murayama:2011hj}.  

\section{MUED}\label{5dUED}
 In the case of the 5D UED, there is a 
compactified flat extra dimension in which all the SM
fields universally propagate in addition to the 4D
Minkowski space-time.
Fields are expanded in the
KK modes (KK particles) in the 4D effective theory, and each zero mode
corresponds to the SM particle. 
The mass of $n$th mode is given by $m_n^2 = m_{SM}^2 + (n/R)^2$. $R$ denotes the radius of the extra
dimension, and $m_{SM}$ denotes a SM particle mass. The fifth dimensional
momentum is the mass in the 4D effective theory, and
this is much greater than $m_{SM}$, because $1/R \sim {\cal O}(\rm TeV) $. 
Therefore, we can neglect $m_{SM}$: $m_n \simeq n/R$, which means
the mass spectrum of each KK level is highly degenerate.

Since the simple  compactified
extra dimension $S^1$ gives vector-like fermions, 
an orbifold compactified extra dimension  $S^1/{\mathbb Z}_2$ with an
identification of $y \leftrightarrow -y$ is considered in order to obtain chiral fermions
in the zero mode. The orbifold compactification results in another
significant characteristic, the KK parity. KK number is conserved by
virtue of the fifth 
dimensional momentum conservation on $S^1$ compactification, but this is
broken down to the KK
parity  by the orbifold compactification.
The KK parity reflects  ``evenness" and ``oddness" of the KK number. 
All the SM particles have the even KK parity. 
The lightest particle
with the odd KK parity, called the lightest
Kaluza-Klein particle (LKP), is stable since it cannot decay into lighter SM
particles due to its oddness.  The stable LKP, typically the first KK
photon $\gamma^{(1)}$, can be a weakly interacting massive particle
(WIMP) and therefore a good Dark Matter candidate.

To discuss collider phenomenology, we have to determine the mass
spectrum. In this paper, we discuss the Minimal Universal Extra
Dimenison model (MUED). 
The MUED is a minimal extension of
 the 4D SM Lagrangian to the 5D UED. At the cutoff scale $\Lambda$
 it contains only SM fields and no other terms,
 especially no localized terms at two fixed points $y=0, \pi R$ led by orbifold compactification. 
The model parameters of MUED are only three: 5D radius $R$, cutoff scale
 of MUED $\Lambda$, and the SM Higgs
 mass $m_h$. 

\subsection{Mass spectrum}
Radiative corrections to masses of the KK modes at the one-loop level were studied in Refs.~\cite{Cheng:2002iz, Georgi:2000ks}.
This correction enlarges mass splitting for each KK level away from the
highly degenerate mass spectrum. The corrected masses are:
\begin{eqnarray}
	 m_{X^{(n)}} ^{2}  =\frac{n^{2}}{R^{2}} + m_{X^{(0)}}^{2}+\delta m_{X^{(n)}}^{2} &	
\end{eqnarray}
where $m_{X^{(0)}}^{}$ is a SM particle (zero mode) mass. 
The neutral gauge bosons of $U(1)_Y$ and $SU(2)_L$ are 
mixed up in the SM, but mass eigenstates of the KK neutral gauge bosons,
$\gamma^{(n)}$ and $Z^{(n)}$, are nearly $U(1)_Y$ and $SU(2)_L$ 
gauge eigenstates,
$B^{(n)}$ and $W^{3(n)}$, respectively because the diagonal components of mass matrix dominates as  
\begin{equation}
 \begin{pmatrix}
 B^{(n)} 
&
W^{3(n)}
 \end{pmatrix}
 \begin{pmatrix}
 \frac{n^{2}}{R^{2}} + {\delta}m_{B^{(n)}}^{2} + \frac{1}{4}g_1^{2}v^{2} & 
 \frac{1}{4}g_1g_2v^{2} 
\\
\frac{1}{4}g_1g_2v^{2}   & \frac{n^{2}}{R^{2}} + {\delta}m_{W^{3(n)}}^{2} + \frac{1}{4}g_2^2v^{2}
 \end{pmatrix}
 \begin{pmatrix}
 B^{(n)} 
\\
W^{3(n)}
 \end{pmatrix}
\end{equation}
where $g_1$ is the gauge coupling of $U(1)_Y$, $g_2$ is that of $SU(2)_L$, and
$v=246 \GEV$ is the vacuum expectation value of the Higgs field. The
radiative corrections to gauge boson masses are given by
\begin{eqnarray}
 \delta m_{B^{(n)}}^{2}&=& -\frac{39}{2} \frac{g_1^{2}\zeta(3)}  {16\pi^{2}}
  \frac{1}{R^{2}}+\frac{n^{2}}{R^{2}}  \biggl(-\frac{1}{6} \frac{g_1^{2}}{16\pi ^{2}}\biggr) 
 \log (\Lambda R)^{2} 
\notag\label{142025_13Jun11} 
\\
\delta m_{W^{(n)}}^{2}&=& -\frac{5}{2} \frac{g_2^2\zeta(3)}  {16\pi^{2}}
 \frac{1}{R^{2}}+\frac{n^{2}}{R^{2}}  \biggl(\frac{15}{2} \frac{g_2^2}{16\pi ^{2}}
   \biggr)  \log (\Lambda R)^{2}
\\
\delta m_{g^{(n)}}^{2} &=& -\frac{3}{2} \frac{g_{s}^{3}\zeta(3)}  {16\pi^{2}} \frac{1}{R^{2}}
+\frac{n^{2}}{R^{2}}  \biggl(\frac{23}{2} \frac{g_{s}^{2}}{16\pi ^{2}}
   \biggr)  \log (\Lambda R)^{2}\notag\label{141717_13Jun11}
  \end{eqnarray}
  where $\zeta(3)=1.20205 ...$ and $g_s$ is the gauge coupling of $SU(3)_C$.
The second terms in the corrections are dominant, so $m_{W^{(n)}}$ is
lifted, and $m_{B^{(n)}}$ is slightly lowered. 

The mixings of the KK quarks (KK
leptons) are also negligible, 
and they become  $U(1)_Y$ and $SU(2)_L$ 
gauge eigenstates. Neglecting $m_{SM}$, radiative corrections to the KK quarks and  KK leptons are given by
\begin{eqnarray}
 \delta m _{Q^{(n)}} &=& \frac{n}{R} \biggl(3 \frac{g_{s}^{2}}{16\pi ^{2}}
  + \frac{27}{16} \frac{g_2^2}{16\pi ^{2}}  + \frac{1}{16}
  \frac{g_1^{2}}{16\pi ^{2}}      \biggr)
  \log (\Lambda R)^{2}
 \notag\label{141824_13Jun11} 
 \\
 \delta m _{u^{(n)}} &=& \frac{n}{R} \biggl(3 \frac{g_{s}^{2}}{16\pi ^{2}}
  +  \frac{g_1^{2}}{16\pi ^{2}}  
  \biggr)
  \log (\Lambda R)^{2}
  \notag\\
\delta m _{d^{(n)}} &=& \frac{n}{R} \biggl(3 \frac{g_{3}^{2}}{16\pi ^{2}}
  + \frac{1}{4} \frac{g_1^{2}}{16\pi ^{2}}  
  \biggr)
  \log (\Lambda R)^{2}
   \\
  \delta m _{L^{(n)}} &=& \frac{n}{R} \biggl(\frac{27}{16} \frac{g_{2}^{2}}{16\pi ^{2}}
  + \frac{9}{16} \frac{g_1^{2}}{16\pi ^{2}}  
  \biggr)
  \log (\Lambda R)^{2}
  \notag \\
 \delta m _{e^{(n)}} &=& \frac{n}{R} \biggl(\frac{9}{4} \frac{g_1^{2}}{16\pi ^{2}}\biggr)
  \log (\Lambda R)^{2}
\notag
\end{eqnarray}
where $Q^{(n)}$ and $L^{(n)}$ denote the $SU(2)_L$ doublet, and $u^{(n)}$, $d^{(n)}$, and $e^{(n)}$ denote
the $SU(2)_L$ singlet.  For the KK top quark we should consider the correction
from its Yukawa coupling, but the production cross section is small. We
do not consider the processes of KK top quark.  

Most KK particles receive positive mass
corrections.  The heaviest particle in each level is $g^{(n)}$   for the
largest correction, and the lightest particle in each level is
typically $\gamma^{(n)}$. Then the LKP is $\gamma^{(1)}$ with the mass $m_{\gamma^{(1)}} \cong 1/R$.  
In the analysis, we fix the Higgs  mass at $m_h = 120\GEV$ for simplicity.

The corrections  are basically proportional to $ \log \Lambda R$, so the degeneracy is
crucial for the smaller $\Lambda R$.
The cutoff scale of the UED was discussed in \cite{Bhattacharyya:2006ym}, and the
appropriate cutoff scale should be  several dozen $1/R$ for a given
$R$. As the energy scale grows, more KK particles appear, and the
 logarithmic running of the gauge coupling changes into power law running above
 the MUED scale $1/R$. The $U(1)_Y$ gauge 
 coupling blows up (Landau pole) at the energy scale 
 $\sim 40/R$, so we should set the cutoff scale below the Landau pole. 
 The very small $\Lambda R$ is also not appropriate
 because we should consider the higher dimensional operators, and the MUED
 framework is not a good effective theory any more.   
In our analysis, we considered $10 \leq \Lambda R \leq 40$. 

\subsection{Production and decay at the LHC}

At the LHC, the first KK particles of the odd KK parity are pair-produced,
and  they eventually decay into the LKP.
The dominant production processes are KK gluon+KK quark ($g^{(1)}+Q^{(1)} / q^{(1)}$) 
and KK quark+KK quark $(Q^{(1)}/q^{(1)}+Q^{(1)}/q^{(1)})$. The cross
sections of the colored particles are shown in Ref.~\cite{Macesanu:2002db}. For
our benchmark point, $\sigma(g^{(1)}+Q^{(1)} / q^{(1)})=12.2$ pb and
$\sigma(Q^{(1)} / q^{(1)}+Q^{(1)} / q^{(1)})=7.4$ pb at $\sqrt{s} =14\TEV$.

The $g^{(1)}$ decays into  $Q^{(1)}Q$ and $q^{(1)}q$ with branching
ratios, BR($g \to Q^{(1)}Q$) $\sim$ 40\% and BR($g \to q^{(1)}q$) $\sim$
60\%, respectively. 
The ratio of inclusive KK quark productions is roughly
$Q^{(1)}Q^{(1)}:q^{(1)}q^{(1)}:Q^{(1)}q^{(1)}=1:1:2$.  
Because $q^{(1)}$ only has the $U(1)_Y$ gauge interaction, 
it directly decays into $\gamma^{(1)}q$. The hard jets mainly come from
this decay. The branching ratios of $Q^{(1)}$ are typically ${\rm
BR}(Q^{(1)}\to Q W^{\pm(1)})\sim 65\%$, ${\rm BR}(Q^{(1)}\to Q
Z^{(1)})\sim 32\%$, and ${\rm BR}(Q^{(1)}\to Q \gamma^{(1)}) \sim 3\%.$
Once $W^{(1)}$ and $Z^{(1)}$ appear from $Q^{(1)}$, they cannot decay
hadronically for kinematical reasons. They democratically decay
into all lepton flavors:
${W^{\pm(1)} \to \gamma^{(1)}l\nu}$ 
and $Z^{(1)}\to \gamma^{(1)}\nu \bar{\nu} \mbox{ or }\gamma^{(1)}l^+ l^-$ through $l^{(1)} \mbox{ or } \nu^{(1)}$.

  This collider signature has been studied in clean channels of
  multilepton such as $4l+ E_{T}^{miss}$
  \cite{Cheng:2002ab,CERN-CMS-CR-2006-062,Bhattacharyya:2009br},
  dilepton, and trilepton  \cite{Bhattacharyya:2009br,Bhatt:2010vm}. 
  The leptons arise only from the KK gauge boson $W^{(1)}$ or $Z^{(1)}$ production.
 The $4l + \Etmiss$ channel has been the most 
promising one because the background is extremely small, but the fraction of
  the MUED events going to this channel is about 1\%:
  from the $Q^{(1)}Q^{(1)}$ production, each $Q^{(1)}$ should decay as
  $Q^{(1)}\to Q Z^{(1)} \to Q l^+l^-\gamma^{(1)}$ with the branching
  ratio of 16\%. 

Multijet channels without requiring multileptons are
  statistically advantageous, so we use the multijet + lepton
  channel. This is accessible to about 65\% of the MUED total
  production. The requirement of one lepton is only to avoid the QCD background.

\section{Definition of $M_{T2}$}
In this section, we briefly review the definition of $\mttwo$. $\mttwo$, an extension of
transverse mass $M_T$, was originally proposed as a mass measurement
variable in the situation with two invisible particles with $\ptvec{\rm inv(1)}$ and $\ptvec{\rm inv(2)}$ \cite{Lester:1999tx,Barr:2003rg}. In each event, we only know the total missing
transverse momentum, $\ptmiss$, but each transverse momentum of the
invisible particle cannot be measured.  
The definition of $\mttwo$ is :

\begin{eqnarray}
 \mttwo &\equiv&  \min_{\ptvec{\rm inv(1)} +\ptvec{\rm inv(2)} = \ptmiss}   \left[    \max \left\{ M_T^{(1)}
, \mbox{ } M_T^{(2)}\right\}  \right] \\\notag
\end{eqnarray}
where $M_T$ is defined by 
\begin{eqnarray}
M_T^{(i)}&=&M_T(m_{{\rm vis}(i)},m_{{\rm inv}(i)}, \ptvec{\rm vis(1)}, \ptvec{\rm inv(1)} )\notag
\\\\
 &\equiv&\sqrt{ m_{{\rm vis}(i)}^{2}+m_{{\rm inv}(i)}^{2}+
2 \left( E_{T}^{{\rm vis}(i)}E_{T}^{{\rm inv}(i)} -\ptvec{{\rm vis}{(i)}} \cdot\ptvec{{\rm inv}{(i)}}  
						    \right) },\notag
\end{eqnarray}
and where $m_{{\rm vis}(i)}$ and $m_{{\rm inv}(i)}$ are visible (observed) particle mass and invisible particle mass in system $i$, respectively.
The transverse energy $E_T^{}$ is given by
\begin{eqnarray}
E_T &\equiv &\sqrt{m^2 + |\ptvec{}|^2}.
\end{eqnarray}
In calculating $\mttwo$, we first construct transverse mass
$M_T^{(i=1,2)}$ and  take the maximum of them for one partition of
$\ptvec{inv{(1)}}$ and $\ptvec{inv{(2)}}$ satisfying $\ptvec{inv{(1)}}+\ptvec{inv{(2)}}=\ptmiss$. 
Then, all the possible partitions are considered, and the minimum value
among them is taken.  

Let us consider the simple case where the same parent
particles are produced and each of them directly decay to a visible
particle and an invisible particle. In the MUED, such a example is $q^{(1)}q^{(1)}\to \gamma^{(1)}\gamma^{(1)}qq$ process: first KK (right-handed) quarks, $q^{(1)}$, are produced and each one directly decays into a quark and the LKP, KK photon, $\gamma^{(1)}$. 
If the invisible
particle mass $m_{\rm inv}$ is known, $M_T$ is bounded by the parent particle mass, $M_T \leq m_{\rm parent}$ in the correct partition. 
Then, as seen from the definition,
$\mttwo$ is also bounded by the parent particle mass, $\mttwo \leq \mttm =m_{\rm parent}$. 
We can also interprete $\mttm$ as the invariant mass.     
Transverse mass $M_T$, essential component of $\mttwo$, actually acted as {\it invariant} mass to successfully measure $W$ boson mass in leptonic decay in UA1 \cite{Arnison:1983zy}. This method worked because invisible particle of the decay is known to be massless neutrino. 

\subsection*{Boost dependence}
In practice,  $m_{\rm inv}$ is not known.
In calculating $\mttwo$, we need to set a test mass for the invisible particle.
Many attempts were made
to simultaneously determine the masses of the parent and
invisible particles \cite{Cho:2007qv, Burns:2008va, Cohen:2010wv}. One of them \cite{Cohen:2010wv} utilizes the effect of Up-stream
Radiations (USR). 
USR is defined as visible particles which are emitted before parent particles of our interest are produced. 
The transverse momentum of USR, $\Ptvec{}$, is given by
\begin{eqnarray}
\ptvec{\rm vis(1)} + \ptvec{\rm vis(2)} + \ptmiss = - \Ptvec{} \hspace{5mm}\mbox{(USR)}.
\end{eqnarray}
\begin{figure}
\begin{center}
  \includegraphics[width=70mm]{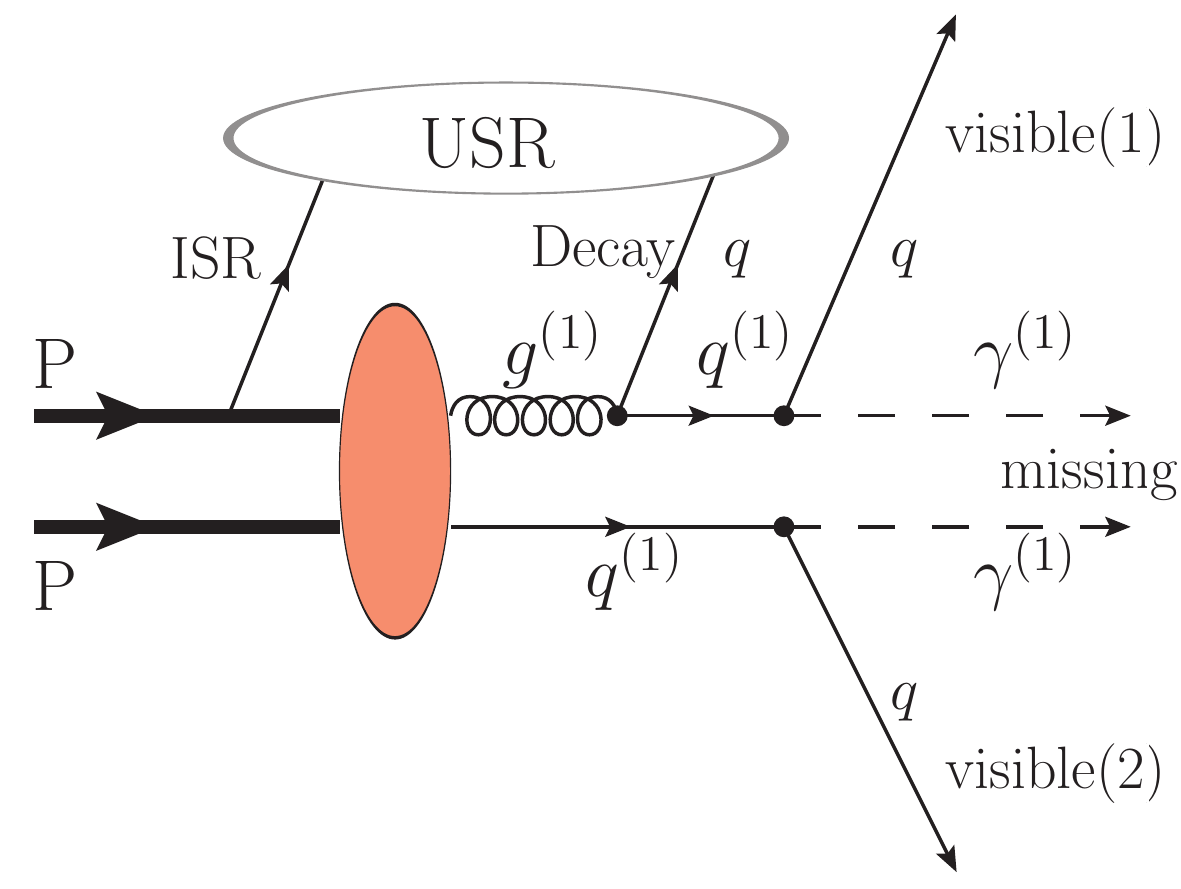}
  \caption{Schematic picture of typical MUED process, $g^{(1)}q^{(1)}$
 production. When two jets from $q^{(1)}\to \gamma^{(1)}q$ and used as
 visible particles to
 construct $\mttwo$, a jet from
 the decay of $g^{(1)}\to q^{(1)}q$ and ISR become USR.}
 \end{center}
\label{USR}
\end{figure}
$\Ptvec{}$ is a measure of the recoil of the parent particles. The
source of USR is mainly initial state radiations (ISR). The decay products can be USR if the decay is before the production of the parent particles of our interest.
Fig.~\ref{USR} illustrates USR in a case of $g^{(1)}q^{(1)}$ production where $g^{(1)}$ is first KK gluon. When we
are interested in $q^{(1)}$, a quark emitted from the decay of $g^{(1)}$
and ISR are considered as USR.

Of course, $\mttm$ has different behaviors depending on whether the test
mass is correct. When we set a correct test mass, $\mttm$
corresponds to the parent particle mass independent of USR. But, when we set a wrong
test mass, $\mttm$ varies with USR, {\it i.e.} it is not Lorentz boost invariant quantity. This is because $M_T$ 
 varies with USR and  is no longer 
bounded by the parent particle mass. This property can be used in the search for compressed models.

\section{Use of $M_{T2}$ as event selection}\label{mt2selection}
\subsection{Properties}
\subsection*{For SM background}
The features of $\mttwo$ for the purpose of event selection were studied in
\cite{Barr:2009wu, Lester:2007fq}. We use the $\mttwo$ cut as an event selection
setting the {\it test mass} to {\it zero}. The set test mass is a correct value for SM because
the only invisible particles of the SM are neutrinos. In this case, $\mttwo$ of
the most background events, especially $\ttbar$ events, is lower than
the top quark mass $m_t$,
	\begin{eqnarray}
	\max_{\rm SM\ events} [\mttwo] \simeq m_{t}	.
	\end{eqnarray}
 This is because we can measure the parent particle mass
with the correct test mass, and the top quark is the heaviest
parent particle in the SM.   If a significant excess of  $\mttwo$ above $m_t$ is observed, it
is likely to be new physics signal.  
An advantage of $\mttwo$ cut is that the value  has physical meaning because this is originally a mass measurement variable unlike most of cut variables.

There are other good features.  It is found that events without missing momentum
or with fake missing momentum which is parallel to a mismeasured jet have very
small values of $\mttwo$ \cite{Barr:2009wu}. The proof is as follows. Suppose one of two energetic jets is mismeasured leading to missing momentum of $\ptmiss ||\ptvec{\rm vis(1)} $, 
and all the jet masses are set to be zero for simplicity.
We can take a partition of $\ptvec{\rm inv(1)}=\ptmiss$ and $\ptvec{\rm inv(2)}=0$, and then $M_T^{(1)}=0$ and $\mttwo=0$ by definition. This result is extended for a case with two mismeasured visible objects, and this proof and other properties are discussed in Ref.~\cite{Barr:2009wu}.

\subsection*{For new physics signal}

$\mttwo$ of new physics
is not bounded by the parent particle mass because the test mass is
wrong for massive invisible particles. The upper bound of
$\mttwo$ is a mass combination of the parent
 particle and
the invisible particle in the absence of USR,
\begin{equation}
 \mttm = \frac{m_{\rm parent}^2 -m_{\rm inv}^2}{m_{\rm parent}} \equiv \mu_0.
\label{muzero} 
\end{equation}
In this case, the signal is extracted from the background for models
with a large mass splitting but not for models with a degenerate mass spectrum particularly for $\mu_0 \leq m_t$ because the signal is buried in the background.

However, considering the recoil momentum of parent particles by USR, 
$\mttwo$ is still a useful variable for the event selection in searching for compressed models. $\mttm$ varies with USR and can exceed $\mu_0$
due to the wrong test mass.  
For example, when parent particles of same
mass are produced and directly decay to invisible particles emitting
visible particles ($q^{(1)}q^{(1)} \to qq \gamma^{(1)}\gamma^{(1)}$), $\mttm$ 
\cite{Burns:2008va, Cohen:2010wv} is given by 
\begin{eqnarray}
 	\mttm &=& \sqrt{\mu_0^2 F^2(P_T)+P_T \mu_0 F(P_T)}\ge \mu_0 ,\label{mttmmu}
\end{eqnarray}
where
\begin{eqnarray}
	F(P_T)&\equiv&
	 \left(\sqrt{1+\left(\frac{P_T}{2m_{\rm parent}}\right)^2}
  	-\frac{P_T}{2m_{\rm parent}}\right)
\\
	&\to& \left(\frac{m_{\rm parent}}{P_T}\right)\hspace{1cm}\mbox{for
 	}P_T\gg m_{\rm parent} .\notag
	\label{endpoint2}
\end{eqnarray}
Here  $P_T$ is the magnitude of the momentum of USR.
There is a rich source of USR because processes of heavy
particles tend to come along with hard QCD radiations including ISR.
Hard ISR gives large recoil of produced particles, and $\mttm$ can have
a large value depending on USR. Note that the background events do not have
$\mttwo$ dependence on USR because the test mass is correct, so most events are kept lower than $m_t$.

\subsection{Comparison in realistic situation}
When analyzing events, we cannot tell the origins of visible particles:
whether the particles come from decays of heavier particles or are
QCD radiations. Practically, leading two jets in $p_T$ are
used as visible particles to construct $\mttwo$. If they
correspond to two ``correct'' particles, namely if each particle is a decay
 product of each pair-produced particle, $\mttwo$ behaves as discussed
above. However, the leading particles can be decay products of one parent
particle, and also hard ISR can be one or both of the leading particles. These
cases are called ``combinatorics.'' 

In many events, $\mttwo$ of the leading particles corresponds to
$\mttwo$ of the correct particles. For instance, the rate of correspondence is about half for  $q^{(1)}q^{(1)}$ or $\ttbar$ as
shown later. 
Combinatorics smears the $\mttwo$ distribution. The
smearing effect is significant for high $\mttwo$, and it is different in each
process.  

\begin{figure}[h!]
\begin{center}
\includegraphics[width=0.49\linewidth]{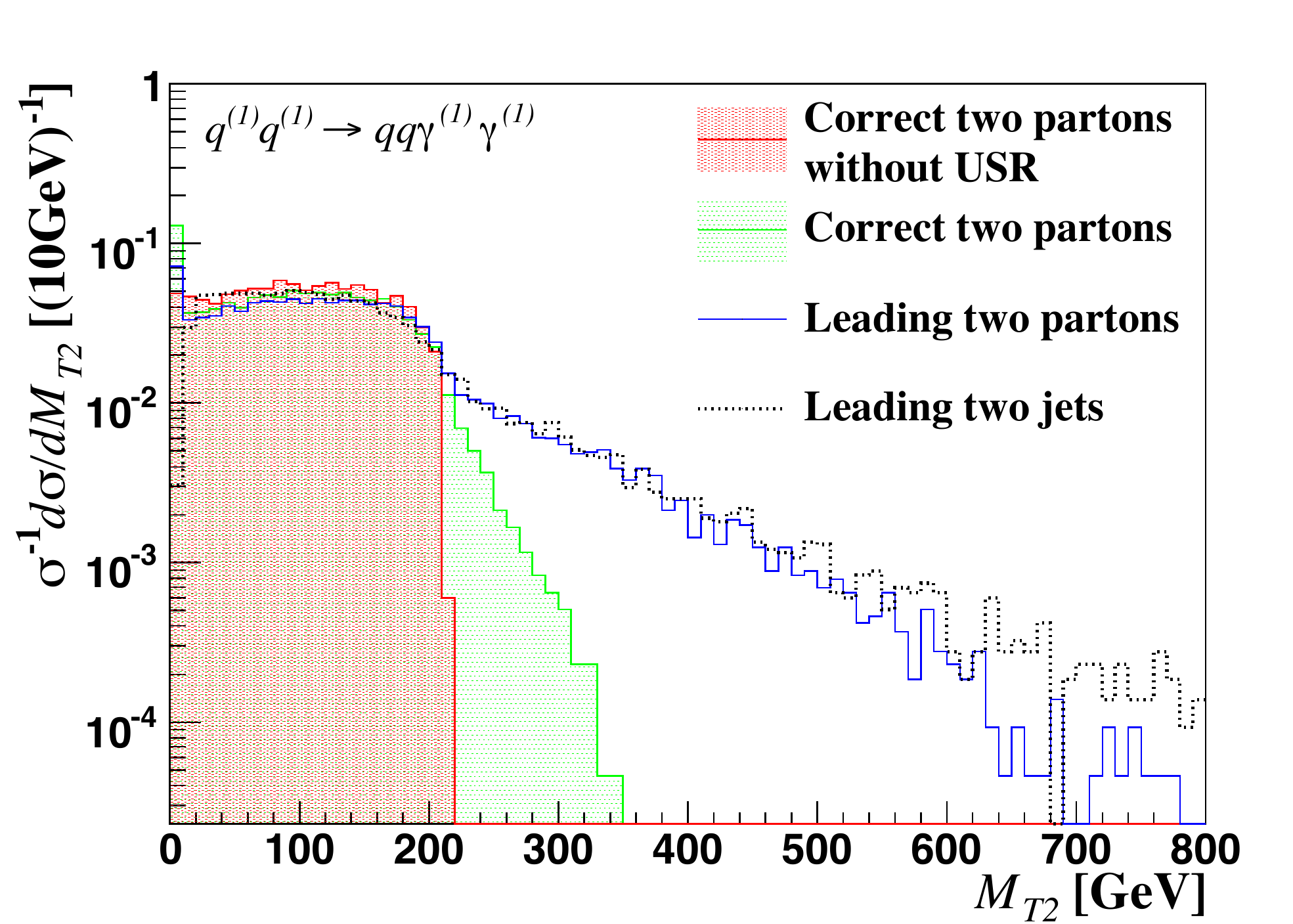}
\includegraphics[width=0.49\linewidth]{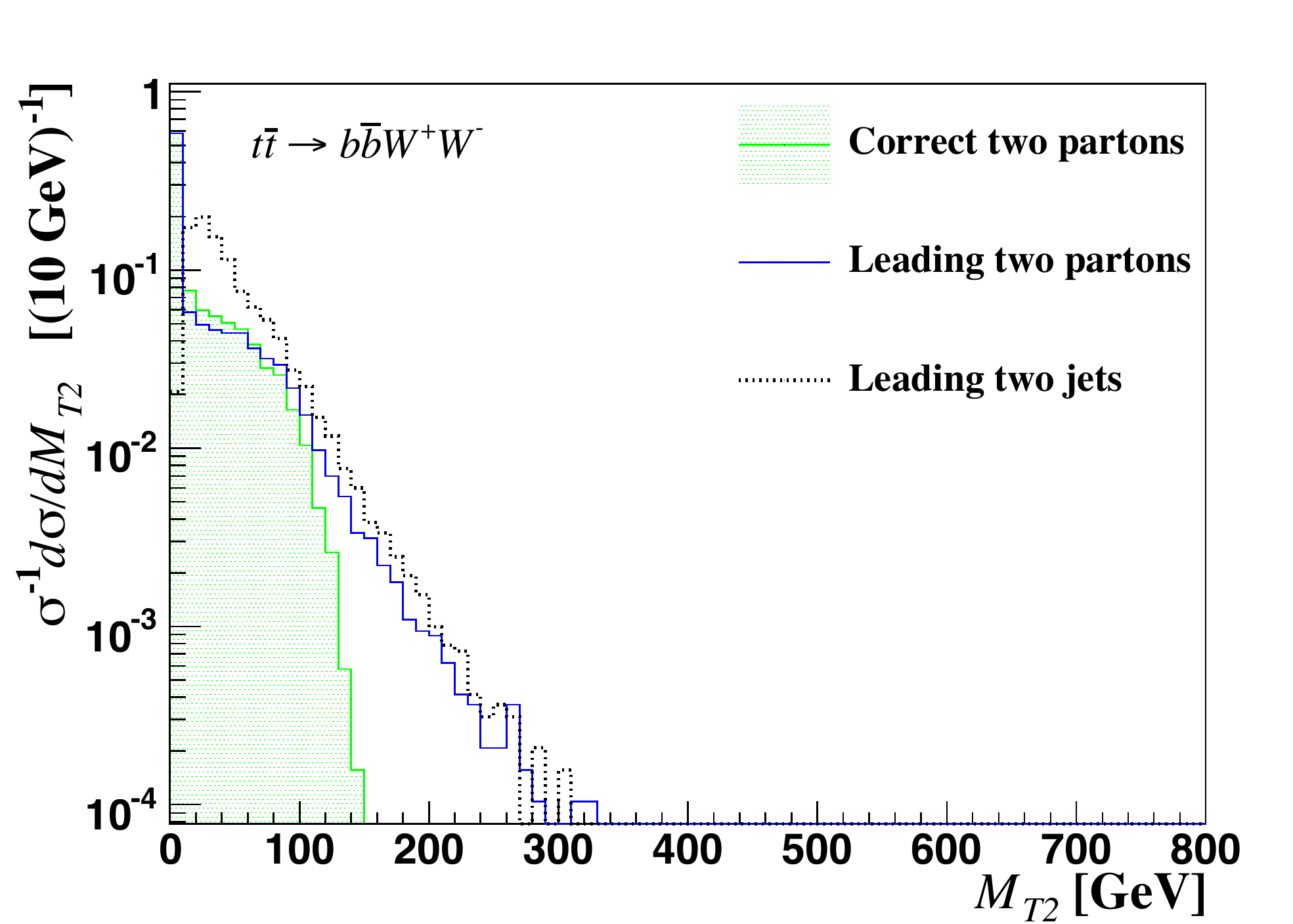}
\label{mt2evsl }
\caption{Distributions of $\mttwo$ for $q^{(1)}q^{(1)}\to qq
 \gamma^{(1)} \gamma^{(1)}$ in the left and $\ttbar \to \bbbar W^+W^-$
 in the right generated by \MG \cite{Alwall:2007st} where $m_{q^{(1)}} = 912.5$
GeV, $m_{\gamma^{(1)}} = 800.1$GeV, and $\mu_0 = 211.0$GeV. $\mttwo$
 is constructed with the correct two
 partons (green shaded area), the leading two partons (blue solid line), and
 leading two jets (dotted line).
 Also, events  were generated without additional jets, that is, without
 USR in the parton level, and $\mttwo$ was calculated with the correct two partons (red shaded area). The distributions are normalized to 1.}
\label{2figmt2}
\end{center}
\end{figure}

In order to see the effects of USR and combinatorics, we generated 
the $q^{(1)}q^{(1)}$ production of the MUED benchmark point  and the $\ttbar$
production adding up to one jet in the parton level \footnote{The events were
generated by \MG \cite{Alwall:2007st} at $\sqrt{s}=14$ TeV. The fragmentation and
hadronization were simulated by \PY,
and the detector effects were simulated by {\tt PGS 4} \cite{PGS}. {The simulation setup is
the same as the simulation described in Sec.~\ref{simulation} except that
here only one jet was added as the Matrix Element correction.} The MLM
matching \cite{Alwall:2007fs} was prescribed.}, and
constructed $\mttwo$ with the correct partons and with the leading partons/jets. 
These $\mttwo$ distributions are shown in Fig.~\ref{2figmt2}.

The green 
(red) shaded area shows $\mttwo$ with (without) an additional jet in
the parton level using correct two partons: quarks from
the direct decay $q^{(1)} \to q \gamma^{(1)}$ for $q^{(1)} q^{(1)}$  and $b$
quarks from $t \to bW $ for $\ttbar$. LKPs and neutrinos produce $\ptmiss$. 
For the signal, $\mttwo$ without USR (red shaded area) is bounded by
the mass combination $\mu_0 =210$ GeV given by (\ref{muzero}).
Including USR (green shaded area) $\mttwo$ varies with $P_T$ and
exceeds $\mu_0$ as (\ref{mttmmu}), and theoretically it reachs about 440
GeV with extremely large $P_T$. 
For the $\ttbar$ background, $\mttwo$ of correct two partons with USR (green
shaded rarea) does not exceed $m_t$ as expected.

\subsection*{Combinatrics}
\begin{table}[htd]
\begin{center}
\begin{tabular}{|c|c|c|}
\hline
 Parton level & $q^{(1)}q^{(1)}\to qq
 \gamma^{(1)} \gamma^{(1)}$ + 0, 1 jet & $\ttbar \to \bbbar W^+W^-$ + 0,
 1 jet \\
\hline
 $\mttwo^{\rm leading} =\mttwo^{\rm correct}$& 61.6\% &49.1\%  \\
\hline
$\mttwo^{\rm leading} >\mttwo^{\rm correct}$ & 30.3\% (78.9\%)&22.4\% (44\%)  \\
\hline
$\mttwo^{\rm leading} <\mttwo^{\rm correct}$ & 8.10\% (21.1\%) &28.5\%  (56\%)\\
\hline
\end{tabular}
\end{center}
\vspace{-15pt}
\caption{The evaluation of combinatrics for $q^{(1)}q^{(1)}$ of the
 benchmark point and $\ttbar$. $\mttwo$ is constructed in the parton level. We compare $\mttwo$ of leading partons and correct partons in each event. }
\label{combi}
\end{table}%

When the leading partons are used for $\mttwo$ (blue solid line),
combinatorics smears the $\mttwo$
distribution. $\mttwo$ of the leading partons spreads over the endpoint of $\mttwo$ of correct partons
to roughly double the value at the endpoint as in Fig.~\ref{2figmt2}. Table \ref{combi} shows the evaluation of
combinatorics. The leading partons correspond to the correct partons for
60\%  of the time for
$q^{(1)}q^{(1)}$ and for 50\% for $\ttbar$. The smearing effect due to
combinatorics is different in each process: $\mttwo^{\rm leading} >\mttwo^{\rm correct}$ for three quarters of
the combinatorics events of $q^{(1)}q^{(1)}$, while $\mttwo^{\rm leading} <\mttwo^{\rm correct}$ for more
than half of the combinatorics events of $\ttbar$. Therefore, combinatorics
assists to enhance the signal to background ratio for high $\mttwo$. 

Also, the detector effects are simulated after the fragmentation and
hadronization, and $\mttwo$ is constructed with the leading two jets
(dotted line). The distribution is similar with the $\mttwo$
distribution with the leading two partons.  

The dependence on USR makes the signal excess in the high $\mttwo$ region
even for compressed models, and the smearing effect of combinatorics enhances the
excess in this particular example. 
It can been seen that events with high $\mttwo$, say above 200 GeV, are
dominated by the $q^{(1)}q^{(1)}$ signal over the $\ttbar$ background. Since for the other
SM background processes the parent particle is lighter than the top quark,
those background events are expeced to have $\mttwo$ lower than
$m_t$. 
Hence, $\mttwo$ is an effective event selection to search for the nearly
degenerate model.

\section{Application for the MUED}
We explicitly perform $\mttwo$ event selection for MUED discovery potential. The MUED has two essential parameters relevant to collider phenomenology: $R$ and the cutoff of the theory, $\Lambda$. It is known the radiative corrections spreads spectrum and the corrections are proportional to $\ln \Lambda R$. The degeneracy is crucial for the smaller $\Lambda R$ and $\Lambda R=10\sim40$ is adequate. 

For the difficulty of compressed spectrum, promising previous study is $4l+\slashed{E}_T$ channel where the background level is quite low  \cite{Cheng:2002ab,CERN-CMS-CR-2006-062,Bhattacharyya:2009br}. Relatively many leptons are emitted in the MUED processes. 
In order to see the use of $\mttwo$ cut, we also study the discovery potential in $4l+\slashed{E}_T$ channel. 

\subsection{Simulation}\label{simulation}

Monte Carlo (MC) samples of MUED signal were generated both by
a private implementation in \MG (MG/ME) \cite{Alwall:2007st} and an implementation
\cite{ElKacimi:2009zj} in \PY ~\cite{Sjostrand:2006za}. CTEQ5.1L
was used as the leading-order (LO)  parton distribution function
(PDF). In the case of MG/ME, the Matrix Element was calculated by {\tt HELAS}
\cite{Murayama:1992gi}, and the fragmentation and hadronization were simulated with \PY. 
 The effects of jet reconstruction and detector smearing were simulated
 through {\tt PGS 4} \cite{PGS}. 

 We consider $1/R$ from 400 GeV to 1.6 TeV in steps of 100 GeV with
 $\Lambda R$ = 10, 20, 30, and 40.
 The remaining parameter, $m_h$, is set to 120 GeV. 
  The MUED spectrum is simplified by neglecting $m_{SM}$ for the first KK level.
  The processes we consider are pair productions of the colored first KK particles, $g^{(1)}$, $q^{(1)}$, and $Q^{(1)}$. The signal events 
 corresponding to  luminosities 
 of more than 10 fb$^{-1}$ at $\sqrt{s}$ =14 TeV were generated
 by \PY. 
 
   Since we use the $\mttwo$ dependence on USR, the ISR has an important
  role. In order to reliably evaluate the hard ISR, we considered the Matrix Element correction in MG/ME adding up to one jet to the pair productions. The MLM
  matching \cite{Alwall:2007fs} was applied to remove the overlap between jets from the Matrix Element and ones from the Parton Shower. 
This prescription was demonstrated for the benchmark point of $\Lambda
  R=20$ and  $1/R =800\GEV$. The spectrum of this point is listed in table \ref{benchmark}. 
However it is very time consuming to generate all of the signal MC
  samples with the Matrix Element correction, so we used \PY{ }rather than MG/ME to generate
  them for the discovery study. 
We will show in Sec.~\ref{EVSelection}. that MC samples generated by ME/ME with
  the Matrix Element correction have larger excess over background than ones generated by
  \PY{ }for the benchmark point. Hence, the event generation by \PY{ }is conservative. 
\begin{table}[htd]
\begin{center}
\begin{tabular}{|c|c|c|c|c|c|c|c|c|c|}
\hline
$m_{\gamma^{(1)}}$ & $m_{ W^{(1)}}$ & $m_{ Z^{(1)}}$& $m_{ e^{(1)}}$ &$m_{ L^{(1)}}$& $m_{ d^{(1)}}$ &$m_{ u^{(1)}}$ &$m_{ Q^{(1)}}$ & $m_{ g^{(1)}}$& \\
800.1 & 847.3 &847.4 & 808.2 & 822.3 & 909.8 & 912.5 & 929.3 & 986.4 & GeV\\
\hline
\end{tabular}
\end{center}
\vspace{-15pt}
\caption{Mass spectrum of first KK level for a benchmark point $(1/R, \Lambda R)=(800,20)$ }
\label{benchmark}
\end{table}%

 MC samples of the SM backgrounds: $\ttbar(+1j,2j)$, $W/Z+{\rm jets} (1j,2j)$, Diboson (+1j,2j),\footnote{Diboson denotes $WW, WZ \mbox{ and } ZZ$.} $Z/W+b\bar{b}$, $\ttbar+Z/\gamma^*(\to ll,\nu\nu)$ and $\ttbar+W(\to l\nu)$, were produced with 
MG/ME  using the PDF set CTEQ6.1L, and fragmentation and
hadronization were simulated with \PY{ }in the same way of the signal. 
For $\ttbar$, $W/Z+\rm jets$, and Diboson, up to two partons were added in
the Matrix Element and the MLM matching was prescribed.
The MC samples were detector-simulated through {\tt PGS 4}. 
The dominant background processes, $\ttbar$ and $W/Z+\rm jets$, were
normalized to the next-leading-order (NLO) cross section consistent
with the inclusive dijet analysis of the ATLAS MC study \cite{Aad:2009wy}. 

For the sake of comparison with the $4l + \slashed{E}_T$ analysis, we generated some multilepton background processes, such as four leptons through off-shell $Z^{\ast}$ or $\gamma^{\ast}$. 
The luminosities of generated SM background are 
 more than 10 \fb at $\sqrt{s}$= 14 TeV, respectively. 

\subsection{Event selection} \label{EVSelection}

The object selection is that an electron and a muon are required to have
$p_T>10 \GEV$ and $|\eta|< 2.5$ and a jet is required to have $p_T>20\GEV$ and $|\eta| <2.5$. 
In order to avoid recognizing a shower from an electron as a jet, a jet within
$\Delta R <0.2$ ($\Delta R = \sqrt{\Delta \eta^2 + \Delta \phi^2}$) from any electron is removed.
Charged leptons from hadronic activity also should be removed. If an electron and a jet are found within $0.2 <\Delta R < 0.4$, the jet is kept
and the electron is rejected 
Similarly, if a muon and a jet are found within $\Delta R <0.4$, the muon is rejected.

We impose event selections as follows,
\begin{itemize}
  \item CUT1: $p_T^{\rm jet} > \{100, 20 \mbox{ GeV}\}$
  \item CUT2: $\slashed{E}_T(=\slashed{p}_T)>100 \mbox{ GeV}$
  \item CUT3: At least one lepton with $p_T^{\rm lep} > 20 \mbox{ GeV}$ 
  \item CUT4: If the number of lepton is one, $M_T^{\rm lep,miss}>100 \mbox{ GeV}$\ \footnote{$M_T^{\rm lep,miss} \equiv \sqrt{2 ( p_T^{\rm lep}\slashed{p}_T- \ptvec{\rm lep}\cdot \ptmiss )}$} 
  \item CUT5: $\mttwo >200 \mbox{ GeV}$. 
\end{itemize}

In order to demonstrate the effectiveness of $\mttwo$ for the MUED, we only
use the basic cuts 1-4 except one on $\mttwo$. The cuts 1-4 are comparable with ones imposed in the ATLAS and CMS new physics searches in  one lepton + jets + $\slashed{E}_T$ with low luminosity at $\sqrt{s}=7 \TEV$ \cite{Bernet:2011rd,Aad:2011hh}.
We do not use the  $M_{\rm eff}$\footnote{$M_{\rm eff} \equiv \sum^4_{\rm jet}p_T+\sum _{\rm lepton}p_T+ \slashed{E}_T.$}
cut and the $\slashed{E}_T/M_{\rm eff}$
 cut which are used to extract the
signal especially by the ATLAS collaboration in the search for SUSY \cite{Aad:2009wy}.  
It is common that the $\Delta \phi_{\rm jet,miss}$ cut is applied to reduce
events with fake missing due to the mismeasurement of jets, but this is 
not necessary because the later $\mttwo$ cut has a similar role \cite{Barr:2009wu}. 

\begin{table}[htdp]
\begin{center}
\begin{tabular}{|c|c|r r r r r| }
\hline
\multicolumn{2}{|c|}{Process} & CUT1 & CUT2& CUT3 & CUT4 & CUT5 {\footnotesize (Optimal)} \\
\hline
 $g^{(1)}+g^{(1)}$ 
  	& MG/ME &1,028 & 832 & 119 & 62& 25\\
\cline{2-2}
	& PYTHIA & 937 &757 &108 &63 &22 \\
\hline
 $g^{(1)}+q^{(1)}/Q^{(1)}$ 
 & MG/ME &9,196 & 7,218& 1,234& 675&  241\\
\cline{2-2}
& PYTHIA & 8,569 & 6,694 &1,344 & 731 &223\\
\hline
 $q^{(1)}/Q^{(1)}$ 
 & MG/ME&5,315 &4,035  &863 &508 &148\\ 
\cline{2-2}
$+q^{(1)}/Q^{(1)}$
& PYTHIA & 4,497 & 3,276  &690 	&436 &84\\
\hline
 $q^{(1)}/Q^{(1)}$ 
 & MG/ME  &1,444 &1,075  &206  &115  &27 \\
\cline{2-2}
$+\bar{q}^{(1)}/\bar{Q}^{(1)}$
& PYTHIA & 1,301 &955 & 163	&112 &20\\
\hline
Total MUED
& MG/ME & 16,983 & 13,160 & 2,422 & 1,360& 441\\
\cline{2-7}
& PYTHIA & 15,304 & 11,682 & 2,305 &1,342 &349 \\
\hline \hline
\multicolumn{2}{|c|}{$\ttbar$ } & 426,074 & 57,533 &23,239	&5,620 & 243\\
\multicolumn{2}{|c|}{$W$ } 	&400,527 & 97,907 &35,386	&1,031 & 85\\
\multicolumn{2}{|c|}{$Z$ } 		&142,368 & 53,801 &916		&107	&12\\
\multicolumn{2}{|c|}{$W/Z+\ttbar /\bbbar$ }& 1,121 & 304 &103 & 49	&10\\
\multicolumn{2}{|c|}{Diboson } &29,141 &4,482 & 1,335	&252	&40\\
\hline
\multicolumn{2}{|c|}{Total Standard Model} &999,231 &214,027 &60,979 &7,059 &390\\
\hline \hline
Total MUED& MG/ME & 0.05 &   0.17 & 0.06& 0.78	&4.10\\
\cline{2-7}
 $Z_B$& PYTHIA & 0.05 & 0.14 & 0.05 &0.77 	& 3.37 (7.57) \\
\hline 
\end{tabular}
\end{center}
\vspace{-15pt}
\caption{Cut flow for 1 \fb at $\sqrt{s}$ = 14TeV. The MUED  benchmark point
 is $\{1/R, \Lambda R \}=\{ 800\mbox{ GeV},20 \}$. The MUED signal generated by
 \MG(MG/ME) with the MLM matching is normalized to one generated by \PY. 
$\mttwo >$ 350 GeV is an optimal cut that maximizes the
 significance $Z_B = 7.57$.  }
\label{800mass}
\end{table}

The cut flow in table \ref{800mass} shows that the $\mttwo$ cut (CUT5) significantly
reduces the SM background to a level comparable to the MUED. Since the Matrix Element correction increases event rates in the high $\mttwo$ region, there
remain more signal events generated by MG/ME after CUT5 than ones
generated by \PY. Therefore, the signal rate in the fast event generation by \PY{ }is 
a little underestimated and hence is conservative.

\begin{figure}[ht!]
\begin{center}
\includegraphics[width=0.49\linewidth]{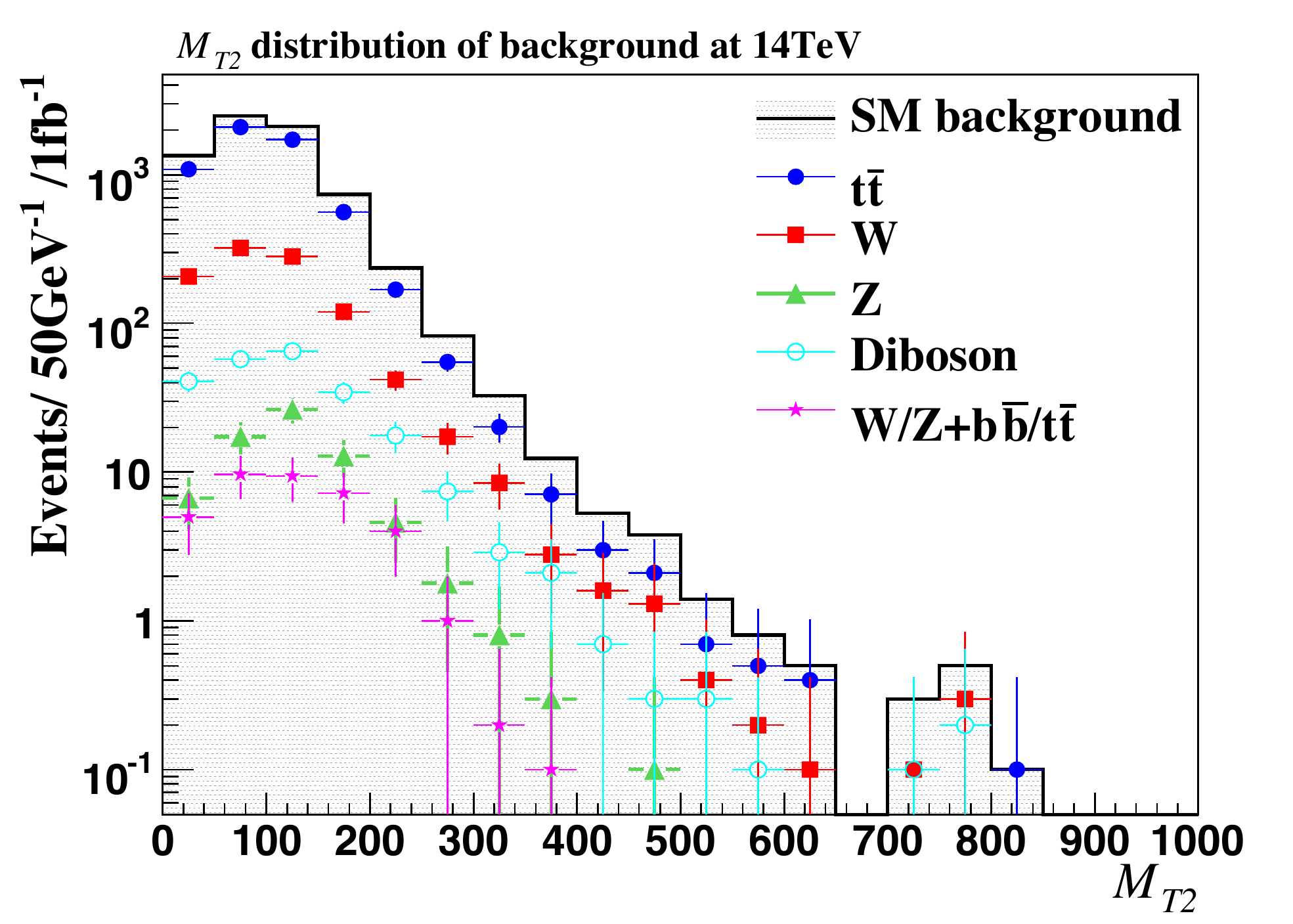}
\includegraphics[width=0.49\linewidth]{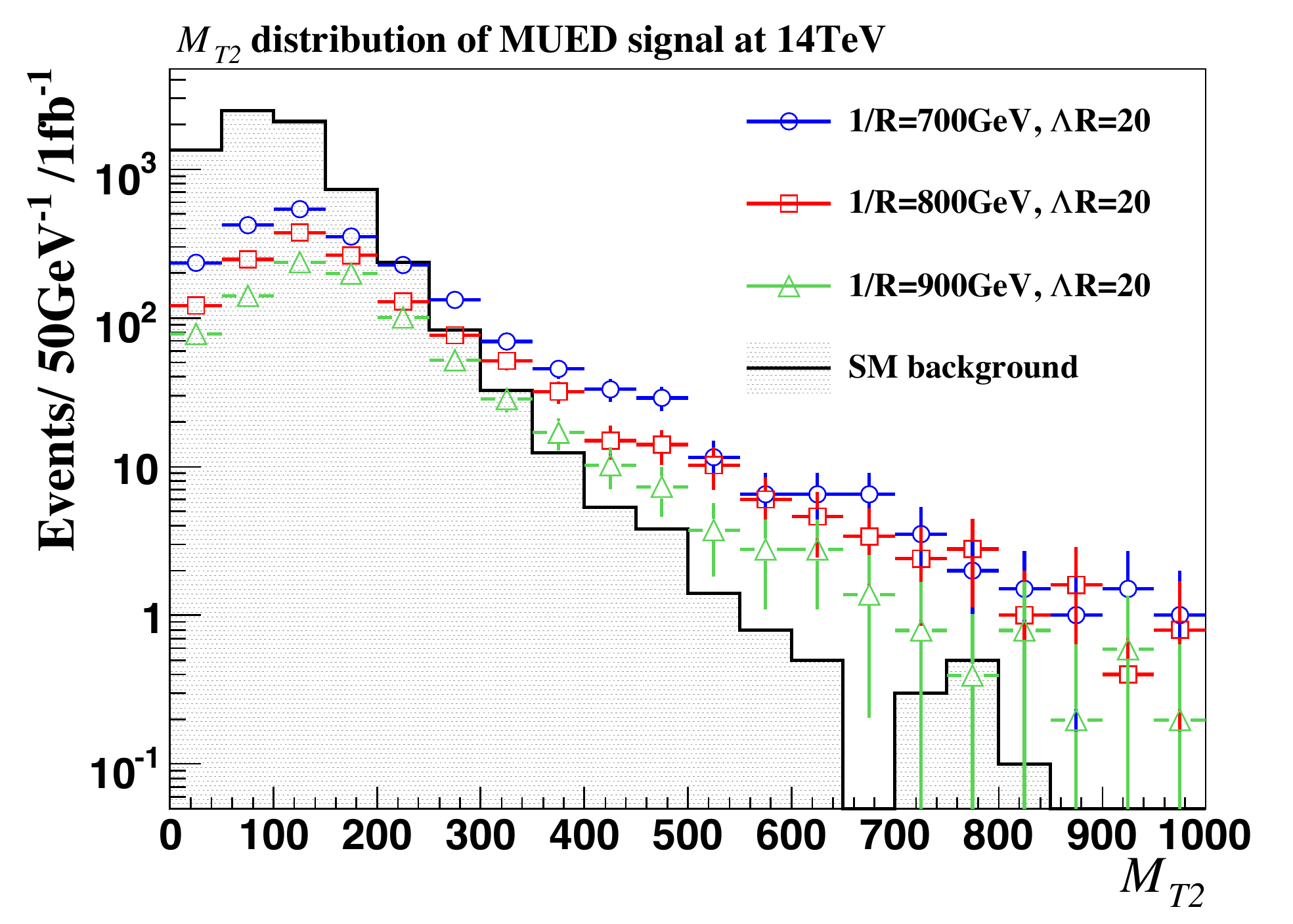}
\caption{Distributions of $\mttwo$ of the leading two jets after CUT4 for each SM background in the left, and
 total SM background and MUED signal points of $1/R= 700, 800, 900 \GEV$
 and $\Lambda R =20$ generated by \PY{ }in the right.  }
\label{BG2}
\end{center}
\end{figure}

Fig. \ref{BG2} shows that the dominant background processes are $\ttbar$,
$W+\rm jets$, and Diboson. 
Background events that remain even after CUT5  mainly come from combinatorics. 
The peak of MUED events is $\mttwo<200\GEV$, but the signal events have
a long
tail which can be understood as a result of  the variant endpoint due to
the wrong test mass discussed in section \ref{mt2selection}. {There is
combinatorics for both the signal and the background, and especially it tends to increase $\mttwo$ of the signal events. Combinatorics help to
enhance the signal excess.}
As a result, we can successfully extract the signal from the
background even based on jets.

\subsection{Discovery potential}
For the study of the discovery potential it is necessary to take
systematic uncertainties into account in addition to statistical uncertainties. 
We use the significance $Z_B$ \cite{Linnemann:2003vw}, which is provided
by the {ROOT} library \cite{Brun:1997pa}, using the same approach as in
 ATLAS discovery study for supersymmetry \cite{Aad:2009wy}. $Z_B$ is calculated using a convolution of a Poisson and a Gaussian term to account for systematic errors. 
For the backgrounds except those from QCD, a reasonable estimate of the systematic
uncertainty is $\pm 20\%$. 
The discovery potential is studied by finding the optimal $\mttwo$ cut (in step of 50GeV) to maximize the significance $Z_B$.
We define ``discovery''
 when $Z_B> 5$ and more than 10 signal events remain after the cuts.

In order to compare the $\mttwo$ analysis in the multijet + lepton mode with the previously studied
$4l+\slashed{E}_T$ analysis, we also check the discovery potential in $4l+\slashed{E}_T$ 
using the same MC samples and using the same definition of discovery. In
the $4l+\slashed{E}_T$ analysis, the following cuts are imposed \cite{Cheng:2002ab}:(1) four isolated leptons with $p_T^{\rm lep} > \{35, 20, 15, 10 \GEV\}$ 
are required, (2) $\slashed{E}_T > 50 \GEV$, and
(3) an invariant mass $M_{ll}$ for all possible pairs of opposite
sign same flavor leptons and remove events if $|M_{ll}-m_Z| <10 \GEV$ to reduce
background from the $Z$ boson. The estimated background from our MC samples is
10 events/100 fb$^{-1}$. The fake leptons should be considered to
evaluate the background level of $4l+\slashed{E}_T$ more appropriately, but
the fake leptons are not considered since they are not important for our analysis based on jets.

\begin{figure}[th]
\begin{center}
\includegraphics[width=0.6\linewidth]{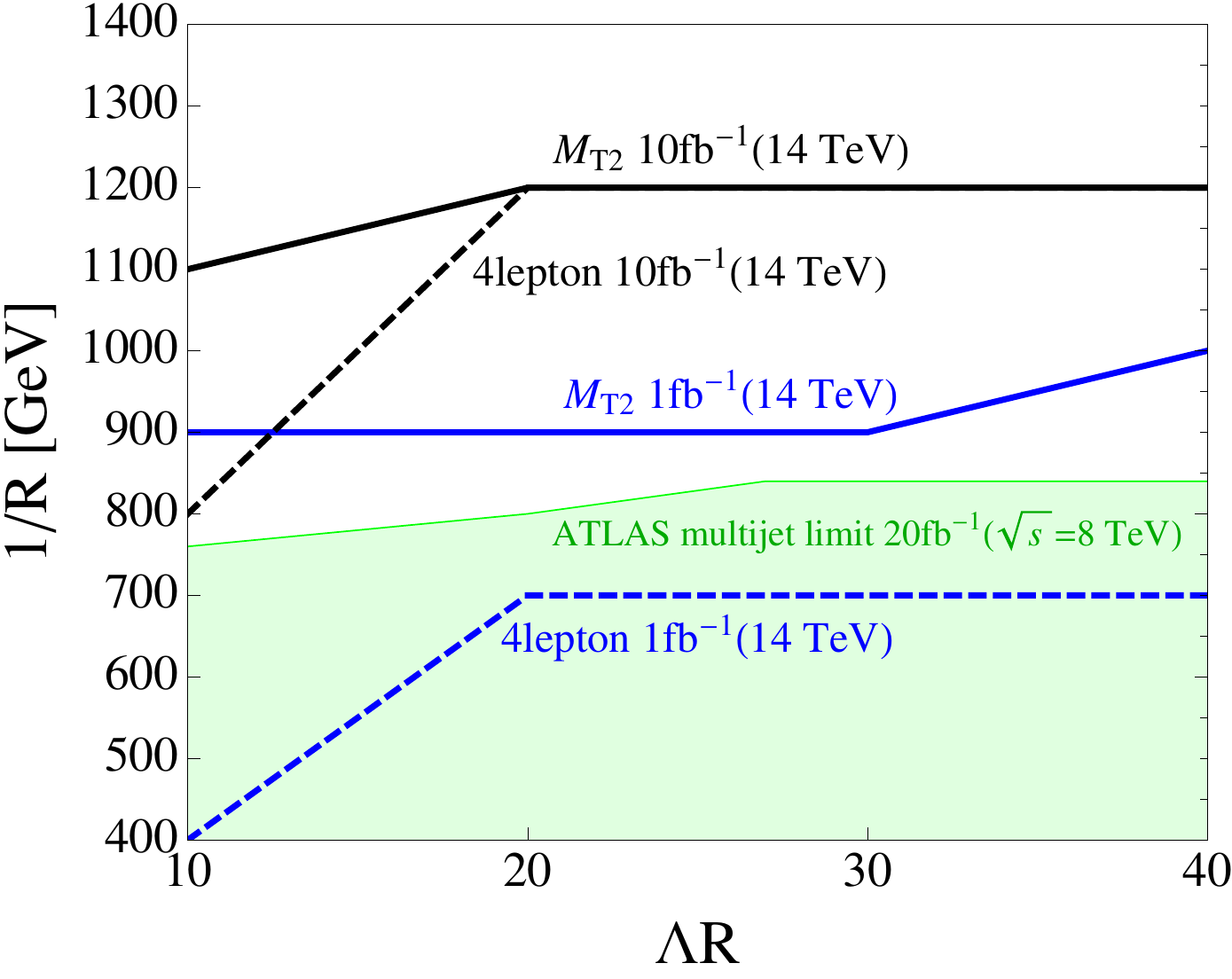}
\caption{Discovery potential of the MUED with 1 \fb and 10 \fb at
$\sqrt{s}$ = 14 TeV in the $4l+\slashed{E}_T$ analysis and the $\mttwo$ analysis. 
  For a given luminosity, the parameter region below the line will be discovered.}
\label{RlambR}
\end{center}
\end{figure}

The spectrum is more degenerate for smaller $\Lambda R$, which is more difficult for discovery in general. 
Note that for fixed $1/R$, the MUED with smaller $\Lambda R$ has a larger
cross section simply because  the KK gluon and the KK quark become lighter as in
Eqs.~(\ref{142025_13Jun11}) and (\ref{141824_13Jun11}).
Fig.~4 shows that the discovery potential does not vary with
changing $\Lambda R$ in
the $\mttwo$ analysis.  
 The second run at 14 TeV will discover up to $1/R \sim 1\TEV$ with 1 \fb
 and $1/R \sim 1.2 \TEV$ with 10 fb$^{-1}$. 
In the $4l+\slashed{E}_T$ analysis, the discovery reach at 14 TeV is $1/R \sim
 700\GEV$
 with 1 \fb and $1/R \sim 1.2 \TEV$ with 10 \fb  for $20 \le \Lambda R \le
 40$, but the sensitivity is very low for $\Lambda R =10$ : the discovery
 reach is only $1/R = 400\GEV$ with 1 \fb and $1/R = 800\GEV$ with 10 fb$^{-1}$.

The result shows that our $\mttwo$ analysis improves the discovery
 potential. In particular, the improvement is so significant for the
 most degenerate parameter $\Lambda R = 10$  that the discovery
 potential improves from $1/R=400\GEV$ to 900 GeV. 

\section{Summary and Discussion}

The problem of the ordinary
multijet $+ E_T^{miss}$ analysis is that the signal of compressed models like the Compact Supersymmetry and the MUED are removed too much because the cuts are optimized for the CMSSM-like models. If the event selections are taken to be weaker, the signal 
is buried in the SM
background. 
This is because the top quark pair production $\ttbar$ generates missing particles, neutrinos,  and  hard jets with a large enough cross section.  

Here, we tackled the search in the multijet channel by using a kinematic variable
$\mttwo$ \cite{Lester:1999tx,Barr:2003rg}. 
$\mttwo$ is bounded by the mass of the produced particles when the true mass of invisible particle is given.
We pointed out that $\mttwo$ is effective in the search for
compressed models utilizing USR dependence of $\mttwo$.

We set a test mass for the invisible particle to zero for $\mttwo$ calculation. 
 The set test mass is wrong for the signal events because it is much
smaller than the mass of invisible particle This leads to the $\mttwo$
dependence on USR which is mainly from ISR. 
$\mttwo$ of the signal can be large depending on USR, although, without
USR, $\mttwo$ is
small in the compressed spectrum.
On the other hand, the test mass is correct for the mass of the SM invisible particle,
neutrino. Then, $\mttwo$ of the SM background does not depend on
USR. As shown in Refs.~\cite{Lester:1999tx,Barr:2003rg}, it is mainly below the mass of the
heaviest particle in the SM, the top quark, 
$  \mttwo^{\rm SM} \lesssim m_{t}$.  
Therefore, an excess in the high $\mttwo$ region beyond $m_{t}$ can be seen as the
signal of new physics, and then, $\mttwo$ is effective
to search for compressed models. 

In the analysis of this paper, leading two jets in $p_T$ are used to calculate
$\mttwo$. They do not always correspond to jets we want, that
is, we have combinatoric issues when choosing jets for defining
$\mttwo$. Combinatorics smears the $\mttwo$ distribution, and the
smearing effect is different in each process. We found combinatorics
makes $\mttwo$ of the signal larger while $\mttwo$ of the background
does not increase as much as that of the signal. 
Therefore, the smearing effect of combinatorics enhances the signal
excess in the high $\mttwo$ region.

We apply $\mttwo$ to the discovery study of the MUED, and we
require at least one lepton in addition to multijet to avoid the QCD
background. Since the ISR
takes an important role in this method, we perform the event generation
with the Matrix Element correction which evaluates the hard ISR
appropriately. 
This way, we show that the $\mttwo$ analysis improves the
discovery potential compared to the $4l + E_T^{miss}$ analysis.  

This technique is useful to apply for a wide class of models with a compressed spectrum. 
Since we have now data, the background can be reliably estimated and the multi-jet discovery analysis becomes easier whereas requiring one lepton in our simulation was not essential. 
Thus the next thing to do is to study discovery potential of the MSSM with a compressed spectrum in multi-jet channel with $\mttwo$. 


\end{document}